\documentclass[preprint,twoside,3p,number,letterpaper]{elsarticle}
\usepackage{graphicx}
\usepackage{amsmath}
\usepackage{amssymb}
\usepackage{epsfig}
\usepackage{graphics,psfrag}
\usepackage{graphicx,psfrag}
\usepackage{xspace}

\newcommand{\nn}{\nonumber\\[1mm]}
\newcommand{\nnn}{\nonumber\\[3mm]}
\newcommand{\beq}{\begin{equation}}
\newcommand{\eeq}{\end{equation}}
\newcommand{\beqa}{\begin{eqnarray}}
\newcommand{\eeqa}{\end{eqnarray}}

\newcommand{\comix}{C\scalebox{0.8}{OMIX}\xspace}
\begin{document} 
\begin{flushright}
Fermilab-PUB-09-406-T
\end{flushright}

\begin{frontmatter}

\title{Efficient Color-Dressed Calculation of Virtual Corrections}
\author[Fermi]{Walter~T.~Giele}
\address[Fermi]{Fermilab, Batavia, IL 60510, USA}
\author[ETH,CERN]{Zoltan Kunszt}
\address[ETH]{Institute for Theoretical Physics, ETH, CH-8093 Z\"urich, Switzerland}
\address[CERN]{Theoretical Physics, CERN, CH-1211 Geneva, Switzerland}
\author[Fermi]{Jan~Winter}

\begin{abstract}
With the advent of generalized unitarity and parametric integration 
techniques, the construction of a generic Next-to-Leading Order 
Monte Carlo becomes feasible. 
Such a generator will entail the treatment of QCD color in the amplitudes.
We extend the concept of color dressing
to one-loop amplitudes, resulting in the formulation
of an explicit algorithmic solution for the calculation of arbitrary
scattering processes at Next-to-Leading order.
The resulting algorithm is of exponential complexity, that is the numerical
evaluation time of the virtual corrections grows by a constant
multiplicative factor as the number of external partons is increased. 
To study the properties of the method, we calculate the virtual corrections 
to $n$-gluon scattering.
\end{abstract}


\end{frontmatter}

\section{Introduction}

Automated Leading Order (LO) generators~
\cite{Stelzer:1994ta,Mangano:2002ea,Draggiotis:2002hm,Boos:2004kh,%
  Gleisberg:2008fv} play an essential role in
experimental analyses and phenomenology in general. However, the theoretical uncertainties
associated with these generators are only understood qualitatively. The 
augmentation of the LO generators with Next-to-Leading Order (NLO) corrections
will give a more quantitative understanding of the theoretical uncertainties. This
is crucial for the realization of precision measurements at the Hadron colliders. 
By calculating NLO corrections using analytic generalized unitarity methods~
\cite{Bern:1994cg,Britto:2004nc,Berger:2006ci},
the one-loop amplitude is factorized into
sums over products of on-shell tree-level amplitudes. This makes the integration of
numerical generalized unitarity methods into the LO generators attractive.
One can use the LO generator as the building block for obtaining
the NLO correction, thereby negating the need for a separate generator
of all the one-loop Feynman diagrams. The generalized unitarity approach reduces 
the complexity of the calculation through factorization. It can reduce 
the evaluation time with increasing number of external particles
from faster than factorial growth to slower than factorial growth.

By utilizing the parametric integration method of Ref.~\cite{Ossola:2006us}
significant progress has been made in the algorithmic
implementation of generalized unitarity based one-loop generators~\cite{Giele:2008ve,Berger:2008sj}
and other non-unitary methods~\cite{vanHameren:2009dr}.\footnote{These methods have matured to the
point where explicit NLO parton generators for specific processes have been 
constructed~\cite{KeithEllis:2009bu,Melnikov:2009dn,Bevilacqua:2009zn,Berger:2009xp,Melnikov:2009wh}.} 
These implementations rely on the color decomposition of the
amplitude into colorless, gauge invariant ordered amplitudes~\cite{Berends:1987cv,Mangano:1987xk}.  
At tree-level these ordered amplitudes can be efficiently calculated by  recursion
relation algorithms~\cite{Berends:1987me}. These algorithms are of polynomial complexity and grow asymptotically
as $n^4$ as the number of external partons, $n$, 
increases~\cite{Bern:2008ef}. By replacing the 4-gluon vertex by an effective 3-gluon
vertex the polynomial growth factor can be further reduced to $n^3$~\cite{Bruinsma,Draggiotis:1998gr,Duhr:2006iq}.

At the one-loop level the ordered amplitudes generalize into 
primitive amplitudes~\cite{Bern:1994zx}. These primitive amplitudes reflect the more complicated
dipole structure of one-loop amplitudes. While the analytic structure of
the factorized one-loop amplitude in color factors and primitive amplitudes is systematic,
the subsequent calculation of the color summed virtual corrections becomes
unwieldy in the algorithmic implementation~\cite{Ellis:2008qc}. The reason for this is the rapid growth
in the number of primitive amplitudes.  This rapid growth is mainly caused by the multiple quark-pairs amplitudes. A further
complication arises from the possible presence of electro-weak particles in the ordered amplitudes.

While in LO generators the analytic treatment of color is more
manageable, alternatives were developed for high parton multiplicity
scattering amplitudes~\cite{Draggiotis:1998gr,Caravaglios:1998yr,Duhr:2006iq}. 
Those alternatives provided a more numerical treatment of the color,
thereby facilitating the construction of tree-level Monte Carlo
programs for the automated generation of high multiplicity parton
scattering amplitudes at LO. This was accomplished by not only choosing the
external momenta and helicities, but also choosing the explicit colors of
the external partons for each scattering event considered. 
In doing so, the tree-level partonic amplitude is a complex number
and the absolute value squared is simply calculated. This numerical treatment
can be done in the context of ordered amplitudes~\cite{Cafarella:2007pc} 
by calculating the explicit color weights of each ordered amplitude. This
method was generalized to one-loop calculations in Ref.~\cite{vanHameren:2009dr}.  
More directly, one can reformulate the recursion relations into
color-dressed recursion
relations~\cite{Caravaglios:1995cd,Draggiotis:1998gr,Duhr:2006iq}.
These color-dressed recursion relations integrate the now explicit
color weights into the recursive formula. The resulting algorithm
is of exponential complexity and grows asymptotically as $4^n$ for
$n$-parton amplitudes; again, a reduction of the growth factor to
$3^n$ can be achieved if the 4-gluon vertex is replaced by the
effective 3-gluon vertex~\cite{Duhr:2006iq}.

In this paper we extend the generalized unitarity method of
Ref.~\cite{Giele:2008ve} as implemented in Ref.~\cite{Winter:2009kd}
to incorporate the color-dressing method.
The algorithm is developed such that it can augment a dressed LO
generator such as \comix~\cite{Gleisberg:2008fv} to become a NLO
generator.\footnote{The LO matrix-element generator needs to be
  upgraded to allow for complex external momenta.}
For the numerical examples presented in this paper, we have used our
own implementation of a color-dressed LO gluon recursion relation to
calculate the virtual corrections for $n$-gluon scattering processes.

The motivation for color dressing at the one-loop level is discussed in Sec.~2.
We outline in Sec.~3 the tree-level dressed recursion relations for
generic theories expressed in terms of Feynman diagrams. 
We optimize the color-sampling performance and study 
the phase-space integration convergence for LO $n$-gluon scattering.
The dressed formalism is extended to one-loop amplitudes in Sec.~4.
The scaling with $n$, the accuracy of the algorithm 
and the color-sampling convergence of the virtual corrections
to $n$-gluon scattering are
studied in some detail. We summarize our results in Sec.~5.
Finally, two appendices are added giving an explicit LO 6-quark example
and details on the color-dressed implementation of the gluon recursion
relation.

\section{Motivation for the Color-Dressed Generalized Unitarity Method}

So far the numerical implementations of generalized unitarity for
the evaluation of one-loop amplitudes make use of color ordering: the ordered one-loop
amplitudes are constructed from tree-level ordered amplitudes through
the $D$-dimensional unitarity cuts. This has the advantage that the
color is factorized off the loop calculation and attached subsequently to each
ordered one-loop amplitude. For the pure gluon one-loop
amplitude, this leads to a particularly simple decomposition in terms
of the adjoint generators $F$\/ of ${\rm SU}(N)$:
\beq
{\cal M}^{(0,1)}(1,2\ldots,n)\ \ \sim\sum_{P(2,3,\ldots,n)} {\rm Tr}\left(F^{a_1}F^{a_2}\cdots F^{a_n}\right) m^{(0,1)}(1,2,\ldots,n)\ .
\eeq
The decomposition is valid for both tree-level~\cite{Berends:1987cv} 
and one-loop amplitudes~\cite{DelDuca:1999rs}.
Once we can calculate the colorless ordered amplitude $m(1,2,\ldots,n)$,
all other ordered amplitudes are obtained by simple
permutations. All kinematic information about the $n$-gluon amplitude is encapsulated in a single ordered amplitude.
However, we also see the drawback of this approach as we are interested in evaluating the amplitude squared. We have to calculate
${\cal M}^{(0,1)}(1,2\ldots,n)\times\left({\cal M}^{(0)}(1,2\ldots,n)\right)^{\dagger}$ summed over all color and spin
states of the external gluons. This immediately leads to a factorial complexity 
when doing the multiplications of the full amplitudes
as we have to sum over the permutations, $P(2,3,\ldots,n)$, of the ordered amplitudes.
Additionally, the color sum has to be performed either analytically or in some numerical manner.

When including quark pairs the situation becomes even more complicated. The reason is that the internal structure of the
one-loop amplitude is not uniquely defined by the external states, thereby affecting the color flow of the 
ordered amplitudes. As a result there exist many types of ordered 
amplitudes depending on the internal configuration of quark and gluon propagators. These
amplitudes are called primitive amplitudes~\cite{Bern:1994zx} and in general cannot be obtained from each other by simple
permutations. For example, the one-loop $q\bar q + n$\/ gluon
amplitude is given by~\cite{DelDuca:1999rs}
\beqa\lefteqn{
{\cal M}^{(1)}(q;1,\ldots,n;\bar q\;\!)\ \ \sim}\nn&&
\sum_{k=2}^n\;\sum_{P(1,\ldots,n)} \left(T^yT^{a_1}\cdots T^{a_k}T^x\right)_{ij}\left(F^{a_{k+1}}\cdots F^{a_n}\right)_{xy} 
m^{(1)}(q,1,\ldots,k,q,k+1,\ldots,n)\ .
\eeqa
where the $T$-matrices are the fundamental generators of ${\rm SU}(N)$.
While for the full amplitude a cut line has an undetermined flavor, each primitive amplitude has
an unique flavor for all the cut lines. Therefore we can apply generalized unitarity to the primitive amplitudes.
However, from a numerical/algorithmic point of view the evaluation 
of this equation becomes tedious as can be seen for instance in
the calculation of the one-loop matrix elements for $W+5$ partons in Ref.~\cite{Ellis:2008qc}. 

It is clear that for an automated generator of one-loop corrections one would like to avoid ordered/primitive amplitudes
altogether. For LO matrix elements, this can be done by applying the color-dressed recursion relations
to evaluate the (unordered) tree-level amplitudes. From these color-dressed
tree-level amplitudes we can build the one-loop color-dressed amplitudes by applying generalized unitarity,
thereby circumventing the need for primitive amplitudes and explicit color summations.  
It is of interest to investigate the feasibility of this approach. 
The $n$-gluon scattering process is good for studying
the behavior of the dressed algorithm. 
The color-ordered approach is most effective for $n$-gluon scattering. 
For processes with quark-pairs, the color-dressed approach will become
even more efficient compared to the color-ordered approach.

An additional advantage of the color-dressed algorithm is that it treats partons 
and color neutral particles on the same footing. Specifically, we can
include electro-weak particles without altering the algorithm.
This is in contrast to the color-ordered algorithm, where the addition of electro-weak
particles would lead to significant modifications in the algorithmic implementation
of the method.

\section{Dressed Recursive Techniques for Leading Order Amplitudes}

In tree-level generators the Monte Carlo sampling over the external 
color and helicity states has become a standard practice
~\cite{Draggiotis:1998gr,Caravaglios:1998yr,Duhr:2006iq}.
Such a color sampling allows for the efficient evaluation of 
large multiplicity partonic processes. A particular efficient
implementation of the color-dressed Monte Carlo method uses the 
color-flow decomposition of the multi-parton amplitudes 
\cite{Draggiotis:1998gr,Maltoni:2002mq,Duhr:2006iq}.

The principle of Monte Carlo sampling over the states of the
external sources generalizes to any theory expressible through Feynman
rules. By explicitly specifying the quantum numbers  of the $n$\/
external sources, one can evaluate the tree-level amplitude squared
and differential cross section using Monte Carlo sampling:
\beqa\label{MCeval}\lefteqn{
d\,\sigma_{\rm LO}(f_1f_2\rightarrow f_3\cdots f_n)\ \ =}\nn
&&\frac{W_{\rm S}}{N_{\rm event}}\times\sum_{r=1}^{N_{\rm event}} 
d\,PS^{(r)}(K_1K_2\rightarrow K_3\cdots K_n)
\left|\;\!{\cal M}^{(0)}\left({\bf f}_1^{(r)},{\bf f}_2^{(r)},\ldots,{\bf f}_n^{(r)}\right)\right|^2\ ,
\eeqa
where
\beq
{\bf f}_i^{(r)}\;=\;\left\{f_i,h_{f_i},C_{f_i},K_{f_i}\right\}^{(r)}
\eeq
denotes the flavor, spin, color and momentum four-vector of external
state $i$\/ for event $r$.\footnote{We will use flavor to indicate the
  particle type, such as e.g.\ gluon, up-quark, $W$-boson, etc.}
The constant $W_{\rm S}$ contains the appropriate identical particle
factors and Monte Carlo sampling weights. For each event $r$, the
external states are stochastically chosen such that when summed over
many events we approximate the correct differential cross section with
sufficient accuracy.

\subsection{The Generic Recursive Formalism}\label{Sec:generictreeformalism}

To calculate the tree-level amplitude ${\cal M}^{(0)}$ in Eq.~(\ref{MCeval}),
we follow the method of color-dressed recursion relations as detailed
in Refs.~\cite{Duhr:2006iq,Gleisberg:2008fv}. A recursion relation
builds multi-particle currents from other currents.
The $m$-particle current $J_{\bf g}\left({\bf f}_\pi\right)$ has
$m$\/ on-shell particles
${\bf f}_\pi=\{{\bf f}_i\}_{i\in\pi}=\{  {\bf f}_{i_1},\ldots,{\bf f}_{i_m} \}$
where $\pi=\{i_1,\ldots,i_m\}$ and one off-shell particle
${\bf g}=\left\{g,L_g,C_g,K_g\right\}$ with $g$, $L_g$, $C_g$ and
$K_g$ denoting the flavor, Lorentz label, color and four-momentum,
respectively. The momentum of the off-shell particle, $K_g$,  is
constrained by momentum conservation: $K_g=-K_{\pi}=-\sum_{i\in\pi} K_i$.

\begin{figure}[t!]
\begin{center}
\includegraphics[angle=0,scale=0.8]{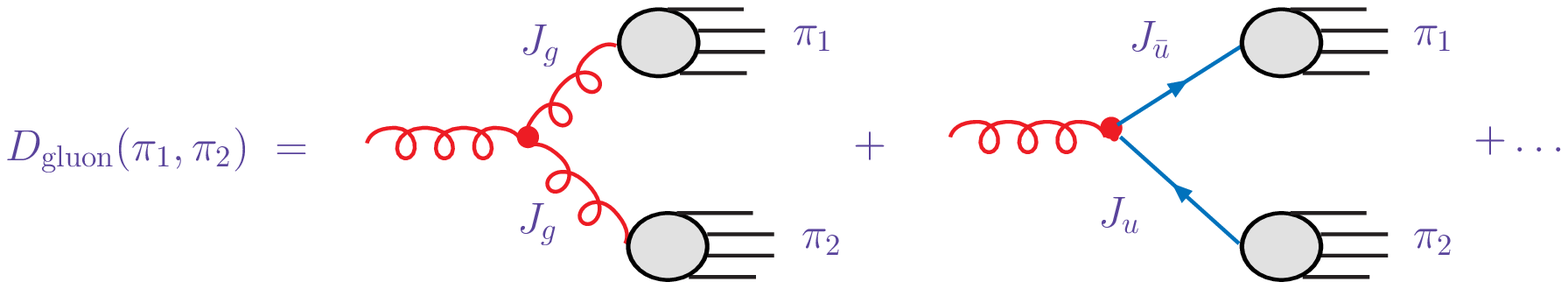}
\caption{\label{Dgblob}
A graphical representation of Eq.~(\ref{vertexblob}) for $k=2$ and an
off-shell gluon in QCD. Because of flavor conservation only one of the
two vertices can contribute for any given partition.}
\end{center}
\end{figure}

The dressed recursion relation generates currents using the
propagators and interaction vertices of the theory. Using standard
tensor notation we can write the propagators as
\beqa
P^{{\bf g}_1{\bf g}_2}(Q)&=&
\delta_{g_1g_2}\delta_{C_{g_1}C_{g_2}}P^{L_{g_1}L_{g_2}}(Q)\ ,\nn
P^{{\bf g}}\big[J({\bf f}_\pi)\big]&=&
\sum_{{\bf g}_1}P^{{\bf g}{\bf g}_1}(K_{\pi})
J_{{\bf g}_1}\big({\bf f}_\pi\big)\ ,\nn
P\big[J({\bf f}_{\pi_1}),J({\bf f}_{\pi_2})\big]&=&
\sum_{{\bf g}_1{\bf g}_2}J_{{\bf g}_1}\big({\bf f}_{\pi_1}\big)
P^{{\bf g}_1{\bf g}_2}(K_{\pi_1})J_{{\bf g}_2}\big({\bf f}_{\pi_2}\big)\ ,
\eeqa
where e.g.\ the gluon propagator is given by
$P^{\mu_1\mu_2}(Q)=-g^{\mu_1\mu_2}/Q^2$. Note that the particle sums
are taken over all quantum numbers of the off-shell particles ${\bf g}_i$.
Furthermore, in all expressions momentum conservation is always
implicitly understood. The on-shell tree-level $n$-particle amplitude
can hence be expressed in terms of an $(n-1)$-current,
\beq\label{treeamp}
{\cal M}^{(0)}\big({\bf f}_1,\ldots,{\bf f}_n\big)\;=\;
P^{-1}\left[J\big({\bf f}_1,\ldots,{\bf f}_{n-1}\big),J\big({\bf f}_n\big)\right]\ .
\eeq
We denote the interaction vertices of the theory as
$V_{{\bf g}_1\cdots {\bf g}_k}(Q_1,\ldots,Q_k)$.
The maximal number of legs for
the allowed vertices of the theory is denoted by $V_{\rm max}$.
The number of legs of the vertex is indicated by the number of its
arguments and the type of vertex is specified by the quantum numbers
of the legs.
The labels ${\bf g}_1,\ldots,{\bf g}_k$ run over the values of all
particles of the theory. Symmetries and renormalizability imply that
many of the vertices are set to zero. The theory is defined by its
particle content and its non-vanishing vertices, which are generalized
tensors:
\beq
V_{{\bf g}_1\cdots{\bf g}_k}(Q_1,\ldots,Q_k)\;=\;
V_{g_1\cdots g_k;C_{g_1}\cdots C_{g_k}}^{L_{g_1}\cdots L_{g_k}}(Q_1,\ldots,Q_k)\ .
\eeq
The sum of all vertices contracted in with currents constitutes the
main building block of the recursion relation. We define it as
\beq\label{vertexblob}
D_{\bf g}\big[J({\bf f}_{\pi_1}),\ldots,J({\bf f}_{\pi_k})\big]\;=
\sum_{{\bf g}_1\cdots {\bf g}_k}
V_{{\bf gg}_1\cdots{\bf g}_k}(K_g=-K_{\Pi_k},K_{\pi_1},\ldots,K_{\pi_k})
\times
J^{{\bf g}_1}\big({\bf f}_{\pi_1}\big)\times\cdots\times
J^{{\bf g}_k}\big({\bf f}_{\pi_k}\big)\ ,
\eeq
where the inclusive list $\Pi_k$ is build up of unions of the
exclusive lists:
\beq\label{Eq:incllists}
\Pi_k=\bigcup_{i=1}^{k}\pi_i\ .
\eeq
Fig.~\ref{Dgblob} is a pictorial representation of
Eq.~(\ref{vertexblob}) when using the example of QCD. For this case,
we will work out the generic vertex blob in detail in the next
subsection.

\begin{figure}[b!]
\begin{center}
\includegraphics[angle=0,scale=0.8]{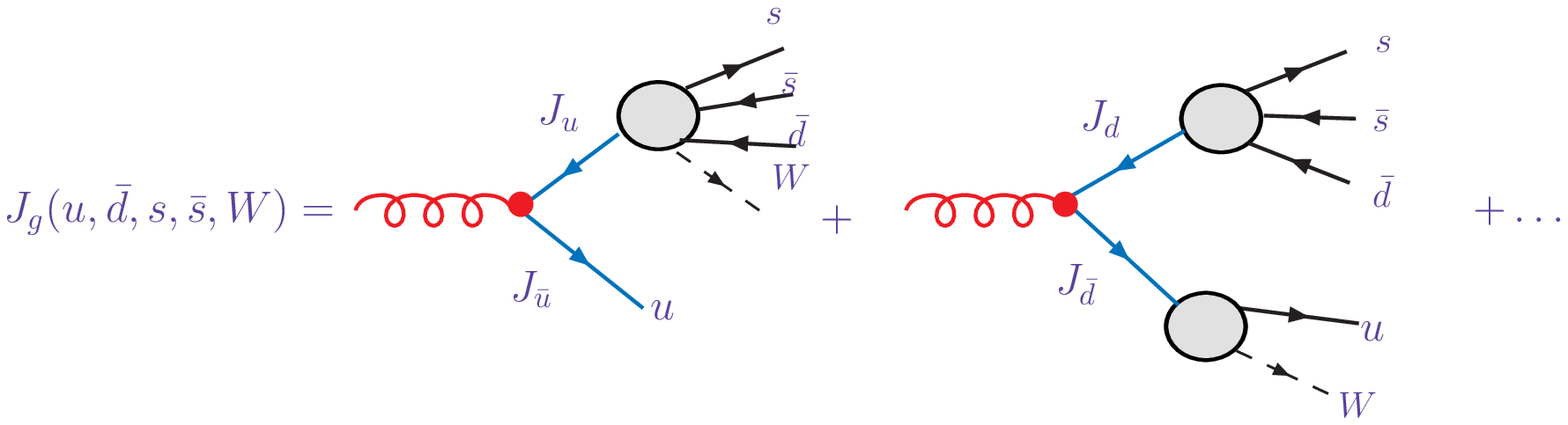}
\caption{\label{Jg5rec}
The first recursion step for the unordered gluon current with
$u,\bar{d},s,\bar{s}$  quarks and a $W^-$ gauge boson in the final
state. There are 15 contributions corresponding to all possible
partitions of the final-state particles into two groups. Because of
flavor conservation there are only 4 non-vanishing contributions for
the ``4+1'' partitions (first term) and 2 non-vanishing contributions
for the ``3+2'' partitions (second term).}
\end{center}
\end{figure}

The recursion relations terminate with the one-particle currents.
A one-particle ${\bf g}$-current is defined in terms of the source
$S^{h_{f_i} C_{f_i}}_{f_i L_{g}}(K_{f_i})$. Hence, we have
\beq
J_{\bf g}\big({\bf f}_i\big)\;=\;\delta^{g f_i}\delta^{C_g C_{f_i}}\,
S^{h_{f_i} C_{f_i}}_{f_i L_{g}}(K_{f_i})\ .
\eeq
For example, the ${\bf g}_1$-gluon one-particle source with helicity
$\lambda_1$, color $c_1$ and momentum $K_1$ is given by
$J_{\bf g}({\bf g}_1)=\delta^{cc_1}\epsilon^{\lambda_1}_{\mu_1}(K_1)$.
I.e.\ the ${\bf g}_1$-gluon source is a matrix in color space
multiplied by the helicity vector.

The $n$-particle currents are now efficiently calculated from a
recursively defined current in the following manner: 
\beq\label{tree-recurrence}
J_{\bf g}\big({\bf f}_1,\ldots,{\bf f}_n\big)\;=\;
\sum_{k=2}^{V_{\rm max}-1}
\sum_{P_{\pi_1\cdots\pi_k}(1,\ldots,n)}^{{\cal S}_2(n,k)}
P_{\bf g}\Big[D\big[J({\bf f}_{\pi_1}),\ldots,J({\bf f}_{\pi_k})\big]\Big]\ ,
\eeq
where ${\cal S}_2(n,k)$ is the Stirling number of the second kind. 
The first recursive step is graphically illustrated in
Fig.~(\ref{Jg5rec}) for the example of
$J_{\bf g}({\bf u},{\bf\bar d},{\bf s},{\bf\bar s},{\bf W^-})$.
The sum over $P_{\pi_1\cdots\pi_k}(1,\ldots,n)$ generates all 
different partitions decomposing the set $\{1,\ldots,n\}$ into the
non-empty subsets $\pi_1,\ldots,\pi_k$. An example for a list of
different partitions is
\beqa
P_{\pi_1\pi_2\pi_3}(1,2,3,4)&=&
\left\{\pi_1^{(i)}\pi_2^{(i)}\pi_3^{(i)}\right\}_{i=1}^{{\cal S}_2(4,3)=6}
\nn&=&\Big\{\
\big\{\{1,2\}\{3\}\{4\}\big\},\,
\big\{\{1,3\}\{2\}\{4\}\big\},\,
\big\{\{1,4\}\{2\}\{3\}\big\},
\nn&&\hphantom{\Big\{\ }
\big\{\{2,3\}\{1\}\{4\}\big\},\,
\big\{\{2,4\}\{1\}\{3\}\big\},\,
\big\{\{3,4\}\{1\}\{2\}\big\}\ \Big\}\ .
\eeqa

The formalism described here fully specifies an automated algorithm of
exponential complexity to calculate the LO differential cross-sections
for any theory defined in terms of Feynman rules. Owing to the
characteristics of the partitioning, the computer resources needed to
calculate the $n$-particle tree-level amplitudes asymptotically grow
in proportion to ${\cal S}_2(n,V_{\rm max})$. The exponential behavior
arises from the large-$n$ limit of the Stirling numbers, i.e.\
${\cal S}_2(n,V_{\rm max})\rightarrow V_{\rm max}^n$~\cite{Bruinsma}.
It may be possible to reduce $V_{\rm max}$ by rewriting higher
multiplicity vertices as sums of lower multiplicity vertices thereby
improving the efficiency of the recursive
algorithm~\cite{Draggiotis:1998gr,Duhr:2006iq}. For the case of the
Standard model this has been fully worked out in
Ref.~\cite{Gleisberg:2008fv} and implemented in the \comix LO
generator.

\subsection{Multi-Jet Scattering Amplitudes}

We specify the generic recursion relations to the perturbative QCD
Feynman rules. This will give an algorithmic description of the
scattering amplitudes at LO for multi-jet production at hadron
colliders.

The external sources are gluons and massless quarks. All these
particles have color and helicity as quantum numbers. Instead of the
traditional color representation in terms of fundamental generators,
we choose the color-flow
representation~\cite{Draggiotis:2002hm,Maltoni:2002mq,Duhr:2006iq},
which is more pertinent to Monte Carlo sampling and easily derivable
from the traditional color representation by making the following two
observation: first, any internal propagating gluon has as a color
factor $\delta^{ab}={\rm Tr}\Big(T^aT^b\Big)$.\footnote{Because of
  this normalization, the structure constants $f^{abc}$ are a factor
  of $\sqrt{2}$ larger than in the conventional definition.}
This color factor can be rewritten as
\beq
{\cal M}\;=\;{\cal A}_a\frac{\delta^{ab}}{K^2}{\cal B}_b
\;=\;{\cal A}_a \frac{{\rm Tr}\left(T^aT^b\right)}{K^2}{\cal B}_b
\;=\;{\cal A}_{ij}\frac{1}{K^2}{\cal B}^{ji}\ .
\eeq
Second, we contract the amplitude with $T^{a_k}_{i_kj_k}$ for
each external gluon:
\beq
|{\cal M}|^2\;=\;{\cal M}^a\,\delta_{ab}\left({\cal M}^b\right)^\dagger
\;=\;{\cal M}^a\,T^a_{ij}T^b_{ji}\left({\cal M}^b\right)^\dagger
\;=\;{\cal M}_{ij}\,{\cal M}_{ji}^\dagger\ .
\eeq
From these observations it follows that we can calculate the interaction
vertices in the color-flow representation
by simply contracting each gluon with $T^{a_k}_{i_kj_k}$ and
summing over $a_k$. 
The three gluon vertex is thus given by
\beqa\label{V3g}
V_{{\bf g}_1{\bf g}_2{\bf g}_3}(K_1,K_2,K_3) &=& V_{i_1j_1i_2j_2i_3j_3}^{\mu_1\mu_2\mu_3}(K_1,K_2,K_3)\nn
&=&T^{a_1}_{i_1j_1}T^{a_2}_{i_2j_2}T^{a_3}_{i_3j_3}\,V_{a_1a_2a_3}^{\mu_1\mu_2\mu_3}(K_1,K_2,K_3)\nn
&=&T^{a_1}_{i_1j_1}T^{a_2}_{i_2j_2}T^{a_3}_{i_3j_3}\,f^{a_1a_2a_3}\,\sqrt{2}\,\widehat{V}_3^{\mu_1\mu_2\mu_3}(K_1,K_2,K_3)\nn
&=&\left(\delta^{i_1}_{j_2}\delta^{i_2}_{j_3}\delta^{i_3}_{j_1}
-\delta^{i_1}_{j_3}\delta^{i_2}_{j_1}\delta^{i_3}_{j_2}\right)\widehat{V}_3^{\mu_1\mu_2\mu_3}(K_1,K_2,K_3)\ ,
\eeqa
with
\beq
\widehat{V}_3^{\mu_1\mu_2\mu_3}(K_1,K_2,K_3)\;=\;\frac{1}{\sqrt{2}}\Big((K_1-K_2)^{\mu_3}g^{\mu_1\mu_2}+(K_2-K_3)^{\mu_1}g^{\mu_2\mu_3}
+(K_3-K_1)^{\mu_2}g^{\mu_3\mu_1}\Big)\ .
\eeq
Similarly, for the four gluon vertex we find
\beq\label{V4g}
V_{{\bf g}_1{\bf g}_2{\bf g}_3{\bf g}_4}\;=\;
V_{i_1j_1i_2j_2i_3j_3i_4j_4}^{\mu_1\mu_2\mu_3\mu_4}\;=
\sum_{C(234)}\left(\delta^{i_1}_{j_2}\delta^{i_2}_{j_3}\delta^{i_3}_{j_4}\delta^{i_4}_{j_1}
+\delta^{i_1}_{j_4}\delta^{i_2}_{j_1}\delta^{i_3}_{j_2}\delta^{i_4}_{j_3}\right)
\widehat{V}_4^{\mu_1\mu_3,\,\mu_2\mu_4}\ ,
\eeq
with
\beq
\widehat{V}_4^{\mu_1\mu_2,\,\mu_3\mu_4}\;=\;2\,g^{\mu_1\mu_2}g^{\mu_3\mu_4}-g^{\mu_1\mu_3}g^{\mu_2\mu_4}
-g^{\mu_1\mu_4}g^{\mu_2\mu_3}\ ,
\eeq
and the sum is over the cyclic permutation of the indices $\{2,3,4\}$.
In the color-flow representation the quark-antiquark-gluon vertex is given by
\beq\label{Vqqbg}
V_{\bf qg\bar q}\;=\;V_{i,\,i_1j_1,\,j}^{s\mu\bar{s}}\;=\;
\left(\delta_{ij_1}\delta_{i_1j}-\frac{1}{N_{\rm C}}\,\delta_{i_1j_1}\delta_{ij}\right) \widehat{V}^{s\mu\bar{s}}\ ,
\eeq
with
\beq
 \widehat{V}_{s\bar{s}}^{\mu}\;=\;\frac{1}{\sqrt{2}}\,\gamma_{s\bar{s}}^{\mu}\ .
\eeq
The external sources are given by
\beqa
J_{\bf g}({\bf g}_1) &=&
\delta^{Ii_1}\delta^{Jj_1}\varepsilon_{\mu}^{\lambda_1}(K_1)\ ,\nn
J_{\bf q}({\bf q}_1) &=&
\delta^{Ii_1} v_s^{\lambda_1}(K_1)\ ,\nn
J_{\bf\bar q}({\bf\bar q}_1)&=&
\delta^{Jj_1} \bar u_{\bar s}^{\lambda_1}(K_1)\ ,
\eeqa
where ${\bf g}=\{g,\mu,(IJ),-K_1\}$,
${\bf g}_1=\{g_1,\lambda_1,(i_1j_1),K_1\}$, 
${\bf q}=\{q,s,I,-K_1\}$, ${\bf q}_1=\{q_1,\lambda_1,i_1,K_1\}$, 
${\bf\bar q}=\{\bar q,\bar s,J,-K_1\}$ and
${\bf \bar q}_1=\{\bar q_1,\lambda_1,j_1,K_1\}$.
The internal propagating particles are given by
\beqa
P^{{\bf g}_1{\bf g}_2}(Q)&=&\delta^{i_1}_{j_2}\delta^{i_2}_{j_1}\left(\frac{-g_{\mu_1\mu_2}}{Q^2}\right)\ ,\nn
P^{{\bf q}_1{\bf q}_2}(Q)&=&\delta^{i_1}_{i_2}\left(Q\!\!\!\!\slash\,-m_{q_1}\right)^{-1}_{s_1s_2}\ ,\nn
P^{{\bf\bar q}_1\bar {\bf q}_2}(Q)&=&\delta^{j_1}_{j_2}\left(Q\!\!\!\!\slash\,+m_{\bar q_1}\right)^{-1}_{\bar s_1\bar s_2}\ .
\eeqa
with ${\bf g}_k=\{g_k,\mu_k,(i_kj_k),Q\}$, ${\bf q}_k=\{q_k,s_k,i_k,Q\}$ and
${\bf\bar q}_k=\{\bar q_k,\bar s_k,j_k,Q\}$.

We can now construct Berends--Giele recursion
relations~\cite{Berends:1987me} using color-dressed multi-parton
currents based on Eq.~(\ref{tree-recurrence}). The result is
\beqa
J_{\bf q}\big({\bf f}_1,\ldots,{\bf f}_n\big)&=&
\sum_{P_{\pi_1\pi_2}(1,\ldots,n)}
P_{\bf q}\Big[D\big[J({\bf f}_{\pi_1}),J({\bf f}_{\pi_2})\big]\Big]\ ,\nnn
J_{\bf g}\big({\bf f}_1,\ldots,{\bf f}_n\big)&=&
\sum_{P_{\pi_1\pi_2}(1,\ldots,n)}
P_{\bf g}\Big[D\big[J({\bf f}_{\pi_1}),J({\bf f}_{\pi_2})\big]\Big]\ ,\nn
&+&\sum_{P_{\pi_1\pi_2\pi_3}(1,\ldots,n)}
P_{\bf g}\Big[D\big[J({\bf f}_{\pi_1}),J({\bf f}_{\pi_2}),J({\bf f}_{\pi_3})\big]\Big]\ ,
\eeqa
where each current violating flavor conservation is defined to give zero. 
The compact operator language can be expanded out to an explicit formula
by adding back in the particle attributes.
For example,
\beqa
P_{\bf g}\Big[D\big[J({\bf f}_{\pi_1}),J({\bf f}_{\pi_2})\big]\Big]
&=&\sum_{{\bf qg}_1{\bf\bar q}}
P_{{\bf gg}_1}(K_{\Pi_2})\,V^{{\bf qg}_1{\bf\bar q}}\,
J_{\bf q}({\bf f}_{\pi_1})\,J_{\bf\bar q}({\bf f}_{\pi_2})\nn
&+&\sum_{{\bf g}_1{\bf g}_2{\bf g}_3}
P_{{\bf gg}_1}(K_{\Pi_2})\,V^{{\bf g}_1{\bf g}_2{\bf g}_3}(-K_{\pi_1\cup\pi_2},K_{\pi_1},K_{\pi_2})\,
J_{{\bf g}_2}({\bf f}_{\pi_1})\,J_{{\bf g}_3}({\bf f}_{\pi_2})\nnn
&=&\frac{1}{K_{\Pi_2}^2}
V^{s_1\mu s_2}_{i,\,IJ,\,j}\times J_{s_1}^i({\bf f}_{\pi_1})\times J_{s_2}^j({\bf f}_{\pi_2})\nn
&+&\frac{1}{K_{\Pi_2}^2}
V^{\mu\mu_1\mu_2}_{IJ i_2j_2 i_3j_3}(-K_{\pi_1\cup\pi_2},K_{\pi_1},K_{\pi_2})\times
J_{\mu_1}^{(ij)_2}({\bf f}_{\pi_1})\times J_{\mu_2}^{(ij)_3}({\bf f}_{\pi_2})\ .
\eeqa
The $n$-parton tree-level matrix element is calculated using
Eq.~(\ref{treeamp}). We exemplify in appendix~\ref{App:6q-example} how
to work out the 6-quark recursion steps using the above formalism.

\subsection{Numerical Implementation of $n$-gluon Scattering}\label{Sec:numtree}

The method of color dressing as discussed in this section relies on
the ability to perform a Monte Carlo sampling over the degrees of
freedom of the external sources. In this subsection we will study in
some detail the properties of such a sampling approach by means of the
color-dressed gluonic recursion relation. We are particularly
interested in the accuracy of the color-sampling procedure and overall
speed of the implementation. The addition of quarks and external
vector bosons is a straightforward extension and will not affect the
conclusions reached in this subsection.

The explicit color-dressed gluon recursion algorithm is given in terms
of colored gluonic currents. The gluonic currents are $3\times3$
matrices in color space and defined as
\beqa\label{CDrecursion}
J_{\bf g}\big({\bf g}_m\big)&=&
\delta^{Ii_m}\delta^{Jj_m}\,\varepsilon_\mu^{\lambda_m}(K_m)\ ,\nn
J_{\bf g}\big({\bf g}_1,\ldots,{\bf g}_m\big)&=&
\sum_{P_{\pi_1\pi_2}(1,\ldots,m)}
P_{\bf g}\Big[D\big[J({\bf g}_{\pi_1}),J({\bf g}_{\pi_2})\big]\Big]\nn&+&
\sum_{P_{\pi_1\pi_2\pi_3}(1,\ldots,m)} 
P_{\bf g}\Big[D\big[J({\bf g}_{\pi_1}),J({\bf g}_{\pi_2}),J({\bf g}_{\pi_3})\big]\Big]\ .
\eeqa
The color-dressed $n$-gluon amplitude is given by
\beq\label{CDLOgluon}
{\cal M}^{(0)}\big({\bf g}_1,{\bf g}_2,\ldots,{\bf g}_n\big)\;=\;
P^{-1}\left[J\big({\bf g}_1,{\bf g}_2,\ldots,{\bf g}_{n-1}\big),J\big({\bf g}_n\big)\right]\ .
\eeq
For this specific example, we have labelled the on-shell gluons by
${\bf g}_i$, the off-shell gluon is denoted by ${\bf g}$ as before.
The operator formulation of the recursive algorithm is particularly
suited for an object oriented implementation of the recursive
algorithm. We have implemented the algorithm presented above in
{\tt C++}. More details including the more explicit recursion equation
are shown in appendix~\ref{App:gluon-recu}.

The first issue to deal with is the correctness of the implemented
algorithm. To this end we want to compare the color-dressed amplitude
to existing evaluations of the gluonic amplitudes based on ordered
amplitudes. To facilitate the comparison, we write the color-ordered
expansion of the amplitude using the color-flow
representation~\cite{Maltoni:2002mq}:
\beqa\label{LOWillen}
{\cal M}^{(0)}\big({\bf g}_1,{\bf g}_2,\ldots,{\bf g}_n\big)&=&
\sum_{P(2,\ldots,n)} 
{A^{(0)}}^{i_1\cdots i_n}_{j_1\cdots j_n}(g_1^{\lambda_1},\ldots,g_n^{\lambda_n})\nn
&=&T^{a_1}_{i_1j_1}\cdots T^{a_n}_{i_nj_n}\sum_{P(2,\ldots,n)}\mbox{Tr}\big(F^{a_1}\cdots F^{a_n}\big)\;
m^{(0)}(g_1^{\lambda_1},\ldots,g_n^{\lambda_n})\nn
&=&\frac{1}{2}\sum_{P(2,\ldots,n)}
\Big(\delta^{i_1}_{j_2}\delta^{i_2}_{j_3}\cdots\delta^{i_{n-1}}_{j_n}\delta^{i_n}_{j_1}
+(-1)^n\,\delta^{i_n}_{j_{n-1}}\delta^{i_{n-1}}_{j_{n-2}}\cdots\delta^{i_2}_{j_1}\delta^{i_1}_{j_n}\Big)\;
m^{(0)}(g_1^{\lambda_1},\ldots,g_n^{\lambda_n})\nn
&=&\sum_{P(2,\ldots,n)} \delta^{i_1}_{j_2}\delta^{i_2}_{j_3}\cdots\delta^{i_{n-1}}_{j_n}\delta^{i_n}_{j_1}\;
m^{(0)}(g_1^{\lambda_1},\ldots,g_n^{\lambda_n})\ .
\eeqa
The $m^{(0)}(g_1^{\lambda_1},\ldots,g_n^{\lambda_n})$ are ordered
amplitudes with the property
$m^{(0)}(1,2,\ldots,n)=(-1)^n\,m^{(0)}(n,\ldots,2,1)$.
From the above formulas it follows that
${A^{(0)}}^{i_1\cdots i_n}_{j_1\cdots j_n}=
{A^{(0)}}^{j_1\cdots j_n}_{i_1\cdots i_n}$.
By choosing the explicit momentum, helicity and color $(ij)_m$ of each
gluon we can compare the numerical values of Eqs.~(\ref{CDLOgluon})
and~(\ref{LOWillen}). We have done the comparison up to
$2\rightarrow12$ gluon amplitudes and found complete agreement,
thereby validating the correctness of the color-dressed algorithm.

An important consideration in calculating the color-dressed amplitudes
is the color-sampling method used in the Monte Carlo program. For a
$2\rightarrow n-2$ gluon scattering amplitude, each of the gluon color
states is stochastically chosen. The full color configuration of the
event is expressed by $\{(ij)_m\}_{m=1}^n$ where $i_m$ and $j_m$ each
denote a color state out of three possible ones that can be labelled
$\{1,2,3\}$. In the ``Naive'' approach one samples uniformly over all
possible color states of the gluons. The number of color
configurations, $N^{\rm Naive}_{\rm col}$, and the color-configuration
weight, $W^{\rm Naive}_{\rm col}$, are given by
\beq
N^{\rm Naive}_{\rm col}\;=\;9^n
\eeq
and
\beq
W^{\rm Naive}_{\rm col}\;=\;1\ ,
\eeq
respectively. About 95\% of the naive color configurations have a
vanishing color factor. This results in a rather inefficient Monte
Carlo procedure when sampling over the color states. As was noted in
Ref.~\cite{Duhr:2006iq}, a significant number of the zero color-weight
configurations can be removed by imposing color conservation. This is
implemented by vetoing any color configuration for which the condition
$\exists\,c\in\{1,2,3\}:
\sum_{m=1}^n\left(\delta_{i_m,c}-\delta_{j_m,c}\right)\neq0$ is true.
In other words, the non-vetoed color configurations can be obtained by
uniformly choosing the colors $i_1,\ldots,i_n$ and subsequently
generating the colors $j_1,\ldots,j_n$ through a permutation of the
list $\{i_1,\ldots,i_n\}$. For the number of color configurations to
be sampled over, this approach, which we name ``Conserved'', then
yields
\beq
N^{\rm Conserved}_{\rm col}\;=
\sum_{n_1,n_2,n_3=0}^n\delta_{n_1+n_2+n_3,n}
\left(\frac{n!}{n_1!\,n_2!\,n_3!}\right)^2
\eeq
where $n_c=\sum_{m=1}^n\delta_{i_m,c}$. As this way of sampling is no
longer uniform, each generated color configuration gets an associated
color weight described by
\beq
W^{\rm Conserved}_{\rm col}\;=\;3^n\,\frac{n!}{n_1!\,n_2!\,n_3!}\ .
\eeq
Yet, there still are non-contributing color configurations left in the
sampling set. We have to augment the selection criteria further by
vetoing any color configuration for which the condition
$\exists\,c\in\{1,2,3\}:[\;\forall m\in\{1,2,\ldots,n\}:
(i_m=c\rightarrow i_m=j_m)\;]$ is true.\footnote{When all colors are
  identical, i.e.\ $i_1=j_1=i_2=j_2=\cdots=i_n=j_n$, every color
  factor in Eq.~(\ref{LOWillen}) is equal to one. We can still veto
  the event because the sum over all ordered amplitudes is identical
  to zero at tree level~\cite{Berends:1987me}.}
In other words, we veto a color configuration if {\it all}\/
occurrences of a particular color $c$\/ come paired: $i_m=j_m=c$. By
adding this veto to the ``Conserved'' generation, we obtain the
``Non-Zero'' Monte Carlo procedure that has removed {\it all}
color configurations with zero color weight. The number of leftover
configurations sampled over is given by
\beqa
N^{\textrm{Non-Zero}}_{\rm col}&=&
\sum_{n_1,n_2,n_3=0}^n\delta_{n_1+n_2+n_3,n}
\left(\frac{n!}{n_1!\,n_2!\,n_3!}\right)\;\times
\nn&&\hspace*{14mm}
\left(\frac{n!-n_1!\,n_2!\,n_3!\,\big[1-\sum_c\Theta(n_c-1)\big]-
\sum_c\Theta(n_c-1)\,n_c!\,(n-n_c)!}{n_1!\,n_2!\,n_3!}\right)\ ,
\eeqa
where the step function $\Theta(x)=1$ for $x\geq0$ and zero otherwise.
\begin{table}[t!]\begin{center}\small
\begin{tabular}{|c||r|r|r|}
\hline\hline
Scattering & Naive & Conserved & Non-Zero \\ \hline
$2\rightarrow2$ & 6,561         & 639         & 378\\
$2\rightarrow3$ & 59,049        & 4,653       & 3,180\\
$2\rightarrow4$ & 531,441       & 35,169      & 27,240\\
$2\rightarrow5$ & 4,782,969     & 272,835     & 231,672\\
$2\rightarrow6$ & 43,046,721    & 2,157,759   & 1,949,178\\
$2\rightarrow7$ & 387,420,489   & 17,319,837  & 16,279,212\\
$2\rightarrow8$ & 3,486,784,401 & 140,668,065 & 135,526,716\\
\hline\hline
\end{tabular}
\caption{\label{MCnumber}
The number of color configurations sampled over when using the
different Monte Carlo color schemes.}
\end{center}
\end{table}
The weight associated with each sampled color configuration has to be
modified and reads
\beq
W^{\textrm{Non-Zero}}_{\rm col}\;=\;\left(3^n-3\right)
\left(\frac{n!-n_1!\,n_2!\,n_3!\,\big[1-\sum_c\Theta(n_c-1)\big]-
\sum_c\theta(n_c-1)\,n_c!\,(n-n_c)!}{n_1!\,n_2!\,n_3!}\right)\ .
\eeq
For up to 10-gluon scatterings, Table~\ref{MCnumber} displays the
resulting number of sampled color configurations in the column
indicated ``Non-Zero''. It is also shown how this number compares to
the numbers found for the ``Conserved'' and ``Naive'' sampling scheme.

\begin{table}[t!]
\begin{center}\small
\begin{tabular}{|c||r|r|r|}
\hline\hline
Scattering & color ordered & color dressed     & color dressed\\
           &               & ($V_{\rm max}=4$) & ($V_{\rm max}=3$)\\\hline
$2\to2$ & 0.0313\ $\hphantom{^{(0.00)}}$& 0.117\ $\hphantom{^{(0.00)}}$
& 0.083\ $\hphantom{^{(0.00)}}$\\
$2\to3$ & 0.169\  $^{(5.40)}$& 0.495\ $^{(4.24)}$ & 0.327\ $^{(3.93)}$\\
$2\to4$ & 0.791\  $^{(4.68)}$& 1.556\ $^{(3.14)}$ & 0.822\ $^{(2.51)}$\\
$2\to5$ & 3.706\  $^{(4.69)}$& 6.11\ \ $^{(3.93)}$& 2.66\ \ $^{(3.23)}$\\
$2\to6$ & 17.83\  $^{(4.81)}$& 25.26\ $^{(4.13)}$ & 7.55\ \ $^{(2.84)}$\\
$2\to7$ & 99.79\  $^{(5.60)}$& 93.43\ $^{(3.70)}$ & 24.9\ \ $^{(3.30)}$\\
$2\to8$ & 557.9\  $^{(5.59)}$& 392.4\ $^{(4.20)}$ & 76.1\ \ $^{(3.05)}$\\
$2\to9$ & 2,979\  $^{(5.34)}$& 1,528\ $^{(3.89)}$ & 228\ \ $^{(2.99)}$\\
$2\to10$&19,506\  $^{(6.55)}$& 5,996\ $^{(3.92)}$ & 693\ \ $^{(3.04)}$\\
$2\to11$&118,635\ $^{(6.08)}$&24,821\ $^{(4.14)}$ &\\
$\vdots$&&$\vdots\ \ \ \ \hphantom{^{(0.00)}}$&\\
$2\to15$&                    &6,248,300\ $^{(3.98^4)}$&\\
\hline\hline
\end{tabular}
\caption{\label{MCsize}
The time (in seconds) to evaluate 10,000 color-dressed tree-level
amplitudes for $2\to n-2$ gluon scatterings. Only color configurations
with non-zero weight are taken into account. Also indicated is the
growth factor (given in brackets) with increasing $n$. To compute the
amplitudes a 2.20 GHz Intel Core2 Duo processor was used.}
\end{center}
\end{table}

Next we examine the execution time of $n$-gluon scattering amplitudes
using the ``Non-Zero'' color sampling. In Table~\ref{MCsize} the CPU
time needed to calculate the color-dressed amplitudes according to
Eq.~(\ref{CDLOgluon}) and Eq.~(\ref{LOWillen}) are compared.

The evaluation of Eq.~(\ref{LOWillen}) employs the ordered recursion
relation~\cite{Berends:1987me}. Naively one would expect this
evaluation to grow factorially with the number of gluons. However this
growth is considerably dampened by sampling over non-zero color
configurations only. Note that for a given event we calculate each
ordered amplitude with non-vanishing color factor independently of the
other ordered amplitudes. One can speed up the computation time by
sharing the calculated sub-currents between different orderings. This,
however, is outside the scope of this paper.

For the evaluation of Eq.~(\ref{CDLOgluon}) we use the color-dressed
recursion relation of Eq.~(\ref{CDrecursion}). To study its time
behavior we apply this recursion as discussed in
appendix~\ref{App:gluon-recu} with and without the 4-gluon vertex. As
can be seen from Table~\ref{MCsize}, the required CPU times scales as
$4^n$ or $3^n$ if the 4-gluon vertex is neglected. This exponential
scaling was derived in Ref.~\cite{Bruinsma,Duhr:2006iq}. The
derivation, following~\cite{Duhr:2006iq}, uses the recursive buildup
of the amplitude. To calculate an $n$-particle amplitude using a
$V$-point vertex, we have to evaluate the $(n-1)$-particle current of
Eq.~(\ref{treeamp}). This current in turn is determined by calculating
all $\left({n-1\atop m}\right)$ $m$-particle sub-currents, where
$n-1\geq m\geq2$. Each $m$-current is constructed from smaller
currents using Eq.~(\ref{tree-recurrence}) thereby employing the
$V$-point vertex. All possible partitions into $V-1$ sub-currents
are given by the Stirling number of the second kind, ${\cal S}_2(m,V-1)$.
This leads to the following scaling of the calculation of the
$n$-particle amplitude
\beq
T_n\;=\sum_{m=2}^{n-1}\left({n-1\atop m}\right) 
{\cal S}_2(m,V-1)\;=\;{\cal S}_2(n,V)\sim V^n\ .
\eeq
Consequently, the $n$-gluon amplitude using the standard 3-gluon and
4-gluon vertex has an exponential scaling behavior $T_n\rightarrow
4^{n}$. This is evident from the results shown in Table~\ref{MCsize}.
As can also be seen in the table, the scaling behaves as expected when
the 4-gluon vertex is left out, i.e.\ $T_n\rightarrow 3^{n}$. As was
shown in Ref.~\cite{Draggiotis:1998gr,Duhr:2006iq}, the 4-gluon vertex
can be avoided and replaced by an effective 3-point vertex. This
results in a significant time gain for the evaluation of high
multiplicity gluon scattering amplitudes.

\begin{figure}[p!]
\psfrag{Yconvergelabel}[b][c][0.84]{
  $\left(\left\langle S^{(0)}_{\rm MC}\right\rangle\pm
  \sigma_{\left\langle S^{(0)}_{\rm MC}\right\rangle}\right)\;\left\langle\,
  \sum\limits_{\rm col}\left|{\cal M}^{(0)}\right|^2\right\rangle^{-1}$}
\psfrag{Xconvergelabel}[t][t][0.84]{$
  100\%\times\sigma\Big(R_{\rm MC}(N_{\rm MC})\Big)\Big/
  \mu\Big(R_{\rm MC}(N_{\rm MC})\Big)
  $}
\vspace*{-4mm}
\centerline{
  \includegraphics[width=0.64\columnwidth,angle=-90]{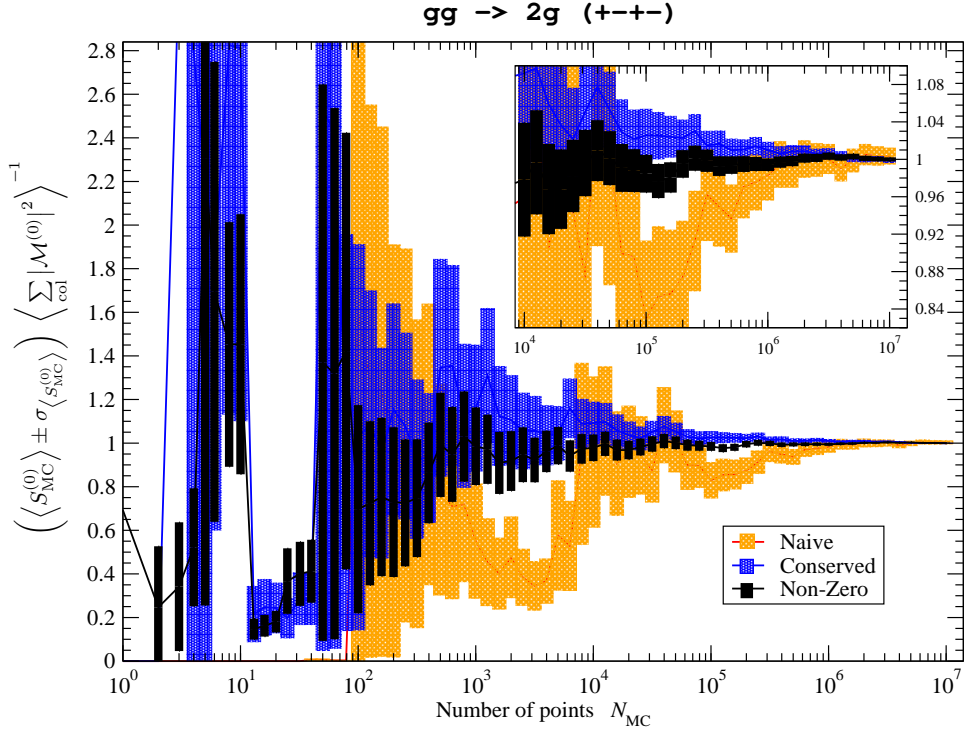}}
\vspace*{-2mm}
\centerline{
  \includegraphics[width=0.64\columnwidth,angle=-90]{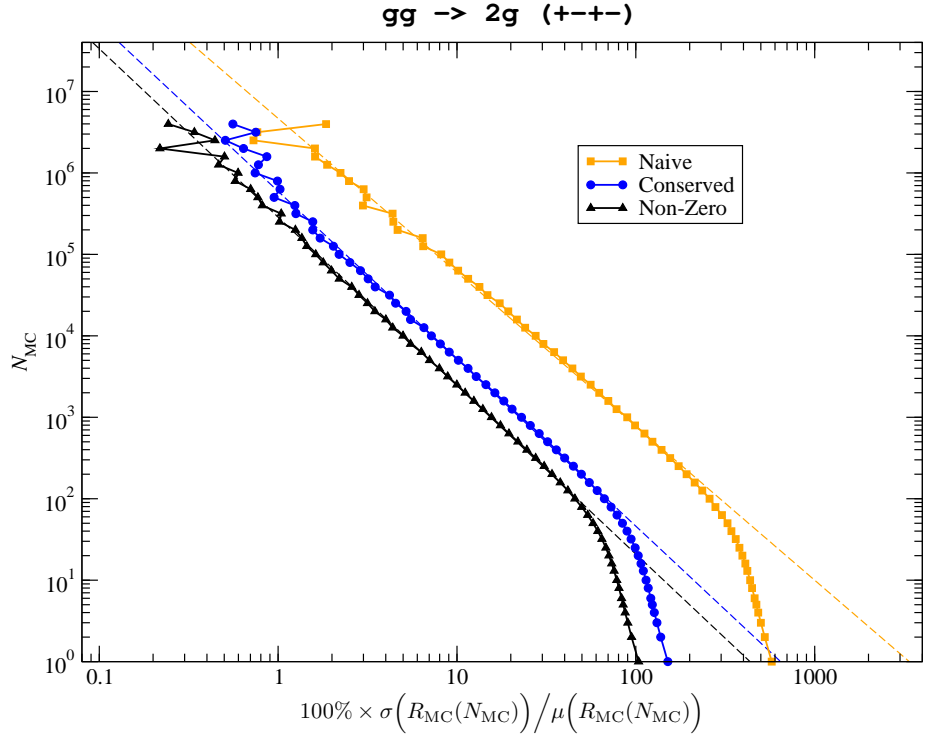}}
\caption{Top panel: comparison of Monte Carlo integrations for the
  various color-sampling schemes, including the standard deviation, to
  the exact color-summed result as a function of the number of
  evaluated phase-space points. One obtains $1.0034\pm0.0091$,
  $0.9989\pm0.0027$ and $0.9999\pm0.0022$ after $10^7$ steps for the
  ``Naive'', ``Conserved'' and ``Non-Zero'' sampling, respectively.
  Bottom panel: number of events required to reach a given relative
  accuracy on the numerical evaluation of the color-sampled
  amplitude. For the definition of $R_{\rm MC}(N_{\rm MC})$ and the
  values of the fit parameters determining the dashed curves, cf.\ the
  text.}
\label{Fig:treeconv4}
\end{figure}

\begin{figure}[p!]
\psfrag{Yconvergelabel}[b][c][0.84]{
  $\left(\left\langle S^{(0)}_{\rm MC}\right\rangle\pm
  \sigma_{\left\langle S^{(0)}_{\rm MC}\right\rangle}\right)\;\left\langle\,
  \sum\limits_{\rm col}\left|{\cal M}^{(0)}\right|^2\right\rangle^{-1}$}
\psfrag{Xconvergelabel}[t][t][0.84]{$
  100\%\times\sigma\Big(R_{\rm MC}(N_{\rm MC})\Big)\Big/
  \mu\Big(R_{\rm MC}(N_{\rm MC})\Big)
  $}
\vspace*{-4mm}
\centerline{
  \includegraphics[width=0.64\columnwidth,angle=-90]{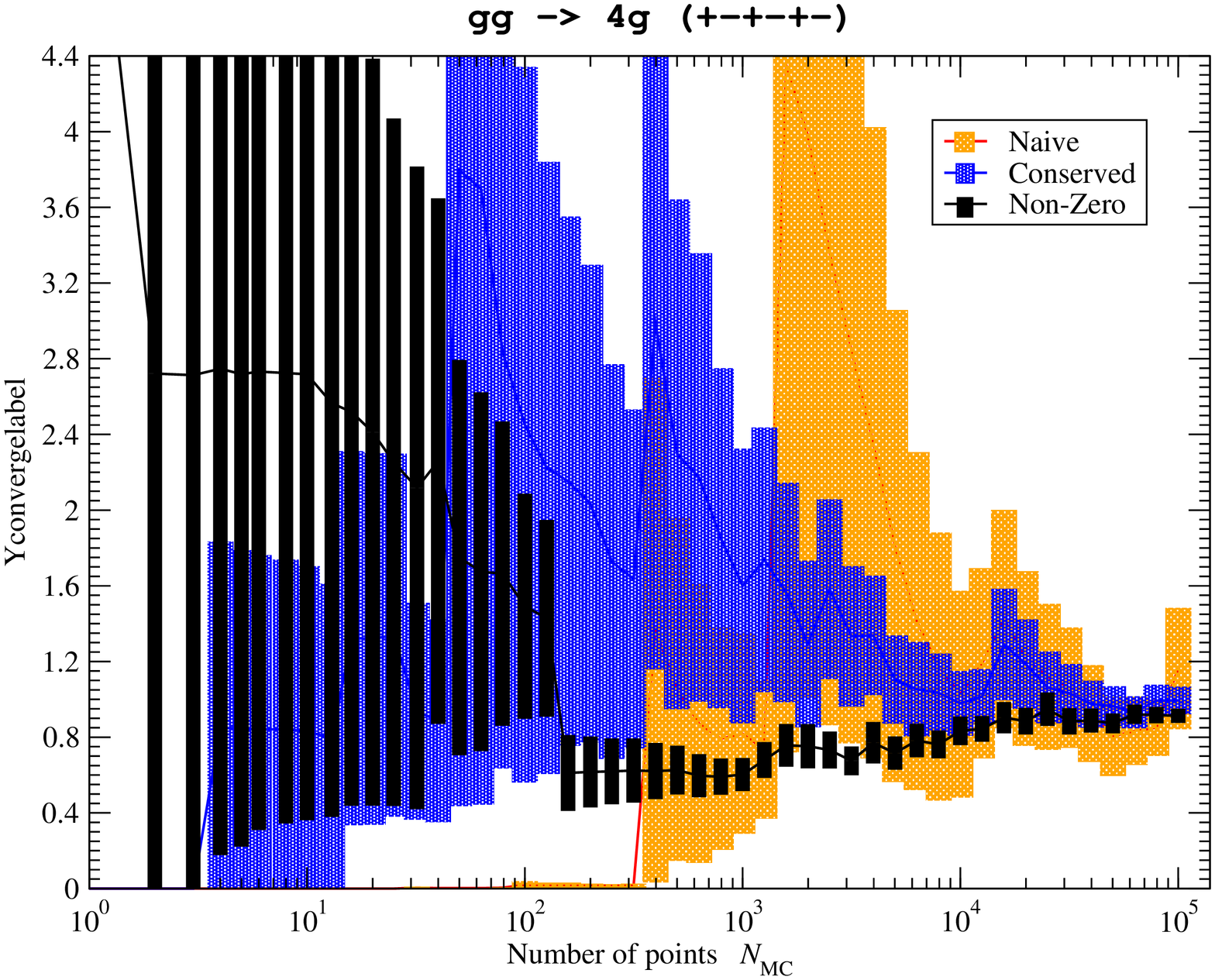}}
\vspace*{-2mm}
\centerline{
  \includegraphics[width=0.64\columnwidth,angle=-90]{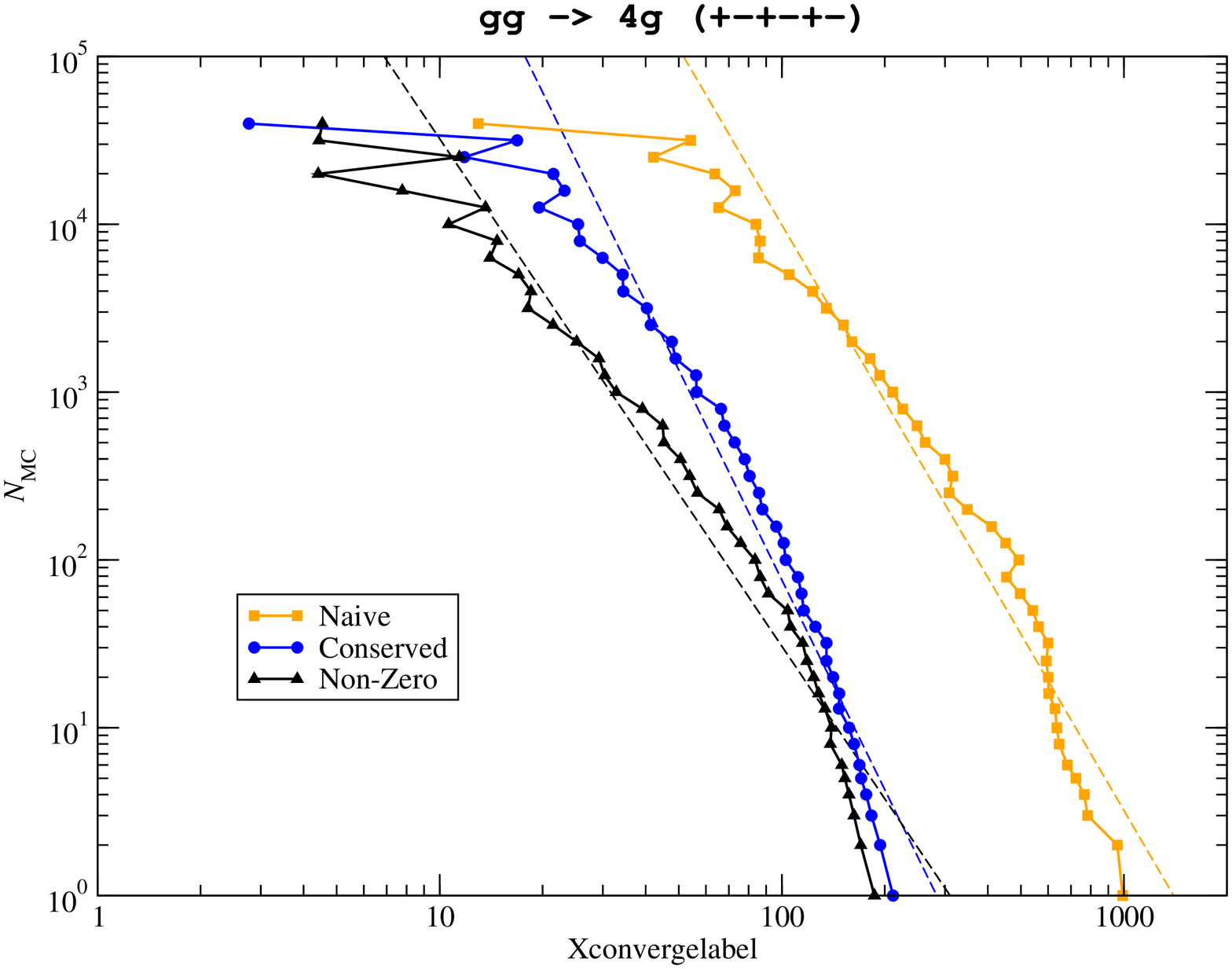}}
\caption{Top panel: comparison of Monte Carlo integrations for the
  various color-sampling schemes, including the standard deviation, to
  the exact color-summed result as a function of the number of
  evaluated phase-space points. One obtains $1.16\pm0.32$,
  $0.995\pm0.071$ and $0.913\pm0.037$ after $10^5$ steps for the
  ``Naive'', ``Conserved'' and ``Non-Zero'' sampling, respectively.
  Bottom panel: number of events required to reach a given relative
  accuracy on the numerical evaluation of the color-sampled
  amplitude. For the definition of $R_{\rm MC}(N_{\rm MC})$, cf.\ the
  text. The fit curves in terms of $\sigma/\mu(N_{\rm MC})$ are
  described by $14.0\,N_{\rm MC}^{-0.287}$, $2.84\,N_{\rm MC}^{-0.241}$ and
  $3.10\,N_{\rm MC}^{-0.331}$ for the ``Naive'', ``Conserved'' and
  ``Non-Zero'' sampling, respectively. The ``Conserved'' and
  ``Non-Zero'' approaches are slower by factors of $f=10.5$ and $f=13.3$,
  respectively (see text).}
\label{Fig:treeconv6}
\end{figure}

An important consideration in the usefulness of the color-sampling
approach is the convergence to the correct answer as a function of the
Monte Carlo sampling size $N_{\rm MC}$. To this end, we compare the
color-sampled result $S^{(0)}_{\rm MC}$ for the tree-level amplitude
squared,
\beq\label{convergenceLO}
S^{(0)}_{{\rm MC},r}\;=\;W_{\rm col}(n_1,n_2,n_3)\times
\left|\,{\cal M}^{(0)}\big({\bf g}^{(r)}_1,\ldots,{\bf g}^{(r)}_n\big)\right|^2\ ,
\eeq
to the color-summed, i.e.\ color-exact, result
\beq\label{Eq:Scol}
S^{(0)}_{{\rm col},r}\;=
\sum_{i_1,\ldots,i_n=1}^3\ \sum_{j_1,\ldots,j_n=1}^3
\left|\,{\cal M}^{(0)}\big({\bf g}^{(r)}_1,\ldots,{\bf g}^{(r)}_n\big)\right|^2\ .
\eeq
We plot the ratio of the average value for the color-sampled amplitude
squared and its standard deviation over the average value of the
color-summed amplitude squared as a function of the number of
evaluated Monte Carlo events:
\beq\label{Eq:Rtree}
R\;=\;\frac{\langle S^{(0)}_{\rm MC}\rangle\,\pm\,
            \sigma_{\langle S^{(0)}_{\rm MC}\rangle}}
           {\langle S^{(0)}_{\rm col}\rangle}\ .
\eeq
We define the ratio this way so that most of the phase-space
integration fluctuations are divided out. The average values are
computed via
\beq\label{Eq:meanMC}
\langle S^{(0)}\rangle\;=\;
\frac{1}{N_{\rm MC}}\,\sum_{r=1}^{N_{\rm MC}}S^{(0)}_r
\eeq
where the index $r$\/ numbers the different events with the only
exception that the gluon polarizations have held fixed:
$\lambda_1,\ldots,\lambda_n=+-\ldots+-(+)$. The standard deviation of
the average is calculated by
\beq\label{Eq:sigmaMC}
\sigma_{\langle S^{(0)}\rangle}\;=\;
\frac{\sqrt{\sum_{r=1}^{N_{\rm MC}}(S^{(0)}_r)^2\,-\,
    N_{\rm MC}\,\langle S^{(0)}\rangle^2}}
     {N_{\rm MC}-1}\ .
\eeq
The 4- and 6-gluon scattering results are shown in the respective top
parts of Figs.~\ref{Fig:treeconv4} and \ref{Fig:treeconv6} for the
three different sampling methods ``Naive'', ``Conserved'' and
``Non-Zero''. The generated phase-space points were subject to the
constraints: $p_{\perp,m}>0.1\sqrt{s}$,\ $|\eta_m|<2$\ and $\Delta
R_{ml}>0.7$, see also Eq.~(\ref{Eq:cuts}). As it can be seen from the
two plots by avoiding sampling over zero-weight color configurations
the convergence is greatly enhanced.

\begin{figure}[t!]
\psfrag{Xconvergelabel}[t][t][0.84]{$
  100\%\times\sigma\Big(R_{\rm MC}(N_{\rm MC})\Big)\Big/
  \mu\Big(R_{\rm MC}(N_{\rm MC})\Big)
  $}
\vspace*{-4mm}
\centerline{
  \includegraphics[width=0.64\columnwidth,angle=-90]{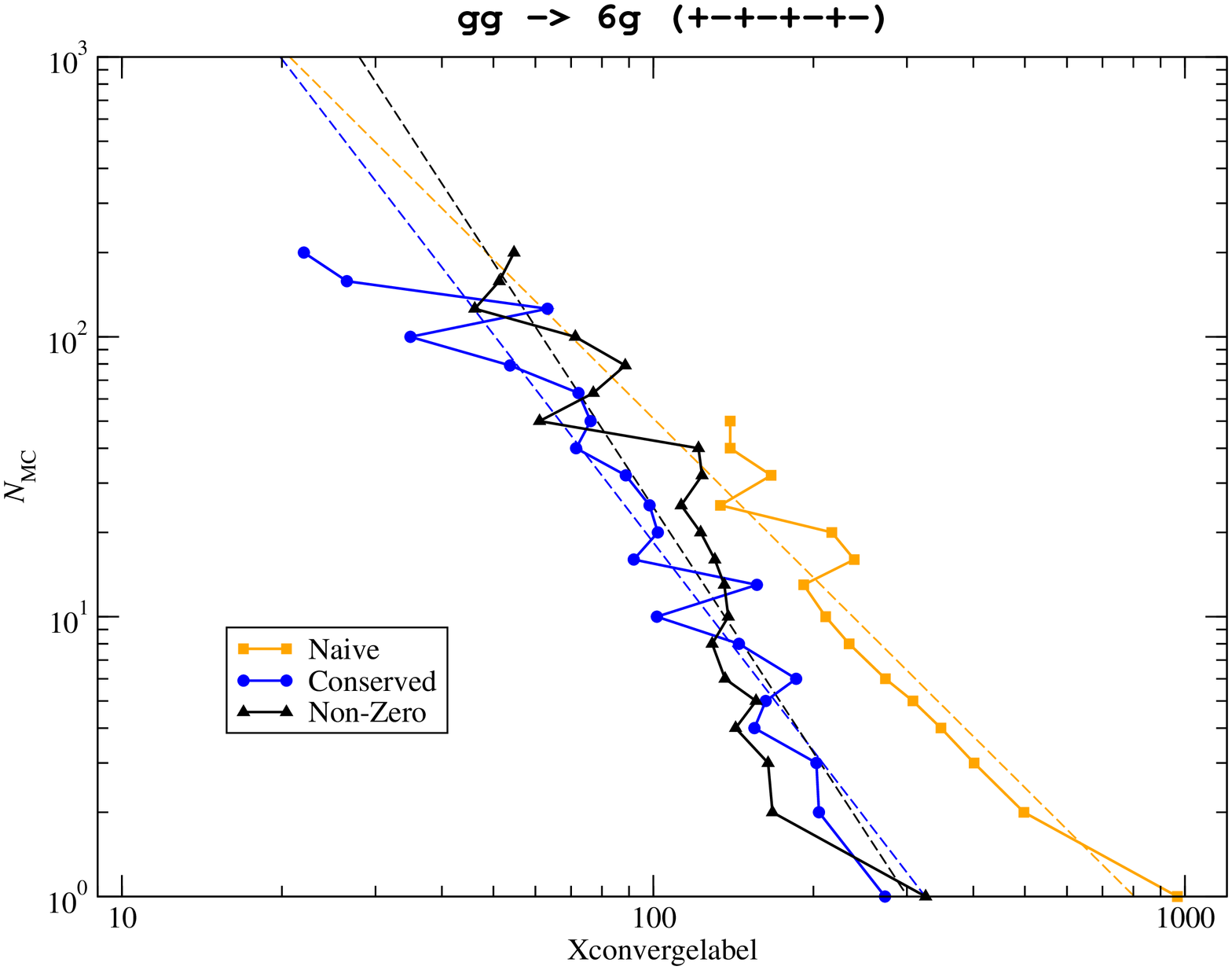}}
\caption{Number of events required to reach a given relative
  accuracy on the numerical evaluation of the color-sampled
  amplitude. For the definition of $R_{\rm MC}(N_{\rm MC})$, cf.\ the
  text. The fit curves in terms of $\sigma/\mu(N_{\rm MC})$ are
  described by $8.04\,N_{\rm MC}^{-0.530}$, $3.25\,N_{\rm MC}^{-0.405}$
  and $3.01\,N_{\rm MC}^{-0.344}$ for the ``Naive'', ``Conserved'' and
  ``Non-Zero'' sampling, respectively. The ``Conserved'' and
  ``Non-Zero'' approaches are slower by factors of $f=9.6$ and $f=10.8$,
  respectively (see text).}
\label{Fig:treeconv8}
\end{figure}

For $N_{\rm MC}={\cal O}(10^5)$, we obtain sufficient accuracy in the
``Non-Zero'' sampling method. To illustrate this more clearly, we show
in the lower panels of Figs.~\ref{Fig:treeconv4} and \ref{Fig:treeconv6}
as well as in Fig.~\ref{Fig:treeconv8} the number of Monte Carlo
events needed to achieve a certain relative precision in the color
sampling. For these plots, we generate $N_{\rm event}$ events, which
are partitioned into trial and sampling events via
$N_{\rm event}=N_{\rm trial}\times N_{\rm MC}$. We define as a
function of $N_{\rm MC}$ the ratio
\beq\label{Eq:RMCtree}
R_{\rm MC}(N_{\rm MC})\;=\;\frac{\sum_{r=1}^{N_{\rm MC}}S^{(0)}_{{\rm MC},r}}
                                {\sum_{r=1}^{N_{\rm MC}}S^{(0)}_{{\rm col},r}}
\;=\;\frac{\langle S^{(0)}_{\rm MC}\rangle(N_{\rm MC})}
	  {\langle S^{(0)}_{\rm col}\rangle(N_{\rm MC})}
\eeq
and plot $N_{\rm MC}$ versus the relative precision
$\sigma(R_{\rm MC})/\mu(R_{\rm MC})$. The mean value
\beq\label{Eq:meanTrial}
\mu(R_{\rm MC})\;=\;
\frac{1}{N_{\rm trial}}\,\sum_{k=1}^{N_{\rm trial}}R_{{\rm MC},k}(N_{\rm MC})
\eeq
and the standard deviation
\beq\label{Eq:sigmaTrial}
\sigma(R_{\rm MC})\;=\;
\sqrt{\frac{\sum_{k=1}^{N_{\rm trial}}\big(R_{{\rm MC},k}(N_{\rm MC})\big)^2
    \,-\,N_{\rm trial}\,\mu^2(R_{\rm MC})}
  {N_{\rm trial}-1}}\ .
\eeq
are computed by using a sufficiently large number of trials, i.e.\
$N_{\rm trial}$ estimates of $R_{\rm MC}(N_{\rm MC})$ are calculated
to obtain the mean value and the standard deviation for $R_{\rm MC}$.
For $N_{\rm trial}>{\cal O}(100)$, we get rather smooth curves. In the
4-gluon case shown in the lower part of Fig.~\ref{Fig:treeconv4} this
gives a reasonable description for $N_{\rm MC}<10^5$. The 6- and
8-gluon scatterings are more involved and require more statistics. The
trend however can be read off the respective plots in
Figs.~\ref{Fig:treeconv6} and \ref{Fig:treeconv8}.

For sufficiently large $N_{\rm MC}$, the expected scaling of the
relative standard deviation $\sigma$\/ with the number of Monte Carlo
events is $\sigma(R_{\rm MC})\sim 1/\sqrt{N_{\rm MC}}$. As can be seen
from the second plot of Fig.~\ref{Fig:treeconv4} the scaling is as
expected and we can fit to the functional form $A\times N_{\rm MC}^{-B}$.
In the 4-gluon case, we find
\beqa
\mbox{\rm Naive}\ &:&\
\frac{\sigma\left(R_{\rm MC}\right)}{\mu\left(R_{\rm MC}\right)}\;=\;
33.8\times N_{\rm MC}^{-0.529}\nn
\mbox{\rm Conserved}\ &:&\
\frac{\sigma\left(R_{\rm MC}\right)}{\mu\left(R_{\rm MC}\right)}\;=\;
6.45\times N_{\rm MC}^{-0.487}\nn
\mbox{\rm Non-Zero}\ &:&\
\frac{\sigma\left(R_{\rm MC}\right)}{\mu\left(R_{\rm MC}\right)}\;=\;
4.35\times N_{\rm MC}^{-0.484}\ .
\eeqa
From these fits we can quantify the enhancements owing to the sampling
strategies. The ``Conserved'' sampling method improves over the
``Naive'' method by a factor of $33.8/6.45=5.2$, while the improvement
of the ``Non-Zero'' method over the ``Conserved'' method yields an
additional factor of $6.45/4.35=1.5$ (or a factor of $33.8/4.35=7.8$
over the ``Naive'' method). The algorithm determines the color
configurations with vanishing color factor before it fully evaluates
the corresponding matrix-element weight. The differences between the
various sampling methods therefore become smaller when we measure the
computer evaluation time to reach a certain relative precision. When
we express this in numbers for the example of 4-gluon scattering, we
notice that the ``Conserved'' and ``Non-Zero''sampling schemes are
slower by factors of $f=2.42$ and $f=3.29$, respectively. This
translates into changing the fit parameter $A\to A'=Af^B$. The
corresponding ratios then read $33.8/9.92=3.4$ and $9.92/7.74=1.3$
when specifying the improvement of the ``Conserved'' versus the
``Naive'' and the ``Non-Zero'' versus the ``Conserved'' method,
respectively. We see using improved sampling over color configurations
is still highly preferred.

\section{Dressed Generalized Unitarity for Virtual Corrections}\label{CDGU}

By using the parametric integration method of 
Ref.~\cite{Ossola:2006us} 
one can implement the generalized unitarity method
of Ref.~\cite{Britto:2004nc} into an efficient algorithmic
solution~\cite{Ellis:2007br}. For the evaluation of
color-ordered amplitudes, the algorithm is of polynomial complexity~\cite{Bern:2008ef}. 
To calculate the  dimensional regulated one-loop amplitude we extend the parametric expressions to $D$-dimensions
and apply the cuts in several integer dimensions to determine all the
parametric coefficients~\cite{Giele:2008ve}.\footnote{If one uses an analytical implementation 
of the $D$-dimensional unitarity method of Ref.~\cite{Giele:2008ve}, 
one can eliminate the penta-cuts~\cite{Badger:2008cm}. 
However, in numerical implementations the removal of the penta-cuts requires 
performing a numerical contour integral in the complex plane~\cite{Berger:2008sj}.}
The algorithm is equally applicable for the inclusion of massive quarks~\cite{Ellis:2008ir}.
The power of this algorithmic solution
was demonstrated in Refs.~\cite{Giele:2008bc,Lazopoulos:2008ex,Winter:2009kd} for pure gluonic scattering.

Given the fully specified external sources and the interaction vertices, both real and virtual corrections
can be evaluated by the recursive formulas. The virtual corrections to the differential cross section are 
given by
\beqa\label{MCvirtual}
\lefteqn{
d\,\sigma^{(V)}(f_1f_2\rightarrow f_3\cdots f_n)\;=\;
\frac{W_{\rm S}}{N_{\rm event}}\times\sum_{r=1}^{N_{\rm event}} 
d\,PS^{(r)}(K_1K_2\rightarrow K_3\cdots K_n)
}\hspace*{42mm}\nn&&
2\,\Re\left({\cal M}^{(0)}\left({\mathbf f}_1^{(r)},\ldots,{\mathbf f}_n^{(r)}\right)^\dagger\times
{\cal M}^{(1)}\left({\mathbf f}_1^{(r)},\ldots,{\mathbf f}_n^{(r)}\right)\right)\ ,
\label{virtual}
\eeqa
where the external sources, including momenta and quantum numbers, are sampled through a Monte Carlo procedure. 
The weight $W_{\rm S}$ is determined by process dependent symmetry factors and sampling weights.

In this section we show how to use the color-dressed tree-level amplitudes discussed in the previous section 
to construct the color-dressed one-loop amplitudes. By color sampling over the
external partons one can calculate the virtual corrections using Eq.~(\ref{virtual}).
The generic algorithm will be outlined and applied to pure gluon scattering.

\subsection{Generic Color-Dressed Generalized Unitarity}

The one-loop amplitude ${\cal M}^{(1)}\left({\bf f}_1,\ldots,{\bf f}_n\right)$ is obtained 
by integrating the un-integrated amplitude denoted by ${\cal A}^{(1)}\left({\bf f}_1,\ldots,{\bf f}_n\mid\ell\right)$
over the loop momentum $\ell$\/:
\beq \label{integrated}
{\cal M}^{(1)}\left({\bf f}_1,\ldots,{\bf f}_n\right)\;=
\int \frac{d^D\ell}{(2\pi)^D}\;
{\cal A}^{(1)}\left({\bf f}_1,\ldots,{\bf f}_n\mid\ell\right)\ .
\eeq
The integrand function can be decomposed into a sum of  a finite number of rational  functions of the loop
momentum with loop independent coefficients~\cite{Ossola:2006us}. The coefficients can  be calculated in terms of tree-level amplitudes.

The parametric form of the integrand is given by the triple sum of
rational functions,
\beq\label{parametricform}
{\cal A}^{(1)}\left({\bf f}_1,\ldots,{\bf f}_n\mid\ell\right)\;=\;
\sum_{k=1}^{C_{\rm max}}\;\
\sum_{RP_{\pi_1\cdots\pi_k}(1,2,\ldots,n)}^{\max\left(1,\frac{1}{2}(k-1)!\right)\,{\cal S}_2(n,k)}\
\sum_{g_{\Pi_1},\ldots,g_{\Pi_k}}
\frac{{\cal P}_k\left(\vec C_{g_{\Pi_1}\cdots g_{\Pi_k}}\mid\ell\right)}
{d_{g_{\Pi_1}}(\ell)\;d_{g_{\Pi_{2}}}(\ell)\cdots d_{g_{\Pi_k}}(\ell)}\ ,
\eeq
where the sum over the propagator flavors $g_{\Pi_1},\ldots,g_{\Pi_k}$ is required as these are not uniquely defined for 
unordered amplitudes.

The maximum number of denominators needed to describe the dimensional regulated one-loop
matrix element is $C_{\rm max}$. The value of $C_{\rm max}$ is given by the dimensionality of the 
loop momentum. For the one-loop calculations in dimensional regularization
the maximum dimension of the loop momentum
is equal to five, i.e.\ $C_{\rm max}=5$. The denominator terms are defined
as
\beq
d_{f_{\Pi_m}}(\ell)\;=\;(\ell+K_{\Pi_m})^2-m_f^2
\eeq
with $\Pi_m$ given through Eq.~(\ref{Eq:incllists}).
The partition sum is over 
$RP_{\pi_1\cdots\pi_k}(1,2,\ldots,n)$ ($\supseteq P_{\pi_1\cdots\pi_k}(1,2,\ldots,n)$)
elements. The total number of elements is given by
$\max\left(1,\frac{1}{2}(k-1)!\right)\times{\cal S}_2(n,k)$.
This extended partition list now also includes non-cyclic and
non-reflective permutations over the
regular partition lists $\{\{\pi_i\}_{i=1}^k\}$; more specifically we
have:
\beqa
RP_{\pi_1\pi_2}&=&\Big\{P_{\pi_1\pi_2}\Big\} \nn
RP_{\pi_1\pi_2\pi_3}&=&\Big\{P_{\pi_1\pi_2\pi_3}\Big\} \nn
RP_{\pi_1\pi_2\pi_3\pi_4}&=&
\Big\{P_{\pi_1\pi_2\pi_3\pi_4}, P_{\pi_1\pi_3\pi_4\pi_2},P_{\pi_1\pi_4\pi_2\pi_3}\Big\} \nn
RP_{\pi_1\pi_2\pi_3\pi_4\pi_5}&=&\Big\{
P_{\pi_1\pi_2\pi_3\pi_4\pi_5},P_{\pi_1\pi_3\pi_4\pi_5\pi_2},P_{\pi_1\pi_4\pi_5\pi_2\pi_3},P_{\pi_1\pi_5\pi_2\pi_3\pi_4},
\nn &&\phantom{\{}
P_{\pi_1\pi_2\pi_4\pi_5\pi_3},P_{\pi_1\pi_4\pi_5\pi_3\pi_2},P_{\pi_1\pi_5\pi_3\pi_2\pi_4},P_{\pi_1\pi_3\pi_2\pi_4\pi_5}, 
\nn &&\phantom{\{}
P_{\pi_1\pi_2\pi_5\pi_3\pi_4},P_{\pi_1\pi_5\pi_3\pi_4\pi_2},P_{\pi_1\pi_3\pi_4\pi_2\pi_5},P_{\pi_1\pi_4\pi_2\pi_5\pi_3}
\Big\}\ .
\eeqa
The  polynomial dependence of the numerator functions ${\cal P}_k $ on the loop momentum is specified with
a vector of parametric coefficients $\vec C_{g_{\Pi_1}\cdots g_{\Pi_k}}$.
The explicit polynomial forms that we are using are given in Ref.~\cite{Giele:2008ve}.
\begin{figure}[tb]
\begin{center}
\includegraphics[angle=0,scale=0.7]{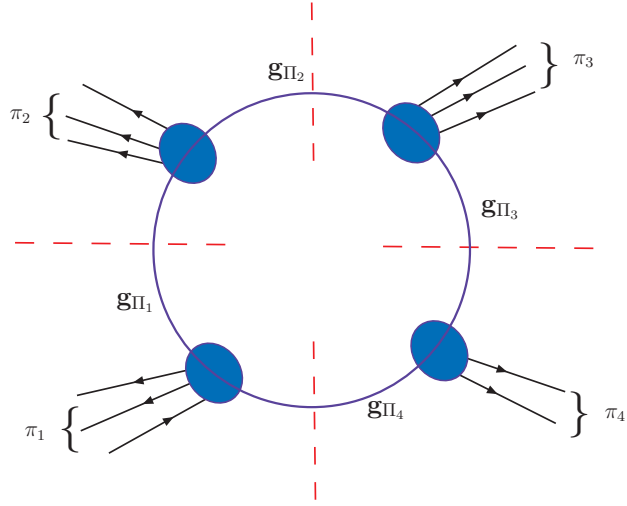}
\caption{\label{PL}
Graphical representation of a quadrupole-cut partitioning of the external legs
into an ordered  set of four unordered subsets ${\pi_1,\pi_2,\pi_3,\pi_4}$ of external  particles.
The corresponding tree-level diagrams are connected with the propagators of particle 
$g_{\Pi_1},g_{\Pi_2},g_{\Pi_3},g_{\Pi_4}$.
}
\end{center}
\end{figure}
The dimensionality of the parameter vector $\vec C_{g_{\Pi_1}\cdots g_{\Pi_k}}$ 
depends on the number of denominators.
In the case of 5 denominators there is only one parameter, for the terms with 4 denominators
we have five parameters, etc.\ .  
The parameters are determined  by
 putting sets of denominators to zero and calculating the residue in terms of tree-level amplitudes.
Setting  denominator factors to zero is on a par with  cutting the corresponding 
propagators as required by    generalized $D$-dimensional unitarity.
Let $\ell_{\Pi_1\cdots\Pi_c}$ be the  
``on-shell'' loop  momentum fulfilling the ``unitarity condition'':
\beq\label{uconstraint}
d_{g_{\Pi_1}}(\ell_{\Pi_1\cdots\Pi_c})\;
=\cdots =\;d_{g_{\Pi_c}} (\ell_{\Pi_1\cdots\Pi_c})\;=0\,;\qquad c=2,\ldots, C_{\rm max}\ .
\eeq
To fulfill the unitarity conditions we allow also complex values
for the components of the loop momenta.
The parametric form of the numerator functions for $c$-cuts becomes
\beqa 
\label{loop-recurrence}
\lefteqn{
{\cal P}_c\left(\vec C_{g_{\Pi_1}\cdots g_{\Pi_c}}
\mid \ell_{\Pi_1\cdots\Pi_c}
\right)\;=\; 
\mbox{Res}_{g_{\Pi_1}\cdots g_{\Pi_c}}
\left({\cal A}^{(1)}\left({\bf f}_1,\ldots,{\bf f}_n
\mid\ell_{\Pi_1\cdots\Pi_c}
\right)\right)} \nn
&-&\sum_{m=c+1}^{C_{\rm max}}\ \
\sum_{PP_{\widehat\pi_1,\ldots,\widehat\pi_m}(1,\ldots,n)}
\delta_{\Pi_1\widehat\Pi_1}\cdots\delta_{\Pi_c\widehat\Pi_c}\
\sum_{g_{\widehat\Pi_{c+1}}\cdots g_{\widehat\Pi_m}}\
\frac{{\cal P}_m\left(\vec C_{g_{\widehat\Pi_1}\cdots g_{\widehat\Pi_m}}
\mid \widehat\ell_{\widehat\Pi_1\cdots\widehat\Pi_c}\right)}
{d_{g_{\widehat\Pi_{c+1}}}( \widehat\ell_{\widehat\Pi_1\cdots\widehat\Pi_c})
\cdots d_{g_{\widehat\Pi_m}}(\widehat\ell_{\widehat\Pi_1\cdots\widehat\Pi_c}) }\ . 
\eeqa
where the sum $PP_{\widehat\pi_1,\ldots,\widehat\pi_m}(1,\ldots,n)$
over all $m!$ permutations of the $m$\/ partitions is
supplemented with the
$\delta$-functions to generate the appropriate subtraction functions.
Each individual subtraction expression has to be evaluated with the
appropriate shift of the loop-momentum,
$\widehat\ell_{\widehat\Pi_1\cdots\widehat\Pi_c}$.
This equation provides us with an iterative procedure starting with
the highest number of cuts.
For a given number of cuts, the numerator function becomes
the residue of  one-loop integrand function minus
the known contributions of terms with higher number of denominator factors.
The residue of the one-loop integrand factorizes into a product of
tree-level amplitudes (see e.g.\ Fig.~\ref{PL}):
\beqa\label{Residue}
\mbox{Res}_{g_{\Pi_1}\cdots g_{\Pi_c}}
\left({\cal A}^{(1)}\left({\bf f}_1,\ldots,{\bf f}_n
\mid \ell_{\Pi_1\cdots\Pi_c}
\right)\right)
&=&
{\left[d_{g_{\Pi_1}}(\ell)\times\cdots\times d_{g_{\Pi_c}}(\ell)\times
{\cal A}^{(1)}\left({\bf f}_1,\ldots,{\bf f}_n\mid\ell\right)\right]}_{\ell=\ell_{\Pi_1\cdots\Pi_c}}
\nn
&=&\sum_{{\bf g}_1\cdots {\bf g}_c}\left\{\prod_{k=1}^c
{\cal M}^{(0)}\big(\,{\bf g}_k^\dagger,\,{\bf f}_{\pi_k},\,{\bf g}_{k+1}\big)\right\}\ ,
\eeqa
where the index $k$\/ is cyclic (i.e.\ ${\bf g}_{c+1}={\bf g}_1$) and
${\bf g}_k$ denotes the particles resulting from the cut lines.

We can determine the parametric vector $\vec C_{g_{\Pi_1}\cdots g_{\Pi_c}}$ in Eq.~(\ref{loop-recurrence})
by evaluating the right hand side for a set of loop momenta 
fulfilling the unitarity constraint of Eq.~(\ref{uconstraint}).
The only physics model input is given through the
tree-level on-shell amplitudes, ${\cal M}^{(0)}$, which we evaluate using 
Eqs.~(\ref{treeamp}) and~(\ref{tree-recurrence}). 
Two of the external lines to the on-shell tree-level amplitudes
are generated by the $D$-dimensional cut lines. These external states have
in general complex, 5-dimensional momenta. 
This extension of the momenta does not modify the general structure of the
tree-level level recursion relations discussed in the previous section.  In this way we obtain
a fully specified algorithm to determine the parameters and thereby
the parametric form on the left hand side of Eq.~(\ref{parametricform}).

It is instructive to illustrate the structure given by  
Eq.~(\ref{parametricform})  for  
a simple example.
Let us consider the  cut-constructible, $D=4$, part of the box terms  in 4-gluon 
scattering ($n=4$, $k=4$).  In this case we have no pentagon terms and the numerator
functions of the box terms are parametrized by two coefficients
\beqa\label{ParametricExample}\lefteqn{
\sum_{RP_{1234}(1,2,3,4)}\ \ \sum_{f=\{g,q\}}\
\frac{
{\cal P}_4\left(\vec C_{f_1f_2f_3f_4}\mid\ell\right)}
{
d_{f_1}(\ell)d_{f_{12}}(\ell)d_{f_{123}}(\ell)d_{f_{1234}}(\ell)}
}\nn
&=&\frac{{\cal P}_4\left(\vec C_{g_1g_2g_3g_4}\mid\ell\right)}
{d_{g_1}(\ell)d_{g_{12}}(\ell)d_{g_{123}}(\ell)d_{g_{1234}}(\ell)}
+\frac{{\cal P}_4\left(\vec C_{g_1g_3g_4g_2}\mid\ell\right)}
{d_{g_1}(\ell)d_{g_{13}}(\ell)d_{g_{134}}(\ell)d_{g_{1342}}(\ell)}
+\frac{{\cal P}_4\left(\vec C_{g_1g_4g_2g_3}\mid\ell\right)}
{d_{g_1}(\ell)d_{g_{14}}(\ell)d_{g_{142}}(\ell)d_{g_{1423}}(\ell)}\nn
&+&\frac{{\cal P}_4\left(\vec C_{q_1q_2q_3q_4}\mid\ell\right)}
{d_{q_1}(\ell)d_{q_{12}}(\ell)d_{q_{123}}(\ell)d_{q_{1234}}(\ell)}
+\frac{{\cal P}_4\left(\vec C_{q_1q_3q_4q_2}\mid\ell\right)}
{d_{q_1}(\ell)d_{q_{13}}(\ell)d_{q_{134}}(\ell)d_{q_{1342}}(\ell)}
+\frac{{\cal P}_4\left(\vec C_{q_1q_4q_2q_3}\mid\ell\right)}
{d_{q_1}(\ell)d_{q_{14}}(\ell)d_{q_{142}}(\ell)d_{q_{1423}}(\ell)}
\eeqa
where
\beq
{\cal P}_4\left(\vec C_{f_1f_2f_3f_4}\mid \ell\right)
=C^{(0)}_{f_1f_2f_3f_4}+C^{(1)}_{f_1f_2f_3f_4}\times \ell
{\bf\cdot} n;\ 
n_{\mu}=\epsilon_{\mu\mu_1\mu_2\mu_3} p_{1}^{\mu_1}p_{12}^{\mu_2}p_{123}^{\mu_3}\ .
\eeq
The parameters are calculated by using  the residue formula of Eq.~(\ref{Residue}).
After the coefficients of  the box functions have been obtained, one
turns to calculate the
coefficients of the triangle contributions.   The numerator function for the triangle
cut of the quark-loop contribution, Eq.~(\ref{loop-recurrence}),  becomes 
\beqa
{\cal P}_3\left(\vec C_{q_1q_2q_{34}}\mid\ell_{\Pi_1\Pi_2\Pi_{34}}\right)&=&
\mbox{Res}_{q_1q_2q_{34}}
\Big({\cal A}^{(1)}({\bf g}_1,{\bf g}_2,{\bf g}_3,{\bf g}_4\mid\ell_{\Pi_1\Pi_2\Pi_{34}})\Big) \nn
&-&\frac{{\cal P}_4\left(\vec C_{q_1q_2q_3q_4}\mid\ell_{\Pi_1\Pi_2\Pi_{34}}\right)}
{d_{q_{123}}(\ell_{\Pi_1\Pi_2\Pi_{34}})}\;-\;
\frac{{\cal P}_4\left(\vec C_{q_1q_2q_4q_3}\mid\ell_{\Pi_1\Pi_2\Pi_{34}}\right)}
{d_{q_{124}}(\ell_{\Pi_1\Pi_2\Pi_{34}})}\ ,
\eeqa
where the residuum of the quark loop 
can be calculated again using Eq.~(\ref{Residue}),
\beqa
\mbox{Res}_{q_1q_2q_{34}}\left({\cal A}^{(1)}({\bf g}_1,{\bf g}_2,{\bf g}_3,{\bf g}_4\mid\ell_{\Pi_1\Pi_2\Pi_{34}} ) \right)
&=&\nn&&\hspace*{-59mm}
=\;\left[\,d_{q_1}(\ell)\times d_{q_{12}}(\ell)\times d_{q_{1234}}(\ell)\times
{\cal A}^{(1)}({\bf g}_1,{\bf g}_2,{\bf g}_3,{\bf g}_4\mid\ell)\,\right]_{\ell=\ell_{\Pi_1\Pi_2\Pi_{34}}}
\nn&&\hspace*{-59mm}
=\sum_{{\bf q}_1{\bf q}_2{\bf q}_{3}} {\cal M}^{(0)}({\bf q}_1^\dagger,{\bf g}_1,{\bf\bar q}_2)
\times {\cal M}^{(0)}({\bf q}_2^\dagger,{\bf g}_2,{\bf\bar q}_3)
\times {\cal M}^{(0)}({\bf q}_3^\dagger,{\bf g}_3,{\bf g}_4,{\bf\bar q}_1)\ .
\eeqa

Finally, we can obtain the one-loop amplitude, Eq.~(\ref{integrated}),
 by integrating  out the parametric forms on the right hand side of Eq.~(\ref{parametricform}) 
over the loop momentum.  In this way  one
finds the master-integral decomposition of the one-loop matrix element for every  specified
scattering configuration point~\cite{Giele:2008ve}:
\beqa\label{integrated-amp}
\lefteqn{
{\cal M}^{(1)}\left({\bf f}_1,\ldots,{\bf f}_n\right)\;=
\int \frac{d^D\,\ell}{(2\pi)^D}\ {\cal A}^{(1)}\left({\bf f}_1,\ldots,{\bf f}_n\mid\ell\right)}\nn
&=&\sum_{k=1}^{C_{\rm max}}\
\sum_{RP_{\pi_1\cdots\pi_k}(1,2,\ldots,n)}\ \sum_{g_{\Pi_1}\cdots g_{\Pi_k}}
S_F^{(g_{\Pi_1}\cdots g_{\Pi_k})}\times
\left(\bar C_{g_{\Pi_1}\cdots g_{\Pi_k}} {\cal I}_{g_{\Pi_1}\cdots g_{\Pi_k}}
+\bar{\bar C}_{g_{\Pi_1}\cdots g_{\Pi_k}} 
{\cal R}_{g_{\Pi_1}\cdots g_{\Pi_k}}\right)
\eeqa
where $S_F^{(g_{\Pi_1}\cdots g_{\Pi_k})}$ is the loop-integral
symmetry factor (e.g.\ for a gluonic self-energy, the symmetry factor
is $\frac{1}{2}$), the ${\cal I}_{g_{\Pi_1}\cdots g_{\Pi_k}}$ denote
the scalar master-integral functions corresponding to the generalized
cut given by the ordered partition list $\{\Pi_1\cdots\Pi_k\}$ 
and flavors of the cut lines $(g_{\Pi_1}\cdots g_{\Pi_k})$.
The terms ${\cal R}_{g_{\Pi_1}\cdots g_{\Pi_k}}$ are the leading terms
of the higher dimensional scalar integrals in the limit $D\to4$,
\beqa
{\cal R}_{f_{\Pi_1}f_{\Pi_2}f_{\Pi_3}f_{\Pi_4}}&=&-\frac{1}{6}\ ,\nn
{\cal R}_{f_{\Pi_1}f_{\Pi_2}f_{\Pi_3}}&=&\frac{1}{2}\ ,\nn
{\cal R}_{f_{\Pi_1}f_{\Pi_2}}&=&
-\frac{\left(K_{\Pi_1}-K_{\Pi_2}\right)^2}{6}+\frac{m_{f_{\Pi_1}}^2+m_{f_{\Pi_2}}^2}{2}\ ,\nnn
{\cal R}_{f_{\Pi_1}}&=&0\ .
\eeqa
The scalar master integrals
\beq
{\cal I}_{f_{\Pi_1}\cdots f_{\Pi_k}}\;=\;I_k\left(K_{\Pi_1},\ldots,K_{\Pi_k},m_{f_{\Pi_1}},\ldots,m_{f_{\Pi_k}}\right)\ ,
\eeq
can be evaluated by e.g.\ using the numerical package developed in
Ref.~\cite{Ellis:2007qk}.
In Eq.~(\ref{integrated-amp}) 
the coefficients $\bar C$\/ and $\bar{\bar C}$ are determined by 
applying Eqs.~(\ref{loop-recurrence}) and~(\ref{Residue}) using a numerical algorithm. 
The $\bar{\bar C}$\/ coefficients are generated due to the dimensional
regularization procedure and are associated with the higher
dimensional terms in the parametric forms.

\subsection{Numerical Results for the Virtual Corrections of $n$-gluon
  Scattering}

We have applied the formalism of the previous sections to multi-gluon
scattering. To this end we have extended the implementation presented
in Ref.~\cite{Winter:2009kd}. Three major changes are required to
alter the generalized-unitarity based algorithm for the evaluation
of color-ordered amplitudes to a numerical algorithm capable of
calculating color-dressed one-loop amplitudes. First, in the
decomposition of the one-loop integrands (cf.\ Eq.~(3) of
Ref.~\cite{Winter:2009kd} and Eq.~(\ref{parametricform})), all sums
over ordered cuts have to be changed into sums over partitions, which
include all configurations obtained by non-cyclic and non-reflective
permutations:
\beq
\sum\limits_{[i_1|i_k]}\quad\to\quad
\sum\limits_{{RP}_{\pi_1\cdots\pi_k}(1\ldots n)}\quad.
\eeq
Note that $[i_1|i_k]=1\le i_1<i_2<\cdots<i_k\le n$. Second, the
tree-level amplitudes occurring in the determination of the integrand's
residues have to be calculated from color-dressed recursion relations.
In addition, one not only has to sum over the internal polarizations
of the gluons but also over their internal colors when computing these
residues. Third, gluon bubble coefficients need to be supplemented by
a symmetry factor of $1/2!$. The appearance of the symmetry factor is
associated with the parametrization ambiguity of the subtraction terms
in the double cuts. For example, Eq.~(\ref{loop-recurrence}) gives for
one of the double cuts in 4-gluon scattering
\beqa
{\cal P}_2(\vec C_{g_{12}g_{34}}\mid\ell)&=&
\mbox{Res}_{g_{12}g_{34}}\Big({\cal A}^{(1)}({\bf g}_1,{\bf g}_2,{\bf g}_3,{\bf g}_4\mid\ell)\Big) \nn
&-&\frac{{\cal P}_3(\vec C_{g_1g_2g_{34}}\mid\ell)}{d_{g_1}(\ell)}
\;-\;\frac{{\cal P}_3(\vec C_{g_2g_1g_{34}}\mid\ell)}{d_{g_2}(\ell)}
\;-\;\frac{{\cal P}_3(\vec C_{g_3g_4g_{12}}\mid\ell)}{d_{g_3}(\ell)}
\;-\;\frac{{\cal P}_3(\vec C_{g_4g_3g_{12}}\mid\ell)}{d_{g_4}(\ell)} \nn
&-&\frac{{\cal P}_4(\vec C_{g_1g_2g_3g_4}\mid\ell)}{d_{g_1}(\ell)d_{g_{123}}(\ell)}
\;-\;\frac{{\cal P}_4(\vec C_{g_2g_1g_3g_4}\mid\ell)}{d_{g_2}(\ell)d_{g_{213}}(\ell)}
\;-\;\frac{{\cal P}_4(\vec C_{g_1g_2g_4g_3}\mid\ell)}{d_{g_1}(\ell)d_{g_{124}}(\ell)}
\;-\;\frac{{\cal P}_4(\vec C_{g_2g_1g_4g_3}\mid\ell)}{d_{g_2}(\ell)d_{g_{214}}(\ell)}\nn
&=&
\mbox{Res}_{g_{12}g_{34}}\Big({\cal A}^{(1)}({\bf g}_1,{\bf g}_2,{\bf g}_3,{\bf g}_4\mid\ell)\Big) \nn
&-&\frac{{\cal P}_3(\vec C_{g_1g_2g_{34}}\mid\ell)}{d_{g_1}(\ell)}
\;-\;\frac{{\cal P}_3(\vec C_{g_1g_2g_{34}}\mid-\ell+K_1+K_2)}{d_{g_1}(-\ell+K_1+K_2)} \nn
&-&\frac{{\cal P}_3(\vec C_{g_3g_4g_{12}}\mid\ell)}{d_{g_3}(\ell)}
\;-\;\frac{{\cal P}_3(\vec C_{g_3g_4g_{12}}\mid-\ell+K_3+K_4)}{d_{g_3}(-\ell+K_3+K_4)} \nn
&-&\frac{{\cal P}_4(\vec C_{g_1g_2g_3g_4}\mid\ell)}{d_{g_1}(\ell)d_{g_{123}}(\ell)}
\;-\;\frac{{\cal P}_4(\vec C_{g_1g_2g_3g_4}\mid-\ell+K_1+K_2)}{d_{g_1}(-\ell+K_1+K_2)d_{g_{123}}(-\ell+K_1+K_2)} \nn
&-&\frac{{\cal P}_4(\vec C_{g_1g_2g_4g_3}\mid\ell)}{d_{g_1}(\ell)d_{g_{124}}(\ell)}
\;-\;\frac{{\cal P}_4(\vec C_{g_1g_2g_4g_3}\mid-\ell+K_3+K_4)}{d_{g_1}(-\ell+K_3+K_4)d_{g_{124}}(-\ell+K_3+K_4)}\ .
\eeqa
We see that each of the four possible parametrized terms is subtracted
twice with a different choice of the loop momentum. The symmetry
factor of $1/2!$ ``averages'' over the double subtractions.

The results of the new formalism can be tested thoroughly beyond
applying the usual consistency checks such as solving for the
master-integral coefficients with two independent sets of loop
momenta. The value of the double pole (dp) can be cross-checked
against the analytic result
\beq\label{Eq:M1dp}
{\cal M}^{(1)}_{\rm dp,th}\;=\;
-\,\frac{c_\Gamma}{\epsilon^2}\,n\,N_{\rm C}\,{\cal M}^{(0)}\ .
\eeq
Moreover, for a given phase-space point, we can use the ordered
algorithm of Ref.~\cite{Winter:2009kd} to compute the full one-loop
amplitude of a certain color and helicity (polarization)
configuration. Following the color-decomposition approach, we can
analytically calculate the necessary color factors and sum up all
relevant orderings to obtain the full result. In particular, we have
employed:
\beqa\label{Eq:NLOgluons}
\lefteqn{
{\cal M}^{(1)}({\bf g}_1,\ldots,{\bf g}_n)\;=\sum_{P(2\cdots n)} 
{A^{(1)}}^{i_1\cdots i_n}_{j_1\cdots j_n}(g_1^{\lambda_1},\ldots,g_n^{\lambda_n})}\nn
&=&\hspace*{-4mm}\sum_{P(1\cdots n-1)}\hspace*{-1mm}\left[
N_{\rm C}\,\Delta_{1\cdots n}+\sum_{k=1}^{{\rm int}(n/2)}\sum_{m_1=1}^{n-k+1}\cdots\hspace*{-5mm}\sum_{m_k=m_{k-1}+1}^n
(-1)^k\Delta_{m_1\cdots m_k}\Delta_{1\cdots \slash\hspace{-0.2cm}m_1\cdots\slash\hspace{-0.2cm} m_k\cdots n}\right] 
m^{(1)}(12\cdots n)
\eeqa
where
\beq
\Delta_{12\cdots n}\;=\;\delta^{i_1}_{j_2}\delta^{i_2}_{j_3}\cdots\delta^{i_{n-1}}_{j_n}\delta^{i_n}_{j_1}\ .
\eeq
For example,
\beqa
{\cal M}^{(1)}({\bf g}_1,{\bf g}_2,{\bf g}_3,{\bf g}_4,{\bf g}_5)&=&\sum_{P(2345)} 
{A^{(1)}}({\bf g}_1,{\bf g}_2,{\bf g}_3,{\bf g}_4,{\bf g}_5)\nn
&=&\sum_{P(2345)}\Big(
N_c\,\Delta_{12345}-\Delta_{1}\Delta_{2345}-\Delta_{2}\Delta_{1345}-\Delta_{3}\Delta_{1245}
-\Delta_{4}\Delta_{1235}-\Delta_{5}\Delta_{1234}\nn
&&\phantom{\sum_{P(2345)}}+\Delta_{12}\Delta_{345}+\Delta_{13}\Delta_{245}+\Delta_{14}\Delta_{235}+\Delta_{15}\Delta_{234}
  +\Delta_{23}\Delta_{145}+\Delta_{24}\Delta_{135}\nn
&&\phantom{\sum_{P(2345)}}+\Delta_{25}\Delta_{134}+\Delta_{34}\Delta_{125}+\Delta_{35}\Delta_{124}+\Delta_{45}\Delta_{123}
\Big)\, m^{(1)}(12345)\ .
\eeqa
Compared to the LO color-ordered decomposition, Eq.~(\ref{LOWillen}),
the NLO color-ordered decomposition leads to many subleading color
factors. The number of one-loop ordered amplitudes with zero color
weight is significantly smaller than the corresponding number for
tree-level ordered amplitudes. As a result, the advantages of color
dressing become more apparent at the one-loop level.

\begin{table}[t!]
\begin{center}\small
\begin{tabular}{|c||r|r|r|r||r|c||r||r|r|r|}\hline
\multicolumn{11}{|c|}{\bf Ordered cuts.}\\\hline
\multicolumn{1}{|c}{\#}
    & 5-gon & box  & triangle & bubble & sum
    & \raisebox{-.4mm}{${\rm sum}_n$} & total $=$ sum $\times$
    & \#orderings & ${\cal N}_{(ab)_k}=$ & ${\cal N}_{(cd)_k}$\\\cline{1-1}
$n$ &  cuts & cuts & cuts     & cuts   &
    & \raisebox{.7mm}{$\overline{{\rm sum}_{n-1}\!\!}$} & \#orderings
    & $=(n-1)!/2$ & $(n-2)!$ &\\\hline
4 &   0 &   1 &   4 &  6 &   11&     &            33&         3&      2&     3\\
5 &   1 &   5 &  10 & 10 &   26& 2.36&           312&        12&      6&     7\\
6 &   6 &  15 &  20 & 15 &   56& 2.15&         3,360&        60&     24&    22\\
7 &  21 &  35 &  35 & 21 &  112& 2.00&        40,320&       360&    120&    40\\
8 &  56 &  70 &  56 & 28 &  210& 1.88&       529,200&     2,520&    720&   144\\
9 & 126 & 126 &  84 & 36 &  372& 1.77&     7,499,520&    20,160&  5,040&   756\\
10& 252 & 210 & 120 & 45 &  627& 1.69&   113,762,880&   181,440& 40,320& 2,688\\
11& 462 & 330 & 165 & 55 & 1012& 1.61& 1,836,172,800& 1,814,400&362,880&\\
12& 792 & 495 & 220 & 66 & 1573& 1.55&31,394,563,200&19,958,400&3,628,800&\\
\hline\hline
\end{tabular}
\caption{\label{Tab:ordrcuts}
The number of cuts required for the calculation of one ordered
$n$-gluon amplitude. The column labelled ``total'' gives the number of
cuts when calculating all $(n-1)!/2$ ordered amplitudes needed to
reconstruct the full virtual correction. The last two columns list the
number of non-zero color-weight orderings for two special color
configurations given in the text.}
\end{center}
\end{table}

\begin{table}[t!]
\begin{center}\small
\begin{tabular}{|c||r|r|r|r||r|c||c|c|c|}\hline
\multicolumn{10}{|c|}{\bf Unordered cuts.}\\\hline
\multicolumn{1}{|c}{\#}
    & pentagon & box  & triangle & bubble & sum
    & \raisebox{-.4mm}{${\rm sum}_n$}
    & \multicolumn{3}{|c|}{ordr total/unordr total}\\\cline{1-1}\cline{8-10}
$n$ & cuts     & cuts & cuts     & cuts   & $\equiv$ total
    & \raisebox{.7mm}{$\overline{{\rm sum}_{n-1}\!\!}$}
    & orderings & $(ab)_k$ & $(cd)_k$\\\hline
4 &         0&         3&     6&     3&         12&     & 2.750& 1.833 & 2.750\\
5 &        12&        30&    25&    10&         77& 6.42& 4.052& 2.026 & 2.364\\
6 &       180&       195&    90&    25&        490& 6.36& 6.857& 2.743 & 2.514\\
7 &     1,680&     1,050&   301&    56&      3,087& 6.30& 13.06& 4.354 & 1.451\\
8 &    12,600&     5,103&   966&   119&     18,788& 6.09& 28.17& 8.048 & 1.610\\
9 &    83,412&    23,310& 3,025&   246&    109,993& 5.85& 68.18& 17.05 & 2.557\\
10&   510,300&   102,315& 9,330&   501&    622,446& 5.66& 182.8& 40.61 & 2.708\\
11& 2,960,760&   437,250&28,501& 1,012&  3,427,523& 5.51& 535.7& 107.1 &\\
12&16,552,800& 1,834,503&86,526& 2,035& 18,475,864& 5.39& 1699 & 308.9 &\\
\hline\hline
\end{tabular}
\caption{\label{Tab:drsscuts}
The number of cuts needed to calculate color-dressed $n$-gluon
amplitudes. The last three columns give ratios of total numbers of
cuts required to compute the virtual corrections in both the
color-decomposition and color-dressed approaches. The first of these
columns shows the ratios for all generic color orderings whereas the
other columns show the ratios for two specific configurations as given
in the text.}
\end{center}
\end{table}

For a more quantitative understanding of the one-loop amplitude
decomposition, we respectively itemize in Tables~\ref{Tab:ordrcuts} and
\ref{Tab:drsscuts} how many cuts need be applied to decompose the
color-ordered and color-dressed one-loop integrands for $n$\/ external
gluons. In both cases we separately list the numbers of pentagon, box,
triangle and bubble cuts and their sum. While for the ordered cuts
these numbers are ruled by combinatorics:
${\cal C}(n,m)=\left({n\atop m}\right)$ with $m=1,\ldots,5$; in the
unordered case they are given by the Stirling numbers%
\footnote{More exactly, the number of bubble cuts is given by
  $2^{n-1}-1-n={\cal S}_2(n,2)-n$, since cuts that isolate one gluon
  do not contribute. For triangle, box, and pentagon cuts, we
  respectively have
  $(3^n-3\cdot 2^n+3)/6={\cal S}_2(n,3)$, $3\,{\cal S}_2(n,4)$ and
  $12\,{\cal S}_2(n,5)$ where, for the determination of the latter
  two, the recurrence relation
  ${\cal S}_2(n,m)={\cal S}_2(n-1,m-1)+m{\cal S}_2(n-1,m)$ is of help.}
of the second kind, ${\cal S}_2(n,m)$, and therefore grow more quickly
with $n$\/ than those of the ordered cuts. This is exemplified in the
``${\rm sum}_n/{\rm sum}_{n-1}$'' columns of the two tables. The
growth factors slowly decrease for larger $n$, approaching the limit
of $5$ for the color-dressed case. As emphasized in
Table~\ref{Tab:drsscuts} the pentagon-cut calculations dominate in
this case over all other cut evaluations. The large-$n$\/ growth of
the total cut number is hence described by that of ${\cal S}_2(n,5)$
leading to the observed large-$n$ scaling of $5^n$. Using the
color-decomposition approach, we have to deal with much fewer cuts per
ordering. However, the total number of ordered cuts is obtained only
after multiplying with the relevant number of orderings. When
considering all possible $(n-1)!/2$ orderings, the final numbers are
given in column ``total'' of Table~\ref{Tab:ordrcuts}. The last three
columns show the number of generic orderings and the numbers $\cal N$\/
of non-vanishing orderings (i.e.\ those having non-zero color factors)
for two color configurations $(ab)_k\equiv(13)(31)(11)\ldots(11)$ and
$(cd)_k\equiv(22)(12)(23)(31)(11)(22)(33)(11)(22)\ldots\ $.%
\footnote{The first four colors are always fixed, supplemented by the
  repeating sequence $(11)(22)(33)$ according to the number of gluons,
  i.e.\ for $n=5$ we have $(cd)_k\equiv(22)(12)(23)(31)(11)$, while
  for $n=9$ we use $(cd)_k\equiv(22)(12)(23)(31)(11)(22)(33)(11)(22)$.}
Of course, for a fair comparison between the ordered and dressed
approach, the latter two columns are of higher interest, since zero
color weights are not counted. Still, the ratios of total numbers of
ordered versus unordered cuts is always larger than one as can be read
off the last three columns of Table~\ref{Tab:drsscuts}. Keeping in
mind the greater cost of evaluating dressed recursion relations, the
color-decomposition approach can be expected to outperform the dressed
method as long as these ratios remain of order ${\cal O}(1)$. This in
particular is true for simple color configurations such as $(cd)_k$.

The analytic knowledge of ${\cal M}^{(1)}({\bf g}_1,\ldots,{\bf g}_n)$
presented in Eq.~(\ref{Eq:NLOgluons}) enables us to perform stringent
tests of our algorithm and its implementation. We consider $2\to n-2$
processes where the gluons have possible polarization states
$\lambda_k\in\{+,-\}$ and colors $(ij)_k$ where $i_k,j_k\in\{1,2,3\}$
and $k=1,\ldots,n$, i.e.\ we make use of the color-flow notation. Our
$n$-gluon results are given in the 4-dimensional helicity (FDH)
scheme~\cite{Bern:2002zk}. In almost all cases, we compare our new
method labelled by
``drss'' with the color-decomposition approach, which -- since it
makes use of the ordered algorithm -- we denote ``ordr''. We will
present all our results for two choices of loop-momentum and
spin-polarization dimensionalities $D$\/ and $D_s$: the ``4D-case'' is
obtained by setting $D=D_s=4$ and sufficient when merely calculating
the cut-constructible part (ccp) of the one-loop amplitude. The
``5D-case'' specified by $D=D_s=5$ allows us to determine the complete
result (all) including the rational part. In NLO
calculations one identifies the momenta of the external gluons with
those of well separated jets. We therefore apply cuts on the generated
$k=1,\ldots,n$ phase-space momenta ($l=3,\ldots,n$):
\beq\label{Eq:cuts}
|\eta_l|\;<\;2\,,\qquad
p_{\perp,l}\;>\;0.1\,|E_1+E_2|\,,\qquad
\Delta R_{kl}\;>\;0.7\,,
\eeq
where $\eta_l$ and $p_{\perp,l}$ respectively denote the
pseudo-rapidity and transverse momentum of the $l$-th outgoing gluon;
$\Delta R_{kl}$ describe the pairwise geometric separations in
pseudo-rapidity and azimuthal-angle space of gluons $k$\/ and $l$.
We perform a series of studies in the context of double-precision
computations: we investigate the accuracies with which the double
pole, single pole (sp) and finite part (fp) of the full one-loop
amplitudes are determined by our algorithm. We also examine the
efficiency of calculating virtual corrections by means of simple
phase-space integrations. To begin with, we will verify the expected
exponential scaling of the computation time for different numbers of
external gluons.

\begin{table}[t!]
\begin{center}\small
\begin{tabular}{|c||rr|c|rr|c|c||rr|c|rr|c|c|}\hline
    & \multicolumn{7}{c||}{\bf 4D-case}
    & \multicolumn{7}{c|}{\bf 5D-case}\\\cline{2-15}
$n$ & \multicolumn{3}{c|}{ordr} & \multicolumn{3}{c|}{drss}
    & \raisebox{-.4mm}{$\underline{\rm ordr}$}
    & \multicolumn{3}{c|}{ordr} & \multicolumn{3}{c|}{drss}
    & \raisebox{-.4mm}{$\underline{\rm ordr}$}\\\cline{2-7}\cline{9-14}
    & $\tau^{({\rm a})}_n$ & $\tau^{({\rm b})}_n$ & $r_n$
    & $\tau^{({\rm a})}_n$ & $\tau^{({\rm b})}_n$ & $r_n$
    & \raisebox{.7mm}{drss}
    & $\tau^{({\rm a})}_n$ & $\tau^{({\rm b})}_n$ & $r_n$
    & $\tau^{({\rm a})}_n$ & $\tau^{({\rm b})}_n$ & $r_n$
    & \raisebox{.7mm}{drss}\\\hline
4   & 0.027 & 0.026 &      & 0.061 & 0.062 &      & 0.43
    & 0.053 & 0.052 &      & 0.139 & 0.140 &      & 0.38\\
5   & 0.159 & 0.161 & 6.04 & 0.368 & 0.364 & 5.95 & 0.44
    & 0.415 & 0.412 & 7.88 & 1.026 & 1.029 & 7.37 & 0.40\\
6   & 1.234 & 1.235 & 7.72 & 2.152 & 2.146 & 5.87 & 0.57
    & 3.887 & 3.928 & 9.45 & 7.137 & 7.124 & 6.94 & 0.55\\
7   & 12.07 & 12.00 & 9.75 & 13.06 & 13.08 & 6.08 & 0.92
    & 41.66 & 41.61 & 10.7 & 49.62 & 49.85 & 6.98 & 0.84\\
8   & 131.2 & 131.3 & 10.9 & 80.22 & 80.53 & 6.15 & 1.6
    & 493.2 & 498.6 & 11.9 & 348.0 & 346.9 & 6.99 & 1.4\\
9   & 1579  & 1563  & 12.0 & 511.6 & 507.8 & 6.34 & 3.1
    & 6316  & 6296  & 12.7 & 2466  & 2470  & 7.10 & 2.6\\
10  & 20900 & 20480 & 13.2 & 3640  & 3629  & 7.13 & 5.7
    & 88320 & 88810 & 14.0 & 21590 & 21620 & 8.75 & 4.1\\\hline\hline
\end{tabular}
\caption{\label{Tab:timepsps}
Computer times $\tau_n$ in seconds obtained from the 4- and
5-dimensional evaluation of $n$-gluon virtual corrections at two
random phase-space points $\rm a$ and $\rm b$ using a 3.00 GHz Intel
Core2 Duo processor. The results are shown for both the color-ordered
and color-dressed method. All virtual corrections were evaluated twice to
check for the consistency of the solutions. The $n$\/ gluons have
colors $(ab)_k$ and polarizations $\kappa_k$ as specified in the text.
Also given are the ratios $r_n=\tau_n/\tau_{n-1}$ where $\tau_n$ is
the time to compute the correction for $n$\/ gluons, in particular
$\tau_n=(\tau^{({\rm a})}_n+\tau^{({\rm b})}_n)/2$. The $\tau_n$
ratios of the ordered versus dressed method are depicted in the
respective last column of the 4- and 5-dimensional case.}
\end{center}
\end{table}

\begin{table}[t!]
\begin{center}\small
\begin{tabular}{|c||rr|c|rr|c||rr|c|rr|c||c|c|}\hline
    & \multicolumn{6}{c||}{\bf 4D-case} & \multicolumn{6}{c||}{\bf 5D-case}
    & \multicolumn{2}{c|}{\bf 4D/5D}\\\cline{2-15}
$n$ & \multicolumn{3}{c|}{ordr} & \multicolumn{3}{c||}{drss}
    & \multicolumn{3}{c|}{ordr} & \multicolumn{3}{c||}{drss}
    & ordr & drss\\\cline{2-13}
    & $\tau^{({\rm a})}_n$ & $\tau^{({\rm b})}_n$ & $r_n$
    & $\tau^{({\rm a})}_n$ & $\tau^{({\rm b})}_n$ & $r_n$
    & $\tau^{({\rm a})}_n$ & $\tau^{({\rm b})}_n$ & $r_n$
    & $\tau^{({\rm a})}_n$ & $\tau^{({\rm b})}_n$ & $r_n$
    &&\\\hline
4   & 0.030 & 0.030 &      & 0.069 & 0.070 &
    & 0.059 & 0.059 &      & 0.156 & 0.157 &      & 0.51 & 0.44\\
5   & 0.180 & 0.179 & 5.98 & 0.418 & 0.413 & 5.98
    & 0.464 & 0.465 & 7.87 & 1.150 & 1.148 & 7.34 & 0.39 & 0.36\\
6   & 1.384 & 1.383 & 7.71 & 2.419 & 2.410 & 5.81
    & 4.370 & 4.340 & 9.38 & 8.036 & 7.996 & 6.98 & 0.32 & 0.30\\
7   & 13.53 & 13.52 & 9.78 & 14.64 & 14.65 & 6.07
    & 46.65 & 46.40 & 10.7 & 56.06 & 55.99 & 6.99 & 0.29 & 0.26\\
8   & 147.2 & 147.5 & 10.9 & 90.48 & 91.60 & 6.22
    & 550.9 & 549.5 & 11.8 & 395.2 & 391.9 & 7.02 & 0.27 & 0.23\\
9   & 1766  & 1764  & 12.0 & 585.9 & 585.0 & 6.43
    & 7013  & 7029  & 12.8 & 2844  & 2845  & 7.23 & 0.25 & 0.21\\
10  & 23100 & 22830 & 13.0 & 4233  & 4208  & 7.21
    & 98760 & 98360 & 14.0 & 24220 & 24410 & 8.55 & 0.23 & 0.17\\\hline\hline
\end{tabular}
\caption{\label{Tab:timepsps2}
Computer times $\tau_n$ in seconds for the same settings as used in
Table~\ref{Tab:timepsps}, this time using a 2.66 GHz Intel Core2 Quad
processor. The rightmost part of the table depicts the ratios of
4- versus 5-dimensional computer times for both approaches.}
\end{center}
\end{table}

The scaling of the computer time can roughly be estimated by
$(f\times C_{\rm max})^n$. The constants $C_{\rm max}=5\,(4)$ and
$1<f\le4$ express the fact that the number of pentagon (box) cuts and
the exponential scaling with $n$\/ of the tree-level color-dressed
recursion relation respectively govern the asymptotic scaling behavior
of the unordered algorithm. Although one naively expects $f=4$, this
factor is reduced by the efficient re-use of gluon currents between different cuts. The
$C_{\rm max}^n$ growth of the number of cuts reflects the large-$n$\/
limit of the Stirling number ${\cal S}_2(n,C_{\rm max})$. We show four
tables summarizing our results for the computation times $\tau_n$ of
obtaining ${\cal M}^{(1)}({\bf g}_1,\ldots,{\bf g}_n)=
{\cal M}_n^{(1)}(\lambda_k,(ij)_k)$ by using two independent solutions
of the unitarity constraints. The time for the re-computation has been
included in $\tau_n$. In real applications such a consistency
check will become unnecessary, thereby halving the evaluation time per
phase-space point.
Table~\ref{Tab:timepsps} lists the times obtained by running the 4-
and 5-dimensional algorithms for the calculation of two random
phase-space points labelled ``$\rm a$'' and ``$\rm b$''. The $n$\/
gluons have colors $(ij)_k=(ab)_k$ and alternating polarizations
$\lambda_k=\kappa_k\equiv+-\ldots+-(+)$. Owing to the absence of
pentagon cuts we find that the ``4D-case'' calculations are faster.
More importantly, the computation time does not vary when the
$n$-gluon kinematics changes. Hence, we can calculate the ratios
$r_n=\tau_n/\tau_{n-1}$ by defining
$\tau_n=(\tau^{({\rm a})}_n+\tau^{({\rm b})}_n)/2$ and show these
ratios in the table. While for the dressed algorithm these ratios
are almost stable, they are larger and increase gradually for the
method based on ordered amplitudes. This reflects the $(n-2)!$
factorial growth of the number of non-vanishing orderings of the color
configuration $(ab)_k$ as given in Table~\ref{Tab:ordrcuts}. For the 
dressed approach, we find constant ratios of $r_n\approx6$ and
$r_n\approx7$ in the ``4D-case'' and ``5D-case'', respectively. This
manifestly confirms our expectation of exponential scaling. The
difference between the 4- and 5-dimensional ratios obviously arises
because of the absence of pentagon cuts in the ``4D-case''.
The $r_{10}$ ratios do not fit the
constant trend. We cannot exclude though that this is a consequence of
the occurrence of large structures of maps to store the vast number of
color-dressed coefficients. The increasing number of
higher-cut subtractions terms may also cause deviations from the expected
scaling, which we derived from our simple arguments stated above.
Also, the conceptually easier way of storing all coefficients and
calculating the largest-$m$ cuts first is by far not the most economic
in terms of memory consumption.\footnote{It is for this reason that
  our calculations are currently limited to $n=12$ in the ``4D-case''
  and $n=10$ in the ``5D-case''.}
For small $n$, the lower complexity of the ordered recurrence relation
facilitates a faster calculation of the virtual corrections through
ordered amplitudes. The turnaround appears for $7<n<8$ and is just
slightly above $n=7$ for the ``4D-case''. With $n\ge8$ the dressed
method becomes superior owing to the different growth characteristics
of the two approaches. This is neatly expressed by the ``ordr/drss''
ratios given in Table~\ref{Tab:timepsps}.

We have cross-checked the measured computation times in a different
processor environment using exactly the same settings. The results are
shown in Table~\ref{Tab:timepsps2} and consistent with those of
Table~\ref{Tab:timepsps}. Instead of the ``ordr/drss'' ratios, here we
list ratios comparing the 4- and 5-dimensional computation for both
approaches. They stress the relative importance of the pentagon-cut
evaluations, which start to dominate the full calculation when $n$
gets large.


\begin{table}[t!]
\begin{center}\small
\begin{tabular}{|c||rr|c|rr|c|c||rr|c|rr|c|c|}\hline
    & \multicolumn{7}{c||}{\bf 4D-case}
    & \multicolumn{7}{c|}{\bf 5D-case}\\\cline{2-15}
$n$ & \multicolumn{3}{c|}{ordr} & \multicolumn{3}{c|}{drss}
    & \raisebox{-.4mm}{$\underline{\rm ordr}$}
    & \multicolumn{3}{c|}{ordr} & \multicolumn{3}{c|}{drss}
    & \raisebox{-.4mm}{$\underline{\rm ordr}$}\\\cline{2-7}\cline{9-14}
    & $\tau^{(\sigma_k)}_n$ & $\tau^{(\kappa_k)}_n$ & $r_n$
    & $\tau^{(\sigma_k)}_n$ & $\tau^{(\kappa_k)}_n$ & $r_n$
    & \raisebox{.7mm}{drss}
    & $\tau^{(\sigma_k)}_n$ & $\tau^{(\kappa_k)}_n$ & $r_n$
    & $\tau^{(\sigma_k)}_n$ & $\tau^{(\kappa_k)}_n$ & $r_n$
    & \raisebox{.7mm}{drss}\\\hline
4   & 0.049 & 0.045 &      & 0.074 & 0.076 &      & 0.63
    & 0.088 & 0.085 &      & 0.153 & 0.155 &      & 0.56\\
5   & 0.186 & 0.185 & 3.95 & 0.364 & 0.364 & 4.85 & 0.51
    & 0.479 & 0.483 & 5.56 & 1.000 & 1.000 & 6.49 & 0.48\\
6   & 1.186 & 1.182 & 6.38 & 2.071 & 2.068 & 5.69 & 0.57
    & 3.629 & 3.586 & 7.50 & 6.805 & 6.752 & 6.78 & 0.53\\
7   & 4.185 & 4.277 & 3.57 & 11.82 & 11.77 & 5.70 & 0.36
    & 14.02 & 13.95 & 3.88 & 44.42 & 44.46 & 6.56 & 0.31\\
8   & 27.12 & 26.96 & 6.39 & 70.34 & 71.10 & 6.00 & 0.38
    & 98.52 & 99.13 & 7.07 & 294.8 & 297.8 & 6.67 & 0.33\\
9   & 245.0 & 242.9 & 9.02 & 443.8 & 445.5 & 6.29 & 0.55
    & 960.3 & 954.8 & 9.69 & 2080  & 2070  & 7.00 & 0.46\\
10  & 1442  & 1446  & 5.92 & 3265  & 3270  & 7.35 & 0.44
    & 5943  & 5968  & 6.22 & 18610 & 18480 & 8.94 & 0.32\\
11  &       &       &      & 28670 & 28690 & 8.78 &&&&&&&&\\\hline
6   &       &       &      & {\sl 2.044}&  & {\sl 5.62}&&&&&&&&\\
7   &       &       &      & {\sl 11.66}&  & {\sl 5.70}&&&&&&&&\\
8   &       &       &      & {\sl 68.85}&  & {\sl 5.90}&&&&&&&&\\
9   &       &       &      & {\sl 420.4}&  & {\sl 6.11}&&&&&&&&\\
10  &       &       &      & {\sl 2972} &  & {\sl 7.07}&&&&&&&&\\
11  &       &       &      & {\sl 26310}&  & {\sl 8.85}&&&&&&&&\\
12  &       &       &      & & {\sl 292000}& {\sl 11.1}&&&&&&&&\\\hline\hline
\end{tabular}
\caption{\label{Tab:timehels}
Computer times $\tau_n$ in seconds obtained for the color-ordered and
color-dressed evaluation of $n$-gluon virtual corrections in 4 and 5
dimensions using a 3.00 GHz Intel Core2 Duo processor. Results are
shown for two different polarization choices $\sigma_k$ and
$\kappa_k$. The virtual corrections were computed at the same random
phase-space point with the $n$-gluon colors set to $(cd)_k$. The
choices are specified in the text. Ratios $r_n=\tau_n/\tau_{n-1}$ are
given where $\tau_n=(\tau^{(\sigma_k)}_n+\tau^{(\kappa_k)}_n)/2$ is
the time to evaluate the correction for $n$\/ gluons two times. The
re-computation is used to check both solutions for their consistency.
The $\tau_n$ ratios of the ordered versus dressed method are depicted
in the respective last column of the 4- and 5-dimensional case.}
\end{center}
\end{table}

\begin{table}[t!]
\begin{center}\small
\begin{tabular}{|c||r|c|r|c|c||r|c|r|c|c||c|c|}\hline
    & \multicolumn{5}{c||}{\bf 4D-case}
    & \multicolumn{5}{c||}{\bf 5D-case}
    & \multicolumn{2}{c|}{\bf 4D/5D}\\\cline{2-13}
$n$ & \multicolumn{2}{c|}{ordr} & \multicolumn{2}{c|}{drss}
    & \raisebox{-.4mm}{$\underline{\rm ordr}$}
    & \multicolumn{2}{c|}{ordr} & \multicolumn{2}{c|}{drss}
    & \raisebox{-.4mm}{$\underline{\rm ordr}$}
    & ordr & drss\\\cline{2-5}\cline{7-10}
    & $\tau_n$ & $r_n$ & $\tau_n$ & $r_n$ & \raisebox{.7mm}{drss}
    & $\tau_n$ & $r_n$ & $\tau_n$ & $r_n$ & \raisebox{.7mm}{drss} &&\\\hline
4   & 0.026 &      & 0.062 &      & 0.42
    & 0.065 &      & 0.151 &      & 0.43 & 0.40 & 0.41\\
5   & 0.222 & 8.54 & 0.394 & 6.35 & 0.56
    & 0.615 & 9.46 & 1.139 & 7.54 & 0.54 & 0.36 & 0.35\\
6   & 1.863 & 8.39 & 2.378 & 6.04 & 0.78
    & 5.544 & 9.01 & 7.970 & 7.00 & 0.70 & 0.33 & 0.30\\
7   & 15.06 & 8.08 & 14.58 & 6.13 & 1.03
    & 50.41 & 9.09 & 56.94 & 7.14 & 0.89 & 0.30 & 0.26\\
8   & 129.2 & 8.58 & 93.09 & 6.38 & 1.39
    & 476.7 & 9.46 & 401.5 & 7.05 & 1.19 & 0.27 & 0.23\\
9   & 1127  & 8.72 & 603.6 & 6.48 & 1.87
    & 4483  & 9.40 & 2800  & 6.97 & 1.60 & 0.25 & 0.22\\
10  & 10980 & 9.74 & 3961  & 6.56 & 2.77
    & 50260 & 11.2 & 25140 & 8.98 & 2.00 & 0.22 & 0.16\\\hline\hline
\end{tabular}
\caption{\label{Tab:timerndm}
Color-configuration averaged computation times $\tau_n$ in seconds obtained from the
4- and 5- dimensional color-ordered and color-dressed evaluations of
$n$-gluon virtual corrections using 2.66 GHz Intel Core2 Quad
processors. Results are shown for random phase- and color-space points
and alternating gluon polarizations $\lambda_k=\kappa_k$, see text.
The respective growth factors $r_n=\tau_n/\tau_{n-1}$ are given where
$\tau_n$ denotes the time that is needed to calculate the $n$-gluon
one-loop amplitude two times. The re-computation is used to check the
two solutions for their consistency. Several time ratios are formed to
compare the ordered with the dressed method and the 4- with the
5-dimensional computation. These ratios are displayed in the columns
indicated accordingly.}
\end{center}
\end{table}

In Table~\ref{Tab:timehels} we detail computation times when varying
the polarizations of the $n$\/ gluons while keeping their color
configuration fixed. We have chosen the two settings
$\lambda_k=\sigma_k\equiv++-\ldots-$ and, as before,
$\lambda_k=\kappa_k$. In terms of colors we consider the
computationally less involved point $(ij)_k=(cd)_k$. Both amplitudes
are calculated at the same random phase-space point ``$\rm c$''
dissimilar from previous points ``${\rm a}$'' and ``${\rm b}$''. For
none of our four calculational options, we notice manifest deviations
between the times $\tau^{(\lambda_k)}_n$ associated with the two
polarization settings. When inspecting the ``ordr/drss'' ratios, we
observe that the ordered approach is advantageous in cases where only
a few orderings contribute to the result of a certain point in color
space. The fluctuations seen in the growth factors mirror the unsteady
increase with $n$\/ in non-zero orderings depicted in the last column
of Table~\ref{Tab:ordrcuts}. For the dressed approach, we get similar,
though somewhat smaller, growth factors compared to the previous test.
In order to validate the dressed algorithm up to $n=12$ external
gluons, we introduced a few more optimizations specific to the
4-dimensional calculations.\footnote{Some parts of the algorithm can
  be speed up when pentagon cuts are completely avoided.}
The lower part of Table~\ref{Tab:timehels} shows the computer times,
which we obtained after optimization. They are consistent with our
previous findings. As mentioned before, $r_{n\ge10}>6$ likely occur
for reasons of increasingly complex higher-cut subtractions and
computer limitations in dealing with large memory structures.

\begin{table}[t!]
\begin{center}\small
\begin{tabular}{|l|c|c||c|c||c|c|}\hline\hline
\rule[-3mm]{0mm}{8mm} configuration:
& \multicolumn{2}{c||}{hard colors $(ab)_k$}
& \multicolumn{2}{c||}{simple colors $(cd)_k$}
& \multicolumn{2}{c|}{random non-zero colors}\\\hline
\rule[-2mm]{0mm}{6mm} fit values:
& $a/10^{-6}$sec & $b$ & $a/10^{-6}$sec & $b$ & $a/10^{-6}$sec & $b$\\\hline
\rule[-1.5mm]{0mm}{5mm} 4D, ordr
& 1.91 & $9.75\ ^{+0.59}_{-0.56}$
& 34.5 & $5.65\ ^{+0.32}_{-0.30}$
& 4.67 & $8.57\ ^{+0.10}_{-0.09}$\\
\rule[-1.5mm]{0mm}{5mm} 5D, ordr
& 2.66 & $10.99\ ^{+0.48}_{-0.46}$
& 45.6 & $6.39\ ^{+0.29}_{-0.28}$
& 7.84 & $9.46\ ^{+0.13}_{-0.12}$\\\hline
\rule[-1.5mm]{0mm}{5mm} 4D, drss
& 39.4 & $6.19\ ^{+0.09}_{-0.08}$
& 28.2 & $6.51\ ^{+0.29}_{-0.28}$
& 38.7 & $6.30\ ^{+0.04}_{-0.04}$\\
\rule[-1.5mm]{0mm}{5mm} 5D, drss
& 50.8 & $7.21\ ^{+0.10}_{-0.10}$
& 62.5 & $6.92\ ^{+0.08}_{-0.09}$
& 53.3 & $7.28\ ^{+0.11}_{-0.09}$\\\hline\hline
\end{tabular}
\caption{\label{Tab:timefits}
Parameter values $a$\/ and $b$\/ obtained from curve fitting of the
computation times $\tau_n$ to the functional form of $\tau_n=a\,b^n$.
The results are given for the three different $n$-gluon color
assignments used in Tables~\ref{Tab:timepsps} (hard),
\ref{Tab:timehels} (simple) and \ref{Tab:timerndm} (random) and for
all four algorithms the 4- and 5-dimensional color-ordered and
color-dressed algorithm.}
\end{center}
\end{table}

For the calculation of the virtual corrections, one might question
whether there exist enough points in color space that occur with many
trivial orderings. If so, the color-decomposition based method would
be more efficient on average. This is not the case for larger $n$\/ as
shown in Table~\ref{Tab:timerndm}. For gluon multiplicities of
$n=4,\ldots,10$ and polarizations set according to $\kappa_k$, we list
mean computation times, growth factors, ``ordr/drss'' and ``4D/5D''
ratios obtained for one-loop amplitude evaluations where the phase-
and color-space points have been chosen randomly. Following the method
outlined in Section~\ref{Sec:numtree}, we only considered non-zero
color configurations. We averaged over many events, for
$n=4,\ldots,10$ gluons, we used ${\cal O}(10^6),\ldots,{\cal O}(10^2)$
points. We observe that the pattern of the results in
Table~\ref{Tab:timerndm} resembles that found in
Tables~\ref{Tab:timepsps} and \ref{Tab:timepsps2} where we have
studied the more complicated color point $(ik)_k=(ab)_k$. The ratios
comparing the ordered and dressed approach are smaller with respect to
those of Tables~\ref{Tab:timepsps} and \ref{Tab:timepsps2}. This
signals that the mean number of contributing orderings is somewhat
lower than for the $(ab)_k$ case. We finally report dressed growth
factors that are consistent with our previous findings confirming the
approximate $6^n$ and $7^n$ growths in computational complexity of the
new method for the 4- and 5-dimensional case, respectively.

\begin{figure}[t!]
\vspace*{-4mm}
\centerline{
  \includegraphics[width=0.71\columnwidth,angle=-90]{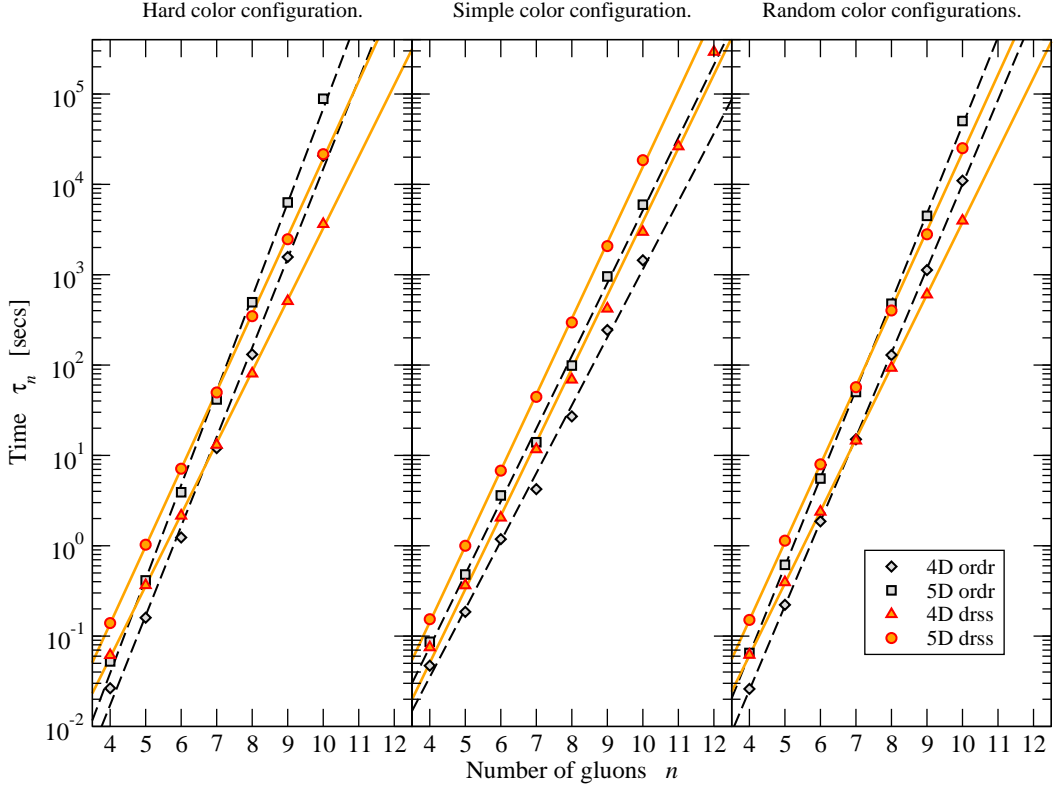}}
\vspace*{-4mm}
\caption{Computation times $\tau_n$ versus the number $n$\/
  of external gluons for the three different gluon color assignments
  used in Tables~\ref{Tab:timepsps} (hard), \ref{Tab:timehels}
  (simple) and \ref{Tab:timerndm} (random). The results reported in
  these tables are shown for the 4- and 5-dimensional color-ordered
  and -dressed algorithms. The solid and dashed curves each represent
  the outcomes of the fits listed in Table~\ref{Tab:timefits} for both
  the dressed and ordered approach, respectively.}
\label{Fig:time}
\end{figure}

Using the results of Tables~\ref{Tab:timepsps}, \ref{Tab:timehels} and
\ref{Tab:timerndm} we have performed fits to the functional form
$\tau_n=a\,b^n$. We show the outcome of the curve fittings in
Table~\ref{Tab:timefits}. Recall that the computation times have been
obtained by using different color assignments for the $n$\/ gluons.
Tables~\ref{Tab:timepsps} and \ref{Tab:timehels} present results where
we have chosen $(ij)_k=(ab)_k$ and $(ij)_k=(cd)_k$ as examples of hard
and simple color configurations, respectively. We have averaged over
non-zero color settings to find the results of
Table~\ref{Tab:timerndm}. Considering the performance of the dressed
algorithm, we conclude that these data are in agreement with
exponential growth for all color assignments. The errors on the fit
parameter $b$\/ are relatively small, only the 4-dimensional case of
simple colors is somewhat worse because we included results up to
$n=12$ where parts of the computation become less efficient as
explained above. The hard- and simple-colors case of the ordered
approach show rather large errors for the $b$-parameter signalling
that the genuine scaling law is not of an exponential kind in both
cases. Interestingly, one observes an effective exponential scaling
when averaging over many non-zero color configurations. The growth
described by the $b$-parameter is however a good two units stronger
for the ordered approach than the growth seen in the color-dressed
approach. To summarize, we have plotted in Fig.~\ref{Fig:time} all
computer times reported in Tables~\ref{Tab:timepsps},
\ref{Tab:timehels} and \ref{Tab:timerndm} as a function of the number
of external gluons in the range $4\le n\le12$. We have included in
these plots the curves $\tau_n=a\,b^n$, which we calculated from the
respective fit parameters stated in Table~\ref{Tab:timefits}.

\begin{figure}[t!]
\centerline{
  \includegraphics[width=0.64\columnwidth]{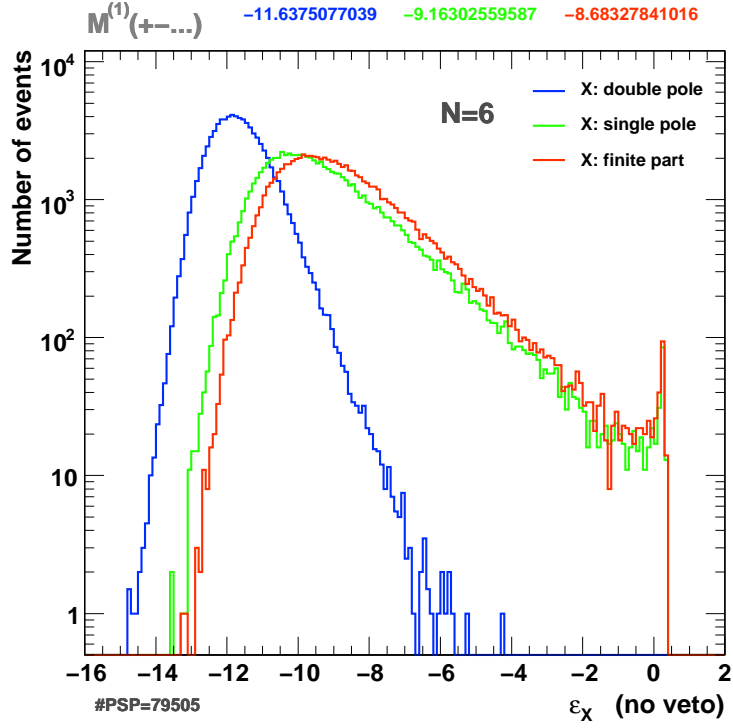}}
\caption{Relative accuracies of the $1/\epsilon^{2,1,0}$ poles of
  $n=6$ gluon one-loop amplitudes as determined by the
  double-precision color-dressed algorithm. The gluon polarizations
  are given by $\lambda_k=+-+-+-$, colors were chosen randomly among
  non-zero configurations. Vetoed events are included, only those
  with unstable ortho-vectors are left out, see text for more
  explanations. The mean accuracies and the number of randomly picked
  phase-space points are displayed in the top row and bottom left
  corner of the plot, respectively.}
\label{Fig:zvaccs6}
\end{figure}

\begin{figure}[p!]
\centerline{
  \includegraphics[width=0.51\columnwidth]{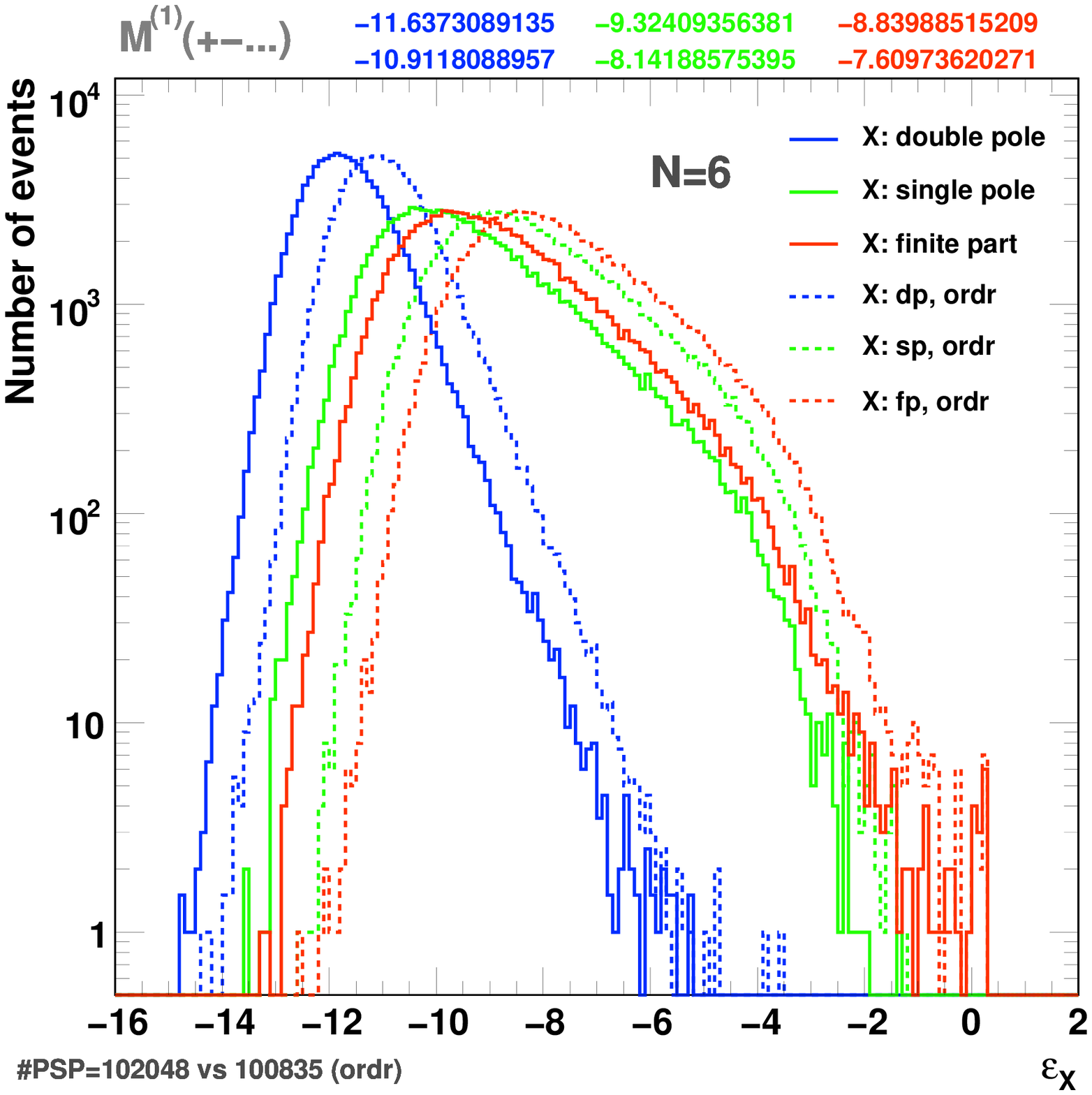}
  \includegraphics[width=0.51\columnwidth]{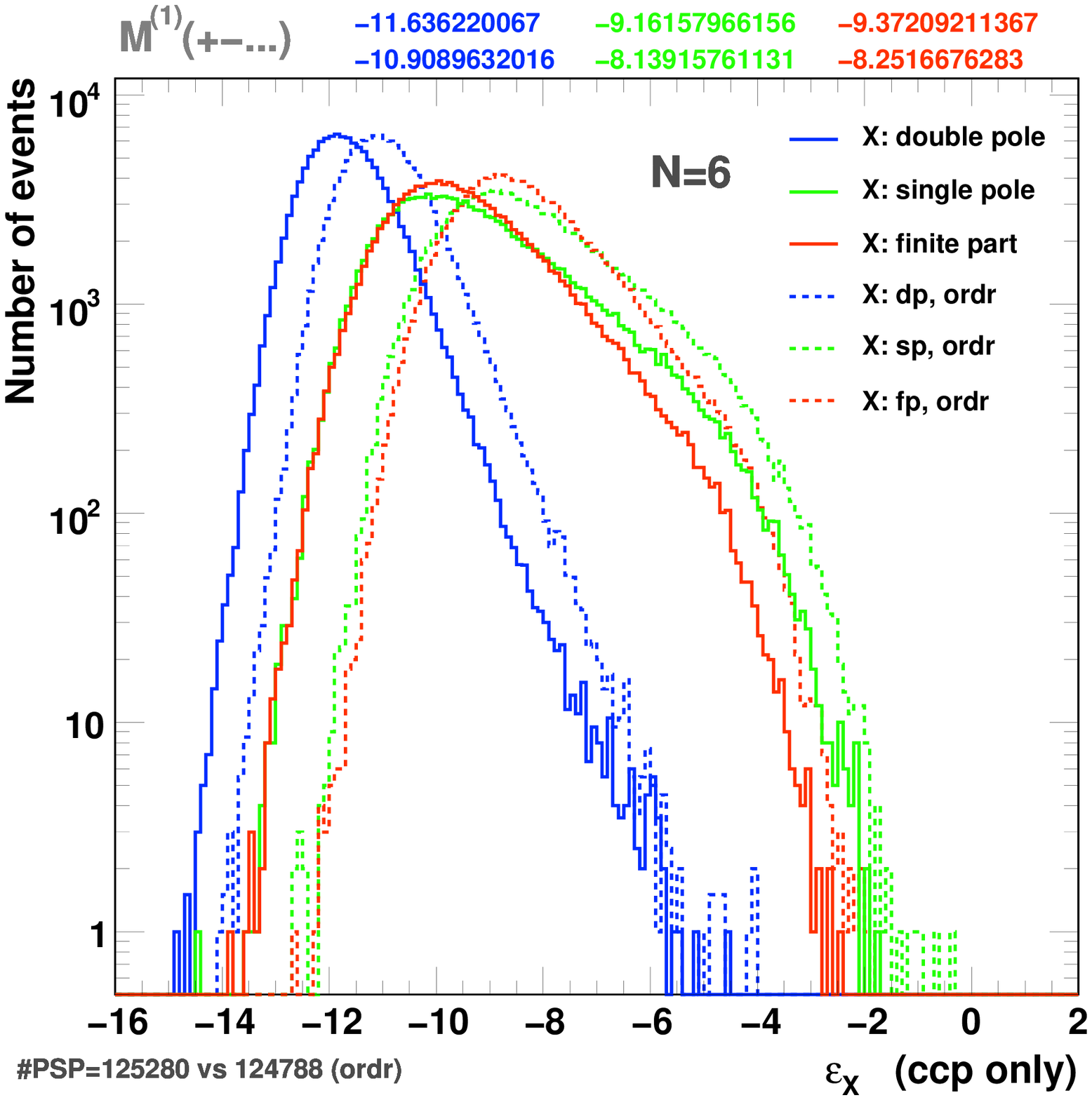}}
\centerline{
  \includegraphics[width=0.39\columnwidth]{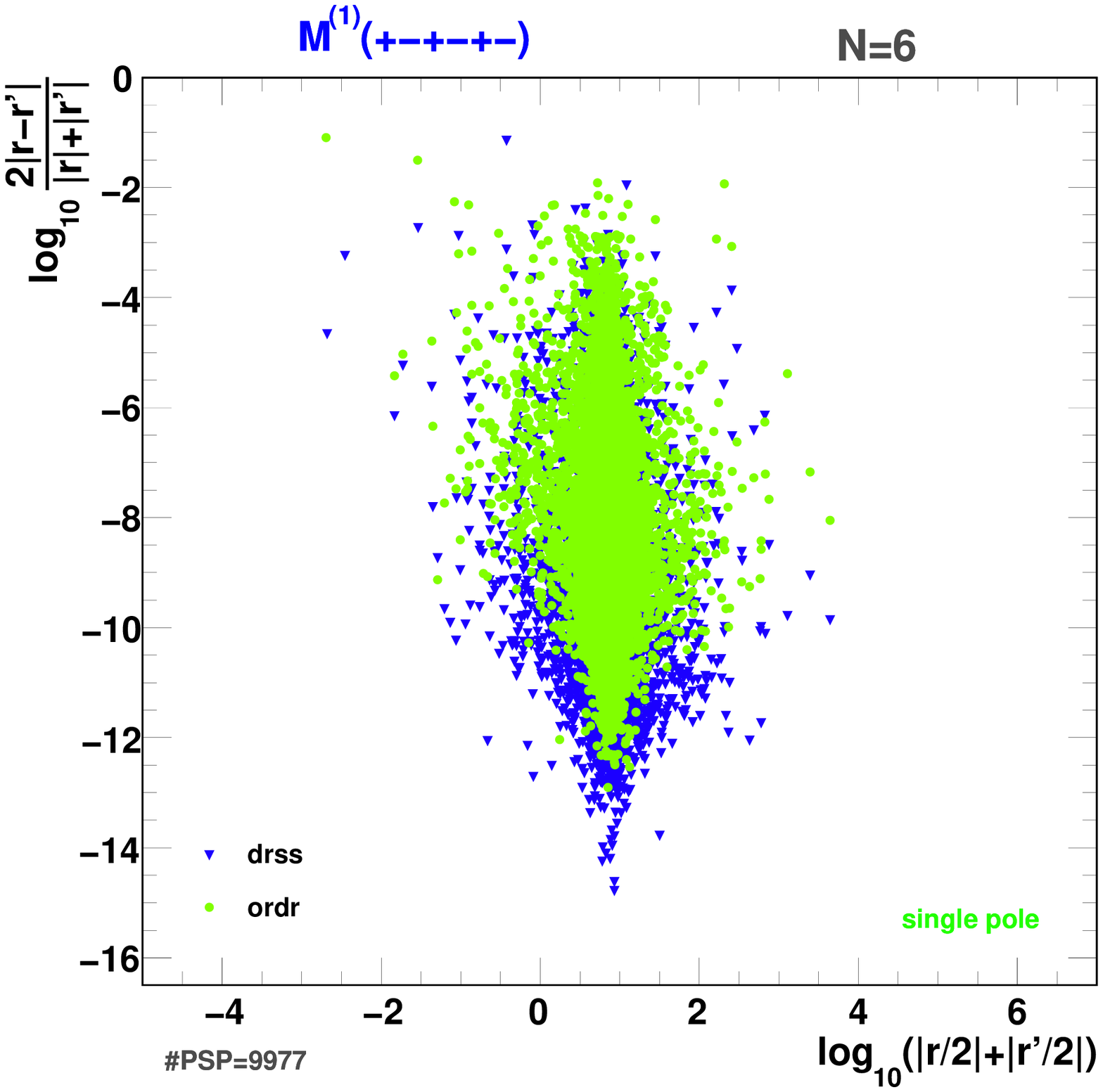}
  \includegraphics[width=0.39\columnwidth]{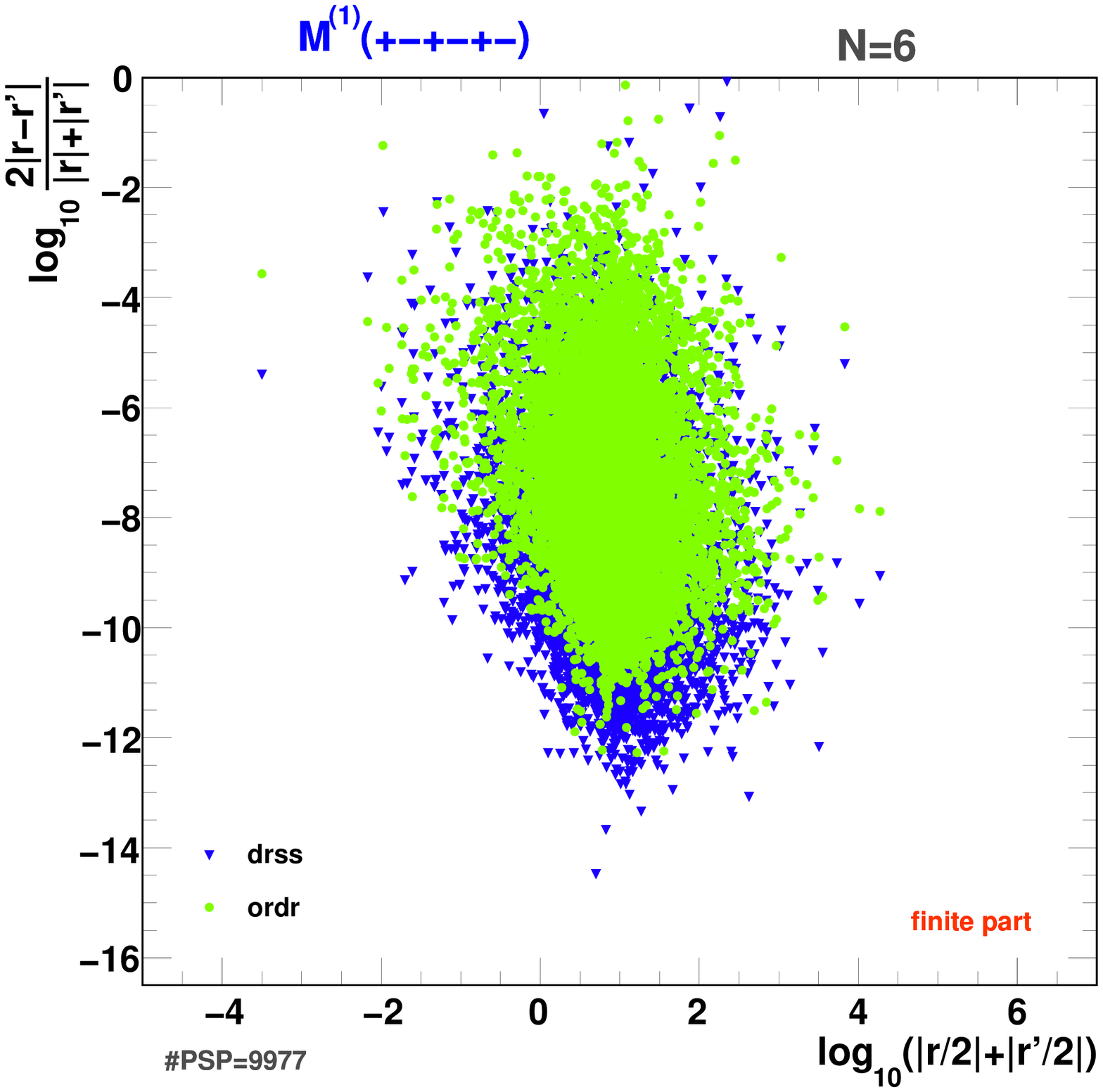}}
\centerline{
  \includegraphics[width=0.39\columnwidth]{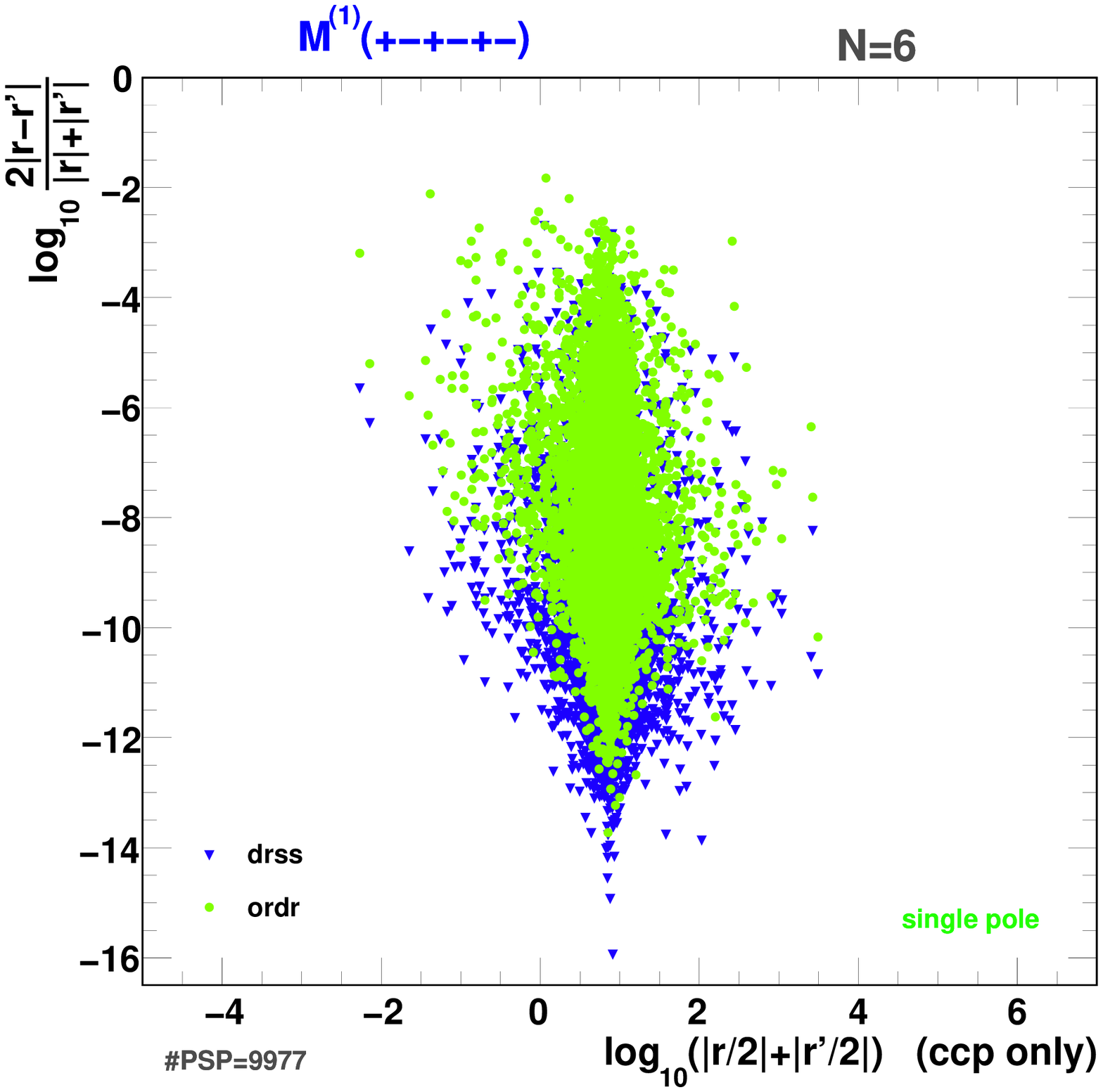}
  \includegraphics[width=0.39\columnwidth]{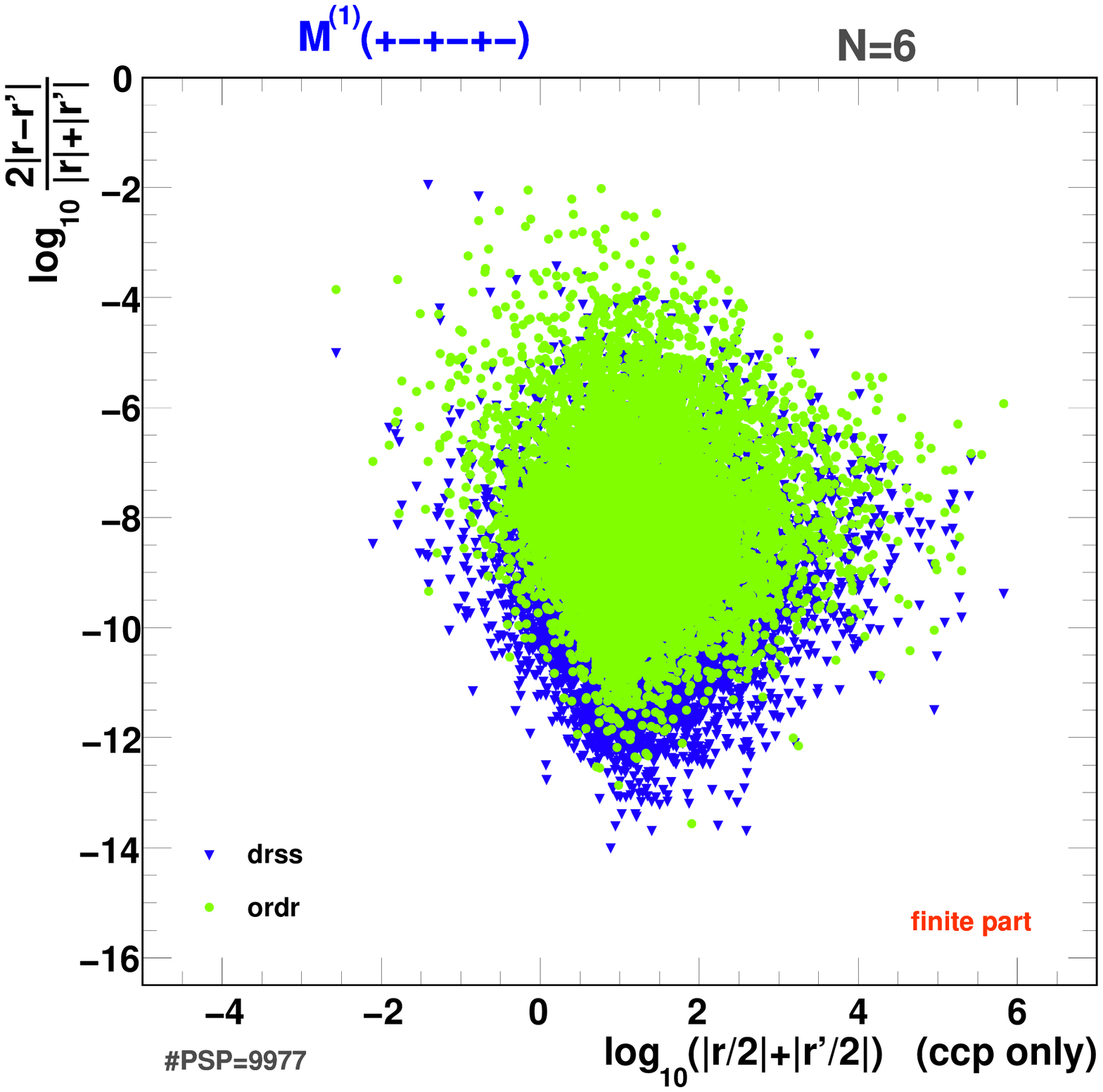}}
\caption{Double-, single-pole and finite-part accuracy distributions
  (upper part) and scatter graphs (lower part) extracted from
  double-precision computations of one-loop amplitudes for $n={\tt N}=6$
  gluons with polarizations $\lambda_k=+-+-+-$ and randomly chosen
  non-zero color configurations. The virtual corrections were
  calculated at random phase-space points satisfying the cuts detailed
  in the text. Unstable solutions were vetoed. Results from the
  color-dressed algorithm are compared with those of the color-ordered
  method indicated by dashed curves and brighter dots in the plots.
  The 5(4)-dimensional case is shown in the top left (right) and
  center (bottom) part of the figure. The definitions of
  $\varepsilon$\/ and $r$\/ are given in the text. All scatter graphs
  contain $2\times10^4$ points.}
\label{Fig:accs6}
\end{figure}

In the following we will discuss the quality of the semi-numerical
evaluations of ${\cal M}_n^{(1)}$ amplitudes for both the
color-ordered and color-dressed approaches. To this end we analyze the
logarithmic relative deviations of the double pole, single pole and
finite part. Independent of the number $n$\/ of gluons, we define them
as follows:
\beq
\varepsilon_{\rm dp}\;=\;
\log_{10}\,\frac{|{\cal M}^{(1)[1]}_{\rm dp,num}-{\cal M}^{(1)}_{\rm dp,th}|}
                {|{\cal M}^{(1)}_{\rm dp,th}|}\,,\qquad
\varepsilon_{\rm s/fp}\;=\;
\log_{10}\,\frac{2\,|{\cal M}^{(1)[1]}_{\rm s/fp,num}-
                     {\cal M}^{(1)[2]}_{\rm s/fp,num}|}
                {|{\cal M}^{(1)[1]}_{\rm s/fp,num}|+
                 |{\cal M}^{(1)[2]}_{\rm s/fp,num}|}\ ,
\eeq
where the structure of the double-poles ${\cal M}^{(1)}_{\rm dp,th}$
is known analytically given by Eq.~(\ref{Eq:M1dp}). We use two
independent solutions denoted by $[1]$ and $[2]$ to test the accuracy
of the single poles and finite parts. All results reported here were
obtained by using double-precision computations. We have run all our
algorithms by choosing color configurations and phase-space points at
random. Colors are distributed according to the ``Non-Zero'' method
presented in Sec.~\ref{Sec:numtree}. The phase-space points are
accepted only if they obey the cuts, which we have specified at the
beginning of this subsection. The gluon polarizations are always
alternating set by $\lambda_k=\kappa_k$.
Figure~\ref{Fig:zvaccs6} shows the $\varepsilon$\/ distributions in
absolute normalization, which we obtain from the 5-dimensional
color-dressed calculation for the case of $n=6$ external gluons. The
number of points used to generate the plots is given in the bottom
left corner, the top rows display the means of the double-,
single-pole and finite-part distributions. Limited to double-precision
computations, we find that the numerical accuracy of our results for
${\cal M}_n^{(1)}$ is satisfying. With $\varepsilon$\/ peak positions
smaller than the respective mean values
$\langle\varepsilon_{\rm d/s/fp}\rangle<-8$, we are able to provide
sufficiently accurate solutions for almost all phase-space
configurations. There is however a certain fraction of events where
the single pole and finite part cannot be determined reliably. These
${\cal O}(100)$ events occur because in exceptional cases small
denominators, such as vanishing Gram determinants made of external
momenta, cannot be completely avoided by the generalized-unitarity
algorithms. We also see accumulation effects where larger numbers get
multiplied together while determining the subtraction of higher-cut
contributions. Owing to the limited range of double-precision
calculations, such effects can lead to insufficient cancellations of
intermediate large numbers that are supposed to cancel out
eventually.\footnote{More detailed explanations can be found in
  Ref.~\cite{Winter:2009kd}.}
The current implementation of the algorithm has no special
treatment for these
exceptional events. One either has to come up with a more
sophisticated method treating these points separately or increase the
precision with which the corrections are calculated. Both of which is
beyond the scope of this paper and we leave it at vetoing these
points. Yet, we need robust criteria that allow us to keep track of
the quality of our solutions: we first test the orthonormal basis
vectors that span the space complementary to the physical space
constructed from the external momenta associated with the particular
cut configuration under consideration. Failures in generating these
basis vectors always lead to the rejection of the event.%
\footnote{We test in particular whether the normalization of the
  orthonormal basis vectors deviates less than $10^{-12}$ units from
  one.}
In the example of Fig.~\ref{Fig:zvaccs6}, such events occurred with a
rate of $0.6\%$ and were not included in the plot. Secondly, and more
importantly, we test the reliability of solving the systems of
equations to determine the master-integral coefficients. To this end
we generate an extra 4-dimensional loop momentum during the evaluation
of the bubble coefficients establishing the cut-constructible part.
Inaccuracies in solving for triangle etc.\ coefficients will be also
detected, since at this level all higher-cut subtractions are
necessary to obtain the correct value of the bubble coefficients. We
use the extra loop momentum to individually re-solve for the
cut-constructible bubble coefficient and compare this solution with
the one obtained in first place. We veto the event, if the deviation
$\Delta_{\rm veto}$ in the complex plane of the two solutions exceeds
a certain amount. We fix the veto cut at $\Delta_{\rm veto}=0.02$ for
this publication. Having this cross-check at hand, we gain nice
control over the events populating the tail of
the accuracy distributions in Fig.~\ref{Fig:zvaccs6}. Applying the
veto, we arrive at the distributions presented in the top left plot of
Fig.~\ref{Fig:accs6} where the steeper tails clearly demonstrate the
effect of the veto. Certainly, both these shortcomings of imprecise
ortho-vectors and inaccurately solved coefficients can be lifted by
switching to higher precision whenever the respective double-precision
evaluations have not passed our criteria. Accordingly,
Table~\ref{Tab:vetoeffects} quantifies
the fractions of events, which are within the
scope of the color-dressed and color-ordered algorithms presented
here. Owing to the more complicated event structures, the fraction of
rejected events increases with $n$, where most of the events fail the
bubble-coefficient test. We observe that the loss of events is more
severe for the ordered algorithm.

\begin{table}[t!]
\begin{center}\small
\begin{tabular}{|c||l|l||l|l|}\hline\hline
\rule[-2mm]{0mm}{6mm} $n\;$ & 4D, ordr & 5D, ordr & 4D, drss & 5D, drss\\\hline
\rule[-1mm]{0mm}{4mm} 4$\;$ & 1.0   & 1.0   & 1.0   & 1.0\\
\rule[-1mm]{0mm}{4mm} 5$\;$ & 0.992 & 0.991 & 0.984 & 0.984 (0.999)\\
\rule[-1mm]{0mm}{4mm} 6$\;$ & 0.960 & 0.960 & 0.964 & 0.972 (0.994)\\
\rule[-1mm]{0mm}{4mm} 7$\;$ & 0.872 & 0.873 & 0.891 & 0.892 (0.982)\\
\rule[-1mm]{0mm}{4mm} 8$\;$ & 0.635 & 0.642 & 0.829 & 0.825 (0.953)\\
\rule[-1mm]{0mm}{4mm}
 9$\;$ & 0.182 (0.84) & 0.205 (0.81) & 0.532 (0.93) & 0.533 (0.903)\\
\rule[-1mm]{0mm}{4mm}
10$\;$ & 0.0\hphantom{00} (0.61) & 0.0\hphantom{00} (0.50)
       & 0.38\hphantom{0} (0.86) & 0.33\hphantom{0} (0.83)\\\hline\hline
\end{tabular}
\caption{\label{Tab:vetoeffects}
Fractions of $n$-gluon events that have a stable set of basis
vectors in orthogonal space and also pass the veto on inaccurate
master-integral bubble coefficients when using $\Delta_{\rm veto}=0.02$.
In brackets, fractions of $n$-gluon events that pass the test for
unstable ortho-vectors.}
\end{center}
\end{table}

\begin{figure}[p!]
\centerline{
  \includegraphics[width=0.51\columnwidth]{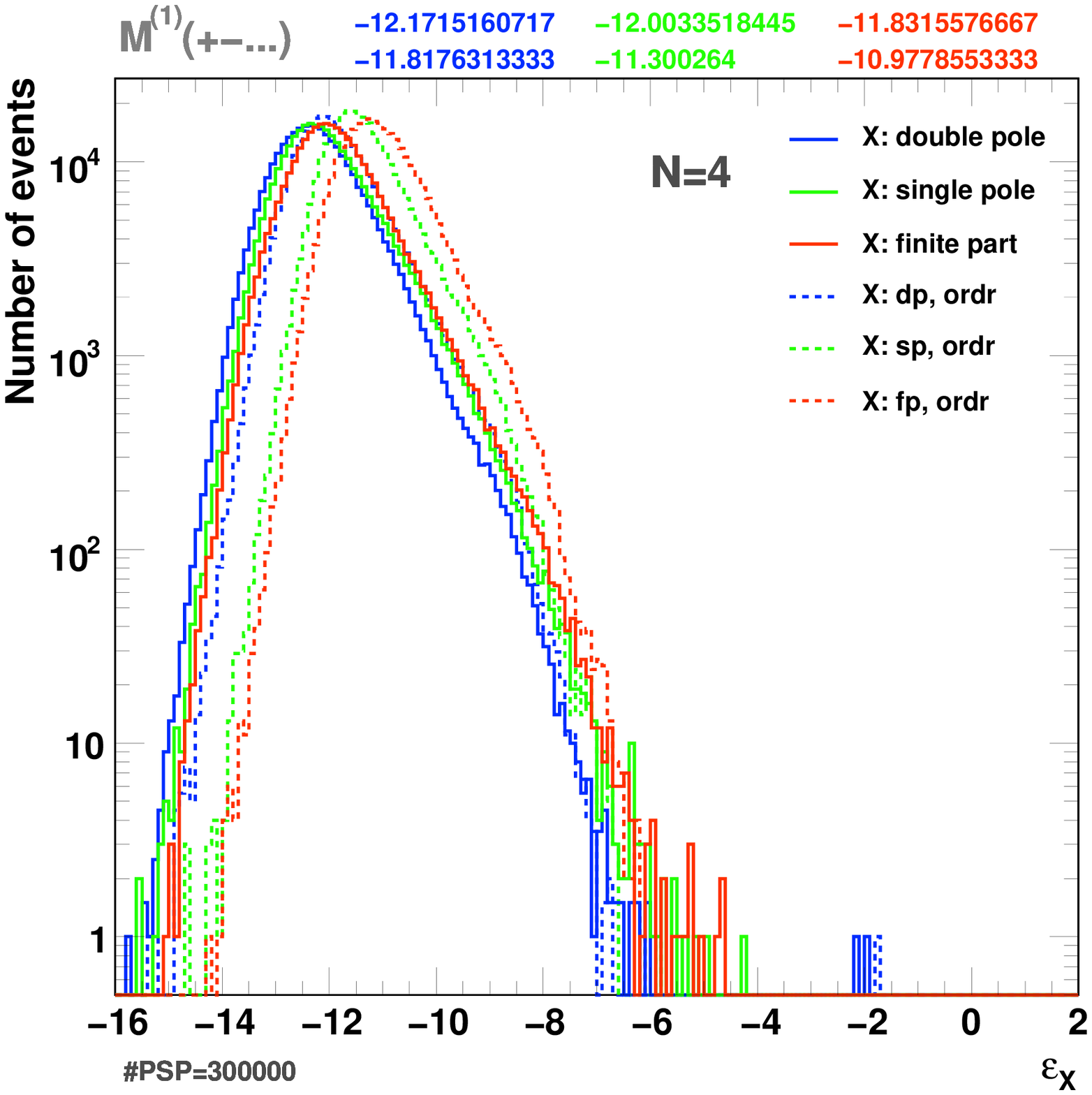}
  \includegraphics[width=0.51\columnwidth]{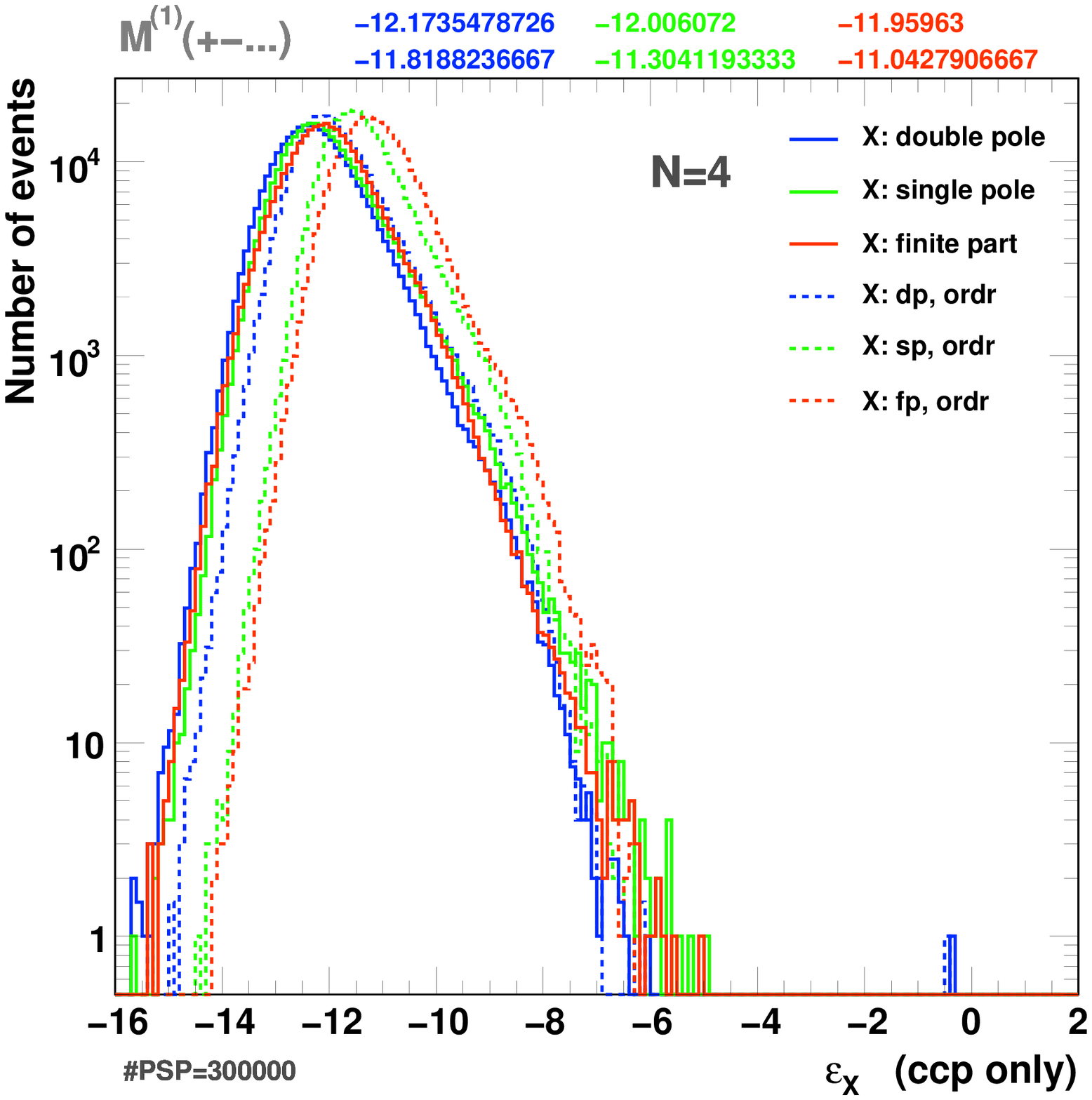}}
\centerline{
  \includegraphics[width=0.39\columnwidth]{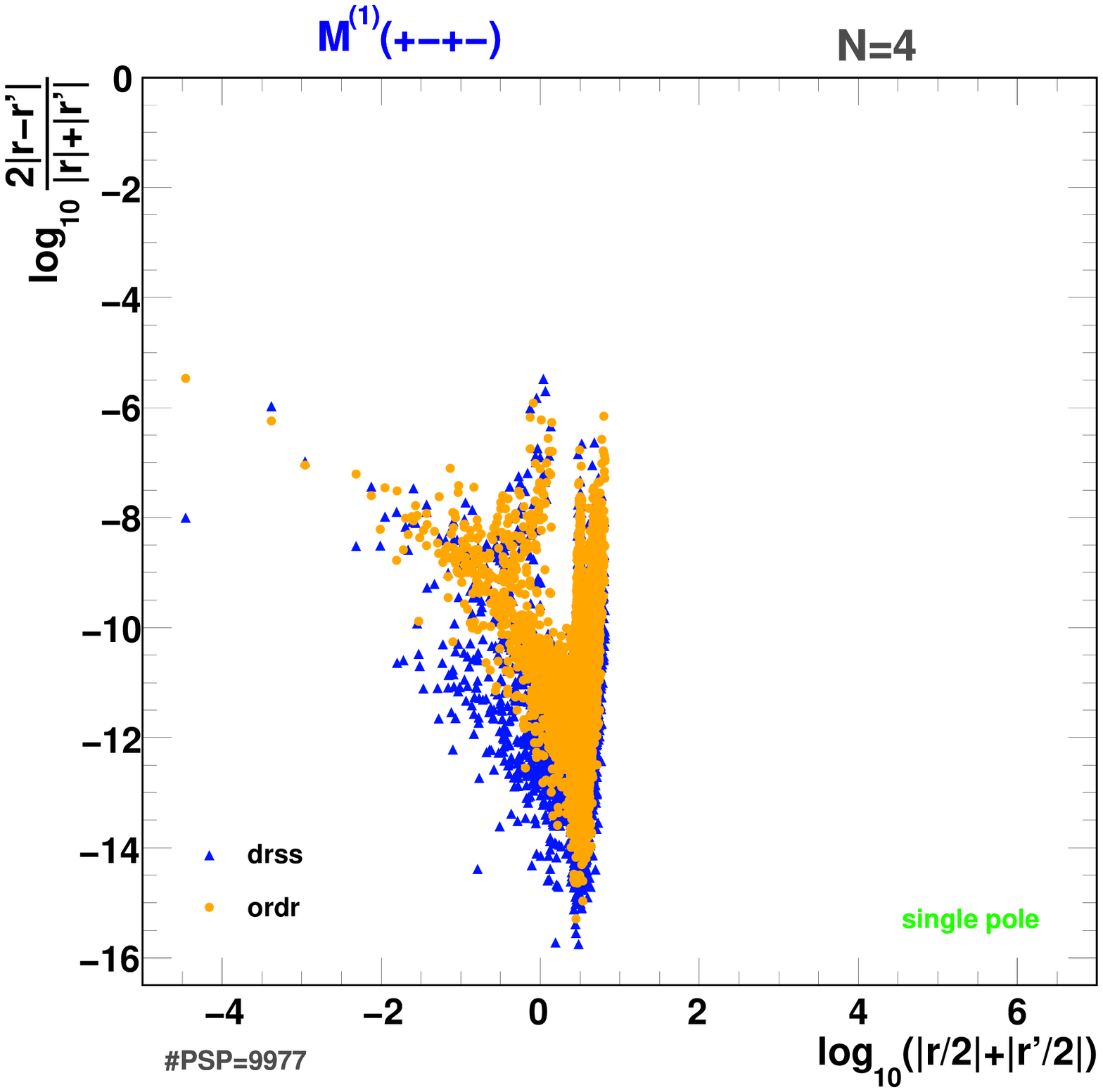}
  \includegraphics[width=0.39\columnwidth]{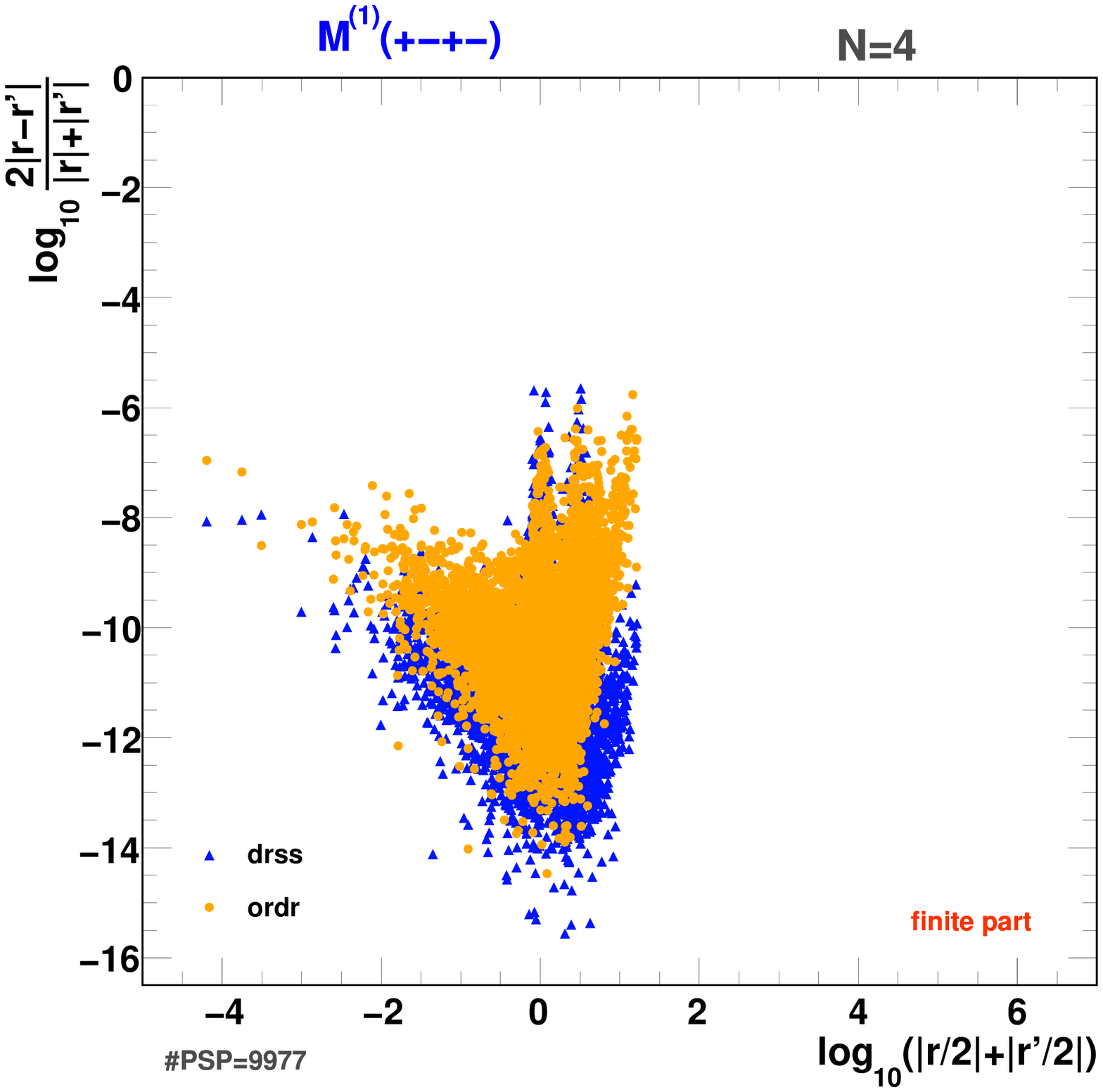}}
\centerline{
  \includegraphics[width=0.39\columnwidth]{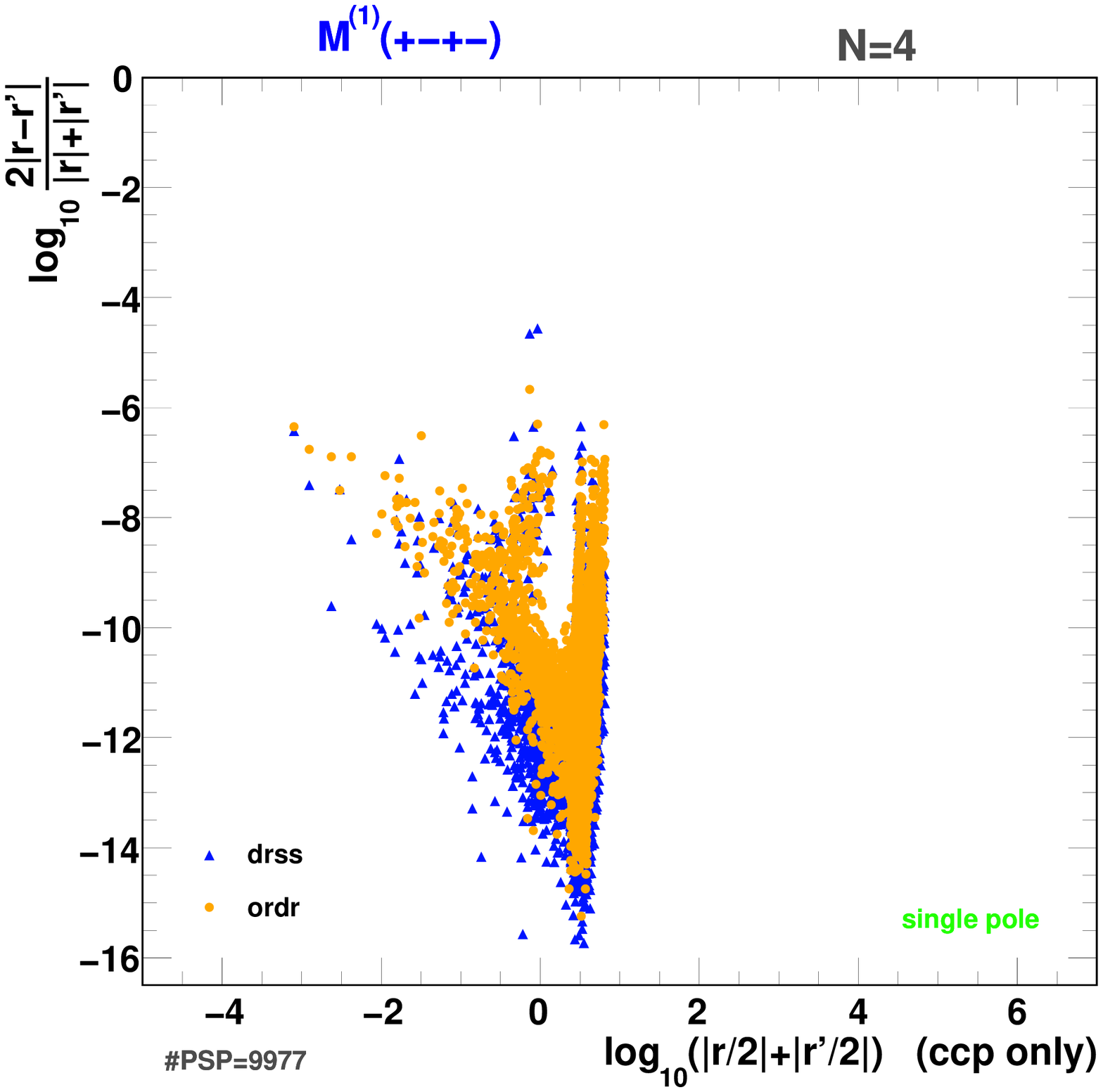}
  \includegraphics[width=0.39\columnwidth]{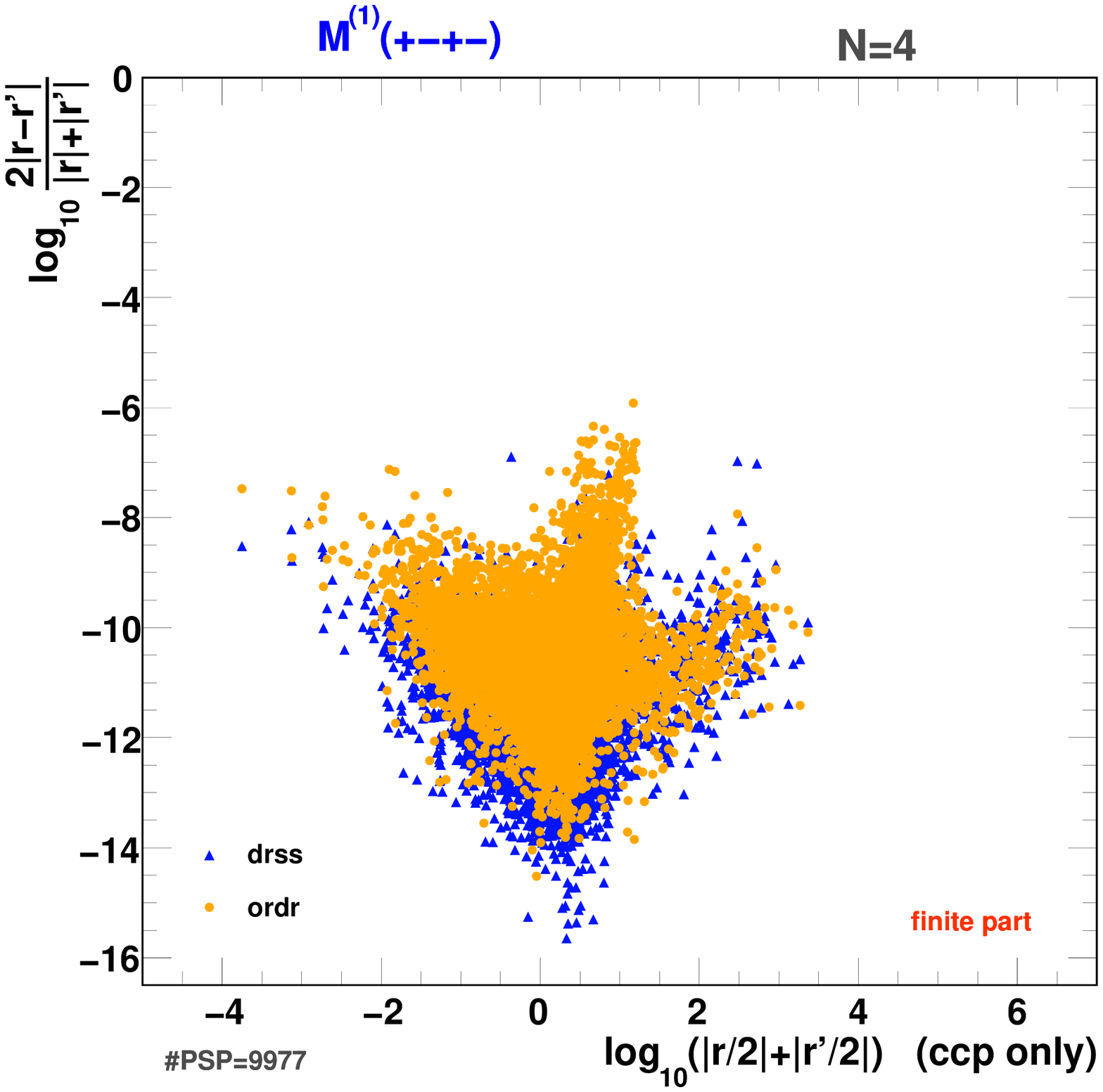}}
\caption{Double-, single-pole and finite-part accuracy distributions
  (upper part) and scatter graphs (lower part) extracted from
  double-precision computations of one-loop amplitudes for $n={\tt N}=4$
  gluons with polarizations $\lambda_k=+-+-$ and randomly chosen
  non-zero color configurations. The virtual corrections were
  calculated at random phase-space points satisfying the cuts detailed
  in the text. Unstable solutions were vetoed. Results from the
  color-dressed algorithm are compared with those of the color-ordered
  method indicated by dashed curves and brighter dots in the plots.
  The 5(4)-dimensional case is shown in the top left (right) and
  center (bottom) part of the figure. The definitions of
  $\varepsilon$\/ and $r$\/ are given in the text. All scatter graphs
  contain $2\times10^4$ points.}
\label{Fig:accs4}
\end{figure}

\begin{figure}[p!]
\centerline{
  \includegraphics[width=0.51\columnwidth]{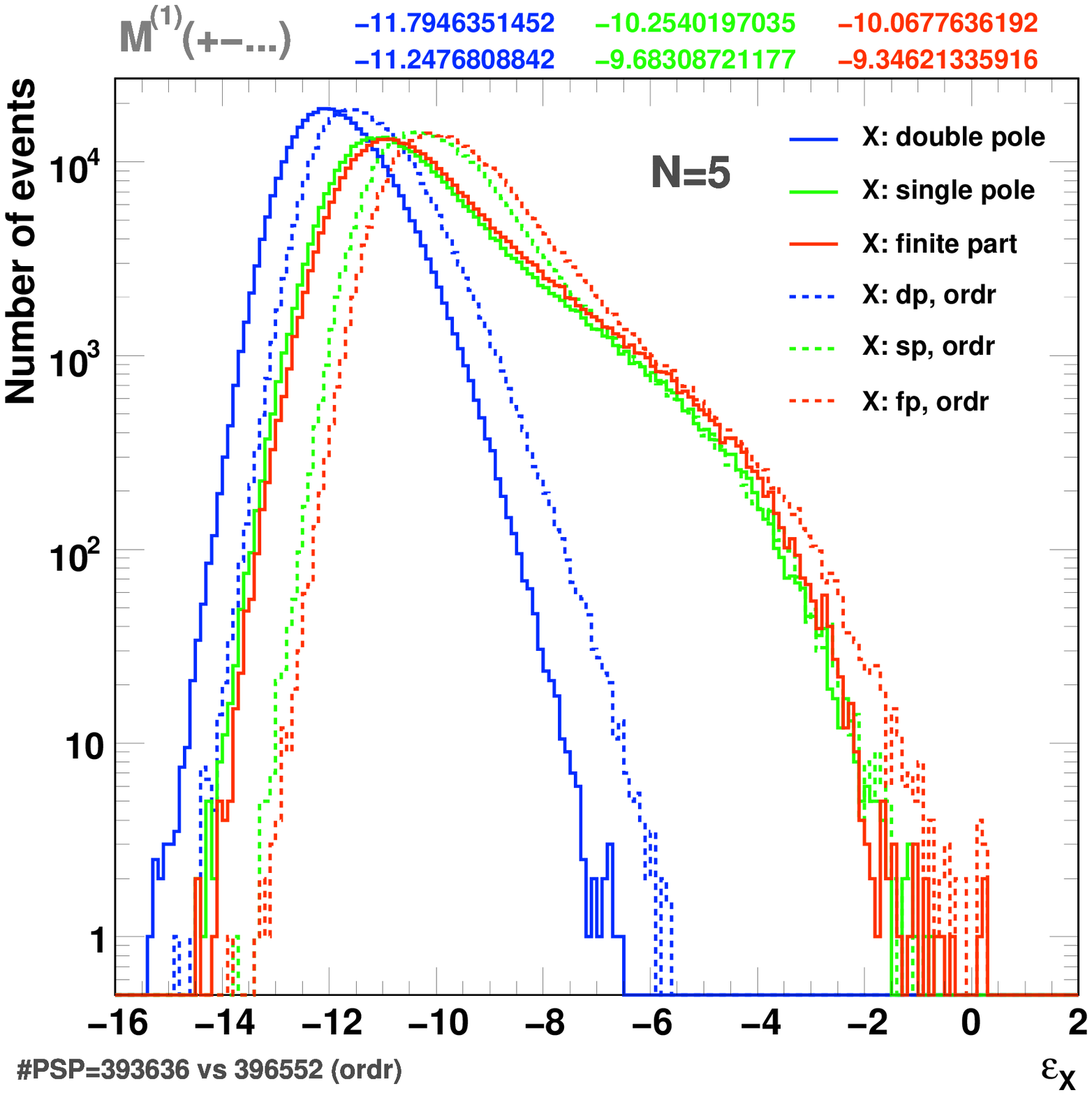}
  \includegraphics[width=0.51\columnwidth]{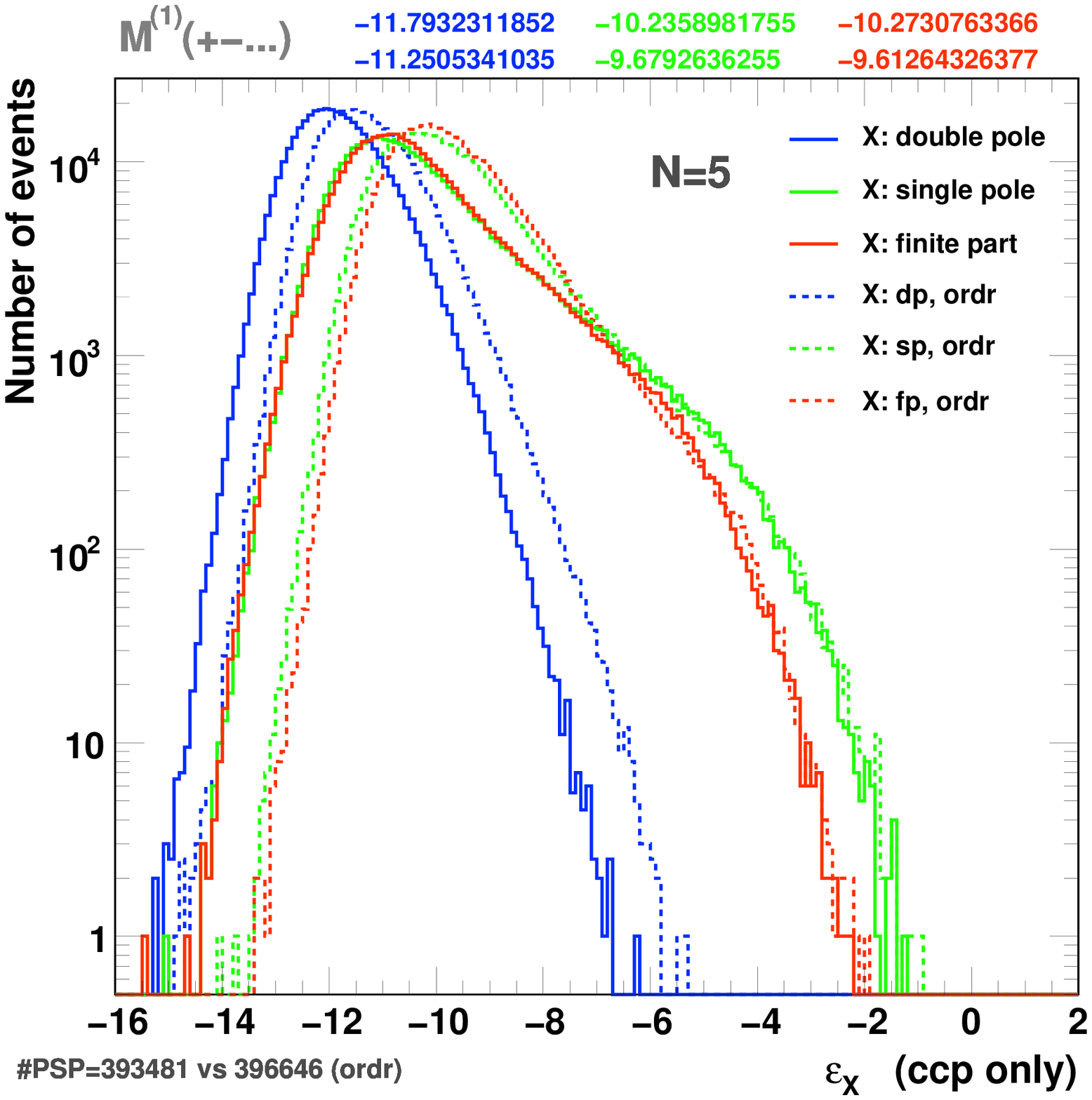}}
\centerline{
  \includegraphics[width=0.39\columnwidth]{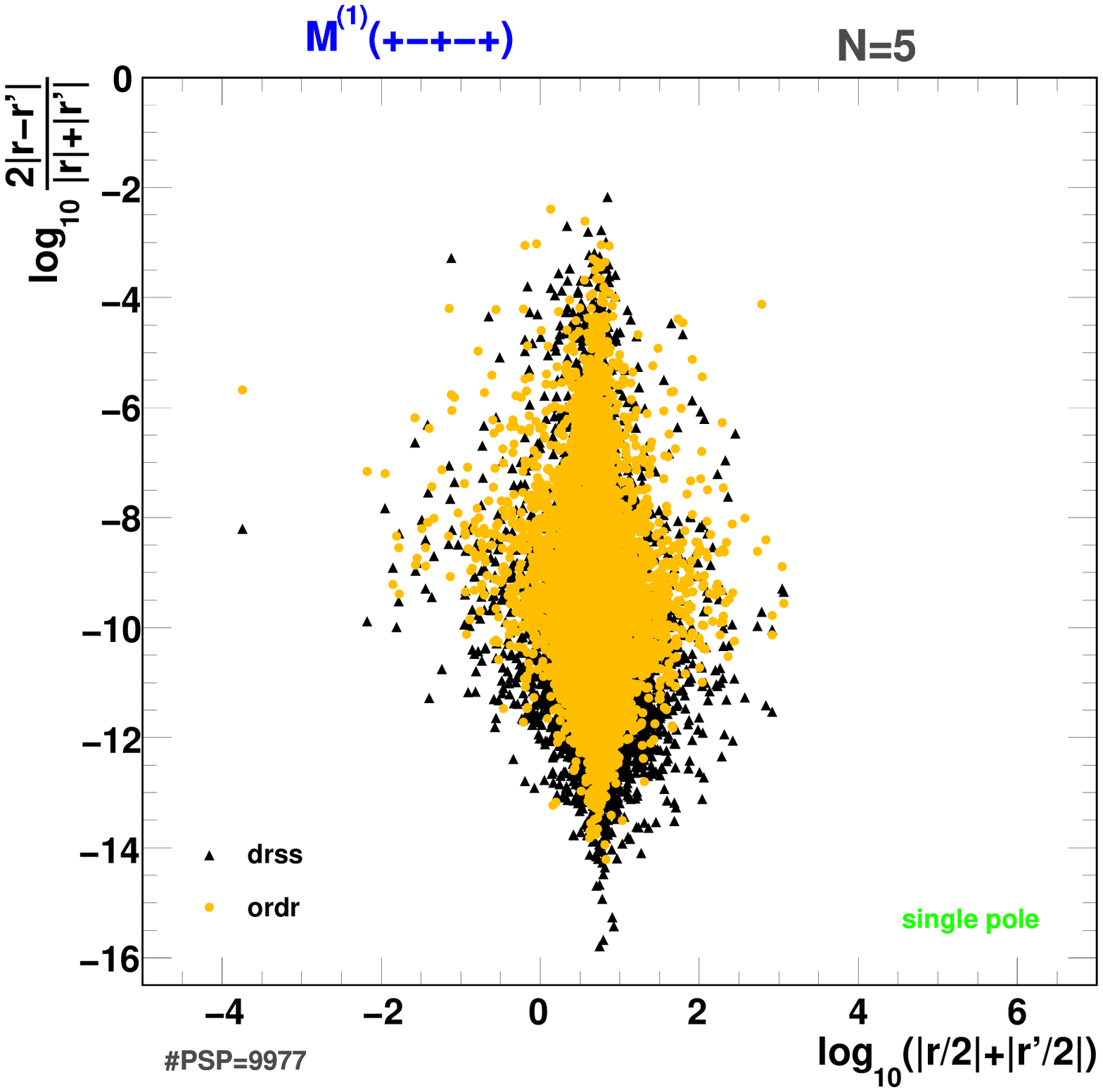}
  \includegraphics[width=0.39\columnwidth]{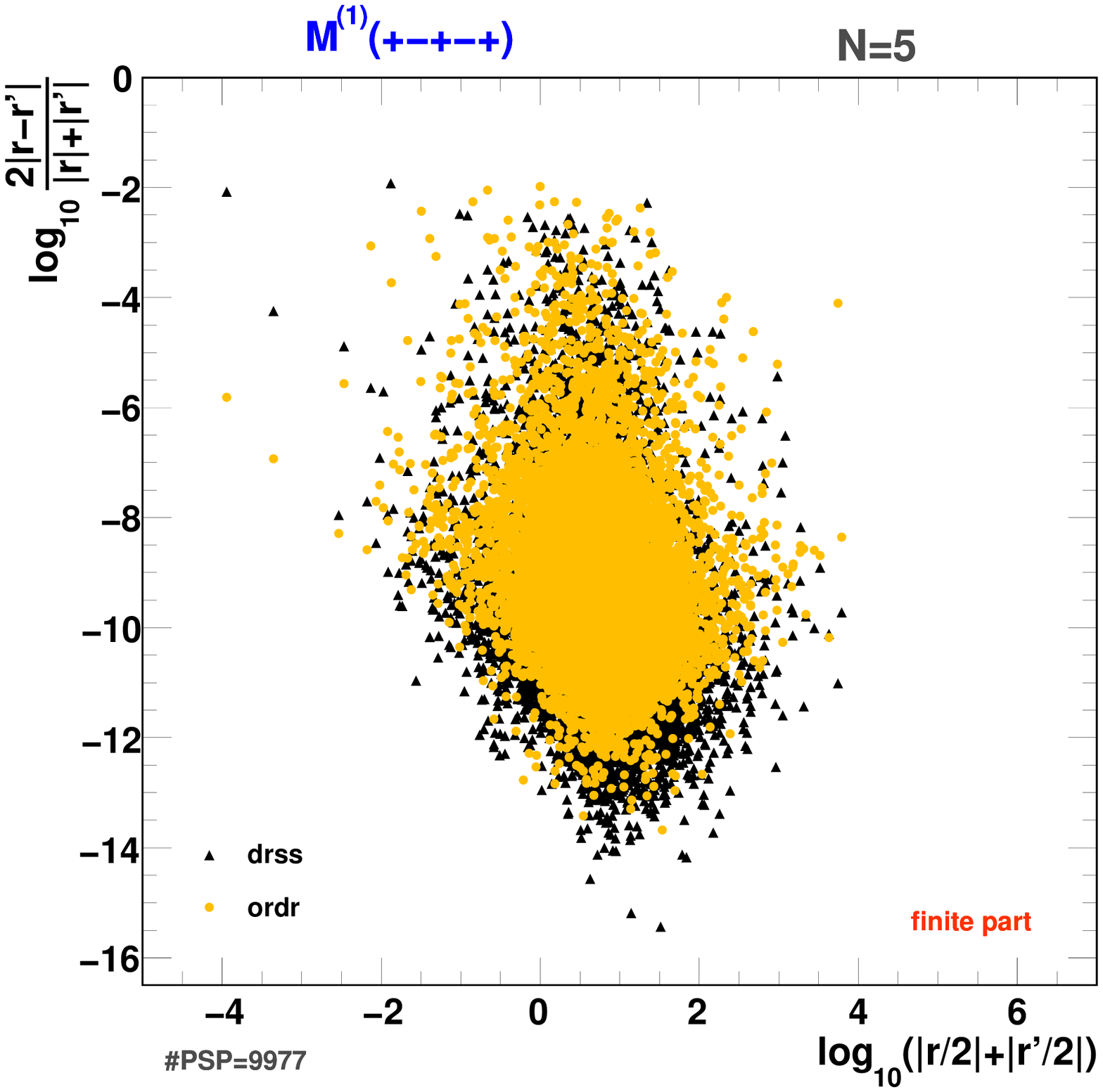}}
\centerline{
  \includegraphics[width=0.39\columnwidth]{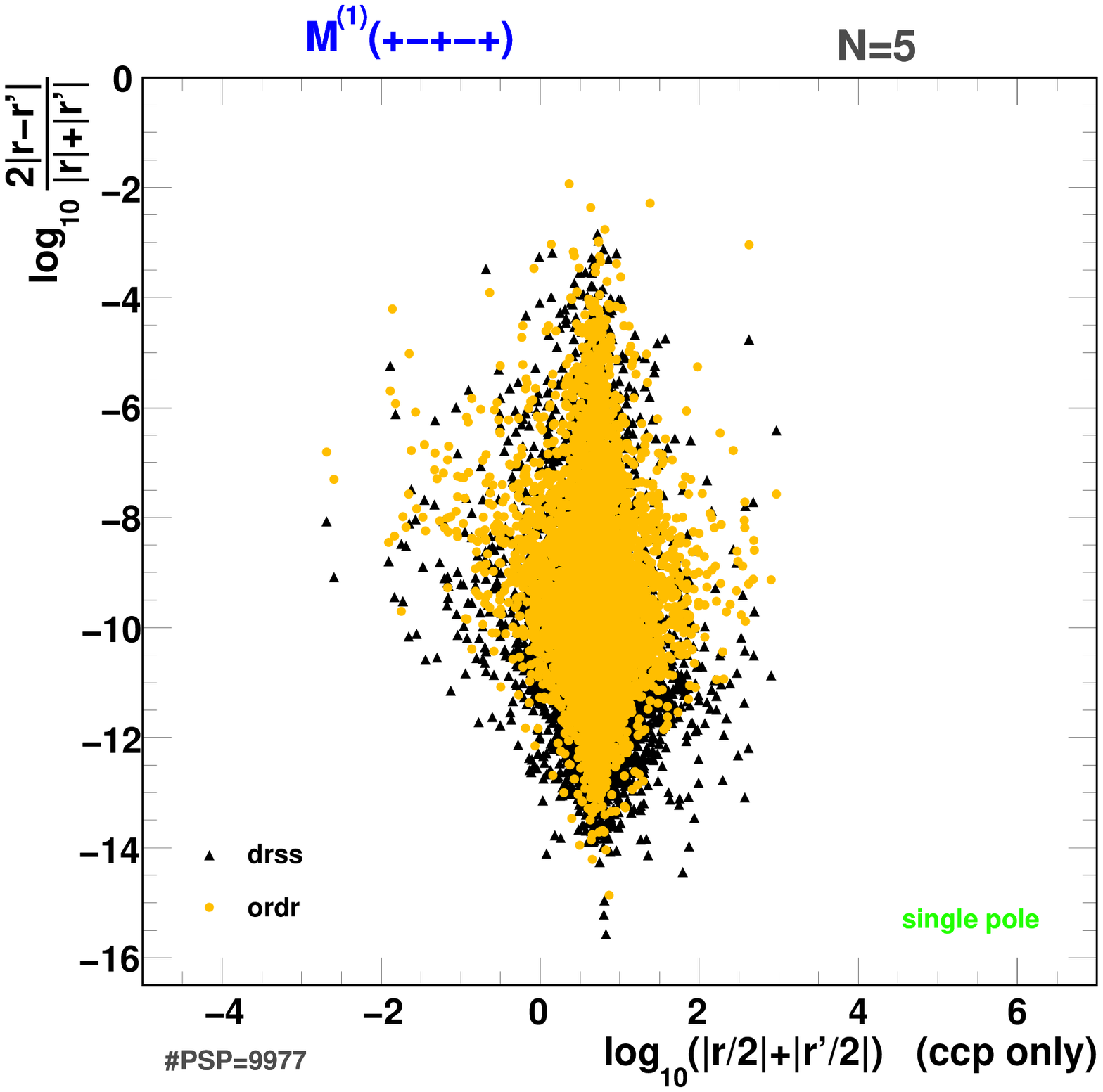}
  \includegraphics[width=0.39\columnwidth]{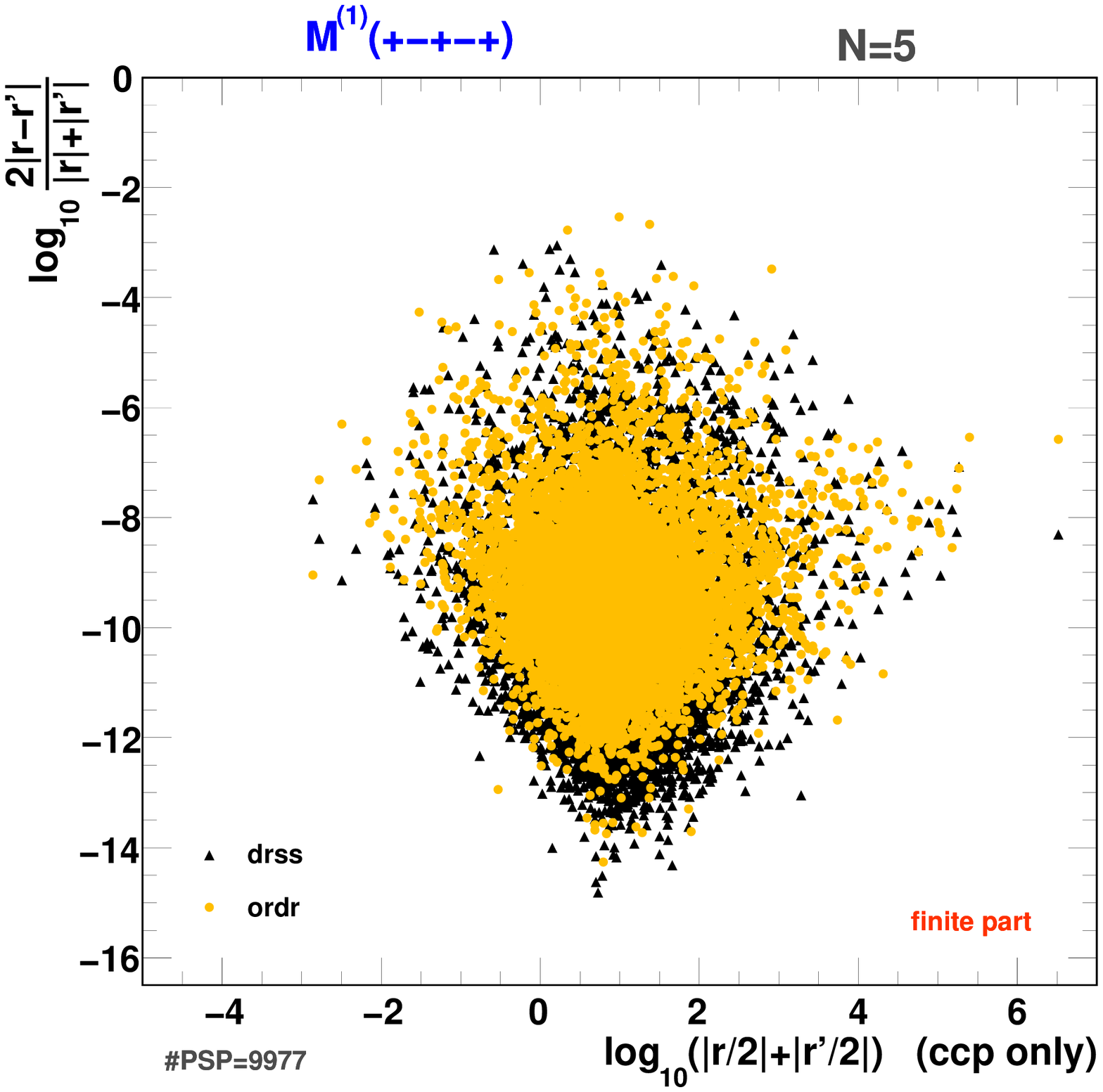}}
\caption{Double-, single-pole and finite-part accuracy distributions
  (upper part) and scatter graphs (lower part) extracted from
  double-precision computations of one-loop amplitudes for $n={\tt N}=5$
  gluons with polarizations $\lambda_k=+-+-+$ and randomly chosen
  non-zero color configurations. The virtual corrections were
  calculated at random phase-space points satisfying the cuts detailed
  in the text. Unstable solutions were vetoed. Results from the
  color-dressed algorithm are compared with those of the color-ordered
  method indicated by dashed curves and brighter dots in the plots.
  The 5(4)-dimensional case is shown in the top left (right) and
  center (bottom) part of the figure. The definitions of
  $\varepsilon$\/ and $r$\/ are given in the text. All scatter graphs
  contain $2\times10^4$ points.}
\label{Fig:accs5}
\end{figure}

\begin{figure}[p!]
\centerline{
  \includegraphics[width=0.51\columnwidth]{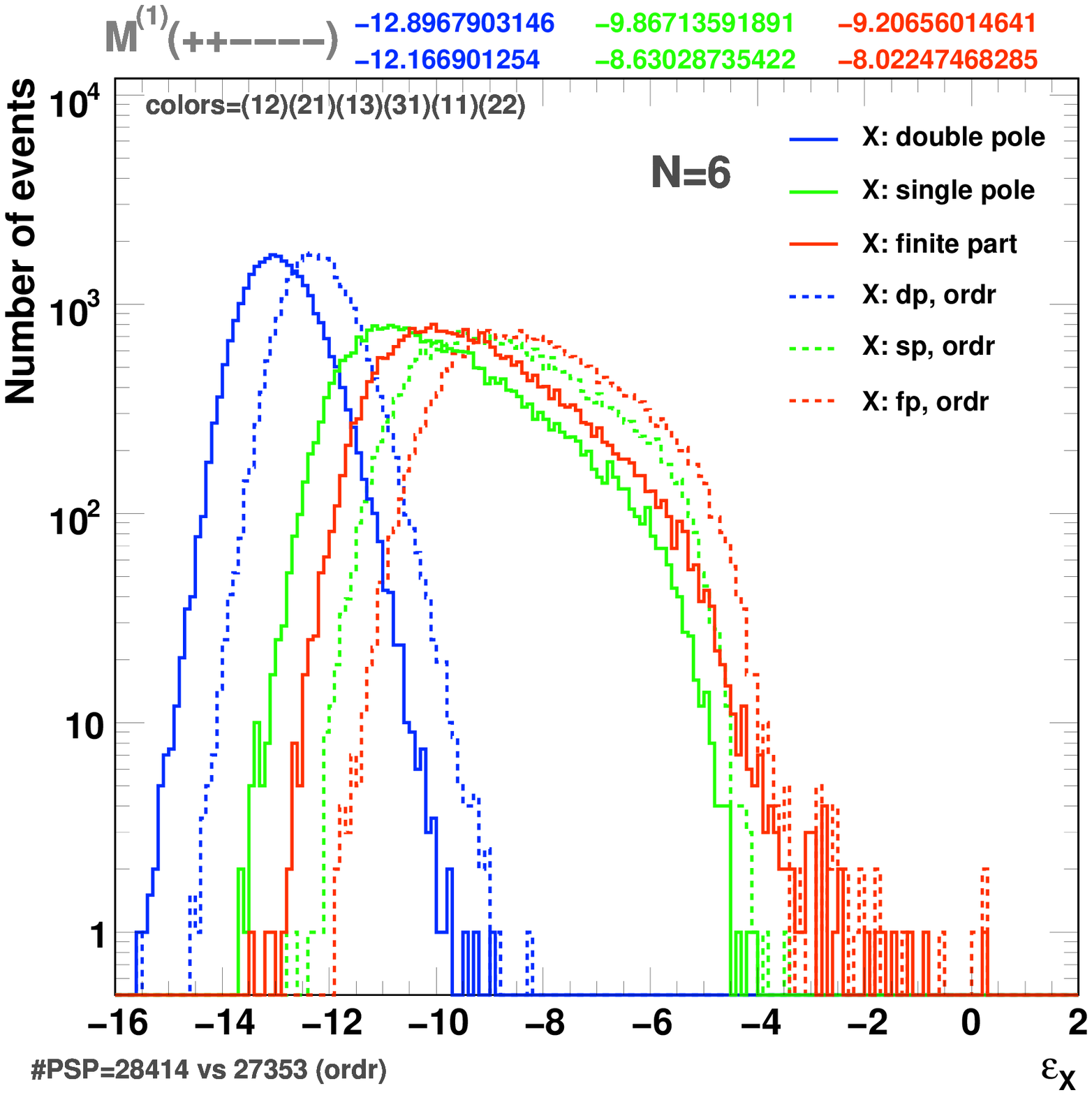}
  \includegraphics[width=0.51\columnwidth]{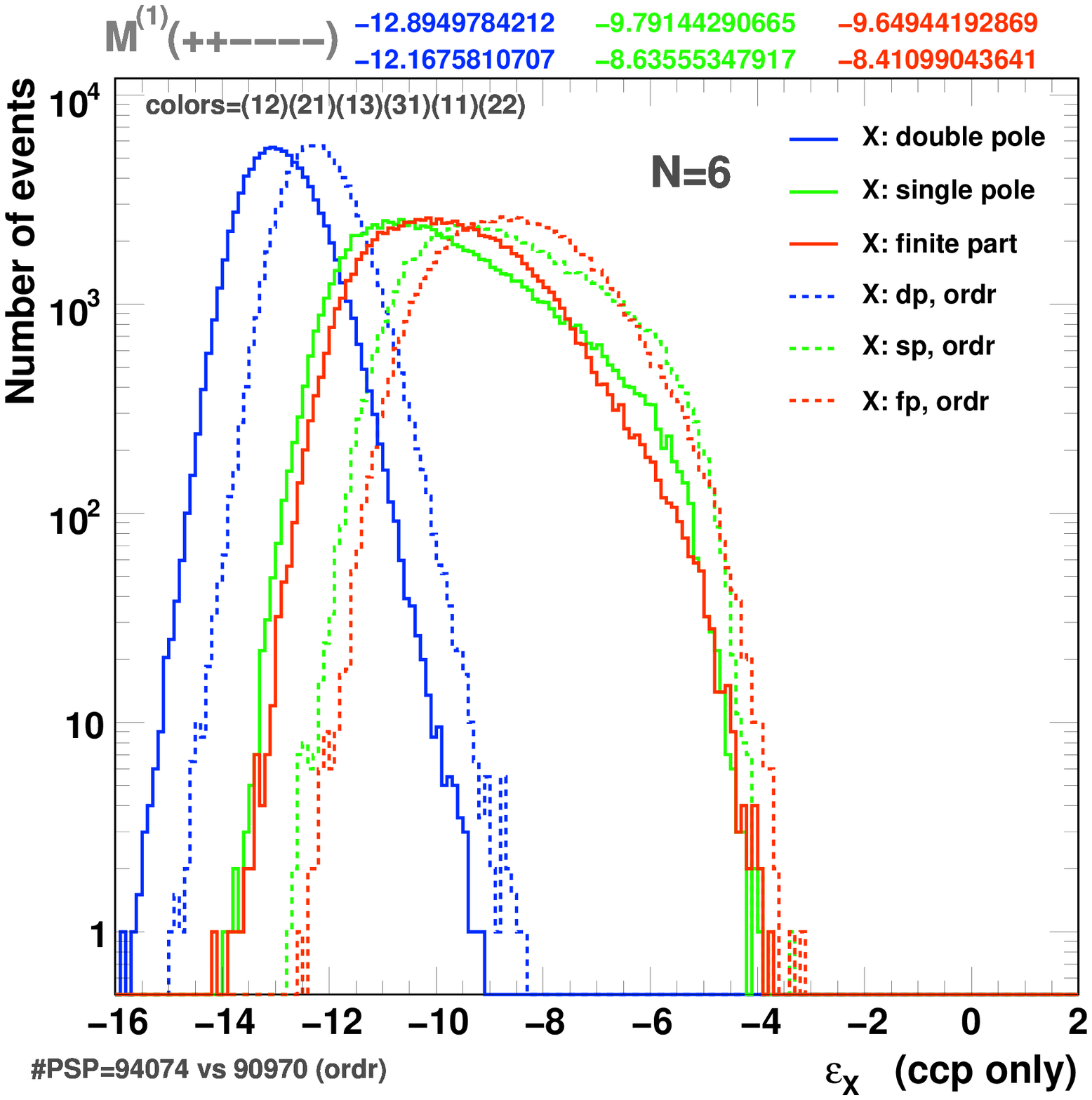}}
\centerline{
  \includegraphics[width=0.39\columnwidth]{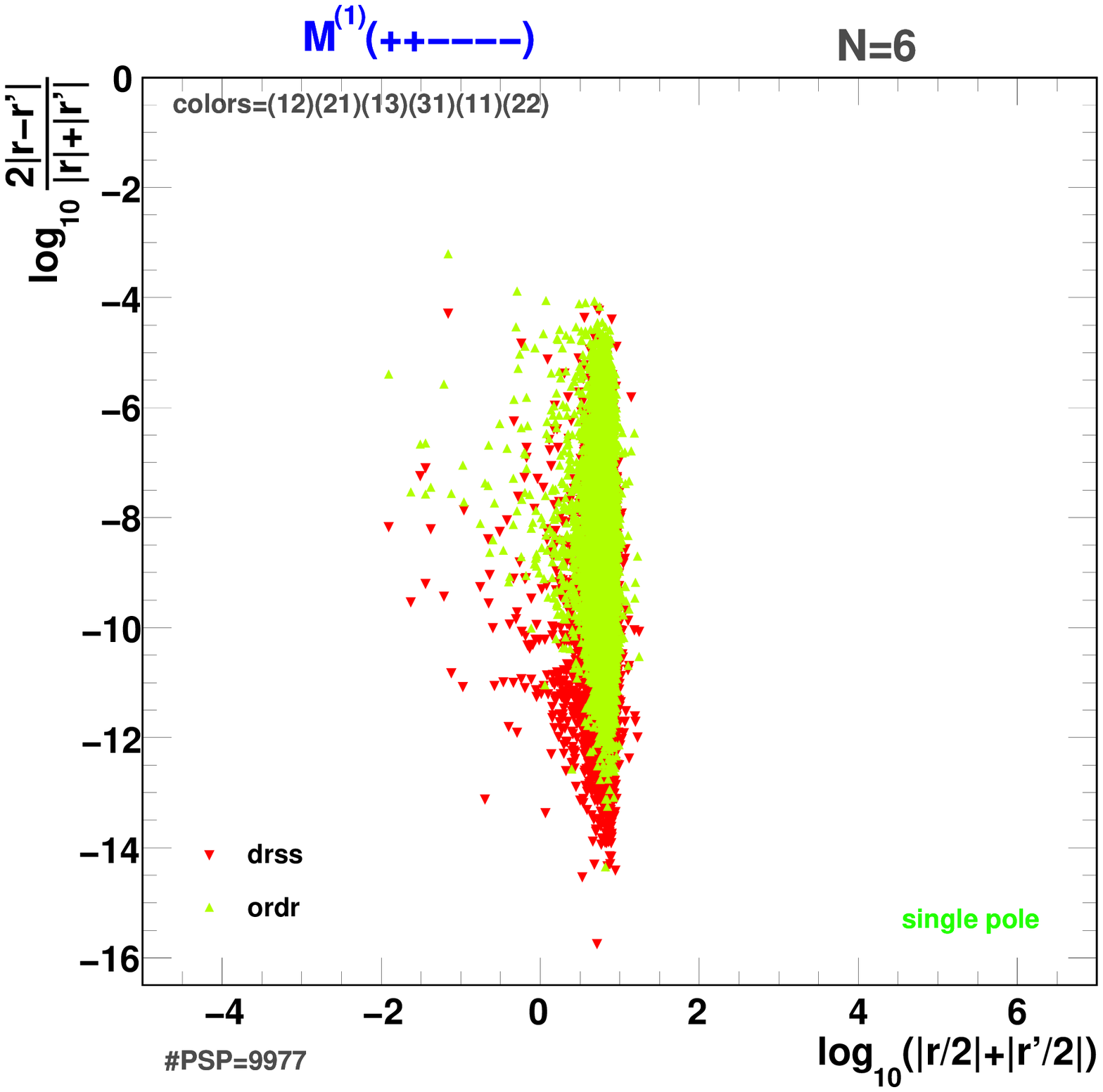}
  \includegraphics[width=0.39\columnwidth]{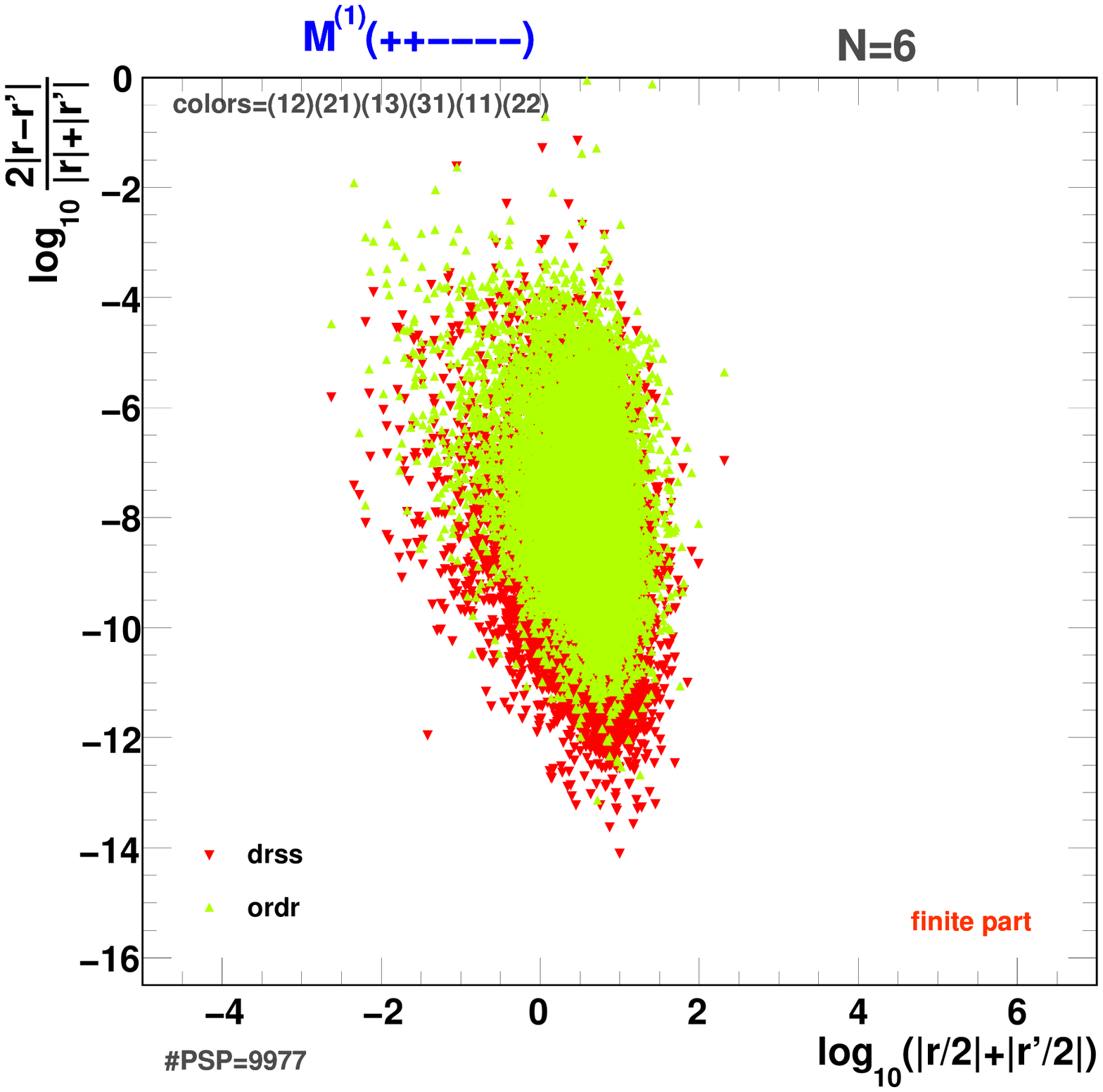}}
\centerline{
  \includegraphics[width=0.39\columnwidth]{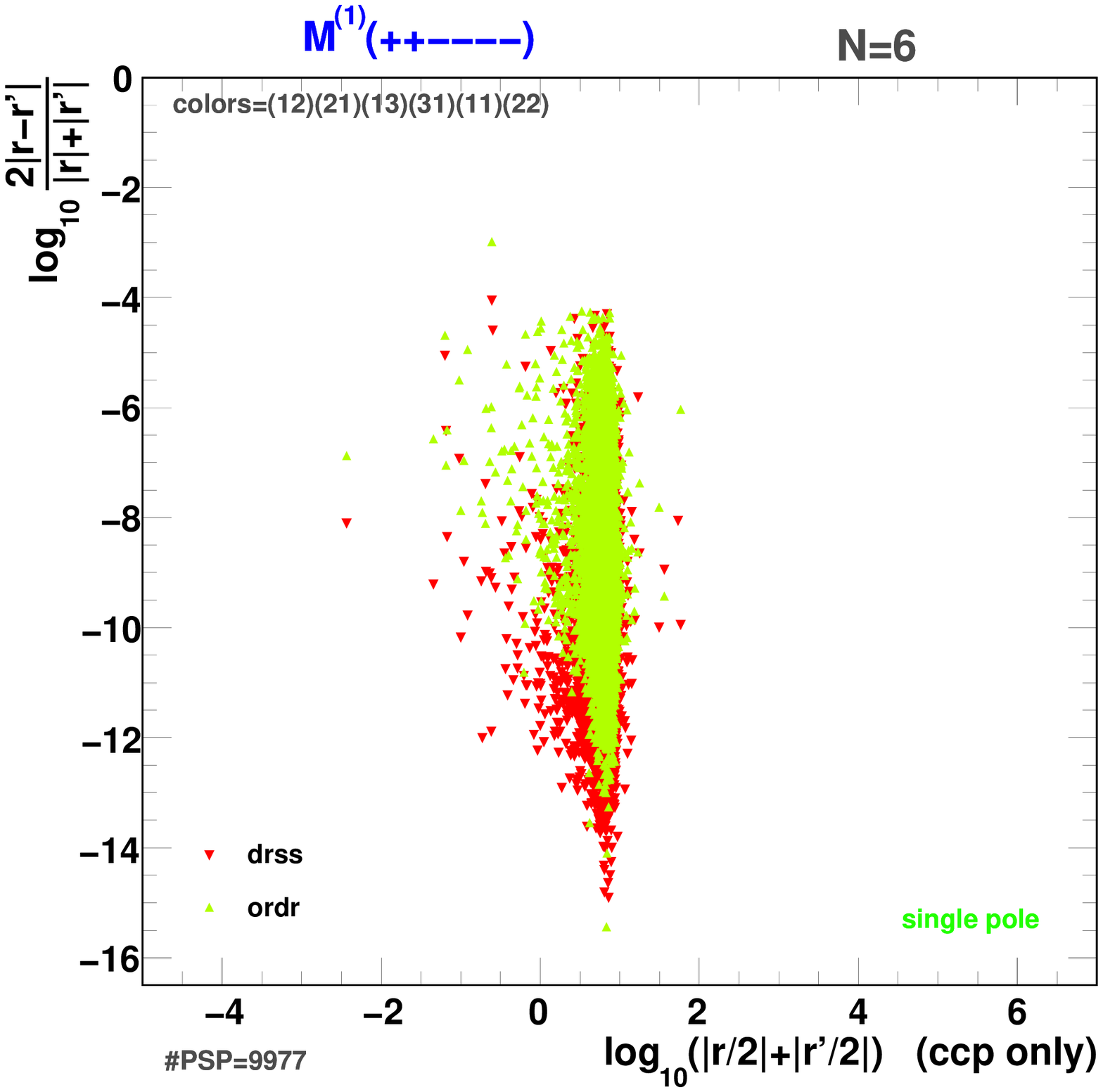}
  \includegraphics[width=0.39\columnwidth]{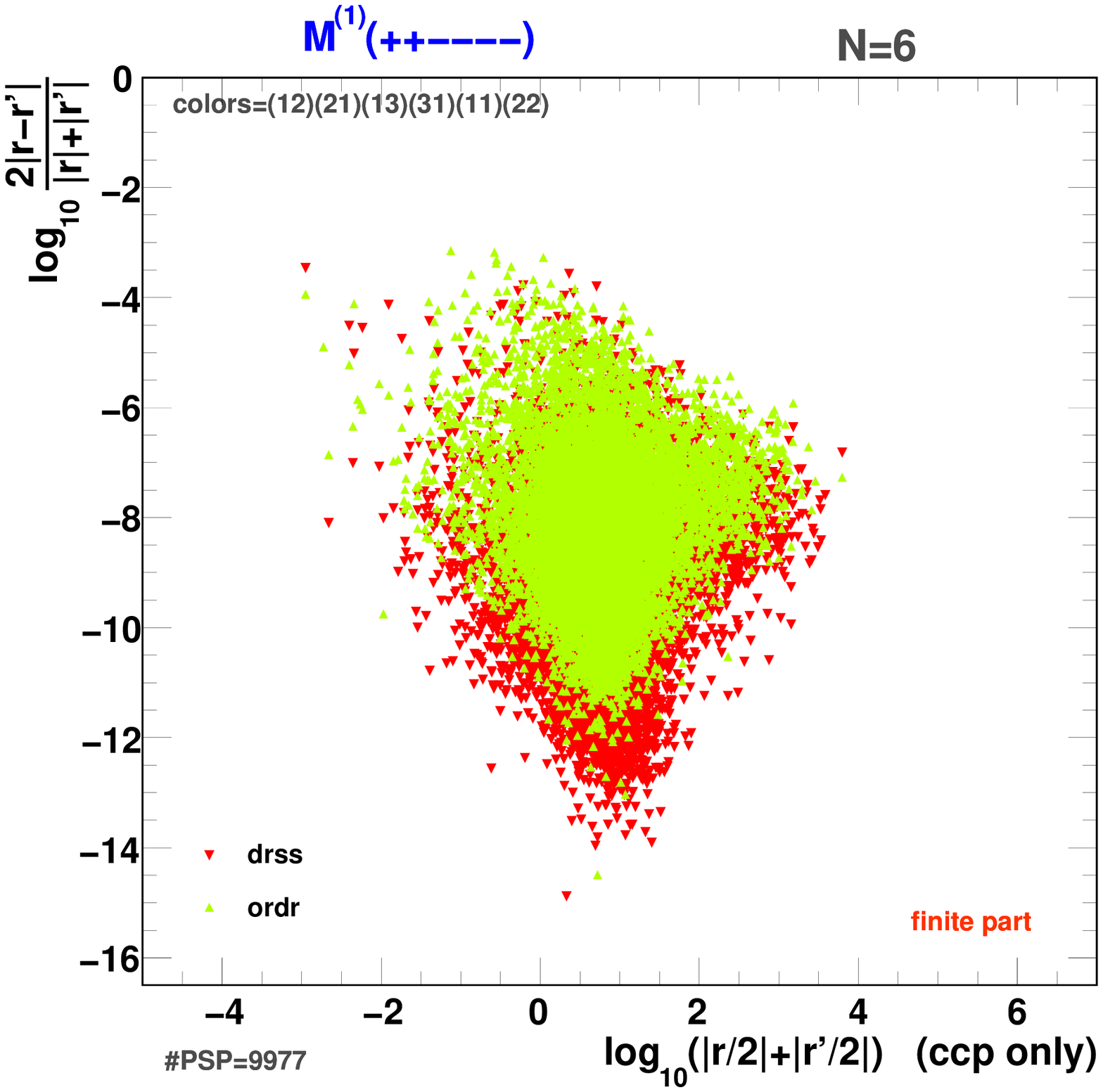}}
\caption{Double-, single-pole and finite-part accuracy distributions
  (upper part) and scatter graphs (lower part) as obtained from
  double-precision evaluations of one-loop amplitudes for $n={\tt N}=6$
  gluons with polarizations and colors set to $\lambda_k=++----$ and
  $(ij)_k=(12)(21)(13)(31)(11)(22)$, respectively. The virtual
  corrections were calculated at random phase-space points satisfying
  the cuts detailed in the text. The veto procedure has been applied
  to reject unstable solutions. The results given by the color-dressed
  algorithm are compared with those of the color-ordered method
  indicated by dashed curves and brighter dots in the plots. The
  5(4)-dimensional case is shown in the top left (right) and center
  (bottom) part of the figure. The definitions of $\varepsilon$\/ and
  $r$\/ are given in the text. Each scatter graph contains
  $2\times10^4$ points. $94.7$($94.1$)\% and $91.2$($91.0$)\% of
  the events pass all tests in the dressed and ordered
  ``5(4)D-case'', respectively.}
\label{Fig:accs6fxcl}
\end{figure}

\begin{figure}[p!]
\centerline{
  \includegraphics[width=0.51\columnwidth]{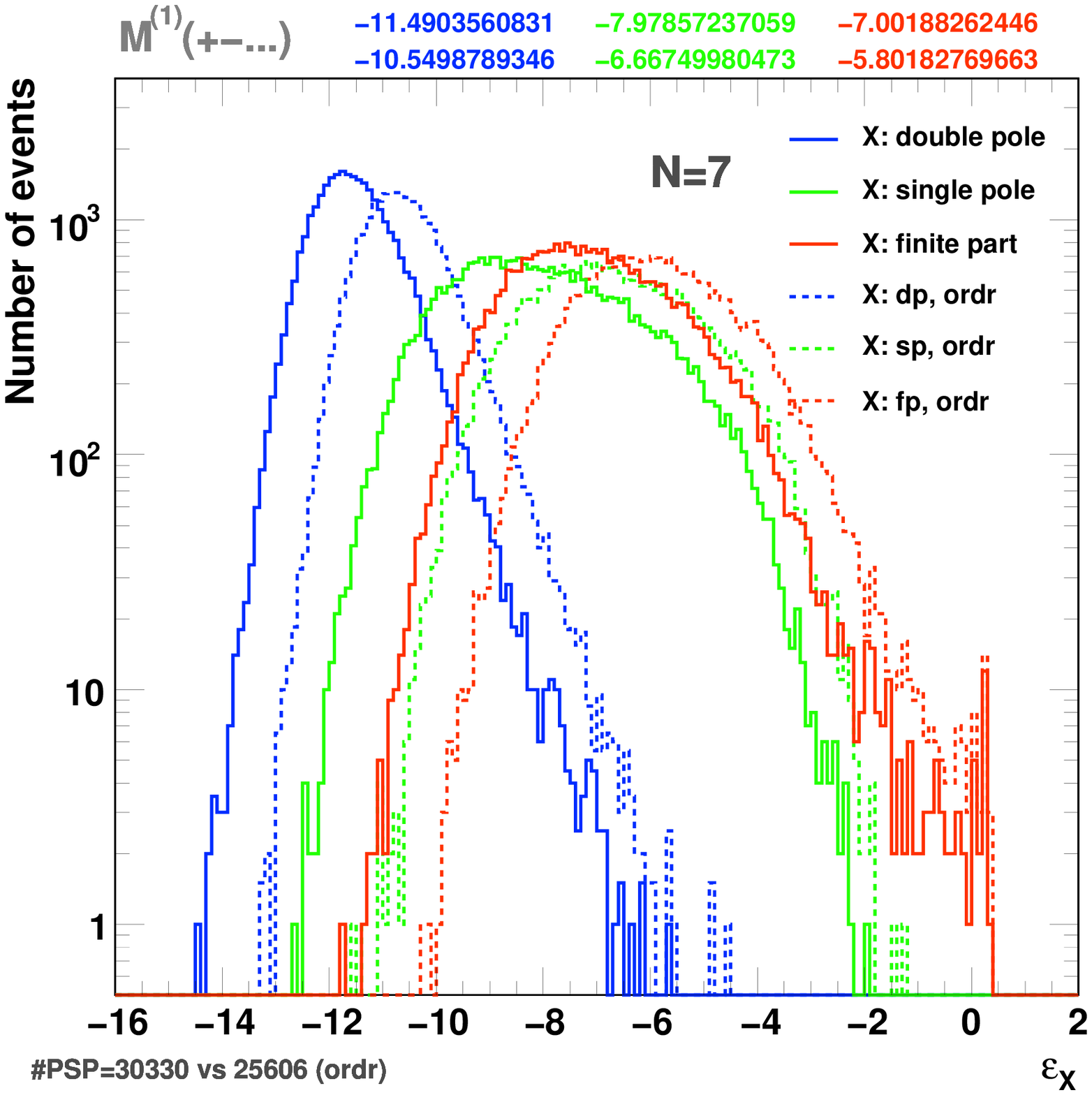}
  \includegraphics[width=0.51\columnwidth]{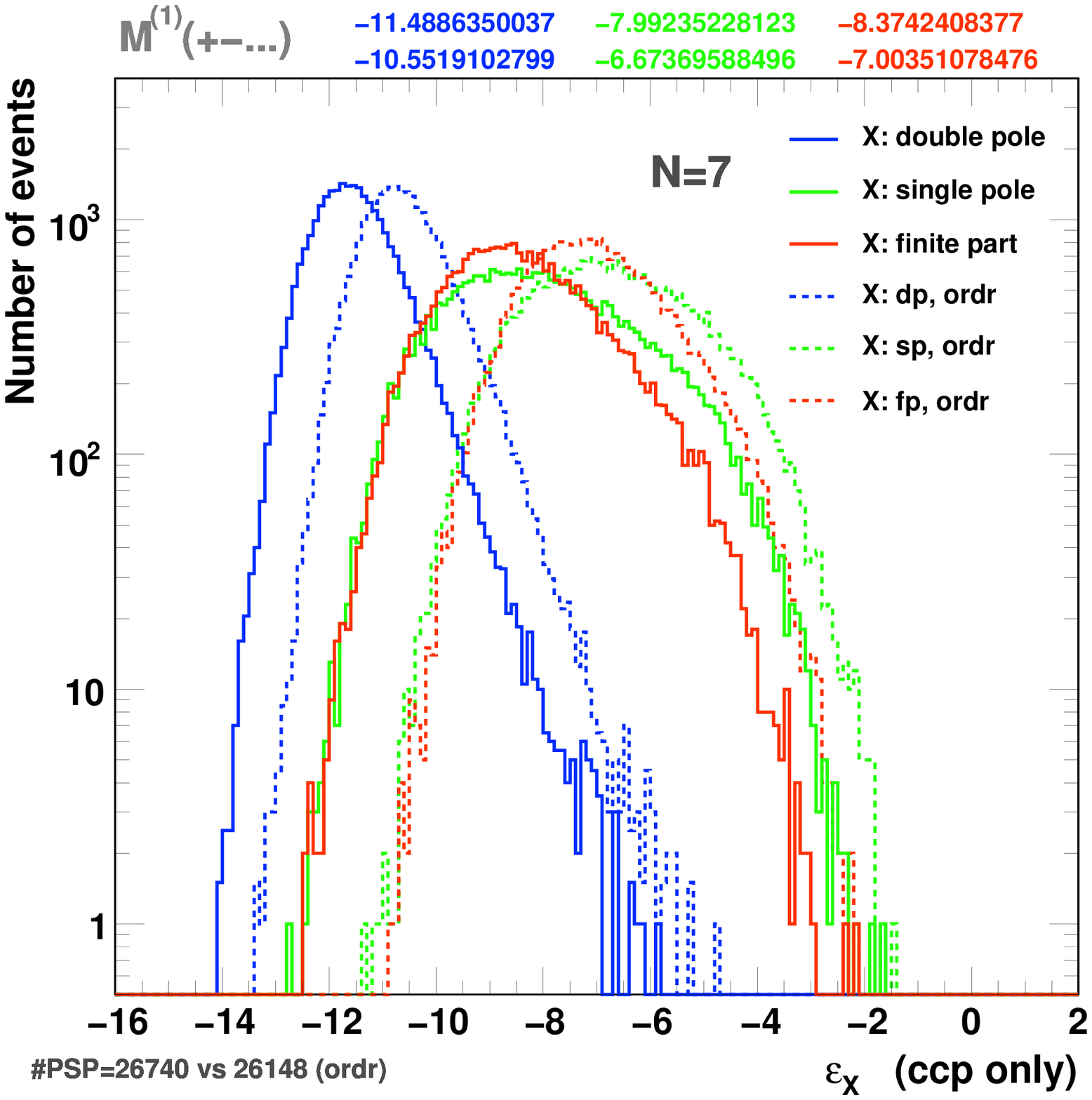}}
\centerline{
  \includegraphics[width=0.39\columnwidth]{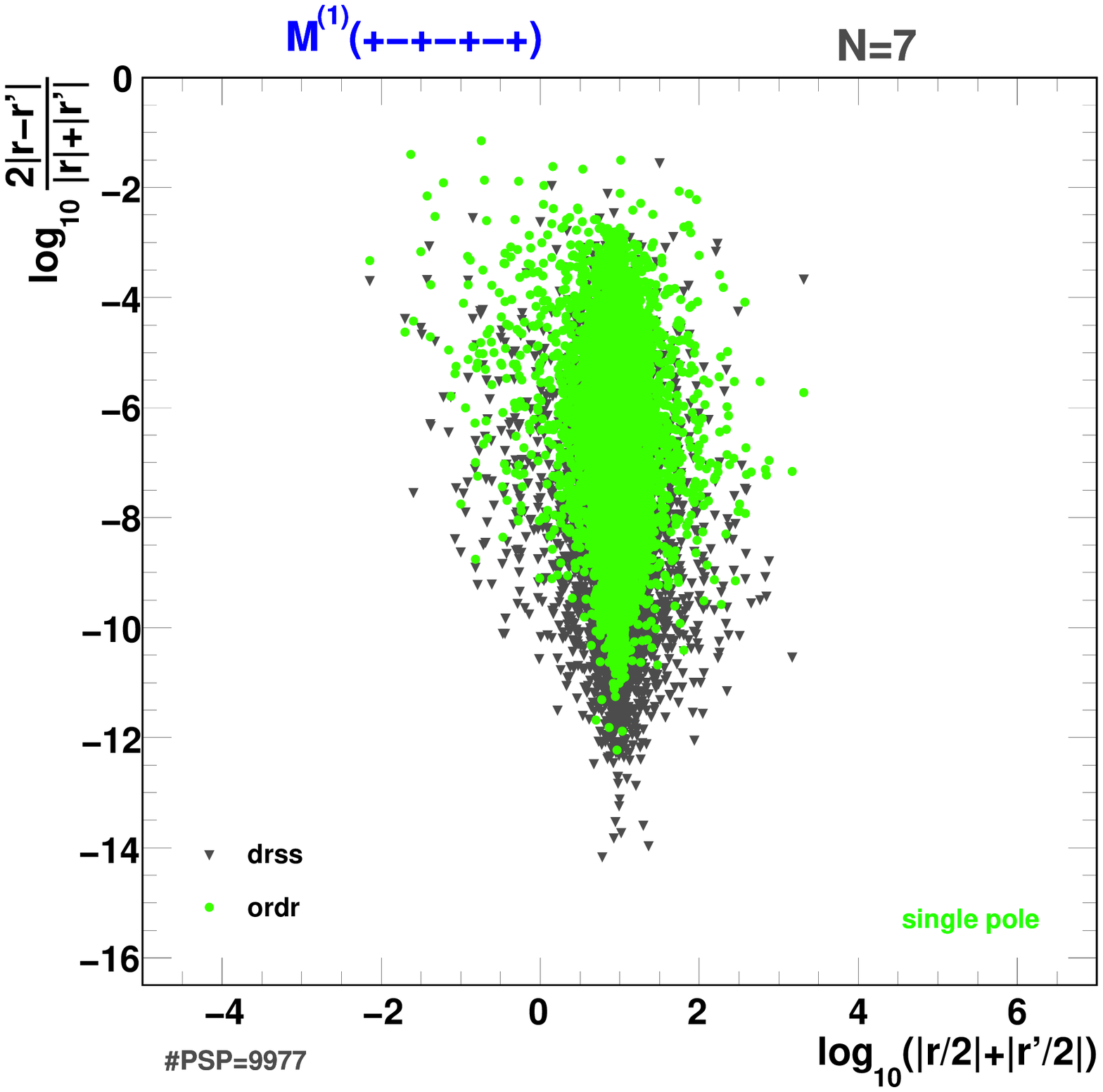}
  \includegraphics[width=0.39\columnwidth]{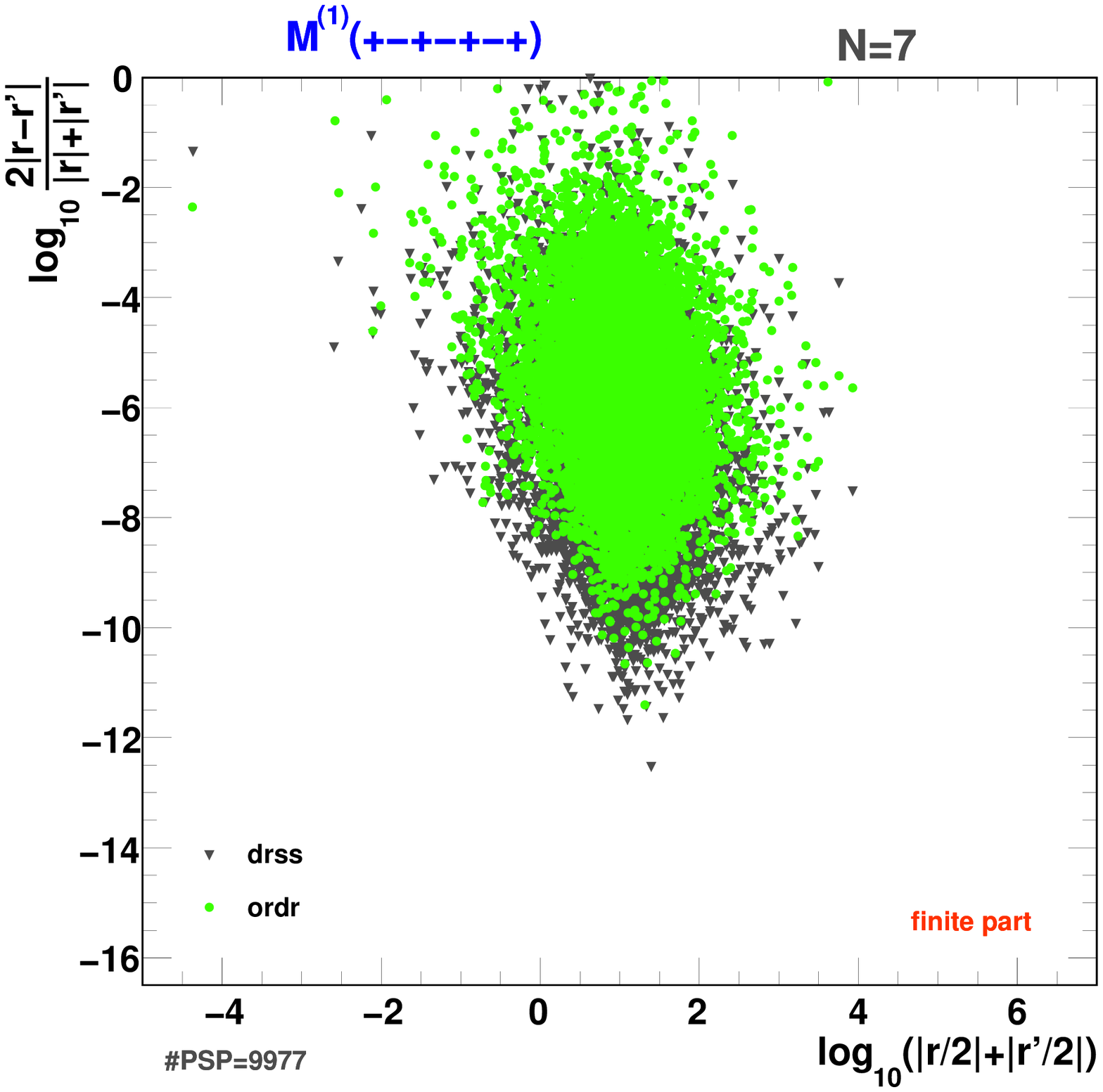}}
\centerline{
  \includegraphics[width=0.39\columnwidth]{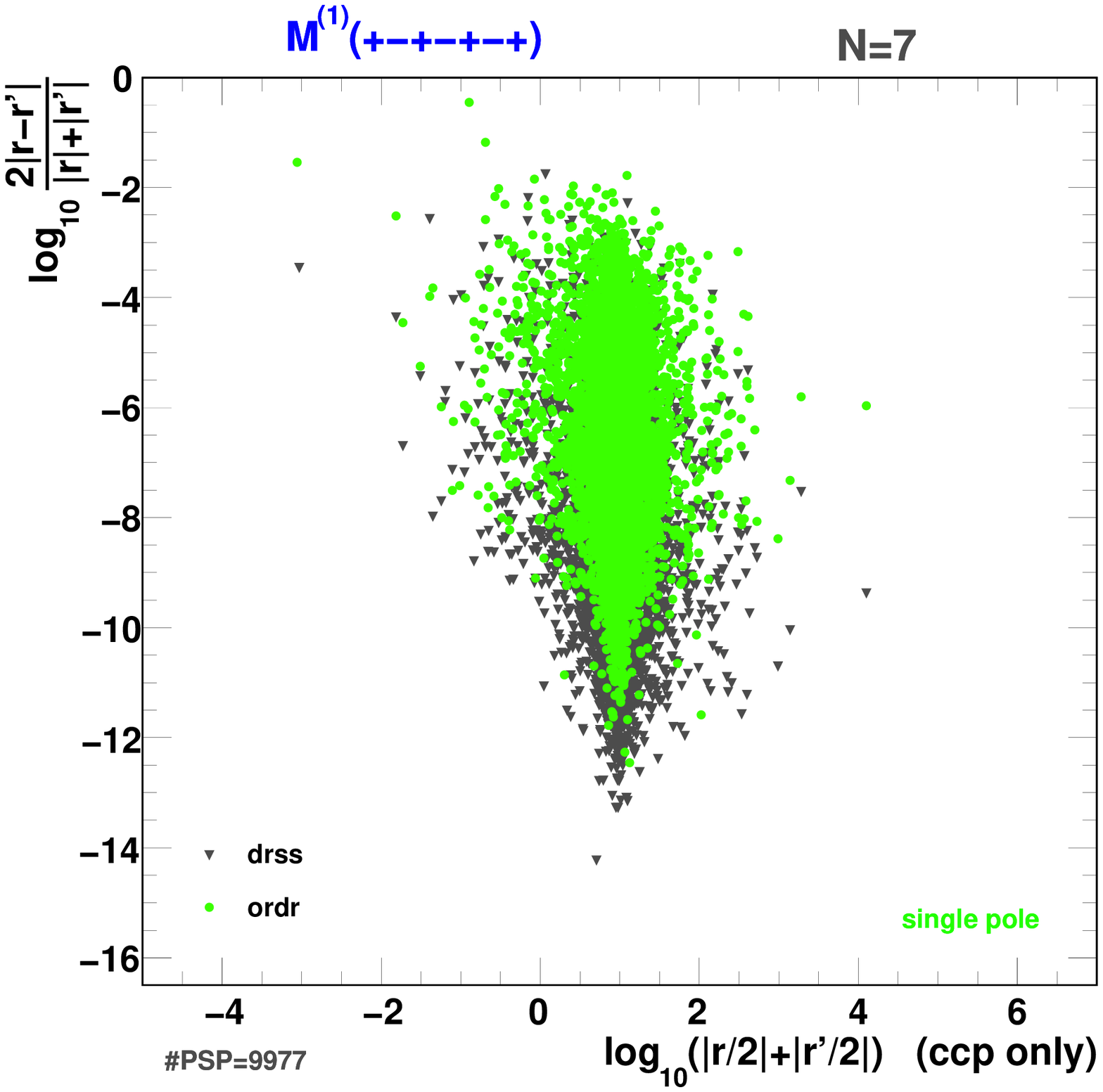}
  \includegraphics[width=0.39\columnwidth]{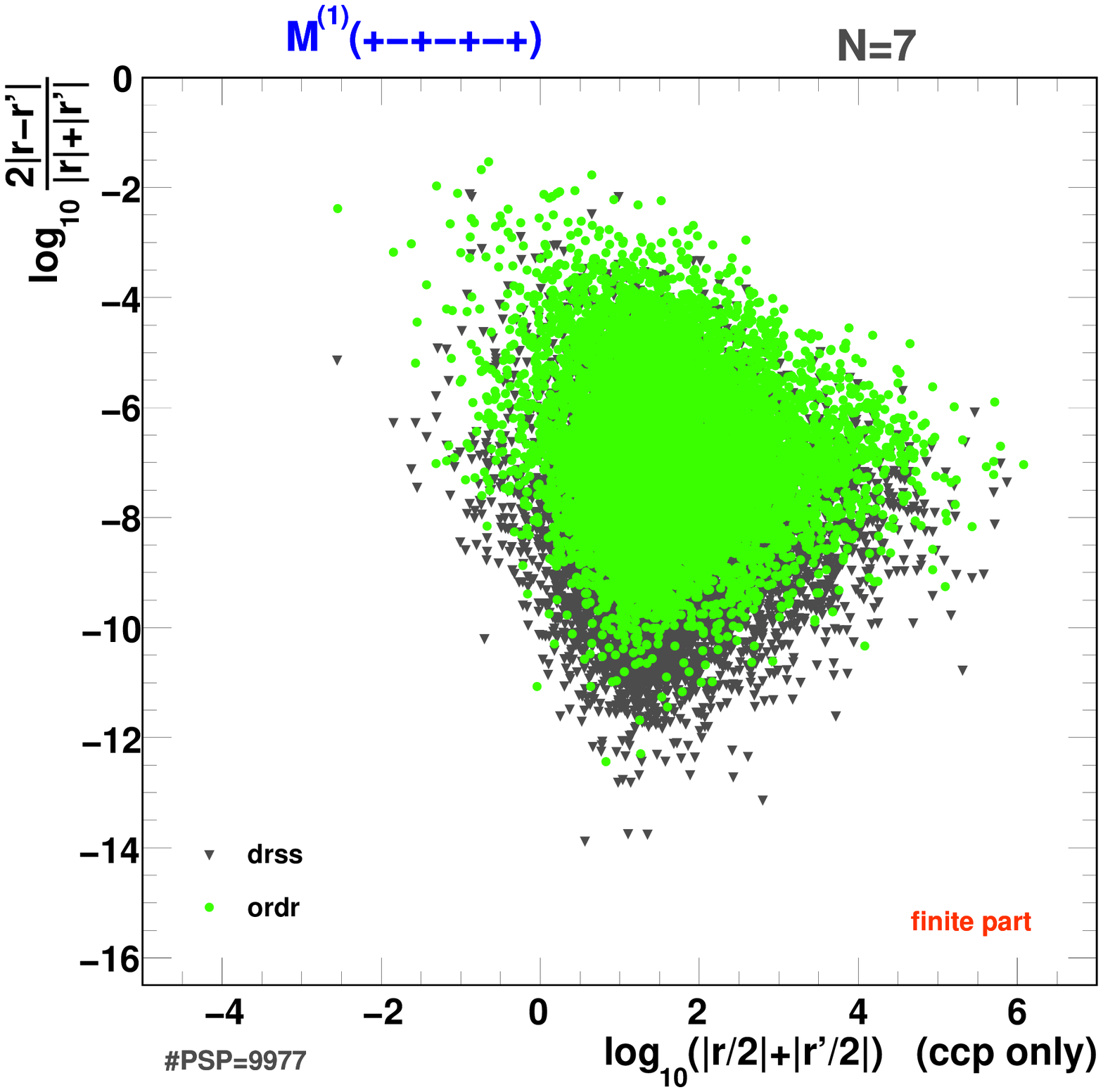}}
\caption{Double-, single-pole and finite-part accuracy distributions
  (upper part) and scatter graphs (lower part) extracted from
  double-precision computations of one-loop amplitudes for $n={\tt N}=7$
  gluons with polarizations $\lambda_k=+-+-+-+$ and randomly chosen
  non-zero color configurations. The virtual corrections were
  calculated at random phase-space points satisfying the cuts detailed
  in the text. Unstable solutions were vetoed. Results from the
  color-dressed algorithm are compared with those of the color-ordered
  method indicated by dashed curves and brighter dots in the plots.
  The 5(4)-dimensional case is shown in the top left (right) and
  center (bottom) part of the figure. The definitions of
  $\varepsilon$\/ and $r$\/ are given in the text. All scatter graphs
  contain $2\times10^4$ points.}
\label{Fig:accs7}
\end{figure}

\begin{figure}[p!]
\centerline{
  \includegraphics[width=0.51\columnwidth]{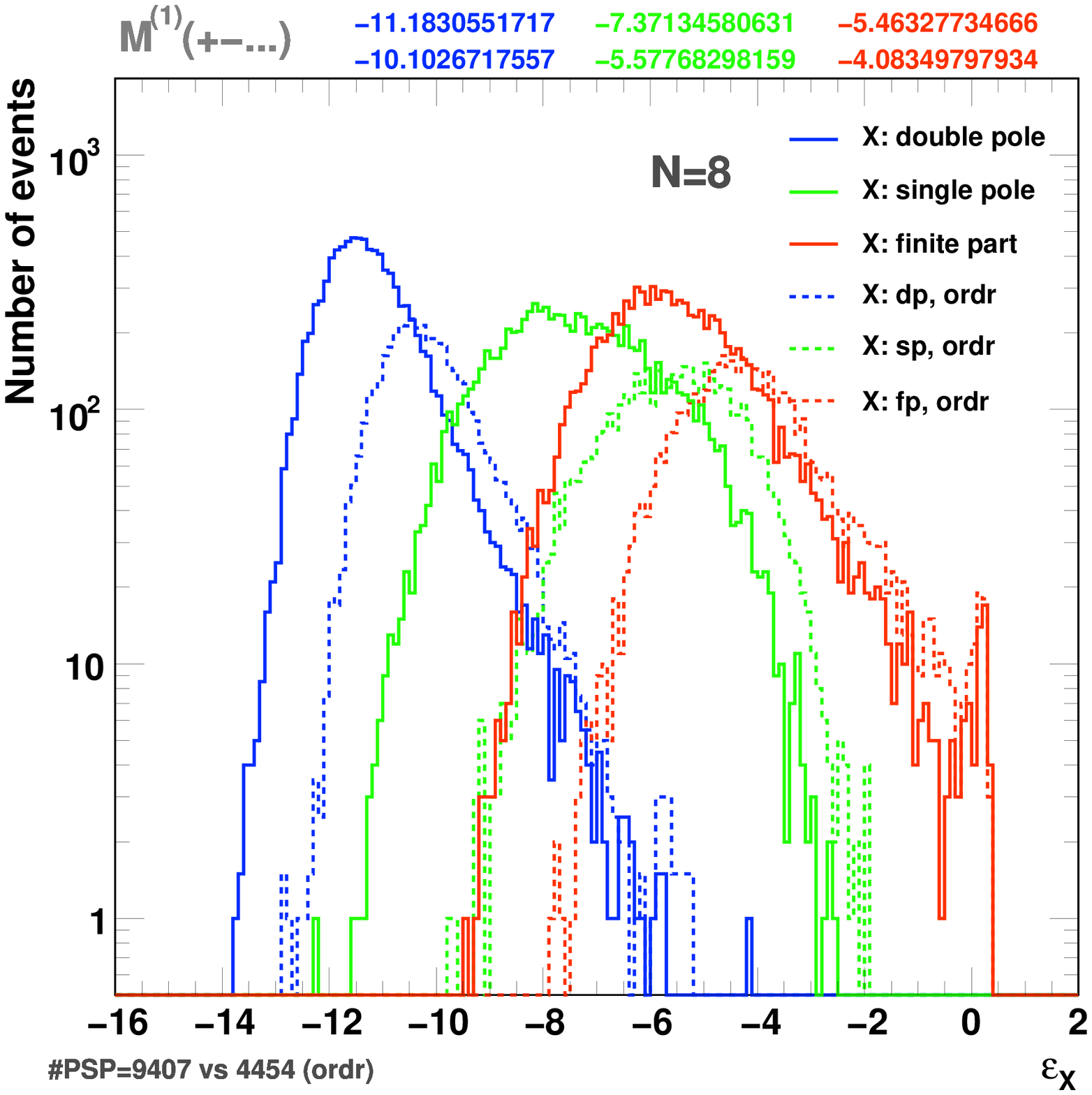}
  \includegraphics[width=0.51\columnwidth]{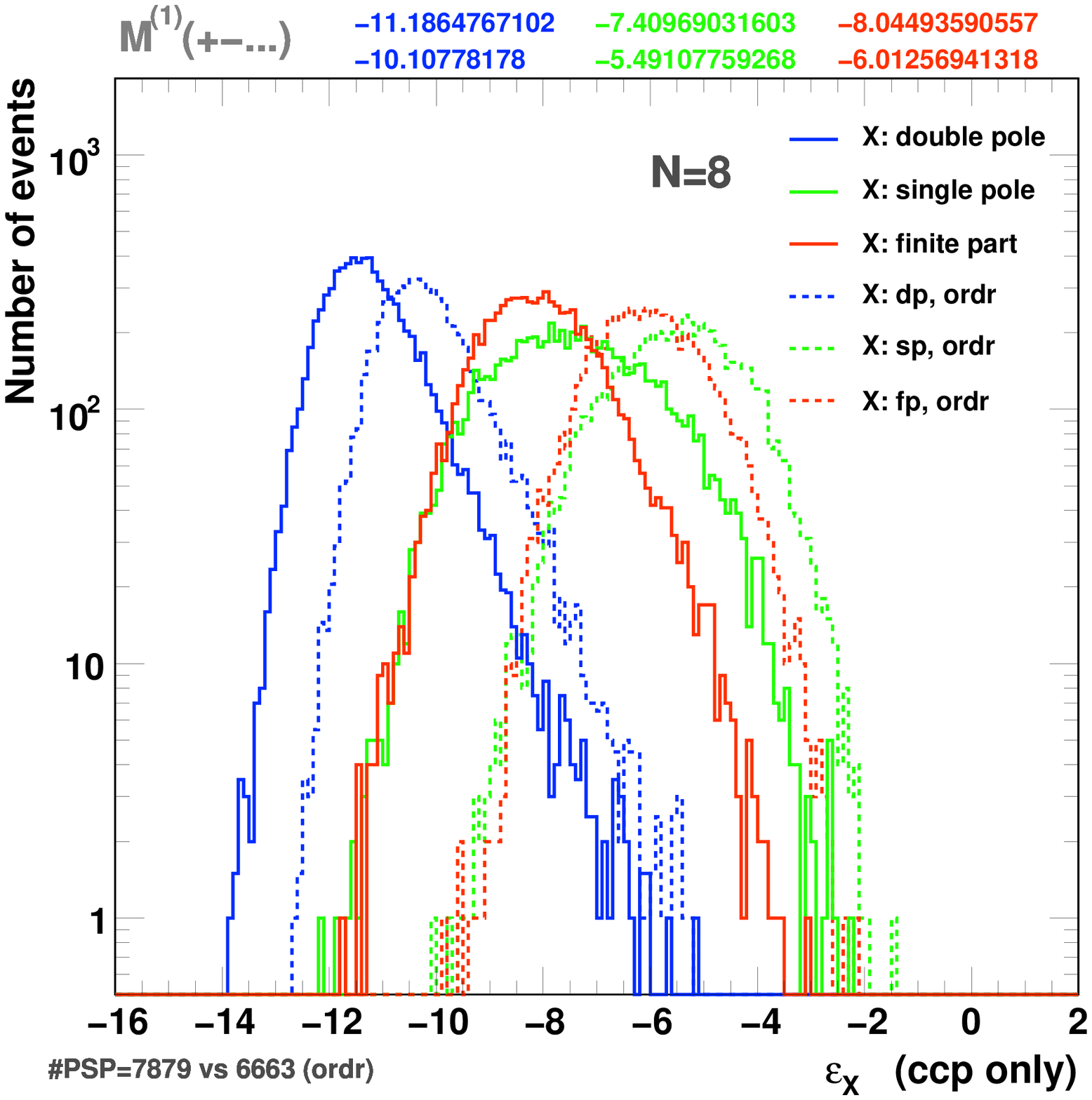}}
\centerline{
  \includegraphics[width=0.39\columnwidth]{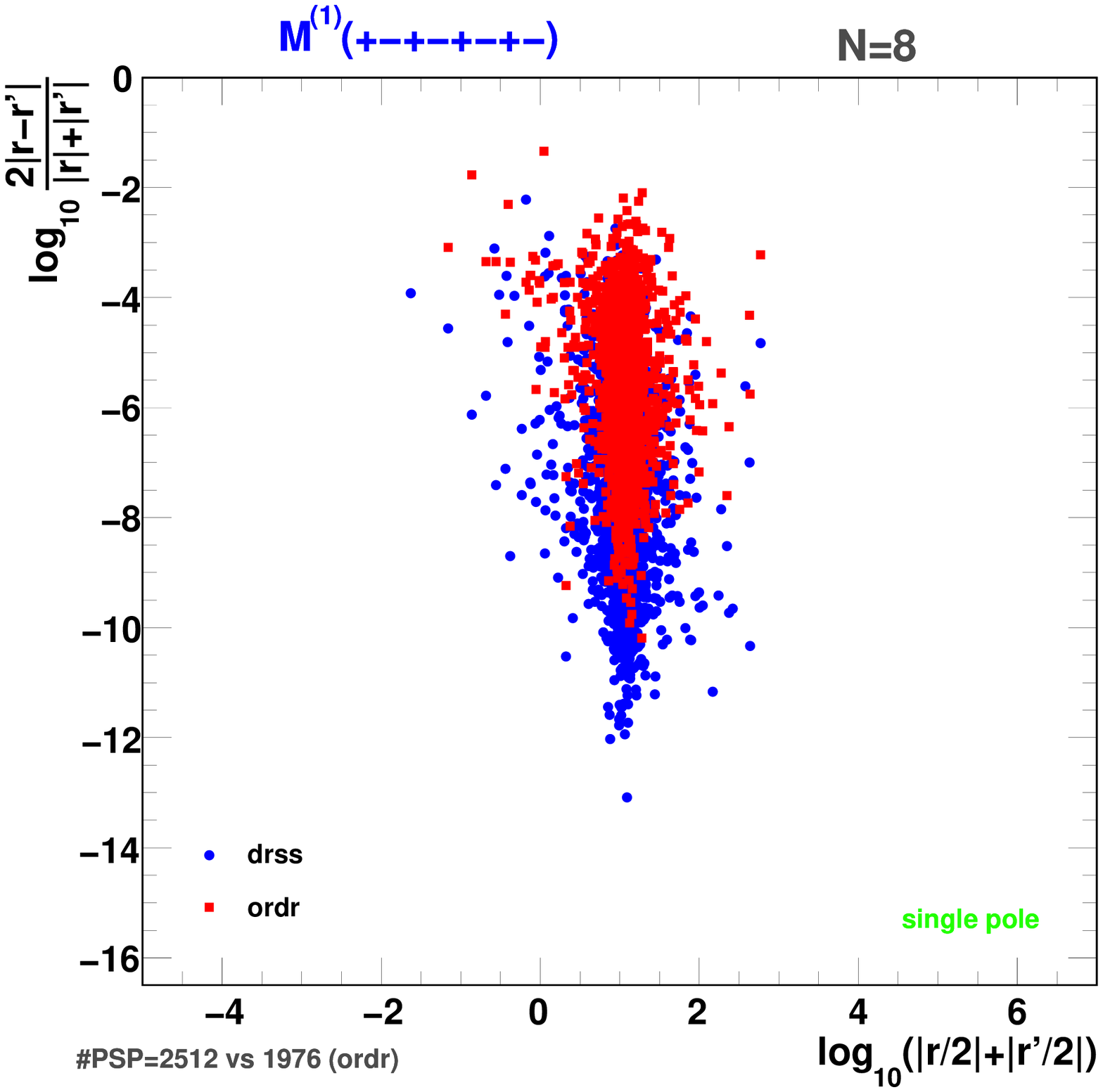}
  \includegraphics[width=0.39\columnwidth]{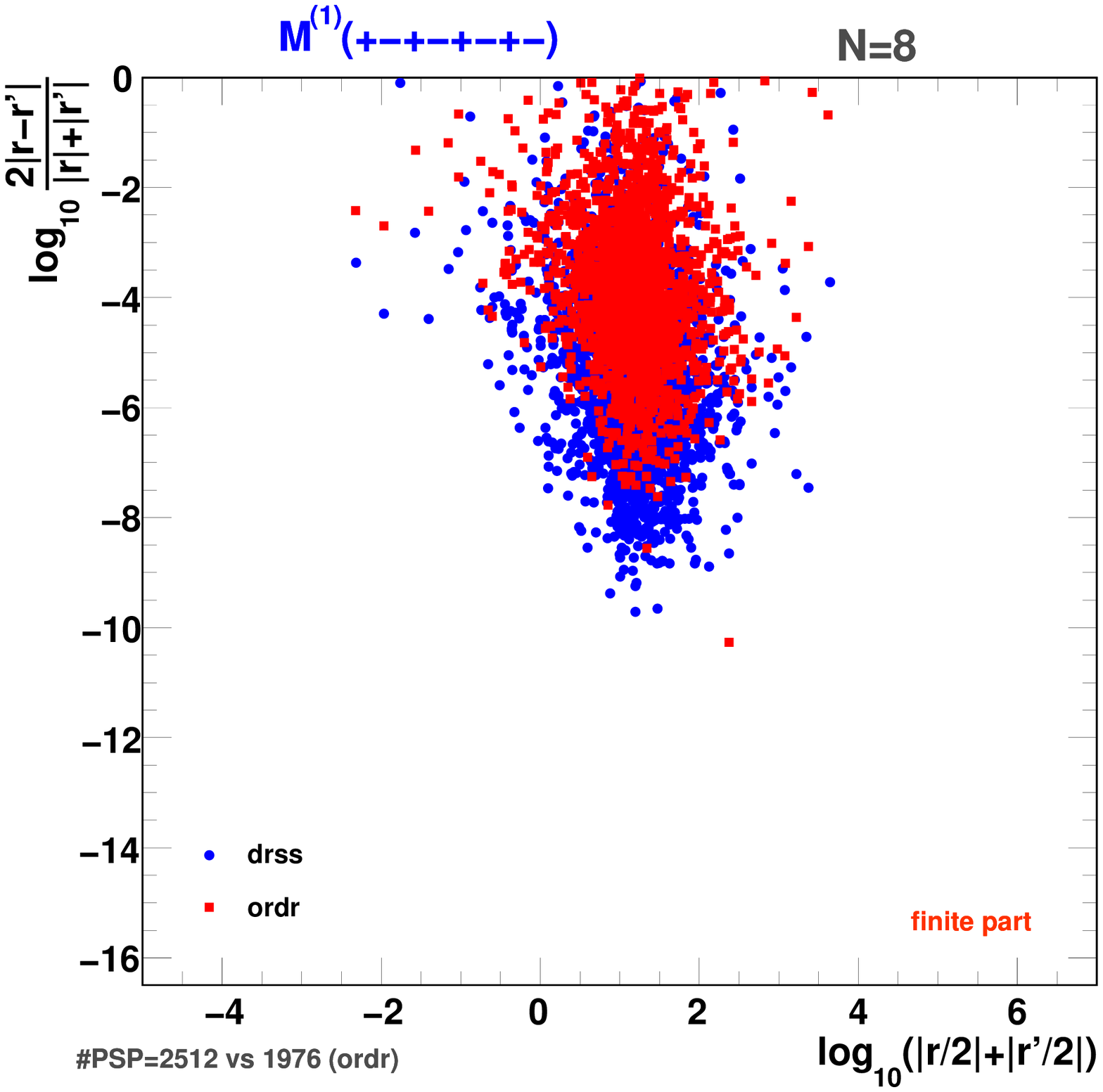}}
\centerline{
  \includegraphics[width=0.39\columnwidth]{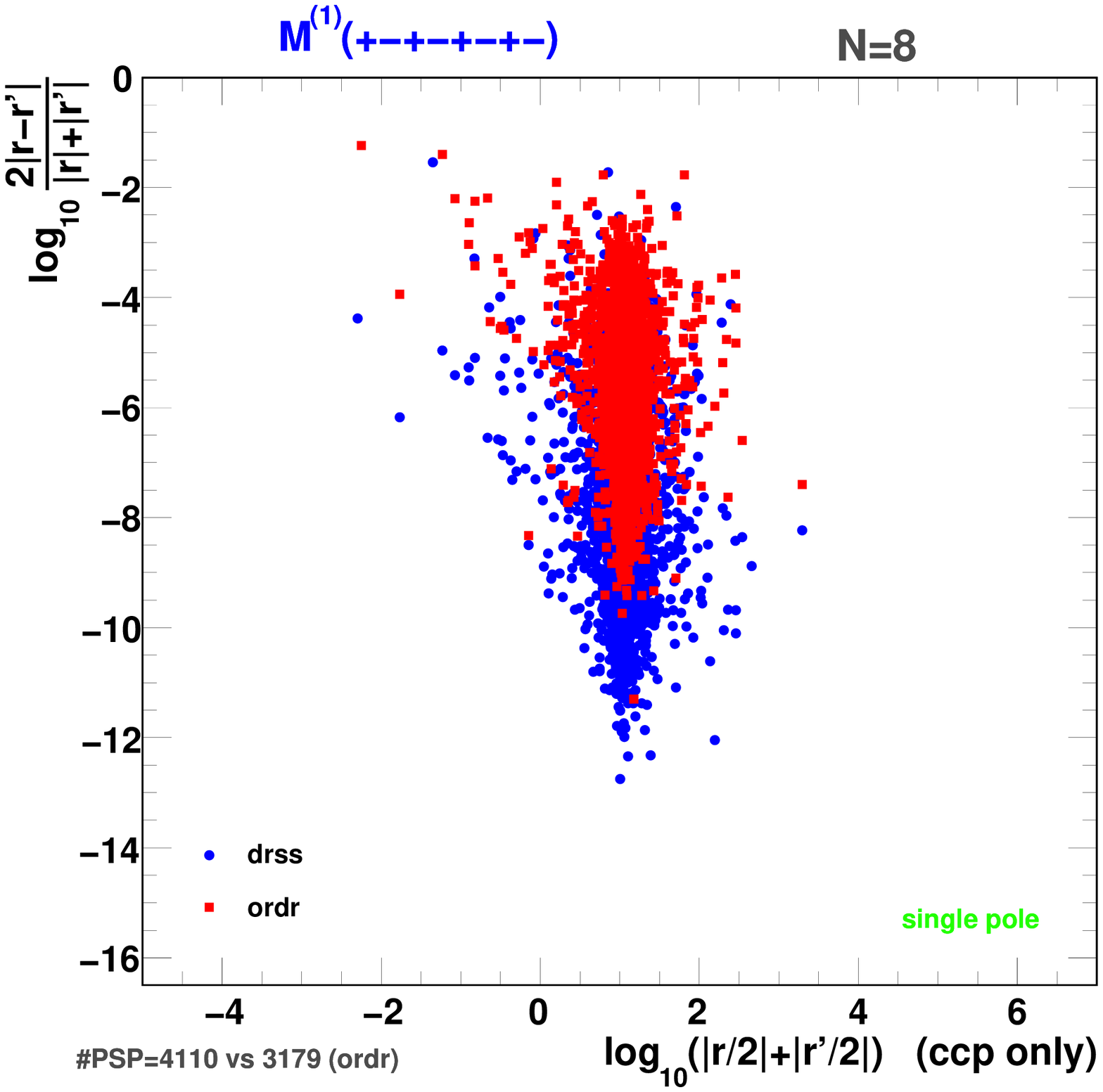}
  \includegraphics[width=0.39\columnwidth]{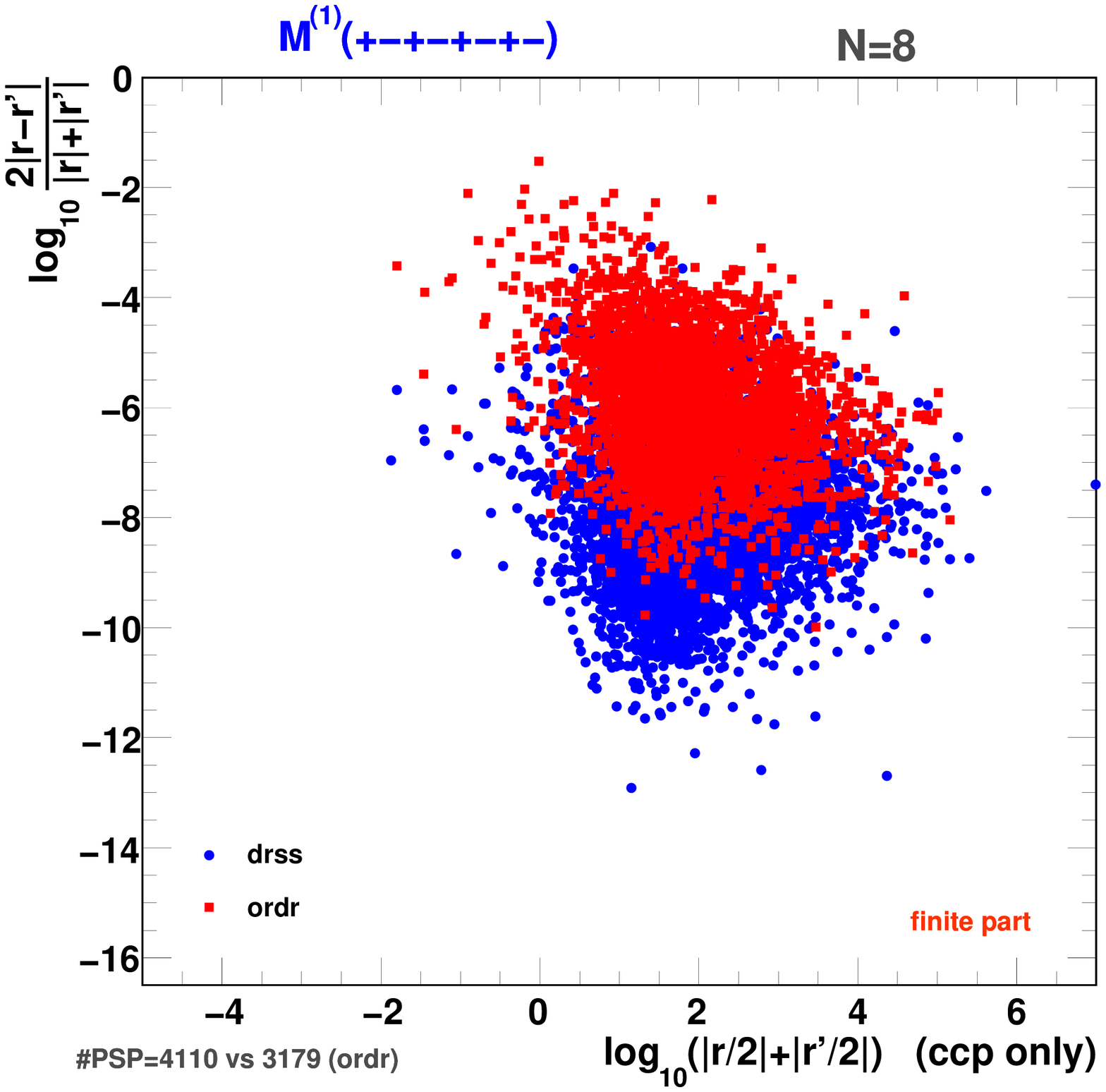}}
\caption{Double-, single-pole and finite-part accuracy distributions
  (upper part) and scatter graphs (lower part) extracted from
  double-precision computations of one-loop amplitudes for $n={\tt N}=8$
  gluons with polarizations $\lambda_k=+-+-+-+-$ and randomly chosen
  non-zero color configurations. The virtual corrections were
  calculated at random phase-space points satisfying the cuts detailed
  in the text. Unstable solutions were vetoed. Results from the
  color-dressed algorithm are compared with those of the color-ordered
  method indicated by dashed curves and brighter dots in the plots.
  The 5(4)-dimensional case is shown in the top left (right) and
  center (bottom) part of the figure. The definitions of
  $\varepsilon$\/ and $r$\/ are given in the text; the number of
  points contained by each scatter graph is found in the lower left.}
\label{Fig:accs8}
\end{figure}

\begin{figure}[p!]
\centerline{
  \includegraphics[width=0.51\columnwidth]{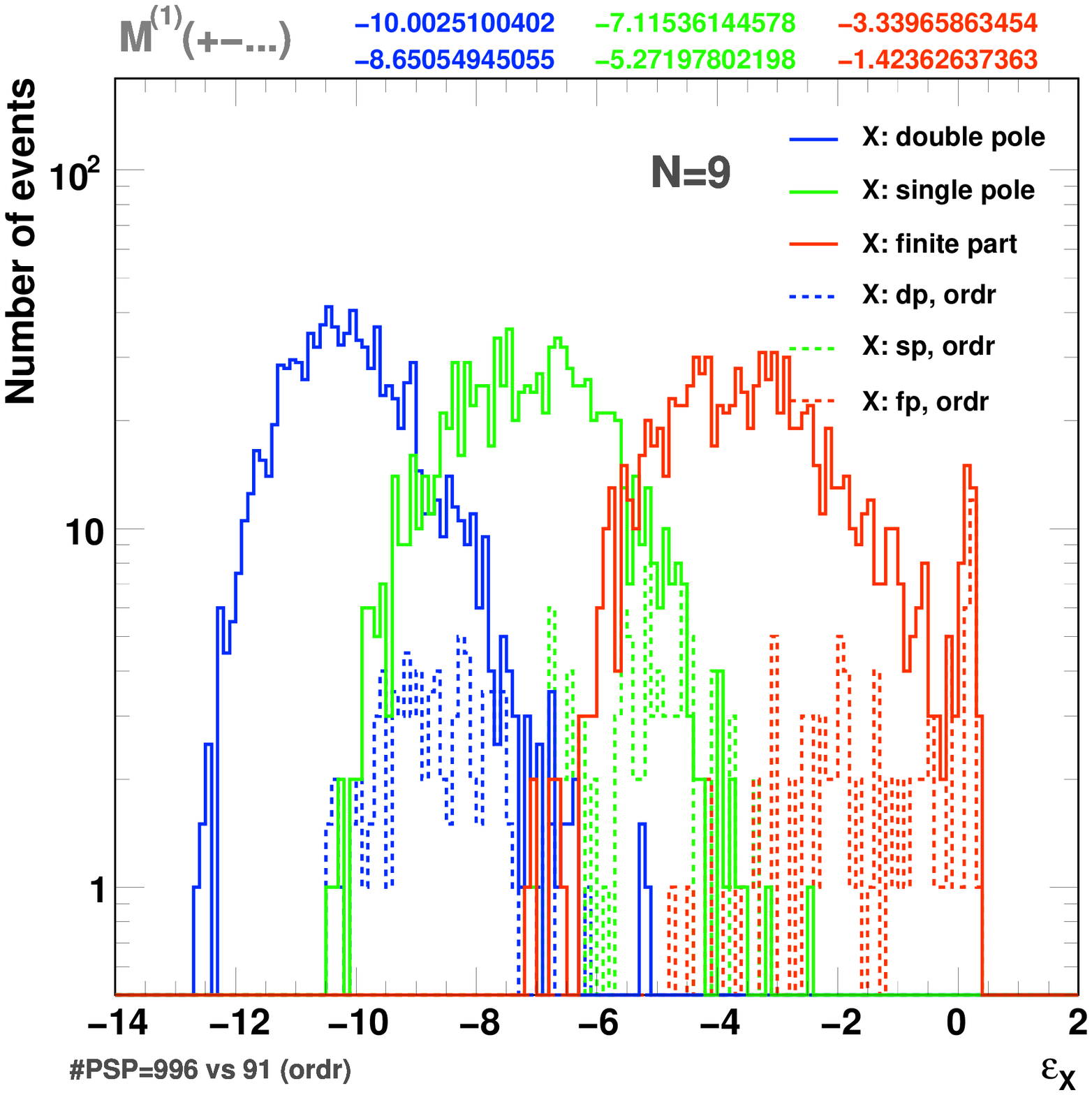}
  \includegraphics[width=0.51\columnwidth]{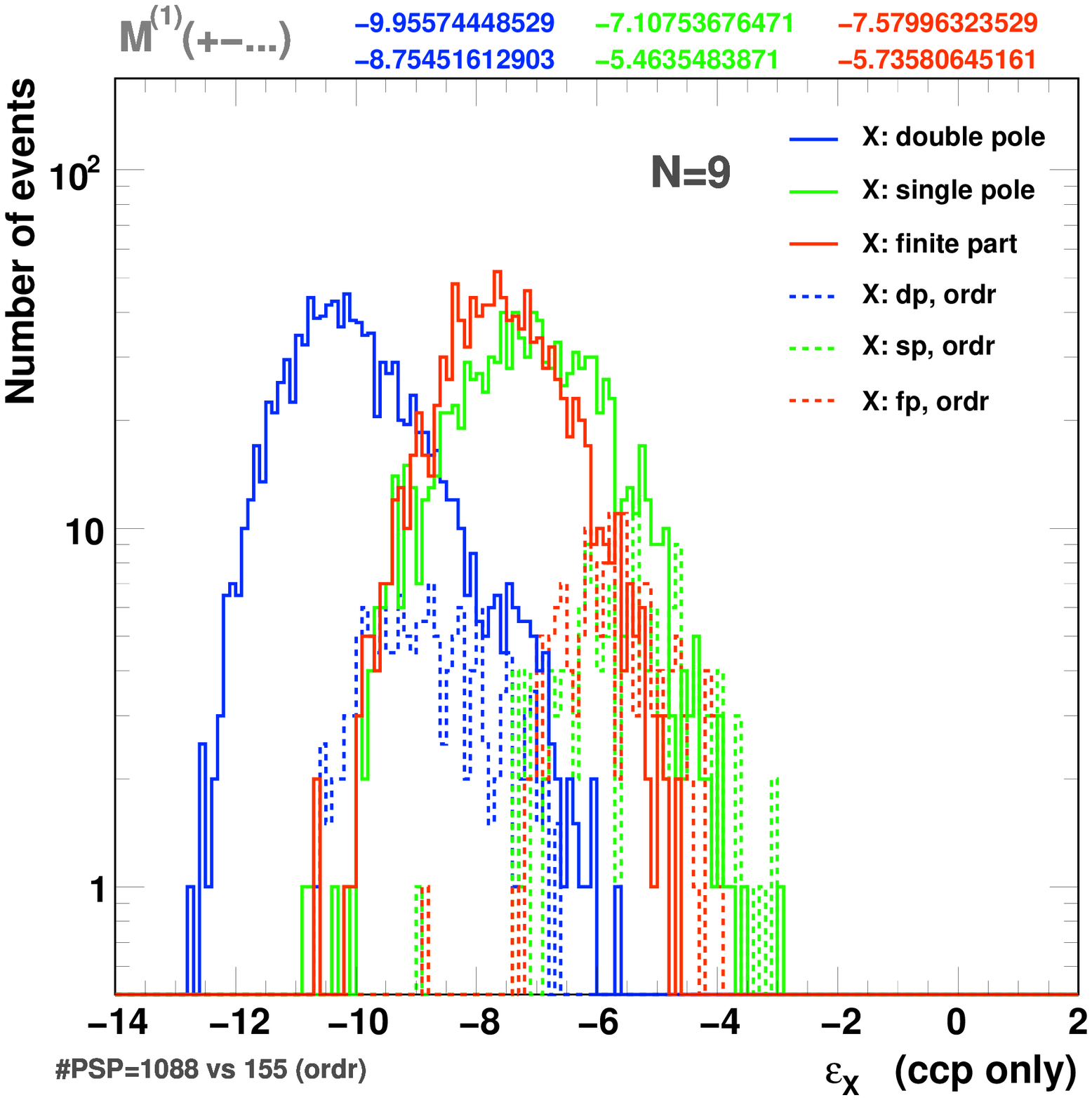}}
\centerline{
  \includegraphics[width=0.39\columnwidth]{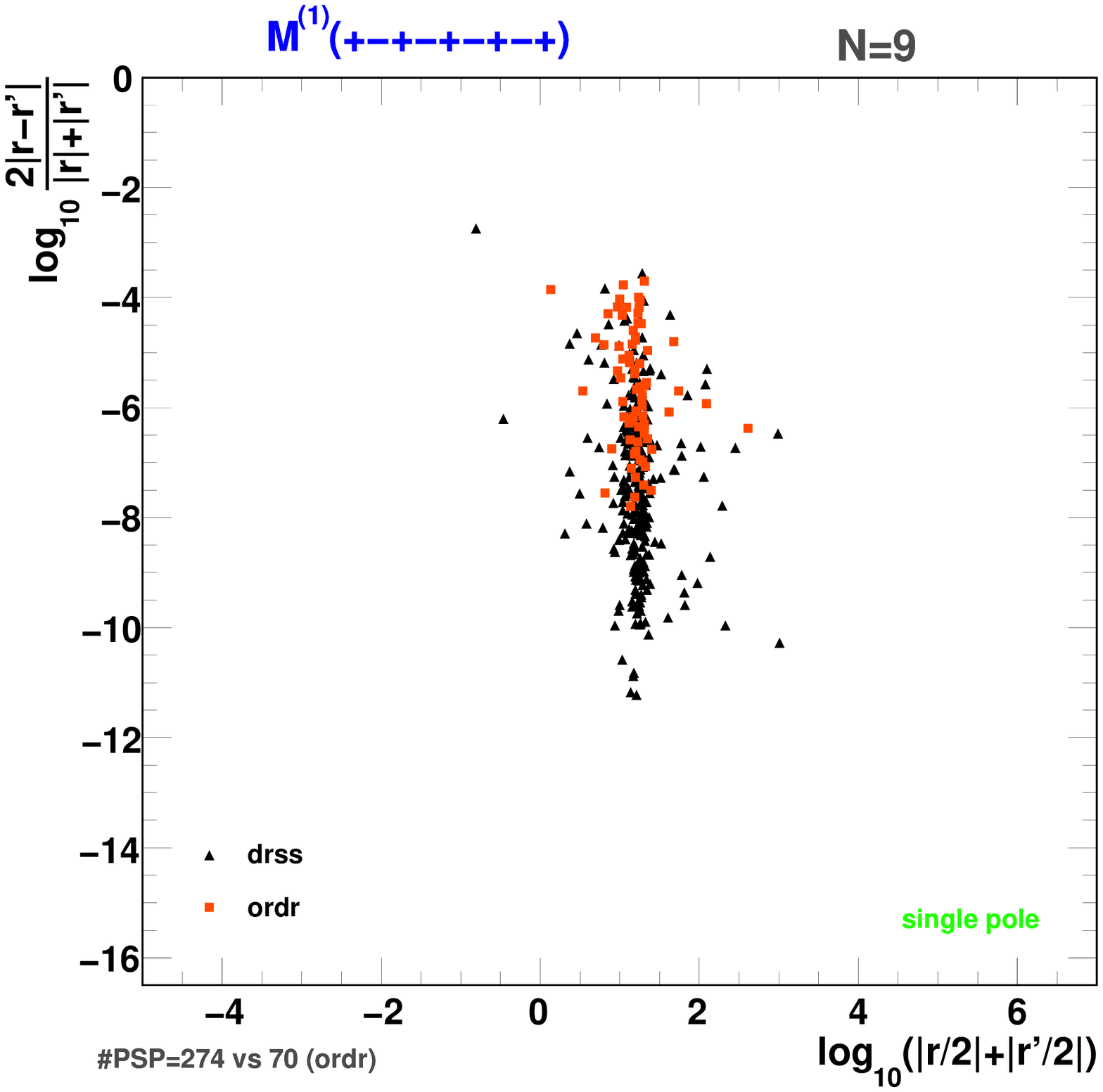}
  \includegraphics[width=0.39\columnwidth]{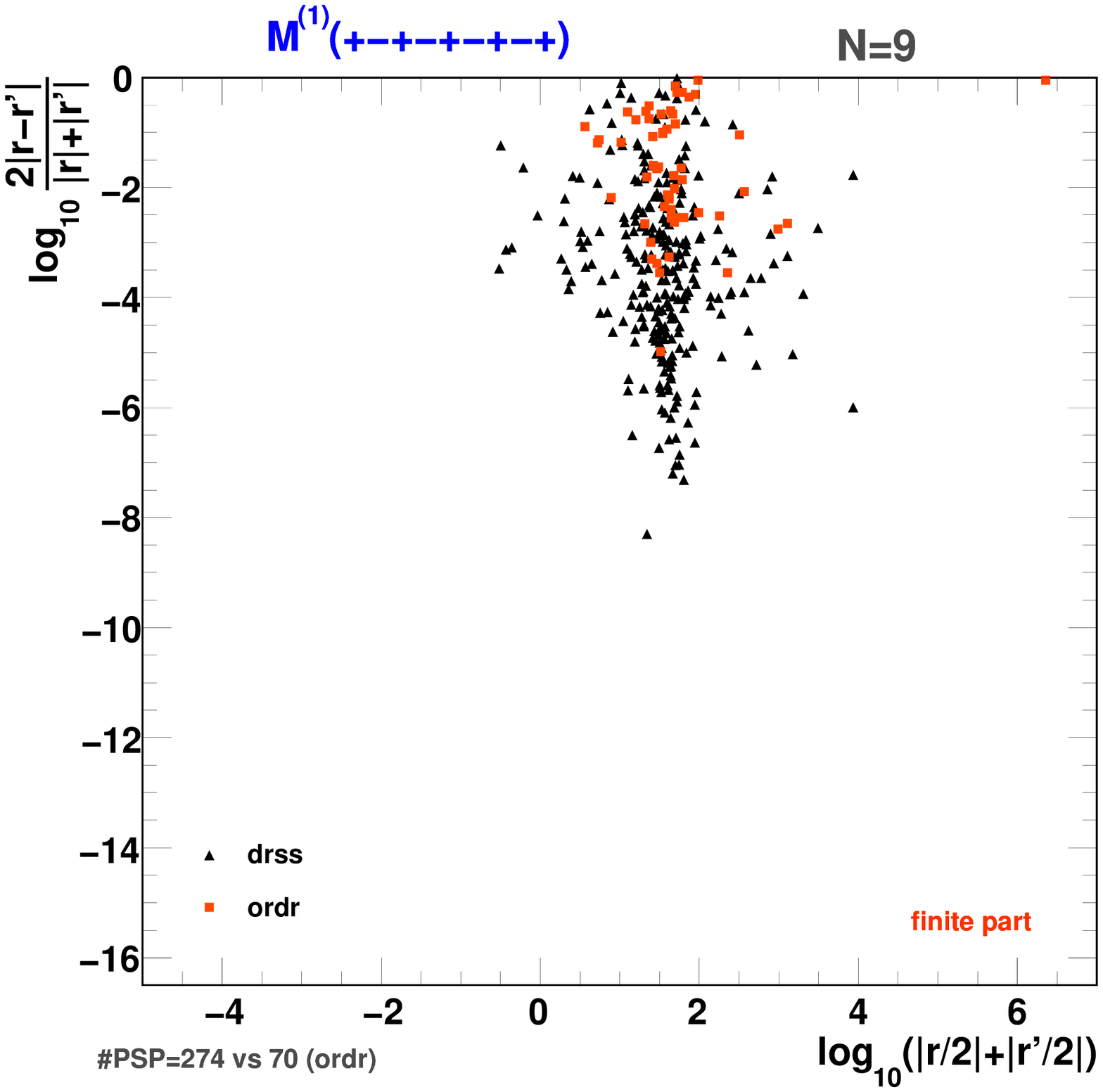}}
\centerline{
  \includegraphics[width=0.39\columnwidth]{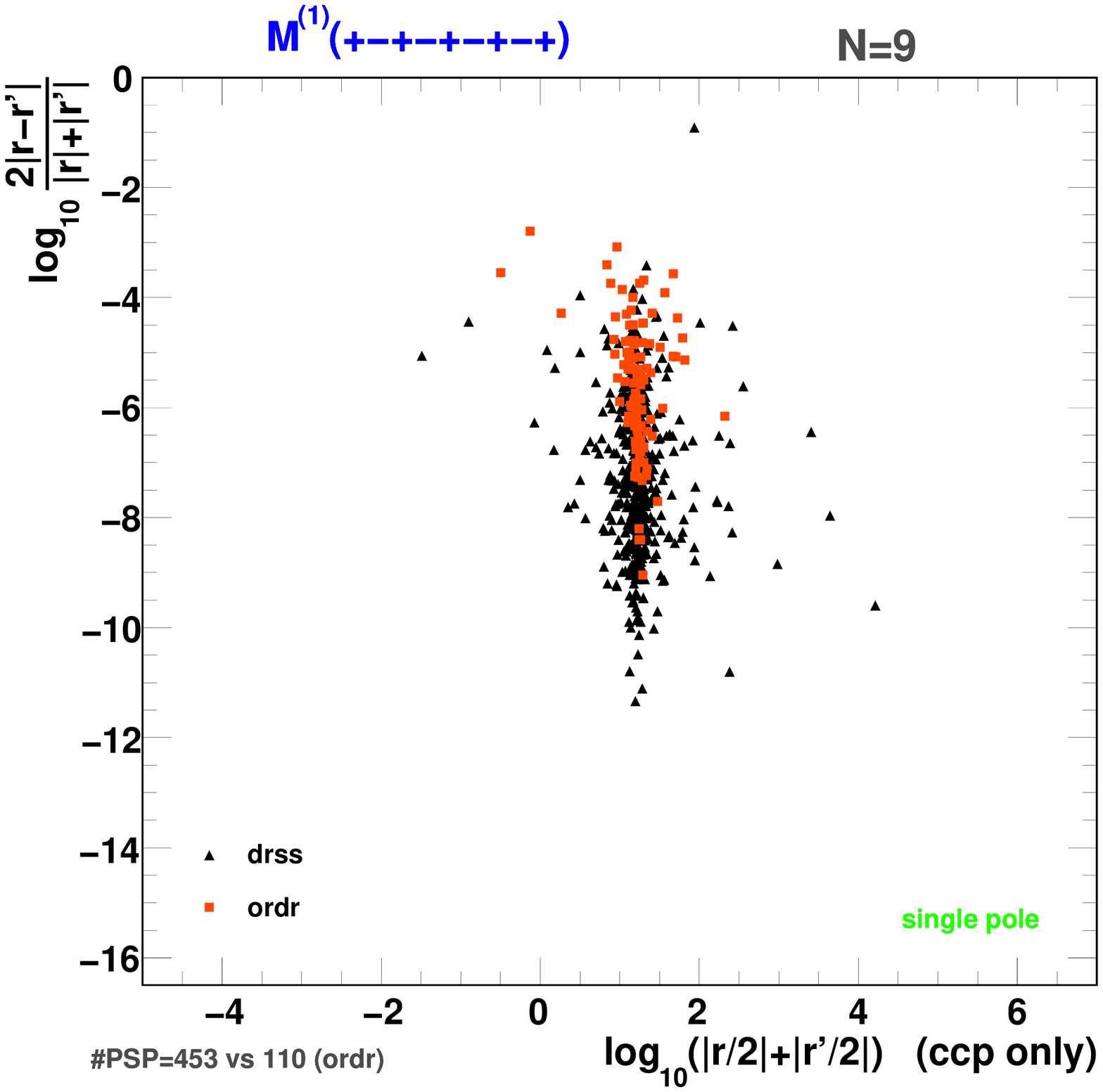}
  \includegraphics[width=0.39\columnwidth]{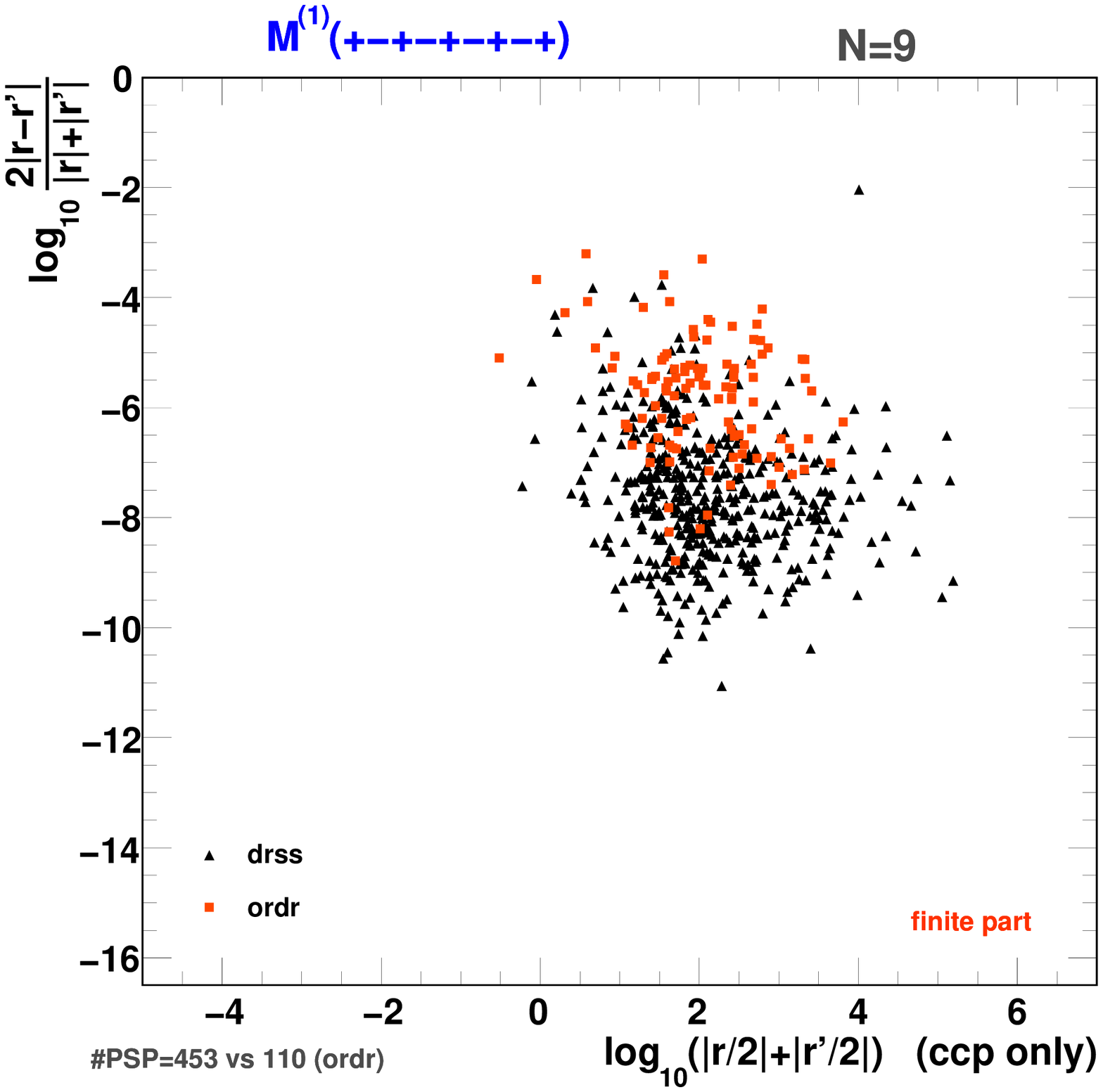}}
\caption{Double-, single-pole and finite-part accuracy distributions
  (upper part) and scatter graphs (lower part) extracted from
  double-precision computations of one-loop amplitudes for $n={\tt N}=9$
  gluons with polarizations $\lambda_k=+-+-+-+-+$ and randomly chosen
  non-zero color configurations. The virtual corrections were
  calculated at random phase-space points satisfying the cuts detailed
  in the text. Unstable solutions were vetoed. Results from the
  color-dressed algorithm are compared with those of the color-ordered
  method indicated by dashed curves and brighter dots in the plots.
  The 5(4)-dimensional case is shown in the top left (right) and
  center (bottom) part of the figure. The definitions of
  $\varepsilon$\/ and $r$\/ are given in the text; the number of
  points contained by each scatter graph is found in the lower left.}
\label{Fig:accs9}
\end{figure}

\begin{figure}[t!]
\centerline{
  \includegraphics[width=0.39\columnwidth]{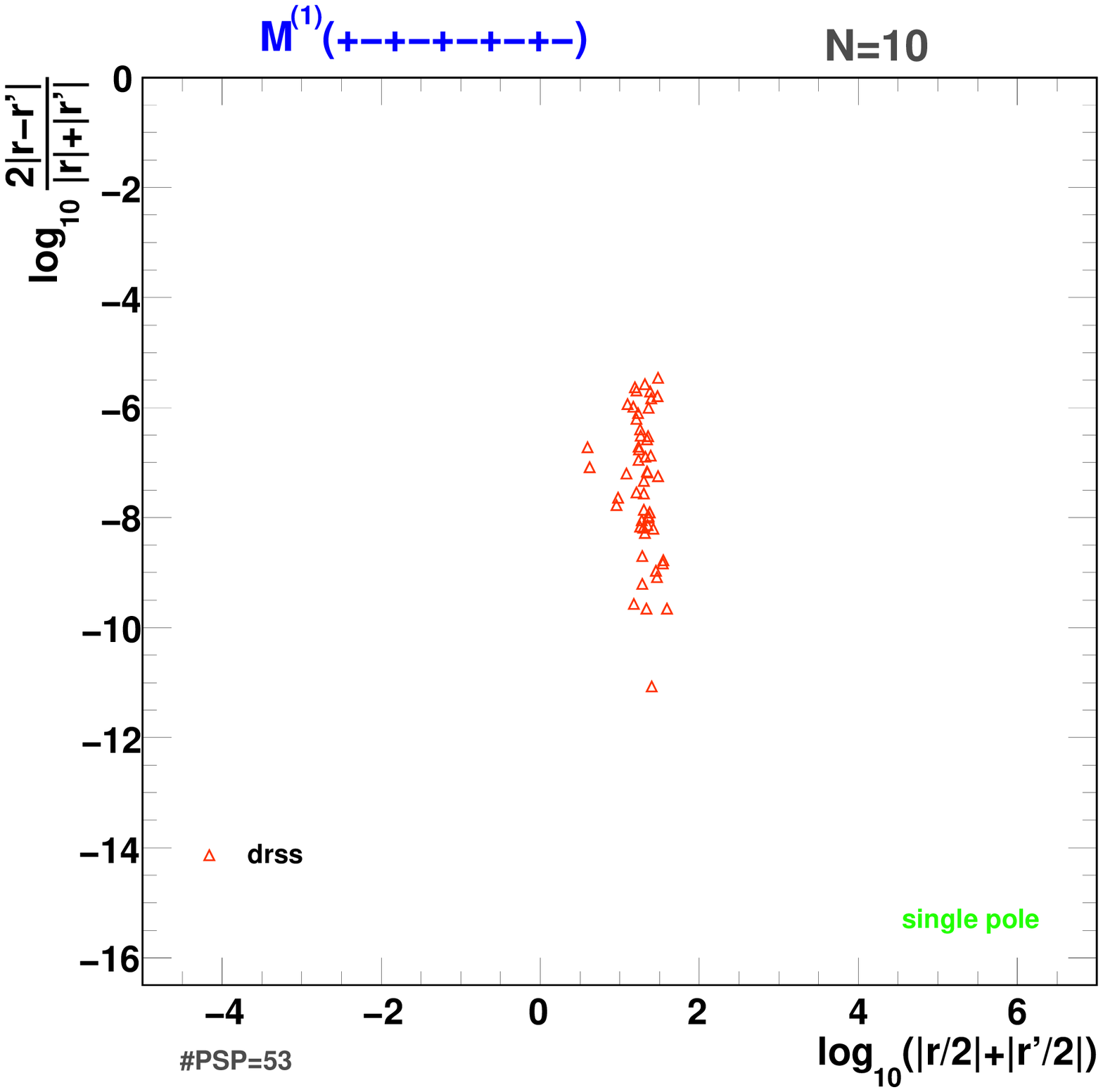}
  \includegraphics[width=0.39\columnwidth]{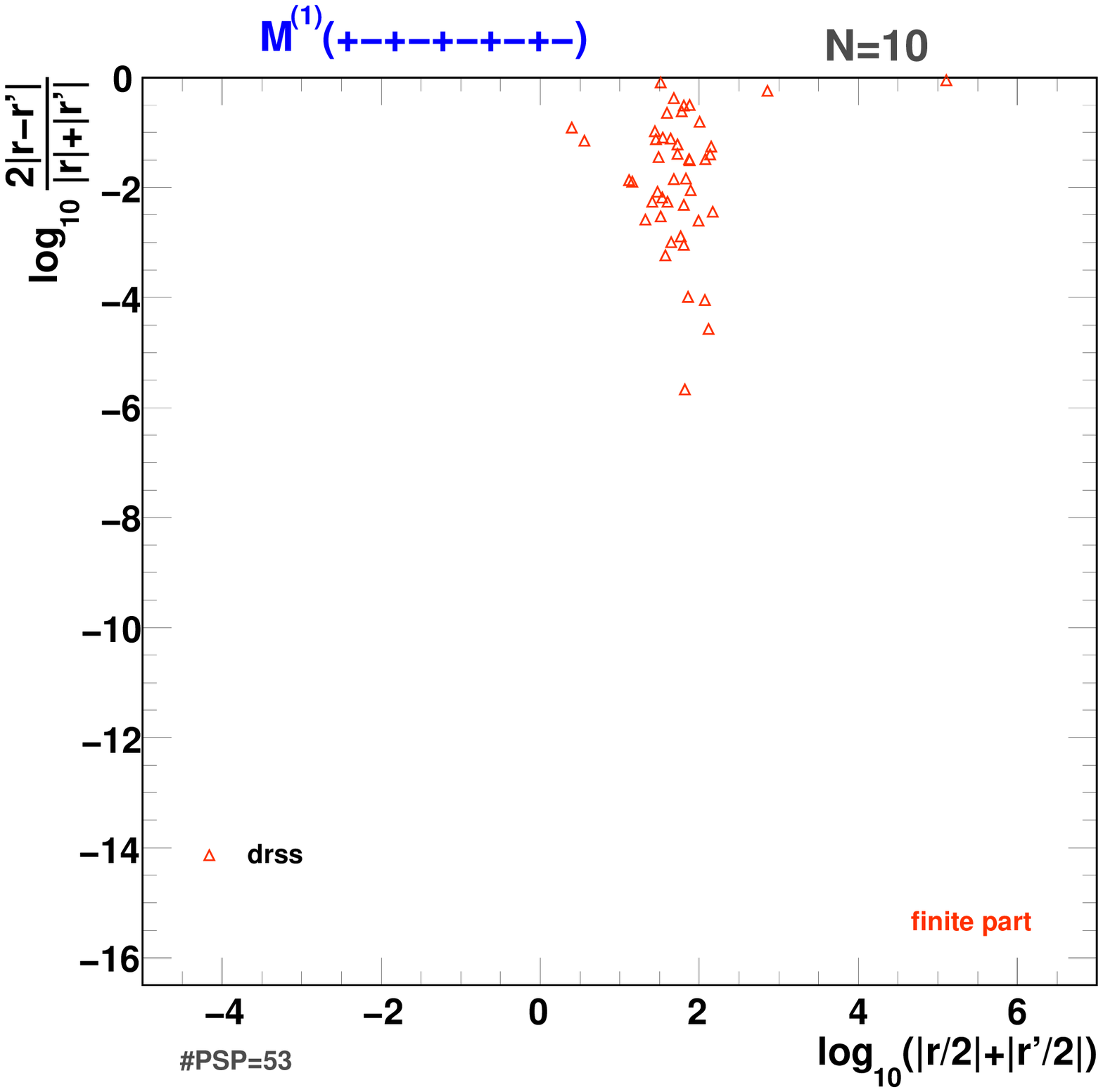}}
\centerline{
  \includegraphics[width=0.39\columnwidth]{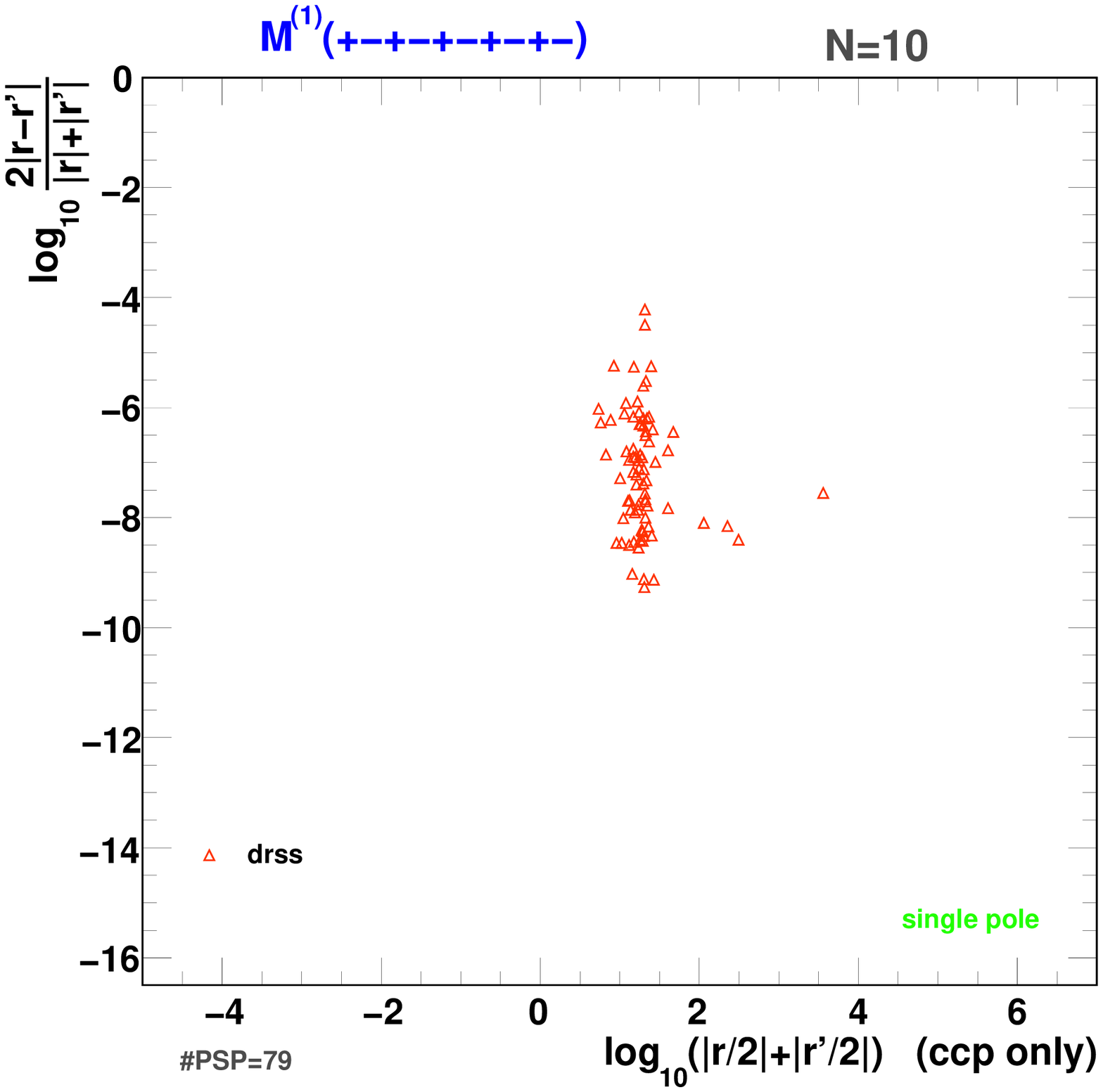}
  \includegraphics[width=0.39\columnwidth]{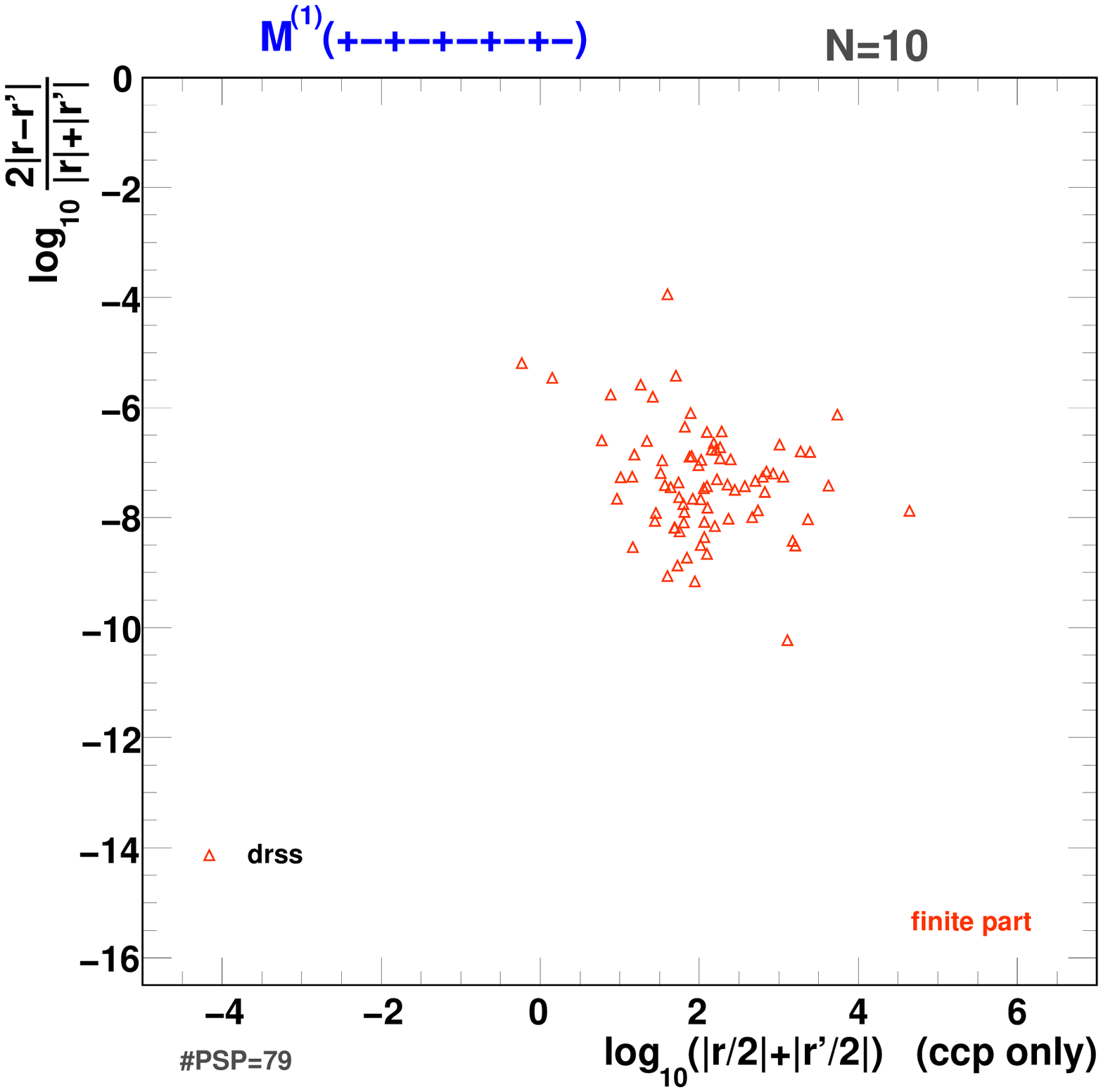}}
\caption{Single-pole and finite-part scatter graphs extracted
  from the double-precision computation of one-loop amplitudes for
  $n={\tt N}=10$ gluons with polarizations $\lambda_k=+-+-+-+-+-$ and
  randomly chosen non-zero color configurations. The virtual
  corrections were calculated at random phase-space points satisfying
  the cuts as described in the text. Unstable solutions were vetoed
  and, therefore, not included in the plots. The upper (lower) row of
  plots shows the results obtained from the 5(4)-dimensional
  color-dressed algorithm. For the definition of $r$, see text. The
  number of points contained by each scatter graph can be found in the
  lower left.}
\label{Fig:accs10}
\end{figure}

\begin{figure}[t!]
\vspace*{-2mm}
\centerline{
  \includegraphics[width=0.4\columnwidth,angle=-90]{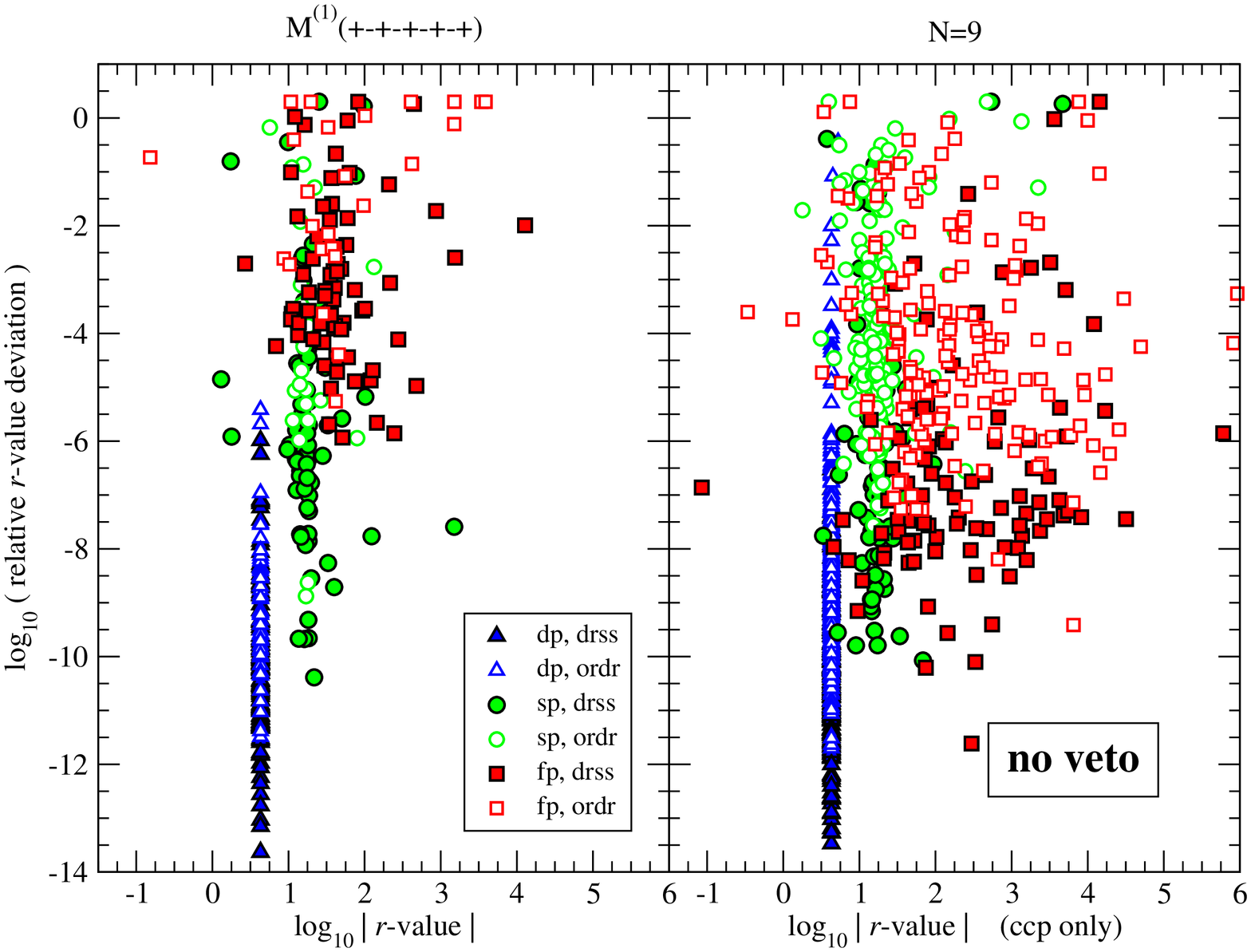}
  \includegraphics[width=0.4\columnwidth,angle=-90]{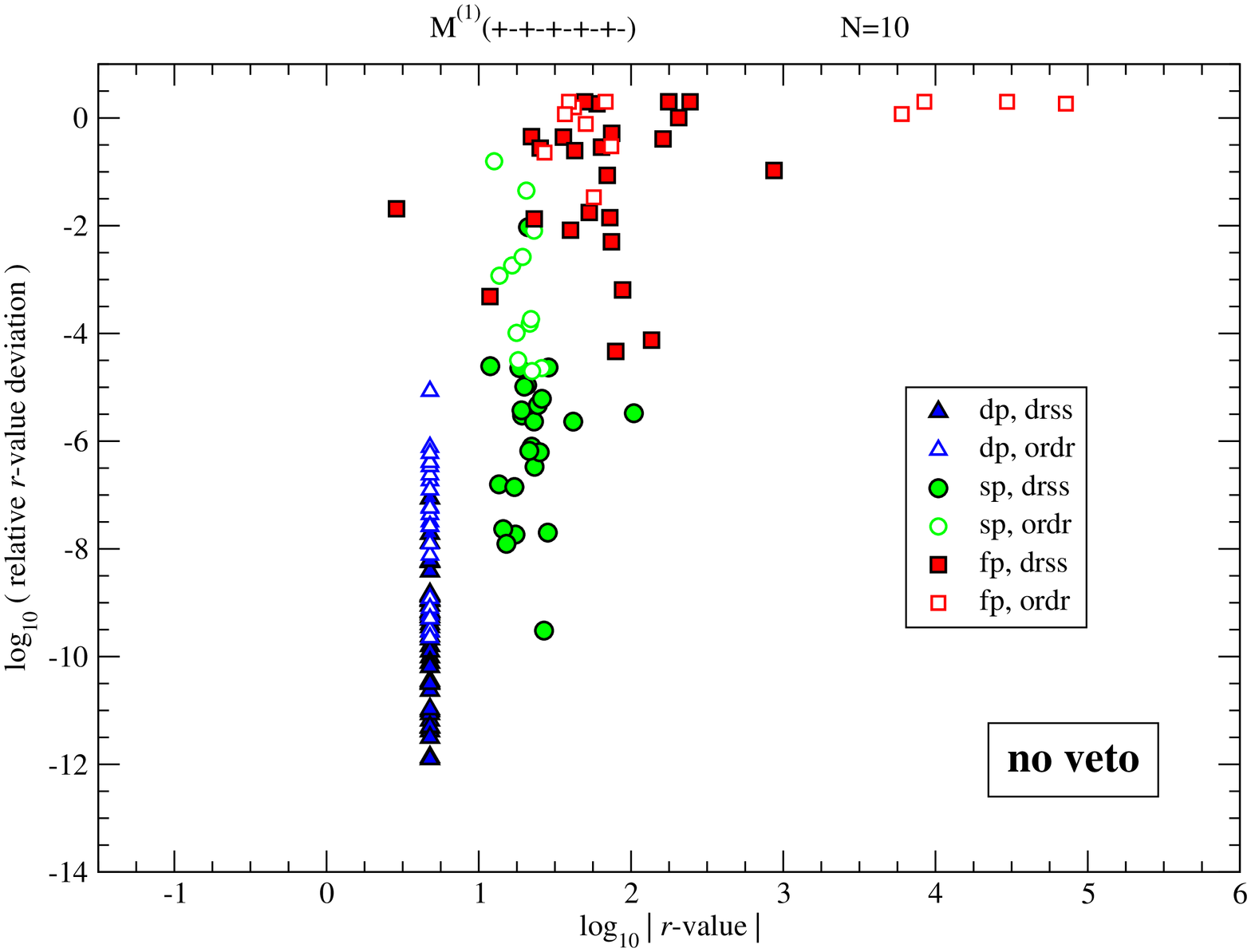}}
\caption{Double-, single-pole and finite-part scatter graphs
  visualizing the accuracy of double-precision evaluations of one-loop
  amplitudes for $n={\tt N}=9$ and $10$ gluons of alternating
  polarizations. Non-zero randomly chosen color
  configurations were used. Note that unstable solutions were not
  vetoed and therefore included in this presentation. The virtual
  corrections were calculated at random phase-space points satisfying
  the cuts as described in the text. Results of the 5-dimensional
  algorithms either based on color ordering (ordr) or color dressing
  (drss) are shown; for $n={\tt N}=9$, the ``4D-case'' results are
  also given (ccp only). For the definition of $r$, see text; axis
  labels as used in Fig.~\ref{Fig:accs10} are understood. The
  rightmost graph contains ${\cal O}(20)$ points per $\epsilon$-pole,
  while the left plot of the ``5(4)D-case'' has approximately
  $50$($120$) points per pole.}
\label{Fig:scttzv}
\end{figure}

In the upper part of Figs.~\ref{Fig:accs4}-\ref{Fig:accs9} we show
the distributions of relative accuracies $\varepsilon$\/ as occurring
in the evaluation of gluon loop corrections with $n=4,\ldots,9$
external gluons. The lower part of these figures and
Figs.~\ref{Fig:accs10} and \ref{Fig:scttzv} themselves depict scatter
graphs visualizing the relative accuracies as a function of the size
of the virtual corrections for the single pole and finite contributions
only, as the double pole contribution has no observable variance. 
This form of presenting the results has
information on whether certain points dominate the uncertainty of the
total correction when averaging over the phase space. The $r$-variables
used in these plots are defined by
\beq
r\;=\;\frac{1}{2\,\pi}\;
      \frac{\Re\left({{\cal M}^{(0)}}^\dagger{\cal M}^{(1)}\right)}
	   {\left|\,{\cal M}^{(0)}\right|^2}
\eeq
and represent corrections of the order of $\alpha_s$. Specifically,
the $r$, $r'$ and $r_{\rm th}$ given in the plots are obtained by
employing ${\cal M}^{(1)}={\cal M}^{(1)[1]}_{\rm d/s/fp,num}$,
${\cal M}^{(1)}={\cal M}^{(1)[2]}_{\rm s/fp,num}$ and
${\cal M}^{(1)}={\cal M}^{(1)}_{\rm dp,th}$, respectively. In all
cases we have rejected events with unreliable basis vectors in
orthogonal space. Except for the results presented in
Fig.~\ref{Fig:scttzv}, we have vetoed all events that led to unstable
solutions of the bubble master-integral coefficient using
$\Delta_{\rm veto}=0.02$. The statistics concerning these rejections
is shown in Table~\ref{Tab:vetoeffects}.

\begin{figure}[t!]
\centerline{
  \includegraphics[width=0.34\columnwidth]{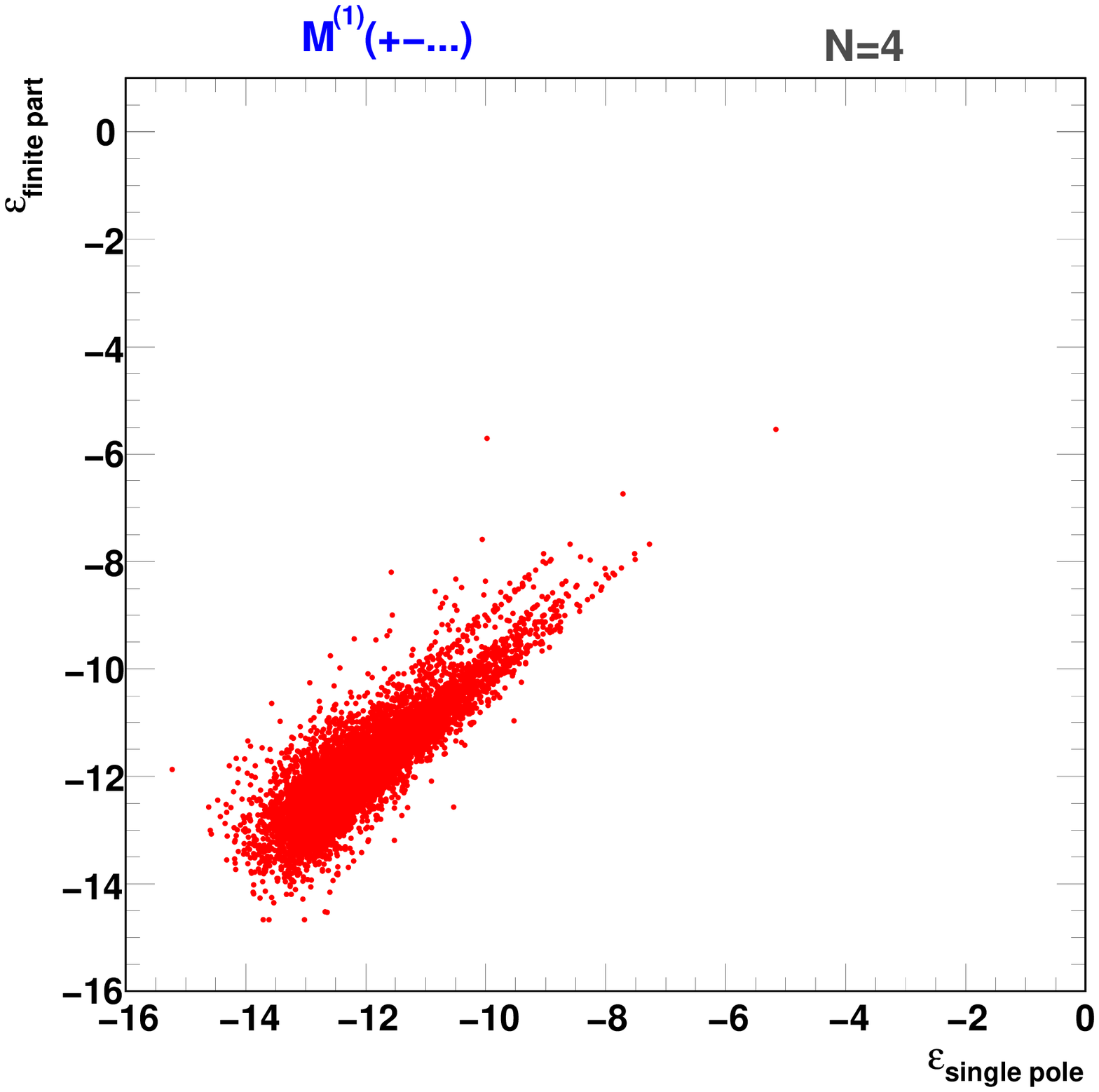}
  \includegraphics[width=0.34\columnwidth]{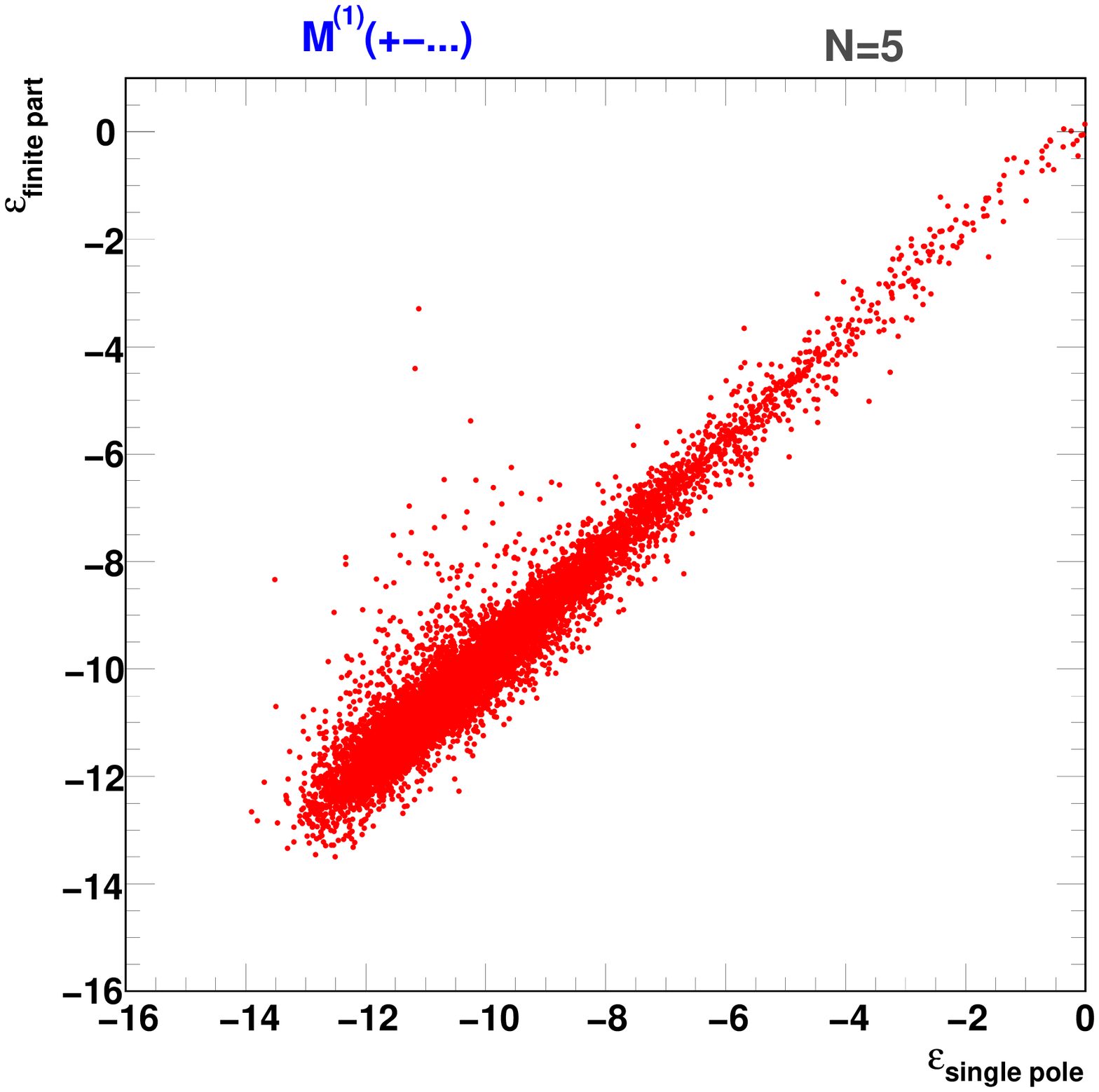}
  \includegraphics[width=0.34\columnwidth]{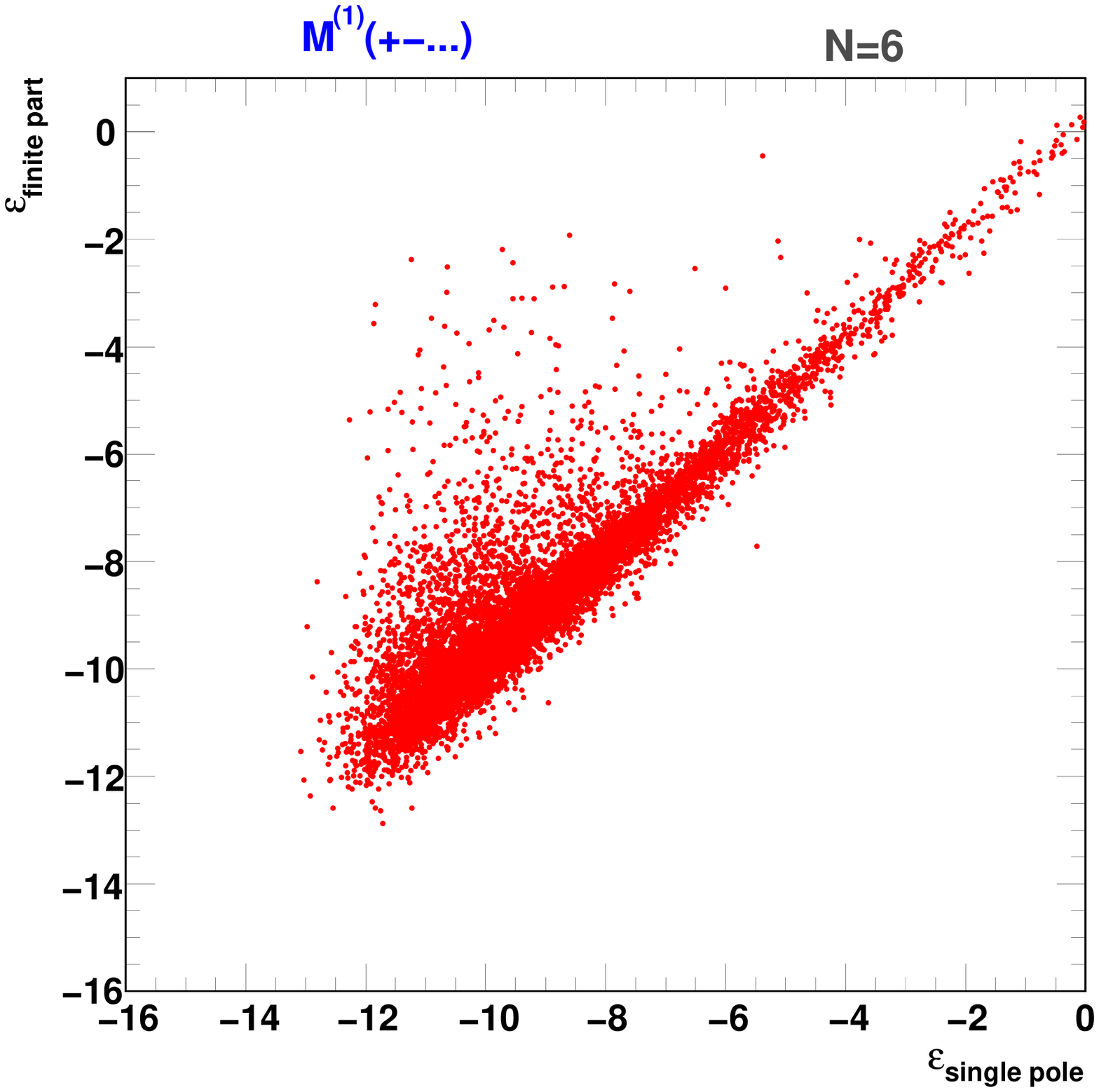}}
\centerline{
  \includegraphics[width=0.34\columnwidth]{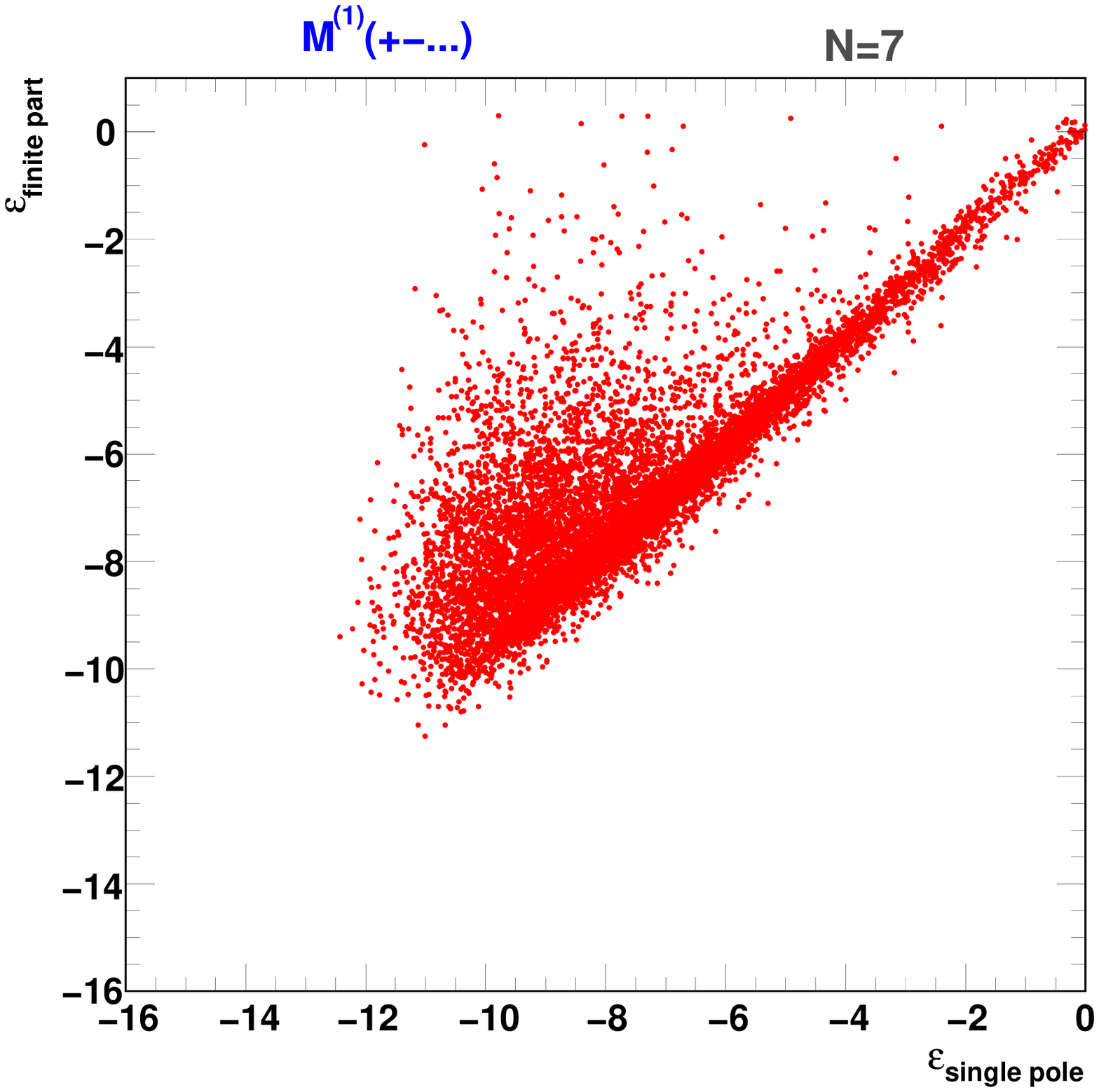}
  \includegraphics[width=0.34\columnwidth]{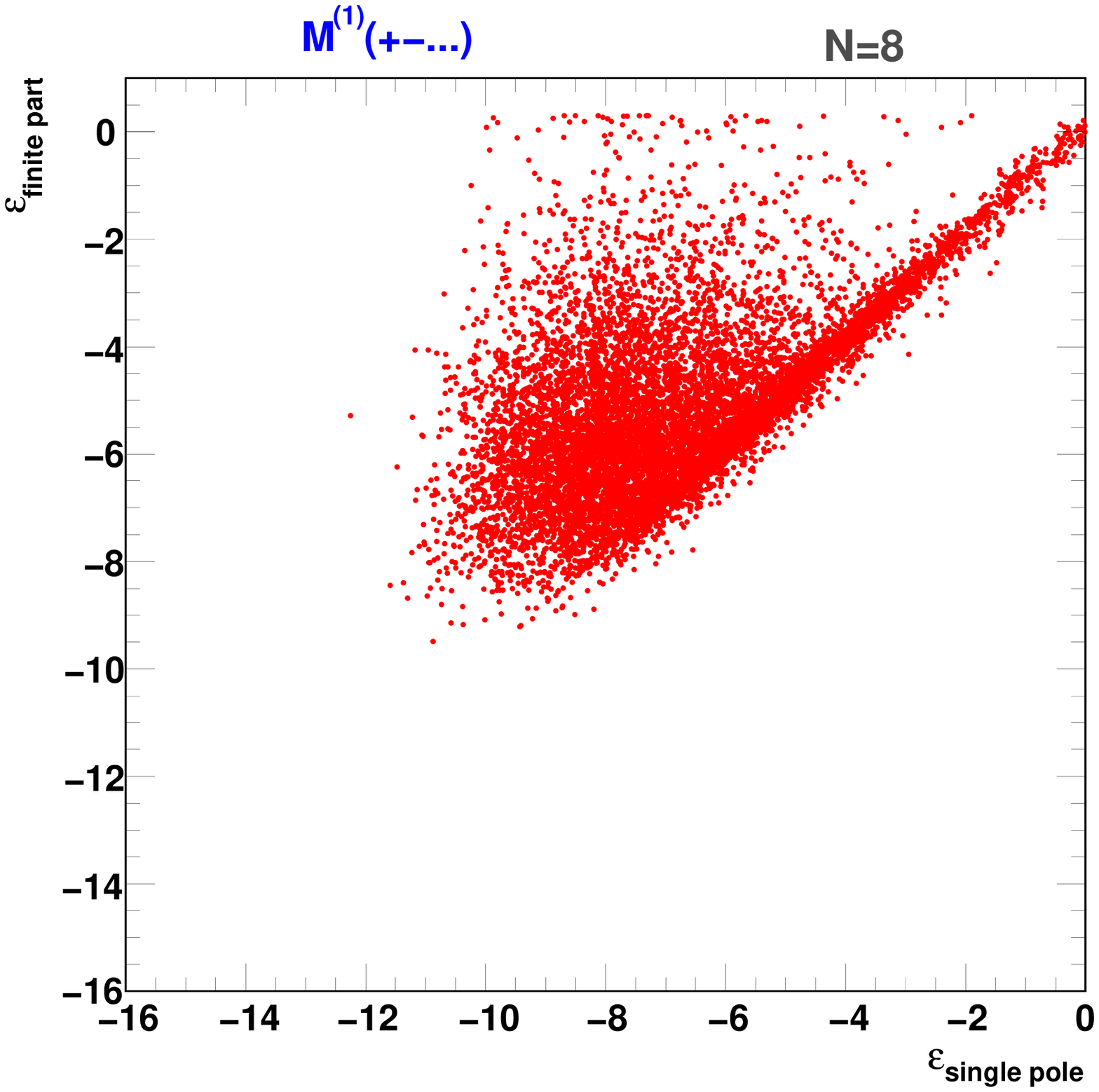}}
\centerline{
  \includegraphics[width=0.34\columnwidth]{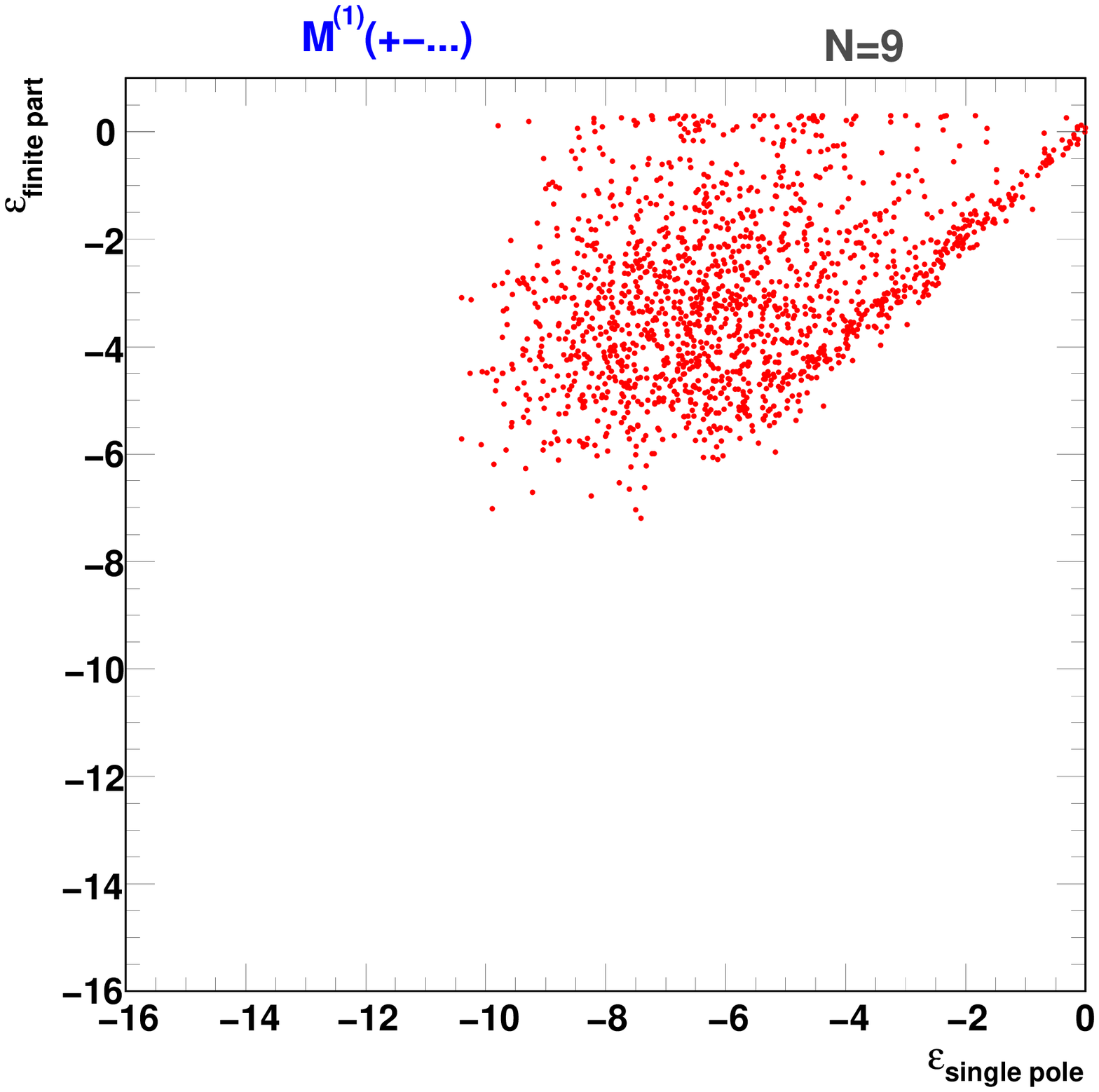}
  \includegraphics[width=0.34\columnwidth]{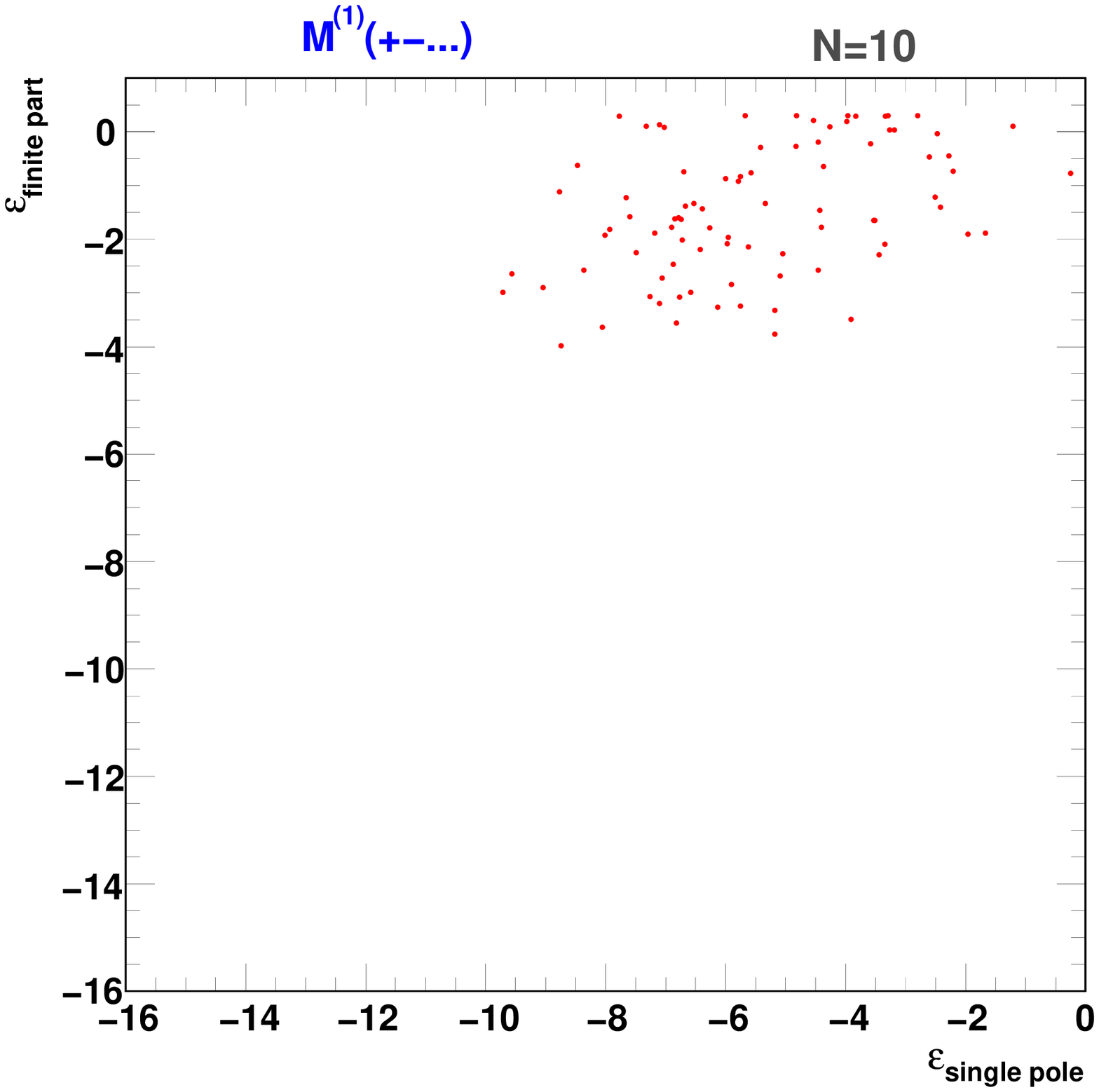}}
\caption{Finite-part versus single-pole accuracy (in double precision)
  as achieved in one-loop amplitude calculations using the
  color-dressed approach for various numbers $n={\tt N}$ of external
  gluons with polarizations $\lambda_k=+-\ldots+-(+)$ and colors
  randomly chosen among non-zero configurations. Note that unstable
  solutions have not been vetoed. The $n={\tt N}=9$ and $n={\tt N}=10$
  graphs only contain $1.6\cdot 10^3$ and $87$ points, respectively,
  whereas all other plots comprise $10^4$ points.}
\label{Fig:fpvssp}
\end{figure}

We compare in all plots of Figs.~\ref{Fig:accs4}-\ref{Fig:accs9} the
color-dressed with the color-ordered approach where the results of the
latter are indicated by dashed curves in the spectra (with the
$\langle\varepsilon\rangle$ given by the lower top row of numbers) and
brighter points in the scatter graphs. The $\varepsilon$\/ spectra of
the ``5D-case'' (``4D-case'') are always shown in the top left (right)
parts of the figures; the associated scatter graphs are compiled in
the center (bottom) parts. In Fig.~\ref{Fig:accs10} we present our
results for $n=10$ gluons where for reasons of limited statistics we
solely show the scatter graphs related to the dressed method. The veto
procedure has a very strong impact on ${\cal M}_{9,10}^{(1)}$
calculations. For the purpose of direct comparisons between vetoed and
non-vetoed samples, we have added in Fig.~\ref{Fig:scttzv} scatter plots
that include vetoed events.

In all cases we notice that the double poles are obtained very
accurately with almost no loss in precision for increasing number of
gluons. The $n$-dependence of the single-pole and finite-part
precisions is not as stable as for the double pole. We see noticeable
shifts of the peak and mean positions towards larger values when
incrementing the number of external gluons. The distribution's tails
are under good control. Because of the introduced veto procedure, they
quickly die off around $\varepsilon\approx-2$. In rare cases worse
accuracies occur, which happens more frequently for the 5-dimensional
calculations. We can avoid these cases, if we extend the veto
criteria by re-solving for and testing the rational bubble coefficient
as well. For $n>9$, the limitations of double-precision computations
unavoidably lead to rather unreliable single-pole and finite-part
determinations. As an interesting fact, we observe that the
color-dressed method yields throughout results of higher precision.
Moreover, the decrease in accuracy for growing $n$\/
is more moderate compared to the method based on
color ordering. Clearly, on the one hand this algorithm has to be run
for many orderings and may therefore lead to an accumulation of small
imprecisions. On the other hand a rather inaccurate determination of
$m^{(1)}$ may appear just for a single ordering, in turn spoiling the
overall result. Both effects make the ordered approach less capable of
delivering accurate results. Turning to the scatter plots, we find
that the most accurate but also inaccurate evaluations occur for
points distributed near the vertical line of ${\cal O}(1)$
corrections. It is very encouraging that all top right quadrants are
rather sparsely populated, dispelling the doubts that insufficiently
determined large corrections may dominate our final results. The
scatter regions of the double-pole solutions remain almost unchanged
for larger $n$, while those of the single poles and finite parts are
slightly growing gradually shifting towards lower relative accuracies.
The scatter patches of the dressed method are displaced with respect
to those of the color-decomposition approach: advantageously, they
cover regions of greater precision, in particular populate the bottom
right quadrants more densely. Due to the simplicity of the 4-gluon
kinematics, the case of $n=4$ gluons stands out from the rest: the
single pole and finite part can be obtained with almost the same
accuracy as the double pole. This feature is preserved even if
rational-part calculations are included. With $5$ gluons or more it is
common that all coefficients contribute to the decomposition of the
one-loop amplitude. The relative accuracies of the single poles and
finite parts therefore develop a much different, less steeper, tail
compared to the double poles. There are almost no differences between
the double- and single-pole results obtained from the 4- and
5-dimensional algorithms. This is no surprise, since the coefficients
necessary to reconstruct these poles can be determined in $4$
dimensions and our algorithms have been set up accordingly. In the
absence of rational-part calculations it turns out that the finite
parts may on average be obtained slightly more precisely than the
single poles. The tails of the $1/\epsilon$\/ spectra reach out to the
largest $\varepsilon$-values occurring in the evaluation of the
cut-constructible part. The behavior is reversed in the 5-dimensional
case owing to the addition of the rational part. For the same reason,
we note increased $\langle\varepsilon_{\rm fp}\rangle$ in the
``5D-case'', furthermore, the 5-dimensional scatter graphs show higher
densities with respect to the 4-dimensional ones at lower accuracies.

As a special case of Fig.~\ref{Fig:accs6} we have displayed in
Fig.~\ref{Fig:accs6fxcl} accuracy distributions and scatter plots for
$n=6$ gluons of polarizations $\lambda_k=++----$ when instead of
random color-space points the fixed non-zero color configuration
$(ij)_k=(12)(21)(13)(31)(11)(22)$ has been selected. We notice that
all $\varepsilon$-spectra are shifted towards smaller accuracies.
Also, as illustrated by the scatter graphs, the magnitude of the
virtual corrections is bound at ${\cal O}(1)$ with the exception of
the finite piece of the cut-constructible part of the one-loop
amplitudes. Interestingly, this is corrected back by adding in the
rational part.

In Ref.~\cite{Winter:2009kd} it was shown that the finite-part
accuracy of the evaluation of ordered amplitudes is mostly correlated
with that of the single poles. We have studied this issue for the
dressed algorithm in the ``5D-case''. The corresponding scatter plots
also include the vetoed events and are presented in Fig.~\ref{Fig:fpvssp}.
The multitude of points is distributed along the diagonal indicating a
strong correlation. As for color-ordered amplitudes the evaluation of
the rational part becomes more involved with increasing gluon numbers.
Therefore, regions of lower finite-part precision start to get
populated distorting the diagonal trend.

Finally, we want to show that the Monte Carlo sampling as defined in
Eq.~(\ref{MCvirtual}) converges sufficiently fast for the
color-dressed calculated virtual corrections. To this end we
generalize the LO discussion following Eq.~(\ref{convergenceLO}) with the
details given in Sec.~\ref{Sec:numtree}. The relevant quantity to
explore in the Monte Carlo averaging is
\beq\label{Saverage}
S^{(0+1)}_{\rm MC}\;=\;
\frac{1}{N_{\rm colpts}}\sum_{k=1}^{N_{\rm colpts}}
W_{\rm col}(n_1,n_2,n_3)\times\left[\,
\left|\,{\cal M}^{(0)}_k\right|^2+\,\frac{\widehat\alpha_s}{2\,\pi}\,
\Re\left({\cal M}_{{\rm fp},k}^{(1)}\,{{\cal M}_k^{(0)}}^\dagger\right)\,\right]
\eeq
where we choose $\widehat\alpha_S=0.12$ and ${\cal M}_{{\rm fp},k}^{(1)}$
is the finite part of the virtual corrections. The sum over the
$N_{\rm colpts}$ color configurations for each phase-space point is an
optional ``mini-Monte Carlo'' over colors for faster convergence as a
function of the number of phase-space point evaluations. By adding the
real corrections to Eq.~(\ref{Saverage}) and performing the coupling
constant renormalization and mass factorization, one obtains the
gluonic contribution to the NLO multi-jet differential cross section.
Therefore, the convergence of Eq.~(\ref{Saverage}) is the relevant
quantity to study.

\begin{figure}[t!]
\psfrag{Yconvergelabel}[b][c][0.84]{
  $\left(\left\langle S^{(0+1)}_{\rm MC}\right\rangle\pm
  \sigma_{\left\langle S^{(0+1)}_{\rm MC}\right\rangle}\right)\;\left\langle\,
  S^{(0+1)}_{\rm col}\right\rangle^{-1}$}
\psfrag{Xconvergelabel}[b][b][0.44]{$
  100\%\times\sigma\Big(R^{(0+1)}_{\rm MC}(N_{\rm MC})\Big)\Big/
  \mu\Big(R^{(0+1)}_{\rm MC}(N_{\rm MC})\Big)
  $}
\vspace*{-4mm}
\centerline{
  \includegraphics[width=0.64\columnwidth,angle=-90]{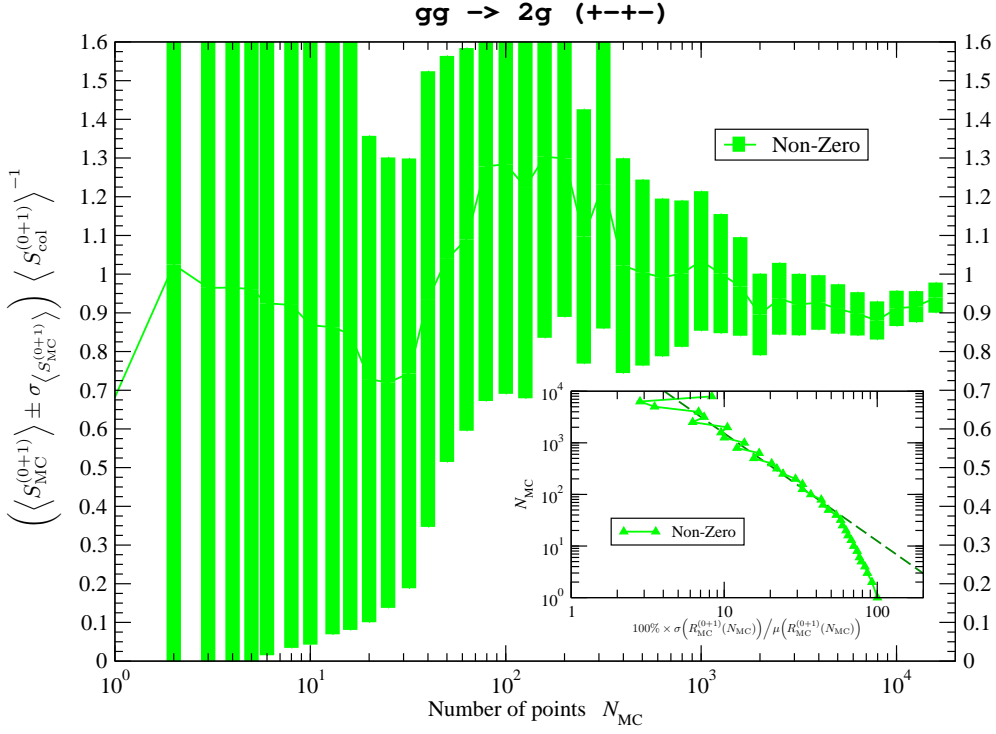}}
\caption{Consistency test for the Monte Carlo integration of 4-gluon
  virtual corrections using ``Non-Zero'' color sampling compared to
  the exact color summing. As a function of the number $N_{\rm MC}$ of
  evaluated phase-space points the $R^{(0+1)}$-ratio is plotted
  converging to one as it should be. The inserted plot shows the
  number of phase-space evaluations needed to reach a given relative
  accuracy in terms of $R^{(0+1)}_{\rm MC}(N_{\rm MC})$ while Monte
  Carlo integrating; for the definitions, see text. The dashed line
  depicts the fit function $\sigma/\mu=A\,N_{\rm MC}^{-B}$, see also
  Table~\ref{Tab:MCfits}.}
\label{Fig:conv4LL}
\end{figure}

By defining the $n$-gluon color-summed counterpart of $S^{(0+1)}_{\rm MC}$,
\beq\label{Scolaverage}
S_{\rm col}^{(0+1)}\;=\sum_{i_1,\ldots,i_n=1}^3\ \sum_{j_1,\ldots,j_n=1}^3
\left[\,
\left|\,{\cal M}_k^{(0)}\right|^2+\,\frac{\widehat\alpha_s}{2\,\pi}\,
\Re\left({\cal M}_{{\rm fp},k}^{(1)}\,{{\cal M}_k^{(0)}}^\dagger\right)
\,\right]\ ,
\eeq
we can form the ratios
\beq\label{Eq:Rvirtual}
R^{(0+1)}\;=\;\frac{\langle S^{(0+1)}_{\rm MC}\rangle\,\pm\,
  \sigma_{\langle S^{(0+1)}_{\rm MC}\rangle}}
{\langle S^{(0+1)}_{\rm col}\rangle}\ ,\qquad
R^{({\rm V})}\;=\;\frac{\langle S^{(0+1)}_{\rm MC}\rangle\,\pm\,
  \sigma_{\langle S^{(0+1)}_{\rm MC}\rangle}}
{\langle S^{(0)}_{\rm col}\rangle}
\eeq
analogously to Eq.~(\ref{Eq:Rtree}). We define the mean values and standard
deviations of the ratios similarly to Eqs.~(\ref{Eq:meanMC}) and
(\ref{Eq:sigmaMC}), respectively. Note that $S^{(0)}_{\rm col}$ is
already defined at LO by Eq.~(\ref{Eq:Scol}). As we increase the
number of Monte Carlo points, $N_{\rm MC}$, the $R^{({\rm V})}$-ratios
quantify the relative importance of the virtual corrections, while
the $R^{(0+1)}$-ratios should converge to one. For the latter, this is
nicely demonstrated in Fig.~\ref{Fig:conv4LL} for the 4-gluon
virtual corrections and the ``Non-zero'' sampling scheme as described
in Sec.~\ref{Sec:numtree}. After $15900$ events we obtain
$R^{(0+1)}=0.939\pm0.039$, which is satisfactory for this consistency
check.

As in the LO discussion we want to illustrate how many events are
needed to achieve a certain relative integration uncertainty when
performing the Monte Carlo color sampling. In analogy to
Eq.~(\ref{Eq:RMCtree}) we can construct the ratio
\beq
R^{(0+1)}_{\rm MC}(N_{\rm MC})\;=\;
\frac{\sum_{r=1}^{N_{\rm MC}}S^{(0+1)}_{{\rm MC},r}}
     {\sum_{r=1}^{N_{\rm MC}}S^{(0+1)}_{{\rm col},r}}
\eeq
as a function of $N_{\rm MC}$. Again, it is interesting to change the
normalization of the ratio and also define
\beq
R^{({\rm V})}_{\rm MC}(N_{\rm MC})\;=\;
\frac{\sum_{r=1}^{N_{\rm MC}}S^{(0+1)}_{{\rm MC},r}}
     {\sum_{r=1}^{N_{\rm MC}}S^{(0)}_{{\rm col},r}}
\eeq
in order to study the impact of the virtual corrections. As before we
partition $N_{\rm event}=N_{\rm trial}\times N_{\rm MC}$ events to
have a certain number of trials to compute the corresponding mean
values $\mu$\/ and standard deviations $\sigma$\/ for $n$-gluon LO and
virtual scattering according to Eqs.~(\ref{Eq:meanTrial}) and
(\ref{Eq:sigmaTrial}), respectively. For the case of
$R^{(0+1)}_{\rm MC}(N_{\rm MC})$ and 4-gluon scattering, the number of
Monte Carlo points versus a given relative accuracy is shown in the
inlaid plot of Fig.~\ref{Fig:conv4LL}. As at LO, the curve bends
behaving as statistically determined after a certain amount of Monte
Carlo integration steps.

\begin{table}[t!]
\begin{center}\small
\begin{tabular}{|l||c|c||c|c|c|c||c|c|c|c||c|c|c|c|}\hline\hline
\multicolumn{1}{|c|}{\rule[-3mm]{0mm}{8mm}:)}
& \multicolumn{2}{c||}{Naive}
& \multicolumn{4}{c||}{Conserved}
& \multicolumn{4}{c||}{Non-Zero}
& \multicolumn{4}{c|}{Non-Zero, $N_{\rm colpts}=4$}\\\hline
\rule[-2mm]{0mm}{6mm} $n$
& $B$ & $A$
& $B$ & $A$ & $A'$ & $f$
& $B$ & $A$ & $A'$ & $f$
& $B$ & $A$ & $A'$ & $f$\\\hline
\rule[-1mm]{0mm}{4mm} 4$^\ast$
& &
& & &&
& 0.479 & 3.36 &&
& & &&\\
\rule[-1mm]{0mm}{4mm} 4
& 0.497 & 22.0
& 0.489 & 5.41 & 17.0 & 10.4
& 0.476 & 3.57 & 13.5 & 16.4
& 0.485 & 2.05 & 15.6 & 65.7\\
\rule[-1mm]{0mm}{4mm} 5
& 0.482 & 59.4
& 0.454 & 13.3 & 43.3 & 13.5
& 0.442 & 9.71 & 36.4 & 19.8
& 0.439 & 5.56 & 37.8 & 79.2\\
\rule[-1mm]{0mm}{4mm} 6
& 0.325 & 7.08
& 0.344 & 5.37 & 14.0 & 16.3
& 0.255 & 1.60 & 3.50 & 21.7
& 0.233 & 0.850& 2.14 & 87.6\\\hline\hline
\end{tabular}
\caption{\label{Tab:MCfits}
Parameter values $B$, $A$\/ and $A'$\/ obtained from curve fitting of
the $\sigma(R_{\rm MC})/\mu(R_{\rm MC})$ to the functional form
$A\times N_{\rm MC}^{-B}$. The results are given for the different
ways of sampling over colors in $n$-gluon scattering. The 4-gluon case
marked by ``$\ast$'' corresponds to the consistency check shown in
Fig.~\ref{Fig:conv4LL}, where $R^{(0+1)}_{\rm MC}$ has been considered.
In all other cases $R^{({\rm V})}_{\rm MC}$ has been used, cf.\
Figs.~\ref{Fig:conv4}, \ref{Fig:conv5} and \ref{Fig:conv6}. Note that
for $n=6$, we have fitted $\sigma(R^{({\rm V})}_{\rm MC})$. The
parameters $A'=Af^B$ take into account that the evaluation of a fixed
number of Monte Carlo events takes longer for the other than ``Naive''
color-sampling methods. The time factors $f$\/ relative to the
``Naive'' case are also displayed.}
\end{center}
\end{table}

To quantify the color-integration performances, we again perform fits
to the functional form $A\times N_{\rm MC}^{-B}$ and show the values
of the fitted parameters in Table~\ref{Tab:MCfits} for the various
cases. As argued in Sec.~\ref{Sec:numtree} for large enough $N_{\rm MC}$,
we expect a scaling of $\sigma/\mu$\/ that is proportional to
$1/\sqrt{N_{\rm MC}}$. The goodness of the sampling schemes is
signified by the $A$- and $A'$-parameters, where the latter is more
important since the time factors are included. Smaller values of these
parameters indicate a better efficiency of the sampling procedure.

\begin{table}[t!]
\begin{center}\small
\begin{tabular}{|l||c|c||c|c||c|c||c|c|}\hline\hline
\multicolumn{1}{|c|}{\rule[-3mm]{0mm}{8mm}:)}
& \multicolumn{2}{c||}{Naive}
& \multicolumn{2}{c||}{Conserved}
& \multicolumn{2}{c||}{Non-Zero}
& \multicolumn{2}{c|}{Non-Zero, $N_{\rm colpts}=4$}\\\hline
\rule[-2mm]{0mm}{6mm} $n$
& $N_{\rm MC}$ & $R^{({\rm V})}$ & $N_{\rm MC}$ & $R^{({\rm V})}$
& $N_{\rm MC}$ & $R^{({\rm V})}$ & $N_{\rm MC}$ & $R^{({\rm V})}$\\\hline
\rule[-1.5mm]{0mm}{5mm} 4
& $4\cdot\!10^6$ & $0.4739\pm0.0054$
& $4\cdot\!10^6$ & $0.4750\pm0.0017$
& $4\cdot\!10^6$ & $0.4724\pm0.0013$
& $1\cdot\!10^6$ & $0.4738\pm0.0020$\\
\rule[-1.5mm]{0mm}{5mm} 5
& 631K & $0.241\pm0.022$
& 631K & $0.2673\pm0.0072$
& 631K & $0.2744\pm0.0058$
& 160K & $0.2790\pm0.0058$\\
\rule[-1.5mm]{0mm}{5mm} 6
& 64K & $-0.10\pm0.12$
& 64K & $-0.059\pm0.094$
& 50.2K & $-0.076\pm0.062$
& 16K & $-0.044\pm0.066$\\
\rule[-1.5mm]{0mm}{5mm} 7
& 4K & $-0.87\pm0.66$
& 4K & $-0.23\pm0.09$
& 4K & $-0.14\pm0.10$
& 2K & $-0.97\pm0.65$\\\hline\hline
\end{tabular}
\caption{\label{Tab:MCintresus}
Monte Carlo integration results for the $R^{({\rm V})}$ ratios as
defined in the text after $N_{\rm MC}$ phase-space point evaluations
for $n$-gluon scattering and different color-sampling schemes using
color-dressed tree-level and one-loop amplitude calculations.}
\end{center}
\end{table}

\begin{figure}[p!]
\psfrag{Yconvergelabel}[b][c][0.84]{
  $\left(\left\langle S^{(0+1)}_{\rm MC}\right\rangle\pm
  \sigma_{\left\langle S^{(0+1)}_{\rm MC}\right\rangle}\right)\;\left\langle\,
  \sum\limits_{\rm col}\left|{\cal M}^{(0)}\right|^2\right\rangle^{-1}$}
\psfrag{Xconvergelabel}[t][t][0.84]{$
  100\%\times\sigma\Big(R^{({\rm V})}_{\rm MC}(N_{\rm MC})\Big)\Big/
  \mu\Big(R^{({\rm V})}_{\rm MC}(N_{\rm MC})\Big)
  $}
\vspace*{-4mm}
\centerline{
  \includegraphics[width=0.64\columnwidth,angle=-90]{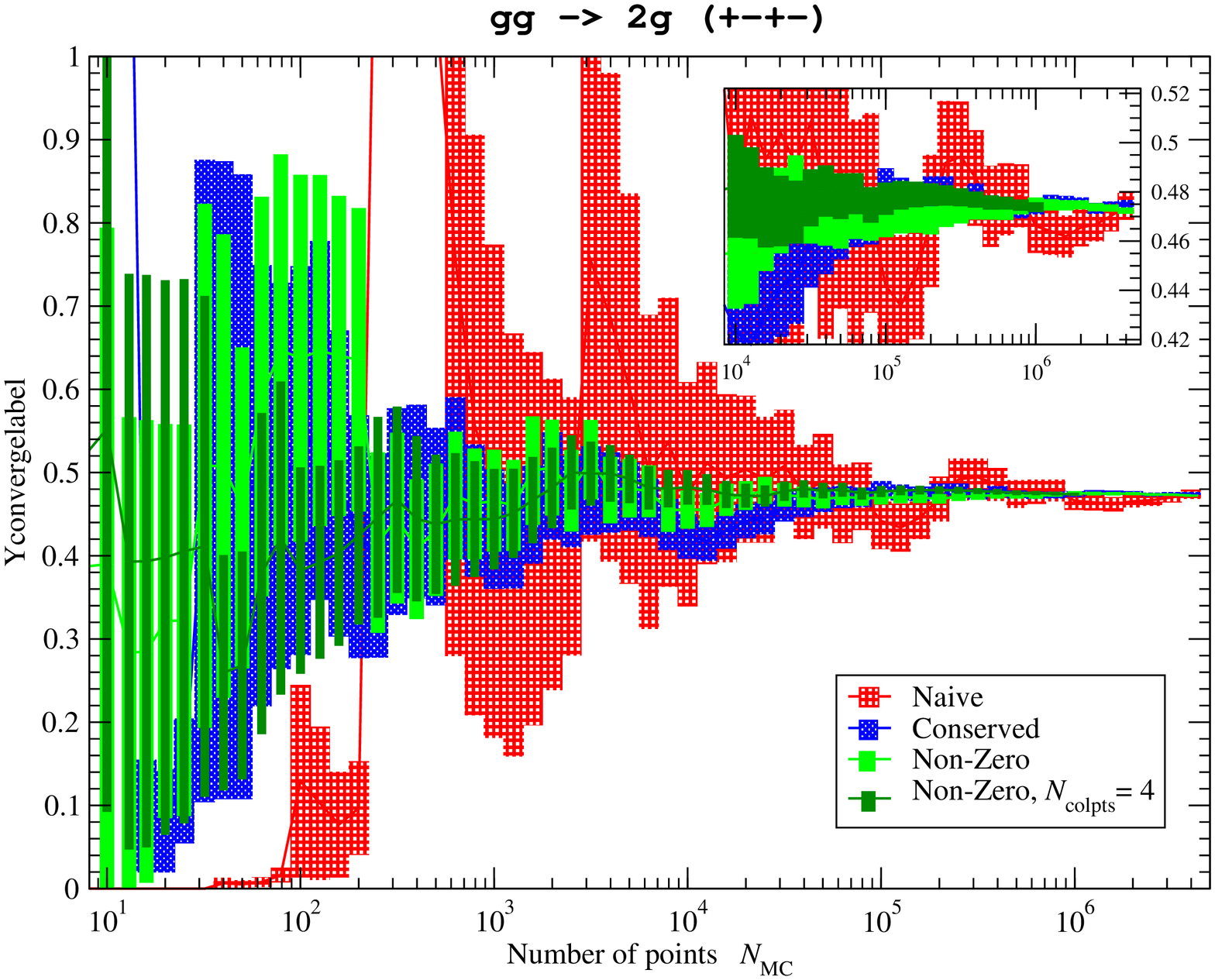}}
\vspace*{-2mm}
\centerline{
  \includegraphics[width=0.64\columnwidth,angle=-90]{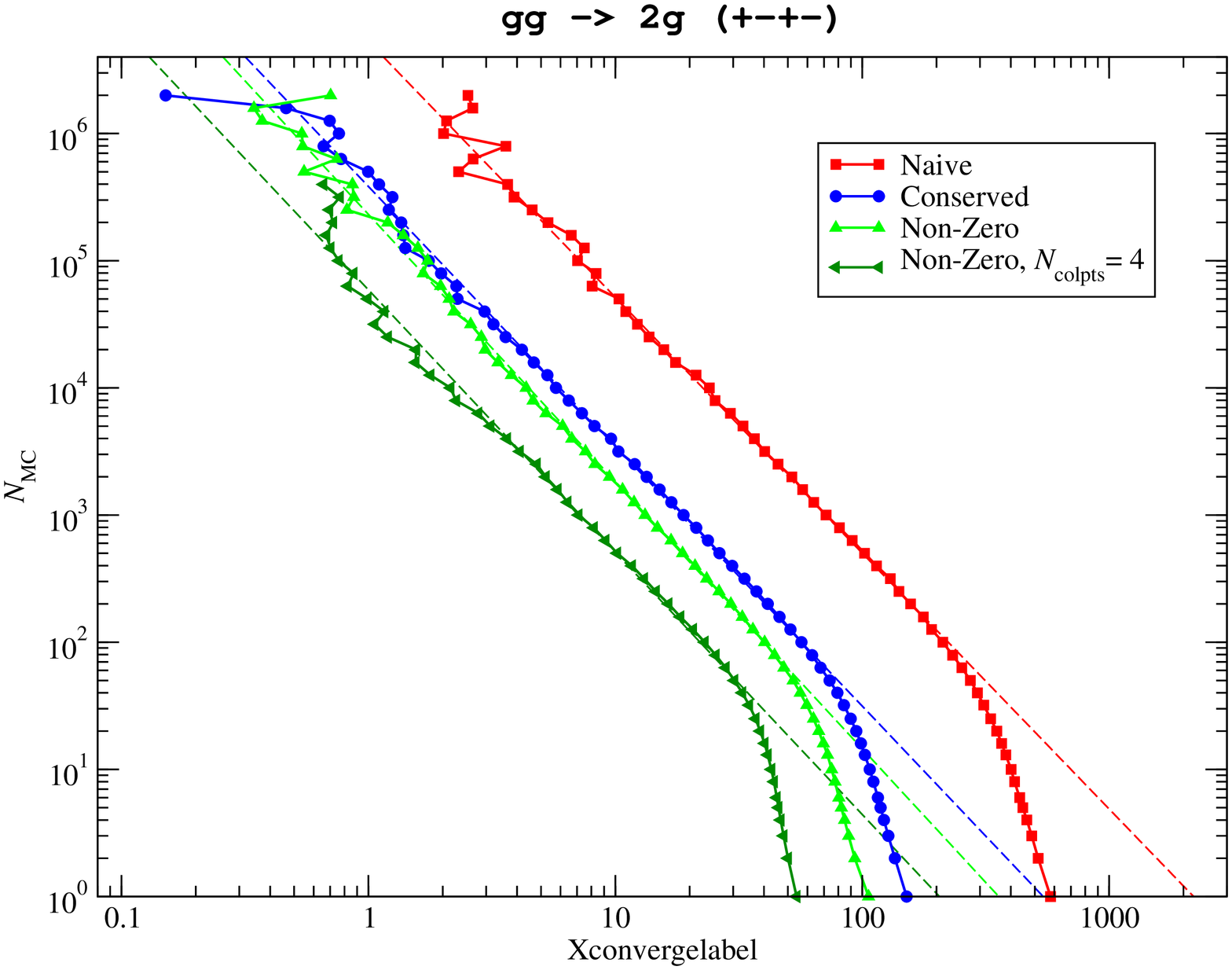}}
\caption{Upper graph: convergence of the 4-gluon virtual corrections
  integration as a function of the number of evaluated phase-space
  points. Also shown is the standard deviation as an estimator of the
  integration uncertainty. Lower graph: convergence of the Monte Carlo
  integration, where the relative integration uncertainty is shown as
  a function of the number of phase-space evaluations. The dashed
  lines describe the fit functions $\sigma/\mu=A\,N_{\rm MC}^{-B}$,
  see also Table~\ref{Tab:MCfits}. The ``Naive'', ``Conserved'' and
  ``Non-Zero'' color-sampling methods are explained in
  Sec.~\ref{Sec:numtree}. The points indicated by ``Non-Zero,
  $N_{\rm colpts=4}$'' average over 4 color configurations per
  phase-space point.}
\label{Fig:conv4}
\end{figure}

\begin{figure}[p!]
\psfrag{Yconvergelabel}[b][c][0.84]{
  $\left(\left\langle S^{(0+1)}_{\rm MC}\right\rangle\pm
  \sigma_{\left\langle S^{(0+1)}_{\rm MC}\right\rangle}\right)\;\left\langle\,
  \sum\limits_{\rm col}\left|{\cal M}^{(0)}\right|^2\right\rangle^{-1}$}
\psfrag{Xconvergelabel}[t][t][0.84]{$
  100\%\times\sigma\Big(R^{({\rm V})}_{\rm MC}(N_{\rm MC})\Big)\Big/
  \mu\Big(R^{({\rm V})}_{\rm MC}(N_{\rm MC})\Big)
  $}
\vspace*{-4mm}
\centerline{
  \includegraphics[width=0.64\columnwidth,angle=-90]{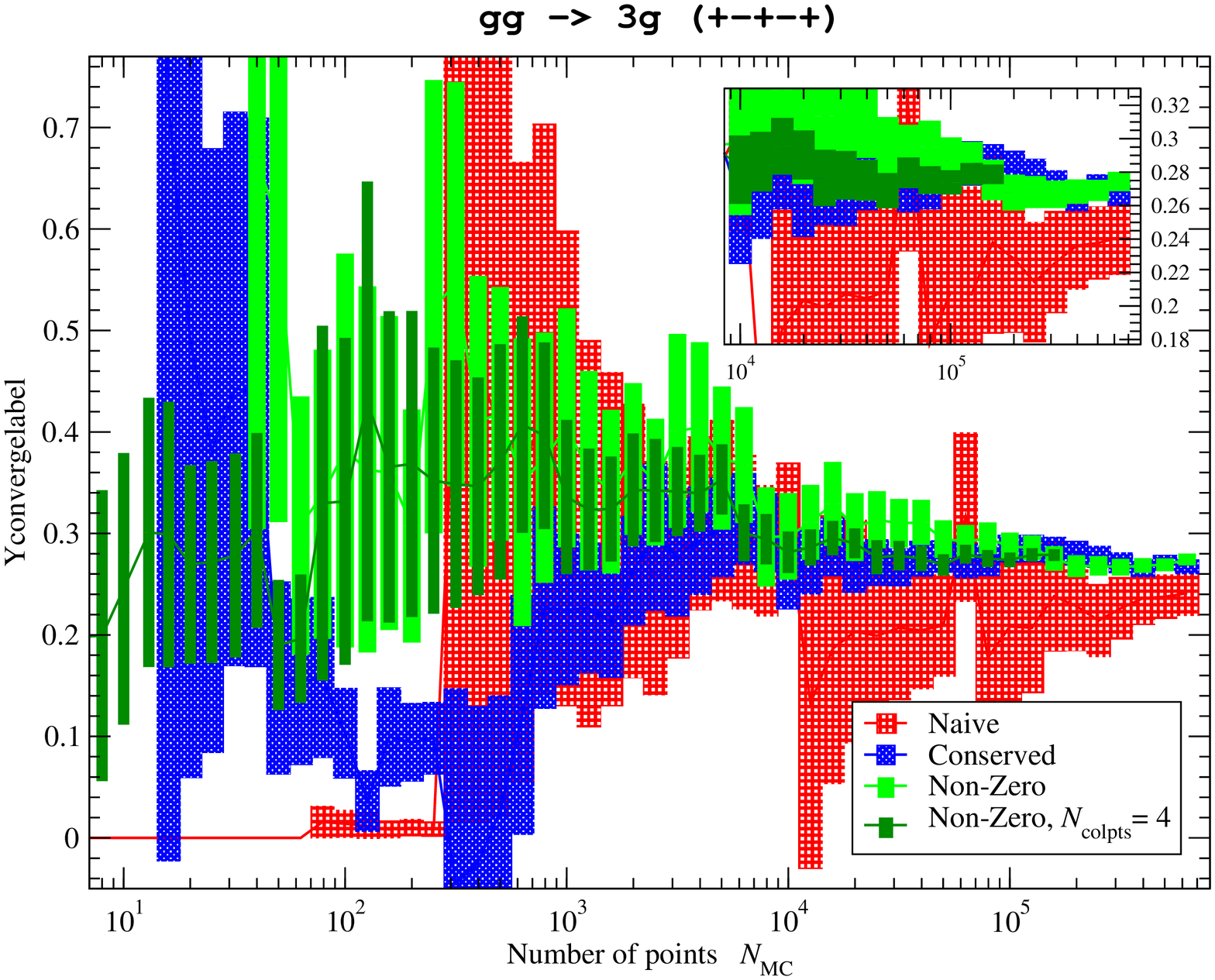}}
\vspace*{-2mm}
\centerline{
  \includegraphics[width=0.64\columnwidth,angle=-90]{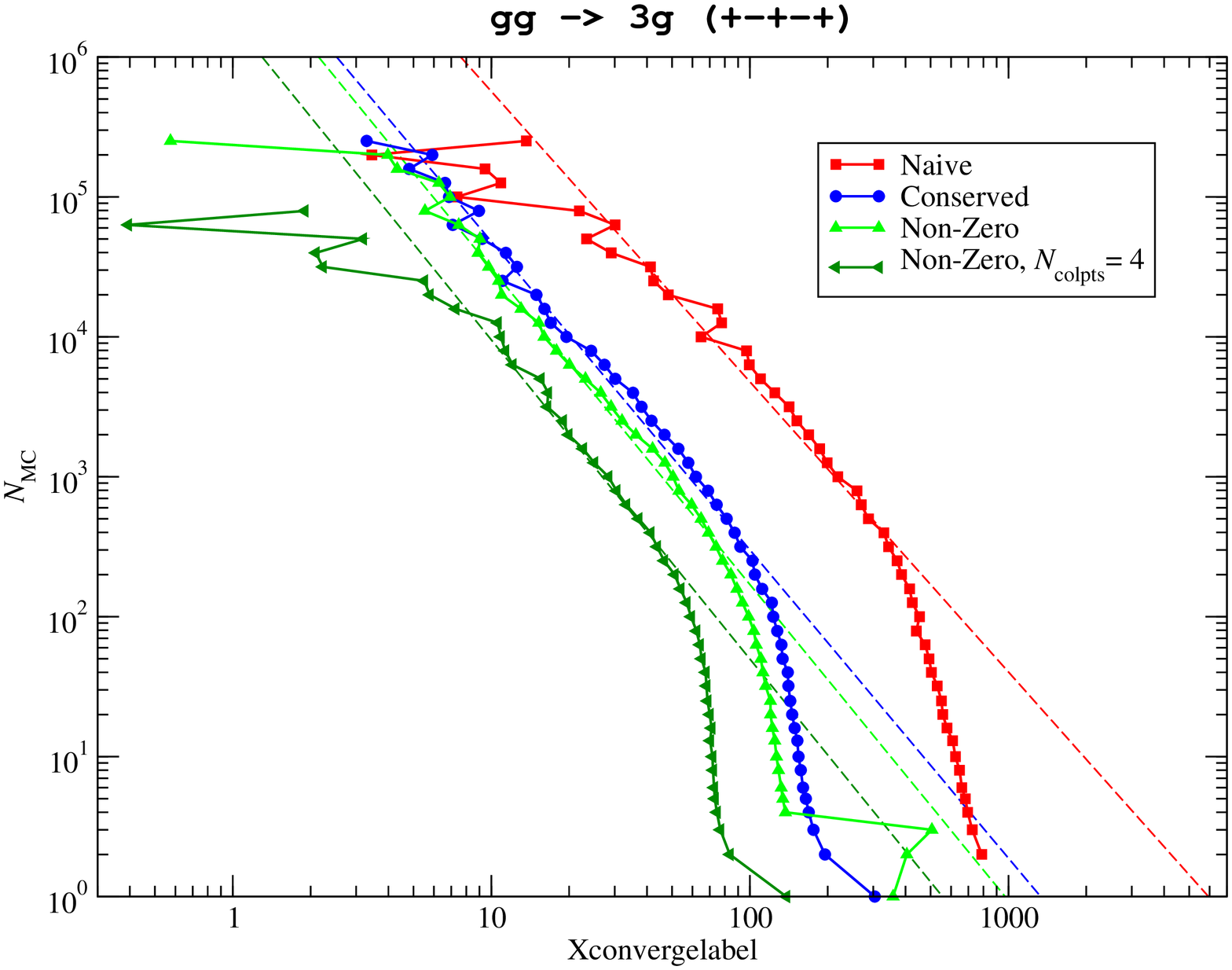}}
\caption{Upper graph: convergence of the 5-gluon virtual corrections
  integration as a function of the number of evaluated phase-space
  points. Also shown is the standard deviation as an estimator of the
  integration uncertainty. Lower graph: convergence of the Monte Carlo
  integration, where the relative integration uncertainty is shown as
  a function of the number of phase-space evaluations. The dashed
  lines describe the fit functions $\sigma/\mu=A\,N_{\rm MC}^{-B}$,
  see also Table~\ref{Tab:MCfits}. The ``Naive'', ``Conserved'' and
  ``Non-Zero'' color-sampling methods are explained in
  Sec.~\ref{Sec:numtree}. The points indicated by ``Non-Zero,
  $N_{\rm colpts=4}$'' average over 4 color configurations per
  phase-space point.}
\label{Fig:conv5}
\end{figure}

\begin{figure}[p!]
\psfrag{Yconvergelabel}[b][c][0.84]{
  $\left(\left\langle S^{(0+1)}_{\rm MC}\right\rangle\pm
  \sigma_{\left\langle S^{(0+1)}_{\rm MC}\right\rangle}\right)\;\left\langle\,
  \sum\limits_{\rm col}\left|{\cal M}^{(0)}\right|^2\right\rangle^{-1}$}
\psfrag{XXconvergelabel}[t][t][0.84]{$
  \sigma\Big(R^{({\rm V})}_{\rm MC}(N_{\rm MC})\Big)
  $}
\vspace*{-4mm}
\centerline{
  \includegraphics[width=0.64\columnwidth,angle=-90]{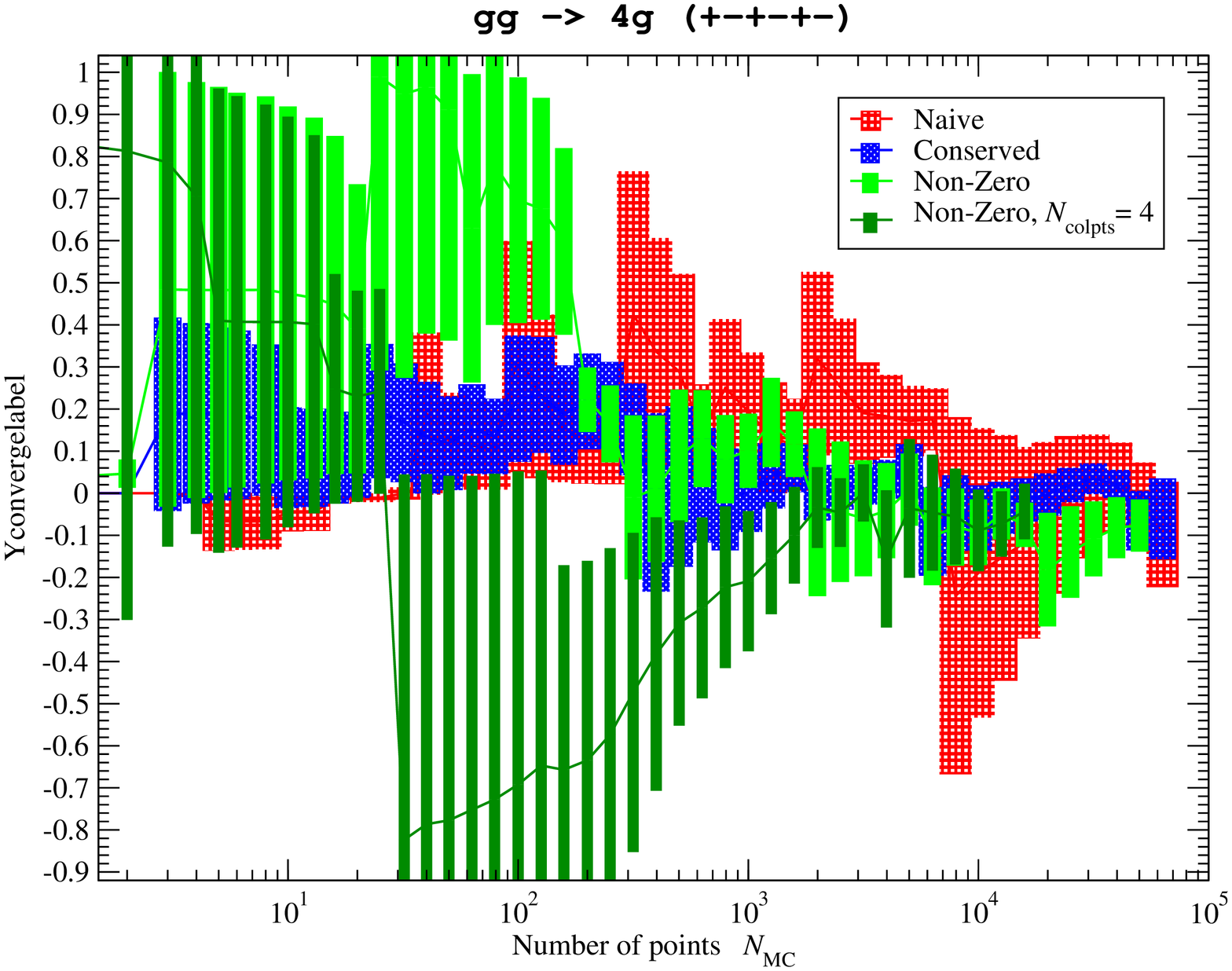}}
\vspace*{-2mm}
\centerline{
  \includegraphics[width=0.64\columnwidth,angle=-90]{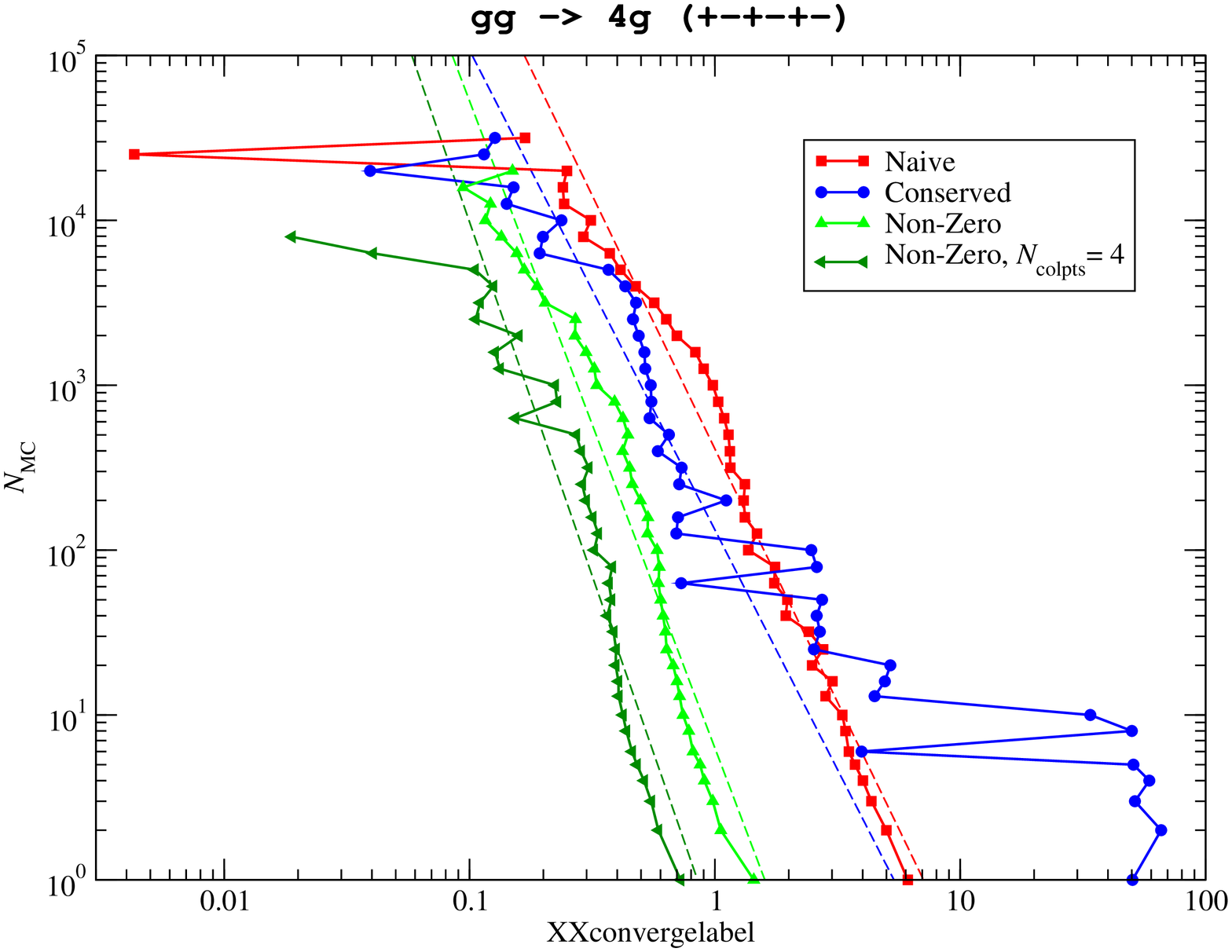}}
\caption{Upper graph: convergence of the 6-gluon virtual corrections
  integration as a function of the number of evaluated phase-space
  points. Also shown is the standard deviation as an estimator of the
  integration uncertainty. Lower graph: convergence of the Monte Carlo
  integration, where this time the standard deviation is shown as a
  function of the number of phase-space evaluations. Note that for
  this case, the virtual corrections are as large as the LO
  contribution so that the full result is close to zero. The dashed
  lines describe the fit functions $\sigma=A\,N_{\rm MC}^{-B}$,
  see also Table~\ref{Tab:MCfits}. The ``Naive'', ``Conserved'' and
  ``Non-Zero'' color-sampling methods are explained in
  Sec.~\ref{Sec:numtree}. The points indicated by ``Non-Zero,
  $N_{\rm colpts=4}$'' average over 4 color configurations per
  phase-space point.}
\label{Fig:conv6}
\end{figure}

\begin{figure}[p!]
\psfrag{Yconvergelabel}[b][c][0.84]{
  $\left(\left\langle S^{(0+1)}_{\rm MC}\right\rangle\pm
  \sigma_{\left\langle S^{(0+1)}_{\rm MC}\right\rangle}\right)\;\left\langle\,
  \sum\limits_{\rm col}\left|{\cal M}^{(0)}\right|^2\right\rangle^{-1}$}
\psfrag{XXconvergelabel}[t][t][0.84]{$
  \sigma\Big(R^{({\rm V})}_{\rm MC}(N_{\rm MC})\Big)
  $}
\vspace*{-4mm}
\centerline{
  \includegraphics[width=0.64\columnwidth,angle=-90]{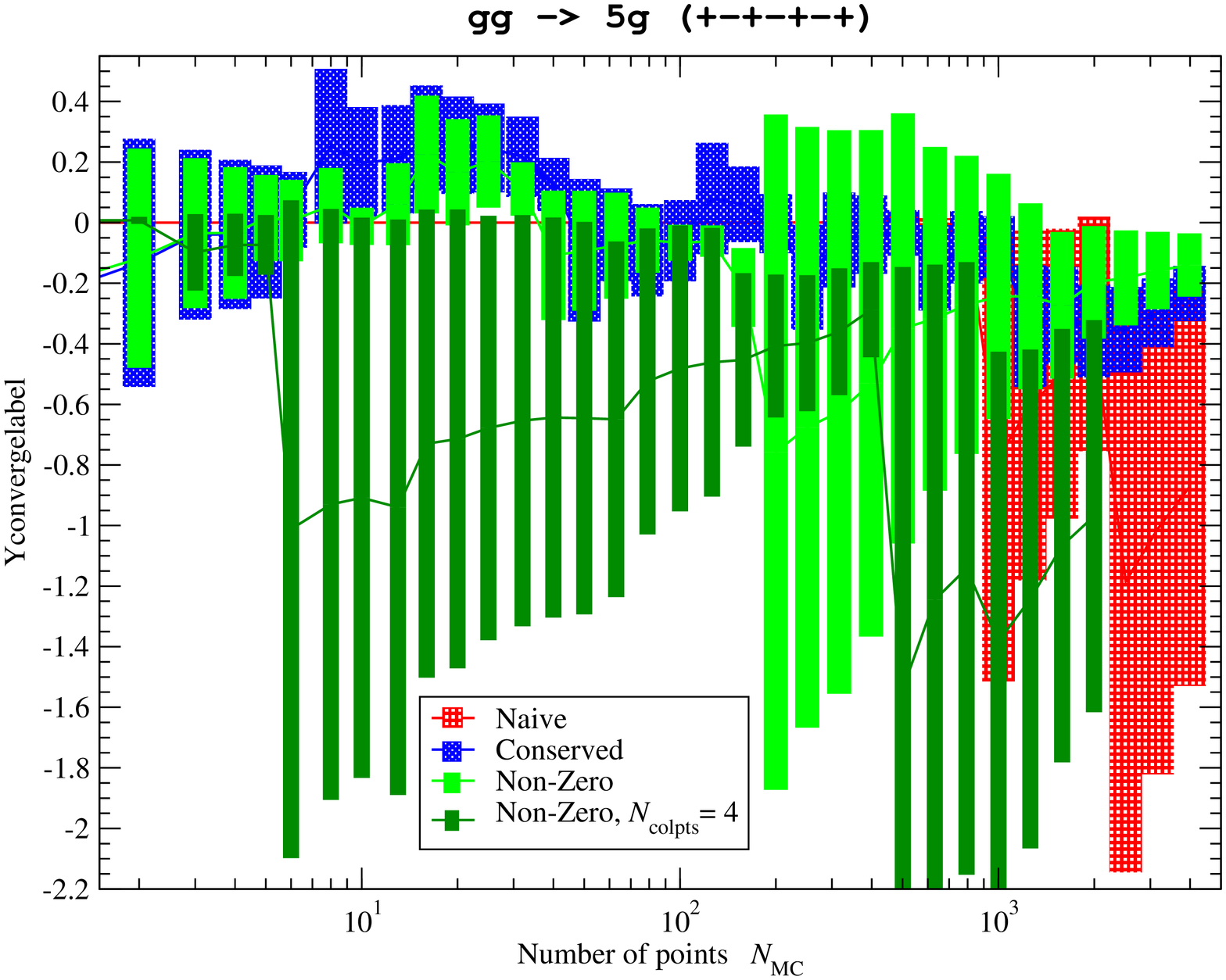}}
\vspace*{-2mm}
\centerline{
  \includegraphics[width=0.64\columnwidth,angle=-90]{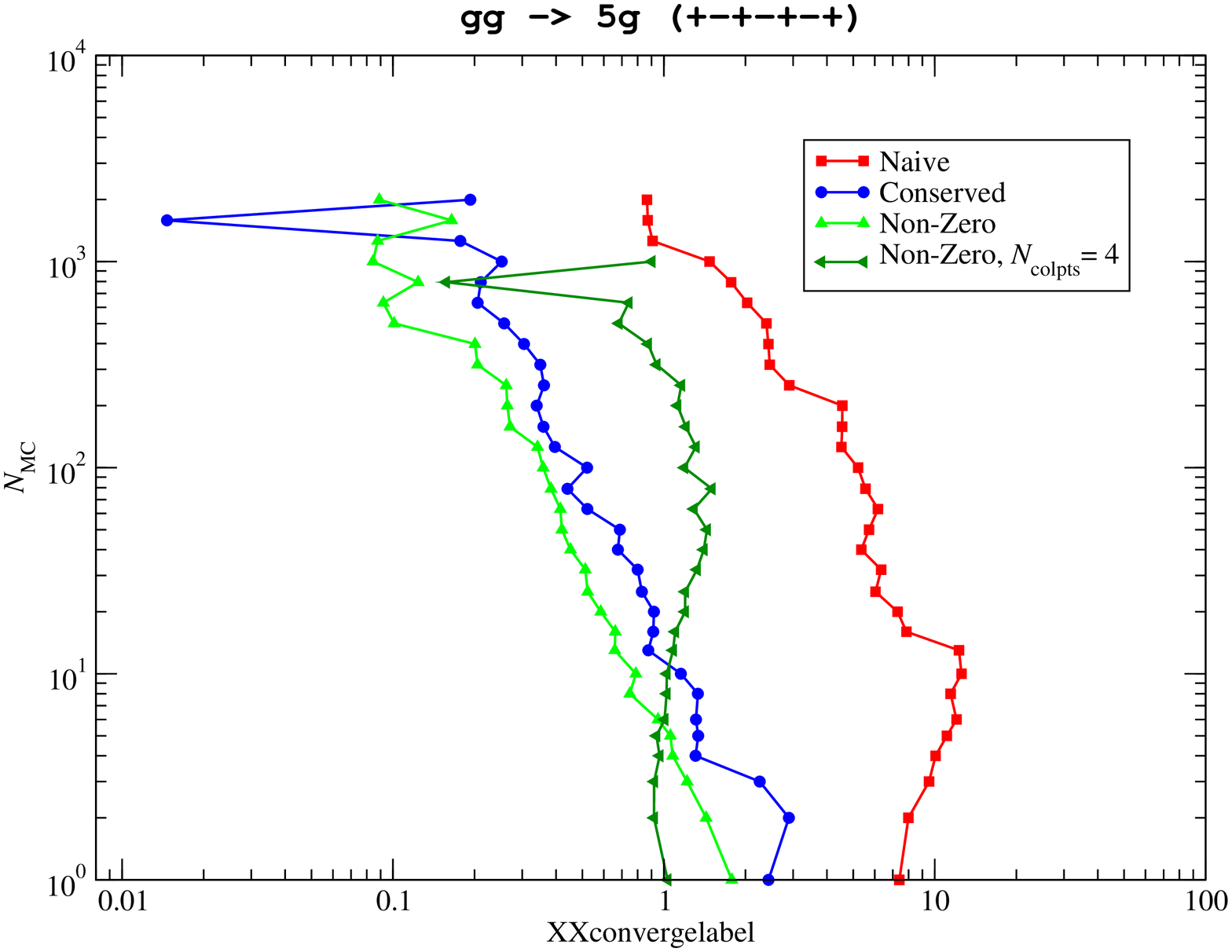}}
\caption{Upper graph: convergence of the 7-gluon virtual corrections
  integration as a function of the number of evaluated phase-space
  points. Also shown is the standard deviation as an estimator of the
  integration uncertainty. Lower graph: convergence of the Monte Carlo
  integration, where for this case, only the standard deviation is
  shown as a function of the number of phase-space evaluations. The
  ``Naive'', ``Conserved'' and ``Non-Zero'' color-sampling methods are
  explained in Sec.~\ref{Sec:numtree}. The points indicated by
  ``Non-Zero, $N_{\rm colpts=4}$'' average over 4 color configurations
  per phase-space point.}
\label{Fig:conv7}
\end{figure}

Using the $R^{({\rm V})}$ and $R^{({\rm V})}_{\rm MC}(N_{\rm MC})$
ratios, we summarize in Figs.~\ref{Fig:conv4}-\ref{Fig:conv7} our
Monte Carlo integration results for $n=4,\ldots,7$ gluon processes and
for the various color-sampling schemes. The upper graphs display the
averaging of $S_{\rm MC}^{(0+1)}$ normalized to the
Monte Carlo average of the color-summed LO contribution as a function
of the number of phase-space evaluations.\footnote{As for the LO
  studies in Sec.~\ref{Sec:numtree}, the gluon polarizations are taken
  alternating and remain fixed while performing the Monte Carlo
  integrations.}
We also indicate the estimate of the integration uncertainty, see
Eqs.~(\ref{Eq:Rvirtual}) and (\ref{Eq:sigmaMC}). To compare all
different test cases, Table~\ref{Tab:MCintresus} list the final values
for $R^{({\rm V})}$. In all these figures we plot in the lower graphs
the number of phase-space point evaluations needed to reach a certain
relative integration uncertainty on $R^{({\rm V})}_{\rm MC}(N_{\rm MC})$.
We show in Table~\ref{Tab:MCfits} the results of the curve fittings
represented by the dashed lines in these plots.

As is clear from these Monte Carlo averaging tests and results, the
convergence is more than satisfactory for future applications of the
color-dressing techniques in NLO calculations. If faster sampling
convergence is required we can evaluate multiple color configurations
per phase-space point. This is shown in the graph, where we have
chosen to evaluate four color configurations at one phase-space point.

\section{Conclusions}

In this paper we explored the possibility of color sampling within the context 
of $D$-dimensional generalized unitarity. Up to now generalized unitarity has
only been used within the context of color-ordered primitive amplitudes. In
the color-ordered approach, color is treated differently from the other quantum numbers such as spin
and flavor. This makes the reconstruction of the full one-loop amplitude rather cumbersome.

We have reformulated the $D$-dimensional generalized unitarity formalism to include
color dressing. That is, we choose the explicit color of each parton, together with all
other quantum numbers, for each Monte Carlo event. In this way all particles, colored
or colorless, are treated on an equal footing. There is no distinction between different
particles as far as the formalism goes. Consequently, the resulting algorithm is independent
of the type and flavor of the external particles. E.g.\ the same algorithm calculates the 6-gluon
virtual corrections, the 6-photon virtual corrections and the $W$+6 parton virtual corrections. 
  
The use of unordered amplitudes requires the partition of the external
legs into unordered subsets.  This is necessary for the calculation of the
tree-level amplitudes as well as for generating all the unitarity
cuts. As a result the complexity
of the resulting algorithm is exponential. That is, the computer time needed
to calculate the virtual corrections grows with a constant multiplicative factor when one adds external particles.
In addition, we have to sum over all color states of the internal lines.
One  may conclude from these general features that the implementation
of the color-dressed  $D$-dimensional generalized unitarity is 
less  efficient in comparison with an implementation based on ordered primitive amplitudes.
As we have explicitly demonstrated for the example of calculating
the virtual corrections to $n$-gluon scattering, this is not the case.
We compared the color-sampling approach for both the color-ordered and color-dressed case.
The calculation of the virtual corrections in the color-dressed case scales as $7^n$,
while in the color-ordered case the effective scaling up to 10 gluons behaves as $9^n$.
Moreover, the color-dressed calculation has a better accuracy in calculating the value
of the one-loop amplitude. The improved accuracy over color-ordered evaluations increases
with $n$.

As we showed for $n$-gluon scattering, the color-dressed approach becomes more efficient
than the color-ordered method for large $n$. One could argue that the differences 
are small and color sampling over the ordered $n$-gluon amplitudes will work as well.
However, when including quarks and other electro-weak particles the color-dressed
approach will easily win out over the color-ordered approach. This is because any
notion of primitive amplitudes is absent. The algorithm simply calculates the virtual
correction.
Moreover, the color-dressed algorithm remains identical when including quarks and electro-weak
particles. It is this algorithmic simplicity that will enable us to employ parallel programming
to significantly improve the computer evaluation time. 

We conclude that the color-dressed formulation is competitive for
calculating one-loop virtual corrections for $n$-gluon scattering.
It is expected that it will be even more efficient in calculating
virtual corrections for processes involving quarks and electro-weak
gauge bosons in addition to the gluons.

\section*{Acknowledgments}
We would like to thank Giulia Zanderighi, Kirill Melnikov,
Stefan H{\"o}che and Tanju Gleisberg for
helpful discussions on the subject.
Fermilab is operated by Fermi Research Alliance, LLC, under contract
DE-AC02-07CH11359 with the United States Department of Energy.

\appendix

\section{The Tree-Level 6-Quark Amplitude}\label{App:6q-example}

As an example we can take a few recursive steps in calculating the
6-quark tree-level matrix element. We start with the definition of the
tree-level matrix element in terms of the 5-quark fermionic current
\beq
{\cal M}^{(0)}\left({\bf u},{\bf\bar u},{\bf d},{\bf\bar d},{\bf s},{\bf\bar s}\right)\;=\;
P^{-1}\left[J\left({\bf u},{\bf \bar u},{\bf d},{\bf \bar d},{\bf s}\right),J\left({\bf\bar s}\right)\right]
\eeq
where we use the shorthand notation 
${\bf u}=u_{i_1}^{\lambda_1}(K_1)$, 
${\bf\bar u}=\bar{u}_{j_1}^{-\lambda_1}(K_2)$,
${\bf d}=d_{i_2}^{\lambda_2}(K_3)$,
${\bf\bar d}=\bar{d}_{j_2}^{-\lambda_2}(K_4)$,
${\bf s}=s_{i_3}^{\lambda_3}(K_5)$
and ${\bf\bar s}={\bar s}_{j_3}^{-\lambda_3}(K_6)$.
The 5-quark fermionic current decomposes into
\beqa
J_{\bf\bar s}\left({\bf u},{\bf \bar u},{\bf d},{\bf\bar d},{\bf s}\right)&=&
P_{\bf\bar s}\Big[D\big[J\left({\bf d},{\bf\bar d},{\bf s}\right),J\left({\bf u},{\bf \bar u}\right)\big]\Big]
+P_{\bf\bar s}\Big[D\big[J\left({\bf u},{\bf\bar u},{\bf s}\right),J\left({\bf d},{\bf \bar d}\right)\big]\Big]\nn
&+&P_{\bf\bar s}\Big[D\big[J\left({\bf s}\right),J\left({\bf u},{\bf\bar u},{\bf d},{\bf\bar d}\right)\big]\Big]\ .
\eeqa
The 3-quark fermionic current decomposes into
\beq
J_{\bf\bar s}\left({\bf q},{\bf\bar q},{\bf s}\right)\;=\;
P_{\bf\bar s}\Big[D\big[J\left({\bf s}\right),J\left({\bf q},{\bf\bar q}\right)\big]\Big]\ ,
\eeq
where ${\bf q}\in\{{\bf u},{\bf d}\}$ and ${\bf\bar q}\in\{{\bf\bar u},{\bf\bar d}\}$.
The 1-quark fermionic current is simply the source term.
Finally the 4-quark gluonic current is given by
\beqa
J_{\bf g}\left({\bf u},{\bf \bar u},{\bf d},{\bf\bar d}\right)&=&
P_{\bf g}\Big[D\big[J\left({\bf u}\right),J\left({\bf d},{\bf\bar d},{\bf u}\right)\big]\Big]
+P_{\bf g}\Big[D\big[J\left({\bf u},{\bf d},{\bf\bar u}\right),J\left({\bf\bar u}\right)\big]\Big]\nn
&+&P_{\bf g}\Big[D\big[J\left({\bf d}\right),J\left({\bf u},{\bf\bar u},{\bf d}\right)\big]\Big]
+P_g\Big[D\big[J\left({\bf d},{\bf u},{\bf\bar u}\right),J\left({\bf\bar d}\right)\big]\Big]\ ,
\eeqa
and the 2-quark gluonic current is written as
\beq
J_{\bf g}\left({\bf q},{\bf\bar q}\right)\;=\;P_{\bf g}\Big[D\big[J\left({\bf q}\right),J\left({\bf\bar q}\right)\big]\Big]\ .
\eeq
The above steps define the 6-quark LO amplitude recursively as would be done
by the algorithm. Note that we have ignored all flavor violating currents.

\section{The Implemented Gluon Recursion Relation}\label{App:gluon-recu}

Making use of the color-flow representation \cite{Maltoni:2002mq}, we
define the color-dressed gluon currents as $3\times3$ matrices of
ordered gluon currents:
\beq\label{Eq:onegcurrent}
J^{(IJ)}_\mu(g^{\lambda_1}_1)\;=\;
\delta^I_{j_1}\delta^{i_1}_J J_\mu(g^{\lambda_1}_1)\ ,
\eeq
where the external gluon $g_1$ has the polarization $\lambda_1$ and
four-momentum $K_1$, its colors are denoted by $(ij)_1$. The
color-flow labels of the dressed current are $(IJ)$ and $\mu$
indicates the Lorentz label. Using this definition, the connection
to the compact notation introduced in
Sec.~\ref{Sec:generictreeformalism} is found as
\beq
J_{\bf g}\big({\bf g}_1\big)\;=\;
\delta^{Ii_1}\delta^{Jj_1}\,\varepsilon_\mu^{\lambda_1}(K_1)\;\equiv\;
\delta^J_{j_1}\delta^{i_1}_I J_\mu(g^{\lambda_1}_1)\;=\;
J^{(JI)}_\mu(g^{\lambda_1}_1)\ .
\eeq
Since we only consider gluons, a plain numbering of the external
particles ${\bf g}_k=\{g_k,\lambda_k,(ij)_k,K_k\}$ is sufficient and
helps simplify the notation such that the color dressing becomes more
emphasized. Hence, in all what follows we write $J_{\bf g}({\bf 1})=
\delta^J_{j_1}\delta^{i_1}_I J_\mu(1)=J^{(JI)}_\mu(1)$. Dressed
$n$-gluon currents are then described by
\beq\label{Eq:ngcurrent}
J^{(IJ)}_\mu(1,2,\ldots,n)\;=\sum\limits_{\sigma\in S_n}
\delta^I_{j_{\sigma_1}}\delta^{i_{\sigma_1}}_{j_{\sigma_2}}\cdots
\delta^{i_{\sigma_{n-1}}}_{j_{\sigma_n}}\delta^{i_{\sigma_n}}_J\
J_\mu(\sigma_1,\sigma_2,\ldots,\sigma_n)\ ,
\eeq
which follows as a consequence of the color decomposition of the
tree-level amplitude into ordered ones:
\beq
{\cal M}^{(0)}(1,2,\ldots,n,n+1)\;=\sum\limits_{\sigma\in S_n}
\delta^{i_{n+1}}_{j_{\sigma_1}}
\delta^{i_{\sigma_1}}_{j_{\sigma_2}}\cdots
\delta^{i_{\sigma_{n-1}}}_{j_{\sigma_{n}}}
\delta^{i_{\sigma_{n}}}_{j_{n+1}}\
m^{(0)}(\sigma_1,\sigma_2,\ldots,\sigma_n,n+1)\ .
\eeq
The vectors $\sigma$\/ describe the elements of the permutations $S_n$
of the set $\{1,2,\ldots,n\}$. With the color-ordered amplitudes
$m^{(0)}(\sigma_1,\sigma_2,\ldots,\sigma_n,n+1)$ expressed through
ordered $J$-currents and the definition of the dressed currents at
hand, we can re-write the last equation and formulate the tree-level
amplitude in terms of the color-dressed currents:
\beqa
{\cal M}^{(0)}(1,2,\ldots,n,n+1)&=&K^2_{\{1,2,\ldots,n\}}\;\times\nn
&&
\sum\limits_{\sigma\in S_n}
\delta^{I}_{j_{\sigma_1}}\delta^{i_{\sigma_1}}_{j_{\sigma_2}}\cdots
\delta^{i_{\sigma_{n-1}}}_{j_{\sigma_{n}}}
\delta^{i_{\sigma_{n}}}_{J}\
J_\mu(\sigma_1,\sigma_2,\ldots,\sigma_n)\
\delta^{J}_{j_{n+1}}\delta^{i_{n+1}}_{I}\ J^\mu(n+1)\nn
&=&K^2_{\{1,2,\ldots,n\}}\
J^{(IJ)}_\mu(1,2,\ldots,n)\;J^{(JI),\;\!\mu}(n+1)\ .
\eeqa
Owing to the simple color structure of the one-gluon current, the
summation over the color indices $(IJ)$ effectively reduces to the
calculation of a single scalar product of the ordered currents
$J^{(i_{n+1}\,j_{n+1})}_\mu$ and $J^{(j_{n+1}\,i_{n+1}),\;\!\mu}$.
The invariant-mass prefactor $K^2$ is determined by the gluon
momenta via $K^2_{\{1,2,\ldots,n\}}=(K_1+K_2+\ldots+K_n)^2$.
The one-gluon current is given in Eq.~(\ref{Eq:onegcurrent}), while
the multi-gluon current is obtained recursively. Starting from
Eq.~(\ref{Eq:ngcurrent}), one incorporates the ordered gluon
recurrence relation to evaluate $J_\mu(\sigma_1,\ldots,\sigma_n)$ and
re-groups accordingly to identify the partitioning. After some
algebra, one finds
\beqa
J^{IJ}_\mu(1,2,\ldots,n)&=&K^{-2}_{\{1,2,..,n\}}\,\Bigg[\;\
\sum_{P_{\pi_1\pi_2}(1,\ldots,n)}
\Big(\delta^{ILN}_{KMJ}-\delta^{INL}_{MKJ}\Big)
\left[J^{(KL)}_\mu(\pi_1),J^{(MN)}_\mu(\pi_2)\right]\ +\nn
&&
\sum_{P_{\pi_1\pi_2\pi_3}(1,\ldots,n)}
\Big(\delta^{ILNP}_{KMOJ}+\delta^{IPNL}_{OMKJ}-\delta^{ILPN}_{KOMJ}-
\delta^{INPL}_{MOKJ}\Big)\ \times\nn
&&\hspace*{21mm}
\bigg(\left\{J^{(KL)}_\mu(\pi_1),J^{(MN)}_\mu(\pi_2),J^{(OP)}_\mu(\pi_3)\right\}
\;+\;\pi_1\leftrightarrow\pi_2\bigg)\;\ \Bigg]
\label{Eq:implrecu}
\eeqa
where we have employed the bracket notation for ordered-current
operations, which was introduced in Ref.~\cite{Berends:1987me}. The
partition sums are explained in Sec.~\ref{Sec:generictreeformalism}
and an implicit summation over the color indices $K,L,M,N,O,P$\/ is
understood. To efficiently compute the dressed currents, the color
factors in front of the operator brackets can be pre-calculated such
that the computation of zero color-weight contributions can be
avoided. We have used the shorthand notation
\beq
\delta^{ik\cdots m}_{jl\cdots n}\;=\;\delta^i_j\delta^k_l\cdots\delta^m_n\ .
\eeq
The recursion relation presented in Eq.~(\ref{Eq:implrecu}) scales
asymptotically as $4^n$, since we kept the 4-gluon vertex as an entity
in our calculation. As a consequence we have to evaluate 3-subset
partitions and the corresponding curly brackets that merge three
different dressed currents.


\end{document}